%% file: blazarzone.tex
\documentclass[fleqn,usenatbib]{mnras}

\usepackage{mathptmx}

\usepackage[T1]{fontenc}
\usepackage{ae,aecompl}

\usepackage{graphicx}
\usepackage{epsfig}
\usepackage{amsmath}
\usepackage{amssymb}
\usepackage{pdflscape}

\usepackage{times}
\usepackage{natbib}

\PassOptionsToPackage{bookmarks=false}{hyperref}

\newcommand\lsim{\mathrel{\spose{\lower 3pt\hbox{$\mathchar"218$}}
     \raise 2.0pt\hbox{$\mathchar"13C$}}}
\newcommand\gsim{\mathrel{\spose{\lower 3pt\hbox{$\mathchar"218$}}
     \raise 2.0pt\hbox{$\mathchar"13E$}}}
\def\ltsima{$\; \buildrel < \over \sim \;$}
\def\lsim{\lower.5ex\hbox{\ltsima}}
\def\gtsima{$\; \buildrel > \over \sim \;$}
\def\gsim{\lower.5ex\hbox{\gtsima}}

\newcommand\ergs{${\rm\thinspace erg \thinspace s^{-1}}$}

\def\gg{$\gamma$-$\gamma$~}
\newcommand{\redchi}{\chi^2_{r}}

\newcommand{\pcms}{\,cm$^{-2}$\,s$^{-1}$~}

\title[Gamma rays from \textit{Fermi} blazars: beyond the BLR]{On the origin of gamma rays in {\it Fermi} blazars: 
beyond the broad line region.}

\author[]{
L. Costamante$^{1,2,3}$ \thanks{E--mail: luigi.costamante@asi.it}, 
S. Cutini$^{4,5}$, G. Tosti$^{3}$, E. Antolini$^{3}$, A. Tramacere$^{6}$ \\  %
$^{1}$ASI -- Unit\`a di Ricerca Scientifica, Via del Politecnico snc, I-00133, Roma, Italy\\
$^{2}$INAF -- Osservatorio Astronomico di Brera, via E. Bianchi 46, I--23807 Merate, Italy \\ 
$^{3}$Dipartimento di Fisica e Geologia, Universit\`a degli Studi di Perugia, via Pascoli snc, I-06123 Perugia, Italy \\
$^{4}$ASI -- Science Data Center, Via del Politecnico snc, I-00133, Roma, Italy\\
$^{5}$INFN -- Sezione di Perugia, Via  Pascoli snc, I-06123, Perugia, Italy\\
$^{6}$ISDC -- Department of Astronomy, University of Geneva, 16, 1290 Versoix, Switzerland\\
}

\date{Accepted 2018 April 4. Received 2018 April 4; in original form 2017 December 7.}
\pubyear{2018}

\begin{document}
\label{firstpage}
\pagerange{\pageref{firstpage}--\pageref{lastpage}}
\maketitle

\begin{abstract}
The gamma-ray emission in broad-line blazars is generally explained as inverse Compton (IC) radiation
of relativistic electrons in the jet scattering optical-UV photons from the Broad Line Region (BLR),
the so-called BLR External Compton scenario.
We test this scenario on the {\it Fermi} gamma-ray spectra of 106 broad-line blazars 
detected with the highest significance  or largest BLR,  
by looking for cut-off signatures at high energies compatible with \gg interactions with BLR photons. 
We do not find evidence for the expected BLR absorption.
For 2/3 of the sources, we can exclude any significant absorption ($\tau_{\rm max} <1$), while for the remaining 1/3
the possible absorption is constrained to be 1.5--2 orders of magnitude lower  than expected. 
This result holds also dividing the spectra in high and low-flux states,
and for powerful blazars with large BLR.
Only 1 object out of 10 seems compatible with substantial attenuation ($\tau_{\rm max} >5$).
We conclude that for 9 out of 10 objects, the jet does not interact with BLR photons.
Gamma-rays seem either produced outside the BLR  most of the time,
or the BLR  is $\sim100\times$ larger than given by reverberation mapping.
This means that  i) External Compton on BLR photons is disfavoured  as the main gamma-ray mechanism,
vs IC on IR photons from the torus or synchrotron self-Compton; 
ii) the {\it Fermi} gamma-ray spectrum is mostly intrinsic, determined by the interaction
of the particle distribution with the seed-photons spectrum;
iii) without suppression by the BLR, broad-line blazars can become copious emitters above 100 GeV, 
as demonstrated by 3C\,454.3.
We expect the CTA sky to be much richer of broad-line blazars than previously thought.
\end{abstract}


\section{Introduction}
The location of the main gamma-ray emitting region in blazars --the so-called ``blazar zone''--
is still rather uncertain, even though
the gamma-ray emission often dominates their entire spectral energy distribution (SED).
Blazars are usually divided into flat spectrum radio quasars (FSRQ) and BL Lac objects, according to the 
presence  or absence of strong broad emission lines in their UV-optical spectra \citep{urrypadovani}.
This often corresponds to a different regime of accretion,  
with most gamma-ray bright FSRQ having a luminous, radiatively efficient accretion disc \citep{gg11,sbarrato12}.

In FSRQ the jet is then expected to propagate through dense regions of radiation, produced 
by gas reprocessing the disc emission  at different distances from the black hole (BH),
like UV-optical emission lines and recombination continua from the broad line region \citep[BLR,][]{peterson}
and thermal IR radiation from the inner parts of the AGN dusty torus.
Reverberation mapping studies in both radio-quiet and radio-loud objects 
indicate that the size of the BLR and dusty torus scales with the disc luminosity $L_{\rm d}$ as $R\propto L_{\rm d}^{1/2}$, namely as  
$R_\text{BLR}\simeq 0.1\,L_{\rm d, 46}^{1/2}$ pc \citep{Bentz2006,Kaspi2007} and $R_\text{IR}\simeq 2.5\,L_{\rm d, 46}^{1/2}$ pc \citep{Cleary2007,Nenkova2008}, respectively
(where we use the notation $L_{\rm d,x}$ for $L_{\rm d}$ in units of $10^x$ erg s$^{-1}$).
The resulting energy density $U_{\rm ext}\propto L_{\rm d}/R^2c$ should therefore be roughly constant irrespective of the disc luminosity,
over regions whose extension depends on the disc luminosity.

These external radiation fields provide effective seed photons for the inverse Compton  
scattering by relativistic electrons in the jet, and are the basis of the External Compton (EC) models 
for the gamma-ray production \citep{sikora94,sikora2009}.
At each distance from the BH, the energy density of the different radiation components and of the magnetic field
changes, as well as their ratio, which determines the IC to synchrotron luminosity ratio 
in the SED (the so-called  ``Compton dominance'').
In a leptonic framework, therefore, the location 
is directly connected with the gamma-ray production mechanism:
for typical bulk-motion Lorentz factors $\Gamma \gtrsim 10-15$, 
gamma rays are mainly produced by EC on UV photons inside the BLR, 
by EC on IR photons between the BLR and the torus sizes,
and by synchrotron self-Compton (SSC) or EC on CMB photons at further distances \citep{ggcanonical,sikora2009}.

So far, EC on BLR photons (i.e. a blazar zone inside the BLR) has been the standard scenario 
to model the SED of the large  majority of gamma-ray detected FSRQ 
\citep[see e.g.][and references therein]{gg98,ggsequence2,ggcanonical,finke13}, providing 
a straightforward way to explain both the difference in Compton dominance and SED peak frequencies between FSRQ and BLLac objects,
and the often observed fast variability (few days to few hours), indicative of compact emitting regions
easier to obtain at the base of a conical jet.

However, the same external photons used for EC  become targets for the \gg collision and pair production process.
The resulting optical depths are large ($\tau>50$)  on paths as short as $10^{16}$ cm.
A gamma-ray region well inside the BLR, therefore, should be accompanied by a strong cut-off in the gamma-ray spectrum
above $\sim$20 GeV (for Ly$\alpha$ photons of ionized Hydrogen), with maximum absorption around 100-200 GeV. 
The densities used for the EC modeling make the BLR essentially opaque to gamma rays beyond tens of GeV.
The opacity is not expected to decrease significantly even with a flat BLR geometry, elongated along the disc,
instead of a spherical geometry (see e.g. \citealt{TavecchioFlat}). 
No emission at Very High Energies (VHE, $\geq$100 GeV) should be detectable, corresponding to the peak of the \gg cross section.

Nevertheless, already in the first years of operation
the {\it Fermi} Large Area Telescope \citep[LAT,][]{LAT} started to provide contrasting evidence,
with some bright FSRQ spectra showing no sign of BLR-induced cut-offs 
and extending up to $\sim$100 GeV rest-frame energy \citep{costamante2010}.
The ``smoking gun'' of a gamma-ray  region outside the BLR was provided 
by the detections at VHE of the FSRQ 4C\,+21.35 \citep{magic4c} and PKS\,1510-08 \citep{hess1510,magic1510}.
The emission measured by MAGIC in 4C\,+21.35 was also rapidly variable ($t_{\rm var}\sim10$ minutes),
indicating that very compact emitting regions of the order of $R\sim1.3\times 10^{14}\,(\delta/10)$ cm
can exist also at pc distances from the BH, beyond the BLR \citep{tavecchio2135}.

At the same time, possible breaks around 3-5 GeV in other objects were interpreted as
evidence of \gg absorption on high-ionization \ion{He}{ii} line photons \citep{poutanen2010}. 
This would locate the dissipation region deep within the BLR.   
Oddly enough, the optical depth caused by the low-ionization Hydrogen complex  
--which provides the dominant contribution to the BLR optical depth, see Fig. 1 in \citet{poutanen2010}--  
was found to be rather low (a few at most), 
not consistent with the deep-within-BLR scenario unless assuming a very large stratification of the BLR 
over a wide range of distances.
The re-analysis of the {\it Fermi} data with new data versions seems now to have reduced
the evidence for such \ion{He}{ii} breaks \citep{stern14}.

Since blazars are typically  highly variable,
these two apparently opposite scenarios might not be in contradiction, after all.
Different states can correspond to different locations along the jet, as well as different emitting blobs.
It is thus natural to expect that the observed gamma rays originate sometimes inside and sometimes
outside the BLR. This should occur most likely in FSRQ with relatively low disc  luminosities,
where the derived BLR size is of the same order of the expected location of the dissipation region 
(e.g. in internal shock scenarios, a few hundred Schwarzchild radii, see \citealt{ggcanonical}).  
However, this type of study has been limited so far to only a few selected objects or single outbursts.

The aim of this paper is to find out the typical behaviour of the population of bright gamma-ray FSRQ,
by looking for the presence or absence of BLR-induced cut-off features in the average GeV spectra 
of the 100 strongest FSRQ detected by {\it Fermi}-LAT.
This study    
is now made feasable by more than seven years of exposure collected by {\it Fermi} in survey mode as well as by the Pass 8 data,
which both enhance the chance of measuring either some source counts or meaningful upper limits at the highest energies,
after integrating over the whole exposure.
We also take advantage  of the systematic spectral study of {\it Fermi}-detected FSRQ in the optical by \cite{Shaw12}, which provides 
the luminosity of the broad emission lines for the majority of bright {\it Fermi} blazars.

Specifically, we want to address the following issues:\\
a) if BLR absorption is a common phenomenon in {\it Fermi}-LAT FSRQ, and at which optical depth;\\
b) if the possible absorption is consistent with the photon densities used by external Compton models;\\
c) if there is a difference in location between high (flaring) and low states.
That is, if  the ``persistent'' (low-flux) emission is mostly produced outside the BLR and the flaring --rapidly variable-- emission 
inside, where the cross-section of the jet is lower, or the other way round 
(see e.g. \citealt{bonnoli454,tavecchio1424,pacciani14,tavani}).

It is important to keep in mind that a cut-off in the gamma-ray spectrum at high energies does not mean {\it per-se} evidence of
absorption due to the BLR: it can also correspond to the end of the emitting particle distribution.
In such case the shape of the cut-off in the gamma-ray band depends on the shape of the cut-off in the 
particle distribution, as well as on the spectrum of the target photons \citep{lefa12}.
This dependence can be used as a powerful diagnostic tool, but it is hindered by the uncertain origin of the cut-off.
The focus of this paper is to test whether the LAT spectra are consistent or not with the blazar zone being inside the BLR, 
under the same assumptions used by EC-modeling of their SEDs. 

\input{tableAa}

\input{tableAb}

\section{Sample Selection and Data Analysis}
We selected the 100 objects detected with the highest significance 
in the Third {\it Fermi}-LAT AGN catalog \citep[3LAC,][]{3LAC}, and classified as FSRQ to guarantee the presence of a BLR.
The list of sources is presented in Table \ref{lista}, in order of decreasing 3LAC significance.
The luminosity $L_{\rm BLR}$ of the BLR is also reported, when available from the literature. 
When  only the luminosity of the observed broad lines was published, for consistency
we calculated $L_{\rm BLR}$ following \citet{celotti97},
using the ratio of the lines to the whole BLR luminosity  based on the relative weights of the lines to the Ly$\alpha$ flux 
in \citet{francis91}  
This is the same recipe used to calculate the BLR luminosity in all the other objects, according to the listed literature.
In Table \ref{lista} we also report the corresponding BLR radius $R_{\rm BLR}$ and maximum optical depth $\tau_{\rm max}$,
defined as the optical depth at the peak of the \gg cross section (details in the following section).
In addition to the most significant 100 FSRQ in the 3LAC, we also considered  6 further objects 
characterized by large values of $L_{\rm BLR}\geq 3\times 10^{45}$ but still detected above 12$\sigma$ in the 3LAC.

The LAT data used in this paper were collected from 2008 August 5 (MJD 54683) to 2014 December 01 (MJD  56992),
corresponding to approximately 7.3 years of data taking.
During this time, the LAT instrument operated almost entirely in survey mode.
The LAT data were analyzed following the standard procedure\footnote{http://fermi.gsfc.nasa.gov/ssc/data/analysis/}, 
using the LAT analysis software ScienceTools v10r01p01. Only events belonging to the ``Source'' class 
(evclass=128, evtype=3) were used. We selected only events within a  maximum zenith angle of 90 degrees 
to reduce contamination from the Earth limb gamma rays, which are produced by cosmic rays interacting 
with the upper atmosphere. The  analysis was performed with the instrument response functions P8R2\_SOURCE\_V6 
using a binned maximum  likelihood method implemented in the Science tool {\it gtlike}.

Events with energies between 0.1 and 300 GeV were extracted from an acceptance cone of 15 degrees centered at the location 
of the source of interest.  Light curves were derived by dividing the data in bins of one week duration.  
The background model for the light-curve extraction included each PGWave source \citep{ciprini07, damiani97}  
detections in each bin, 
to account for possible transients on short timescales that may remain
below the detection threshold on timescales of years.  
For the spectral analysis, the background model included sources from the 3rd LAT catalog 
\citep[hereafter 3FGL]{acero15} within the ROI of 20 degrees.
3FGL sources within a radius of 10 degrees from the source of interest were left free to vary.

Each bin of the light curves and of the spectra was obtained using a power-law model for the source of interest. 
To derive the spectral points for the SED, the spectral parameters of background sources 
were fixed to the results from the fit of the complete time-integrated data. The spectral index of the source of interest was fixed 
to $\Gamma=2$ for the spectral extraction, and left free to vary for the light curve extraction. The models included 
the  isotropic and Galactic diffuse emission 
components\footnote{iso\_P8R2\_SOURCE\_V6\_v06.txt and gll\_iem\_v06.fits}.  
Both background components were left free 
to vary in the time integrated analysis and frozen to the average values in the SED's bin extraction. 

The gamma-ray SEDs were constructed using 4 bins per decade  between 178 MeV and 300 GeV,
as the best compromise between energy resolution in the cut-off region 
and photon statistics in each bin, for the majority of sources.
We excluded the first bin (100-178 MeV) in order to reduce the chance of issues with the Galactic background,
and we enlarged the last bin to increase the statistics (100-300 GeV).
Each SED is then composed of 12 bins in total.
When the Test Statistic  \citep[TS,][]{mattox} 
was TS$<$4 or the number of predicted photons $Npred<3$ or the percentage error on the flux was greater then 50\%,  
we calculated  the upper limit at 2$\sigma$ confidence,
using the Bayesian computation if TS $<$1 \citep{helene84}.
The statistical uncertainty on the fluxes are typically larger than the systematic uncertainty \citep{ackermann12} and only
the former is considered in this paper.

\subsection{BLR and gamma-ray spectra fitting procedure}
For a wide range of ionization levels, the BLR spectrum causes an absorption feature in the gamma-ray spectrum
which starts to be significant above $\sim$20 GeV rest-frame \citep[see e.g.][figure 3]{stern14}.
The spectrum below $\sim$13 GeV in the blazar's rest-frame is therefore essentially unabsorbed, and can thus be 
taken as representation of the intrinsic source spectrum interacting with 
the Hydrogen-induced radiation of the BLR (Ly$\alpha$ emission and continuum). 
The presence or absence of a possible BLR absorption feature 
can thus be determined by comparing  the extrapolation of the spectral fit performed 
in the unabsorbed part of the spectrum with the actual observed data at the highest energies.
This corresponds to a ``best-case scenario'' for BLR absorption, that is without considering possible 
intrinsic cut-offs (and excluding ad-hoc fine-tuned additional emission pile-ups in the spectrum).

To model the effects of \gg absorption, we adopted the same assumptions and values
used for the SED modeling with the External Compton process on BLR photons \citep{ggcanonical,sikora2009}.
Namely, we assume that the BLR intercepts and reprocesses about 10\% of the disc luminosity into an isotropic
radiation field with uniform energy density $U_{\rm BLR}$ and  radius R$_{\rm BLR}$ from the BH. 
Following \citet{tavecchio08} and in part the double-absorber approximation in \citet{poutanen2010},
we approximate the BLR spectrum with a Planckian spectrum peaked at 10.2 eV 
(corresponding to a temperature of $\sim 42000$ K) renormalized to match the BLR energy density.
This provides a good approximation  
to the Hydrogen Ly$\alpha$ emission and recombination continuum complex which
provides the strongest contribution to the BLR gamma-ray opacity \citep{poutanen2010}.
The actual BLR spectrum contains typically many lines and recombination continua,
but their effect is generally smeared out by the broadness of the \gg cross section and the averaging over angles.
The precise shape of the absorption feature depends also on the ionization state of the gas, 
but its main effects can be well approximated by an absorption feature peaking around 100-200 GeV \citep[see][figure 3]{stern14}.
Note that for our study it is irrelevant if a break in the spectrum below 10 GeV is due 
to high-ionization \ion{He}{ii} lines or is intrinsic: our focus is on the interaction with the low-ionization, 
Hydrogen-based part of the BLR emission, 
since it is the one typically used for EC models (yielding the largest energy density)
and it is unavoidable in any scenario inside the BLR, 
even more so if within the smaller high-ionization radius of the BLR. 
The normalization of the Planckian spectrum is obtained by imposing that its integrated energy density $U$ is equal to $U_{\rm BLR}$,
where
\begin{equation}
U_{\rm BLR} \, = \, { L_{\rm BLR} \over 4\pi R_{\rm BLR}^2 c}  \,\,  \quad R_{\rm diss}< R_{\rm BLR}
\label{ublr1}  
\end{equation}
$R_{\rm diss}$ is the distance from the BH of the dissipation region in the jet producing the gamma-rays.
The actual shape of the BLR spectrum longward of this peak (which might be significantly broader than 
the blackbody approximation if the illuminating spectrum is a multicolor black-body, see e.g. \citealt{tavecchio08}) 
is not important in our case, since it affects photons at energies above $\sim$100 GeV.  
Assuming the scaling relation $R_{\rm BLR}=10^{17} L_{\rm d,45}^{1/2}$cm \citep{ggcanonical,sikora2009},
the energy density inside the BLR is constant at $U_{\rm BLR}\simeq0.22$ erg\,cm$^{-3}$.

In this framework, the only free parameter determining the 
optical depth $\tau(E)$ is the path length 
of the gamma rays inside the BLR sphere, from the production region 
(at $R_{\rm diss}$ from the black hole) to the border of the BLR, at $R_{\rm BLR}$.
We thus parameterized the absorption in terms of path length $\ell$, in units of $10^{16}$ cm. 
The BLR luminosity  changes the optical depth by determining the size of the BLR (and thus the photon path). 
%
The spectral shape of the absorption feature is determined by the convolution of the Planckian approximation
with the \gg cross section averaged over angles for an isotropic field \citep{felix04}.
The corresponding maximum optical depth $\tau_{\rm max}(E)$ is reached at $E\simeq$110 GeV.
For each value of the path $\ell$, therefore, there is a corresponding value of $\tau_{\rm max}$.

\input{tableBa}
\input{tableBb}
\input{tableBc}

\begin{figure*}
\vspace*{3cm}
\includegraphics[width=1.5\columnwidth]{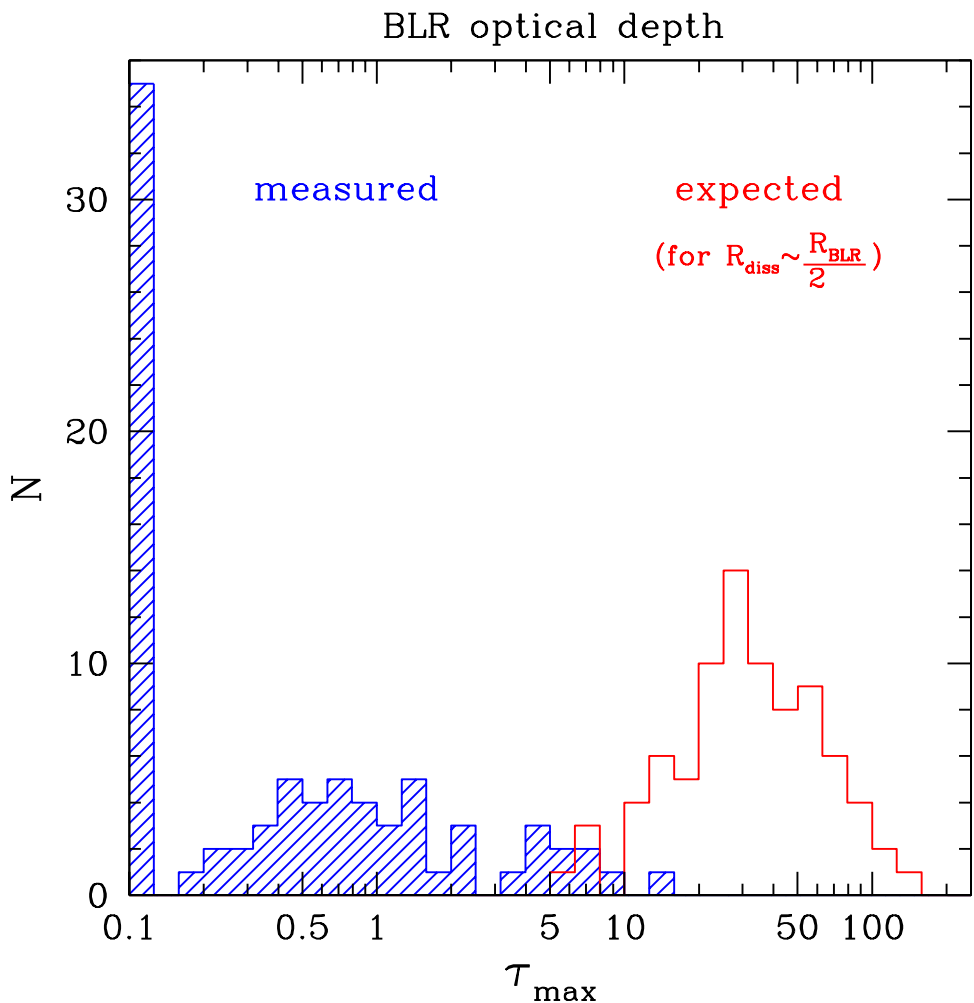}
\caption{Histogram of the distribution of the possible (as measured) vs expected maximum optical depths $\tau_{\rm max}$ 
(i.e. at the peak of the \gg cross-section, see text), for the 83 objects
for which $L_{\rm BLR}$ is available. Histograms are calculated with 10 bins per decade, in logarithmic space.
Where $\tau_{\rm max}\leq0.1$, objects are counted in the first bin. 
The optical depths measured with the fits can be considered upper limits to the possible amount of BLR absorption.
The LAT data indicate that the  possible optical depth
due to BLR photons is at most a factor $\sim30-100\times$ \textit{lower} than typically used in EC(BLR) models, and 
it is less than 1 in about 2/3 of the sample. }
\label{histo}
\end{figure*}

The fitting procedure was performed in two steps: 
first we fitted the data in the unabsorbed part of the LAT spectrum (up to 13 GeV rest-frame) 
with both a powerlaw and a log-parabolic model, choosing the best model as the one with the lowest $\redchi$.
In the second step, we extrapolated this model to the whole LAT band (typically one decade further in energy)
and fitted the optical depth  $\tau(E)=\ell \tau_{16}(E)$ to the data 
(where $\tau_{16}(E)$ is the optical depth for a path of $10^{16}$ cm), 
keeping fixed the spectral parameters to the unabsorbed values.
In addition, we fitted the log-parabolic model plus BLR absorption leaving all parameters free,
to check the robustness of the $\ell$ values when all parameters could be optimized at the same time.

The resulting photon path $\ell$ and optical depth $\tau_{\rm max}$ are to be compared with
the values for an emitting region  deep inside the BLR, $\ell_{\rm BLR}$ and $\tau_{\rm max}^{\rm BLR}$.
As benchmark for such case, we assumed that $R_{\rm diss}$ is located at $R_{\rm BLR}/2$ 
(i.e. halfway between the BH and the BLR), with a lower limit of $2\times10^{16}$ cm 
(the lowest value reported in \citealt{gg2010}).
Therefore $\tau_{\rm max}^{\rm BLR}$ is the optical depth corresponding to a photon path $\ell_{\rm BLR} = R_{\rm BLR}-R_{\rm diss}$.

Our procedure gives the best-fit values of the photon path $\ell$ inside the BLR 
under the assumption that the intrinsic gamma-ray spectrum at higher energies follows the same shape 
measured in the unabsorbed range.
This does not mean that the assumed intrinsic model plus absorption provides the best fit to the data:
it simply gives the largest possible optical depth allowed by the LAT data in the best-case scenario 
of no intrinsic cut-off in the spectrum. 
To check for alternative scenarios with \textit{no} BLR absorption, we also fitted the full-band LAT data
with a pure log-parabolic model and with a power-law model with a high-energy cut-off, of the form:
 
\begin{equation}
{dN \over dE}\, = \, N_{0}\,E^{-\Gamma} \, exp \left( -\left({E \over E_{c}}\right)^{\beta_c} \right)
\label{cut-off}  
\end{equation}

where the exponent $\beta_{c}$ regulates the strength of the cut-off: super-exponential if $\beta_{c}>1$
and under-exponential if $\beta_{c}<1$.
The importance of the shape of the spectrum in the cut-off region is determined by its direct connection 
with the shape of the underlying particle distribution and the spectrum of the seed-photon field: 
in general an exponential cut-off in the particle spectrum generates an under-exponential cut-off in the emitted gamma-ray spectrum,
in the Thomson regime \citep{lefa12}. 
To limit the number of free parameters, we have kept $\beta_{c}$ fixed at $\beta_{c}=1/3$, 
corresponding to an electron spectrum with exponential cut-off interacting with synchrotron photons 
\citep[see][table 1]{lefa12}.

The whole procedure was implemented in a python script, and the fit was performed 
on the gamma-ray SED data using the routines in NumPy (curvefit and lmfit).
This procedure is not as accurate as a direct maximum likelihood fit to the LAT event data,
but it is much faster and from our tests it gives comparable results within the errors.
A disadvantage is that it does not make full use of the upper limit information in the fit,
but in no case of our sample does it play a decisive role in the fit results.
Given the larger systematic uncertainties brought by the still limited knowledge of the BLR itself 
(in geometry, extension and spectrum), and the possible accomodation of small differences with slight 
changes in the position of the emitting region,  
we do not deem  a more detailed analysis necessary for the scope of our population study.

\section{Results: average spectra}
We first fitted the average spectrum integrated over the whole exposure time.
Table \ref{fitblr} lists the results of the BLR absorption fits. 
The model spectral parameters and the comparison with non-absorption scenarios are reported in the Appendix (Table A1),  
together with the gamma-ray SED of each single source and the corresponding fitting models (Fig. \ref{spectra1}).

Table \ref{fitblr} provides also the following information: 
a) the ratio (in logarithm) between the observed flux of the highest-energy data point 
and the expected flux assuming a location deep inside the BLR; i.e. how many orders of magnitude
the observed LAT flux is above the expected flux from deep within the BLR;
b) the p-value of the F-test between the model with free (measured) or fixed (at $\ell_{\rm BLR}$) BLR absorption, 
using the same intrinsic model determined in the unabsorbed band.
This gives the probability of measuring the path value $\ell$   if the null-hypothesis  $\ell=\ell_{\rm BLR}$ (i.e. deep within the BLR)
were true.

For the large majority of sources, there is no evidence for strong BLR absorption (e.g. $\tau_{\rm max}\gtrsim10$).
In 83\%   of the objects, the LAT spectrum extends well beyond 20 GeV, in the ``forbidden zone'' according to the 
BLR photon densities.  
The  possible photon path $\ell$ inside the BLR is  constrained to be  $\ell<3\times10^{16}$ cm 
in 95\%  of the objects,
with an average $\tau_{\rm max}\simeq 0.74$ and a path length of $\ell_{\rm avrg}\simeq 0.38\times10^{16}$ cm.
That is comparable with the typical sizes of the gamma-ray emitting region used in SED modeling.
Even considering the small sub-sample of objects with $\ell >3\times10^{16}$ cm (6\% of the total),
the average photon path 
is still $\ell_{\rm avrg}\simeq 4.3\times10^{16}$ cm, with a corresponding $\tau_{\rm max}\simeq 8.3$,
significantly less than expected from the deep-within-BLR scenario.

The same trend emerges considering the additional sample of 6 sources characterized by large BLRs:
three of them are not compatible with strong BLR absorption, while for the other three 
the data are not really  constraining. 
For these, some BLR absorption --though slightly lower than assumed for $\tau_{\rm BLR}$--
might be compatible (see Appendix, Fig. A2).  
Remarkably, in none 
of the 106 objects considered here  BLR absorption is strictly required: 
all the LAT spectra can be fitted equally well (or better) by models with no BLR absorption.

Figure \ref{histo} shows a comparison between the expected and measured optical depths for all the objects 
with an estimate of the BLR luminosity (83 among 106 objects in total).
The optical depths compatible with the LAT data  are much smaller than expected,
by 1.5 to 2 orders of magnitude, 
with 73\% of the objects having $\tau_{\rm max}<1$, and 88\% less than 3.
Of the total sample of 106 objects, only 8 FSRQ  ($\sim 8\%$) have $\tau_{\rm max}>5$.

In summary, considering all 106 {\it Fermi}-LAT FSRQ of our sample,
only approximately {\it one out of ten} seems compatible with significant absorption, either for
lack of constraints at higher energies (flux too low) or for having an estimate of  $\tau_{\rm max}>5$.
The maximum optical depth $\tau_{\rm max}$ is less than 1  in 73\% of the total sample, and less than 3 in 87\% of the cases.  
This means that, in general,  for about 2/3 of the objects  the LAT data do not seem consistent 
with any BLR absorption,
and for the remaining 1/3 the data are compatible at most  with a very low optical depth.
Even in such case, however, the smooth cut-offs are equally or better reproduced with intrinsic models,
either a log-parabolic shape or an intrinsic, under-exponential cut-off at high-energy.

\section{Results: High vs Low States}
For 21 objects with the highest overall statistics and/or  a lightcurve characterized by clearly recognizeable 
flares of high brightness  (on a 7-days time bin),  
we divided the spectra in a ``high'' and ``low''  state by making a specific cut in flux or epoch,  adapted to each source's lightcurve. 
The results of the fits are reported in Table \ref{tabHL},
together with the cut value for each source.
Lightcurves and flux cuts are shown in the Appendix, Fig. A3, 
together with the corresponding gamma-ray spectra on the side.

With the exception of 3C\,273, for which the high/low data are not really constraining,
all objects show absent or very low BLR absorption in {\it both}  high and low states.
This means that, on average, the emitting region in these sources seems located outside the BLR most of the time,
regardless of variability. 
Changes by a factor of 2 in the limiting flux do not affect the results.
There seems to be no relevant difference between the two states, with the possible exception 
of three cases where the attenuation seems slightly stronger during the high state than in the low state
(4C\,+28$+$07, PKS\,0454-234 and PKS\,2326-502).
This might indicate an emitting region closer to the BLR during flares, 
but it could also be caused by a different cut-off in the emitting particle spectrum.
Given that the allowed absorption even in such cases is rather low, 
with $\tau_{\rm max}\sim2-4$ and photon paths $\ell\sim 1-2\times10^{16}$ cm, the latter seems a more likely scenario.

\input{tableD}

\begin{figure*}
\includegraphics[width=7.5cm]{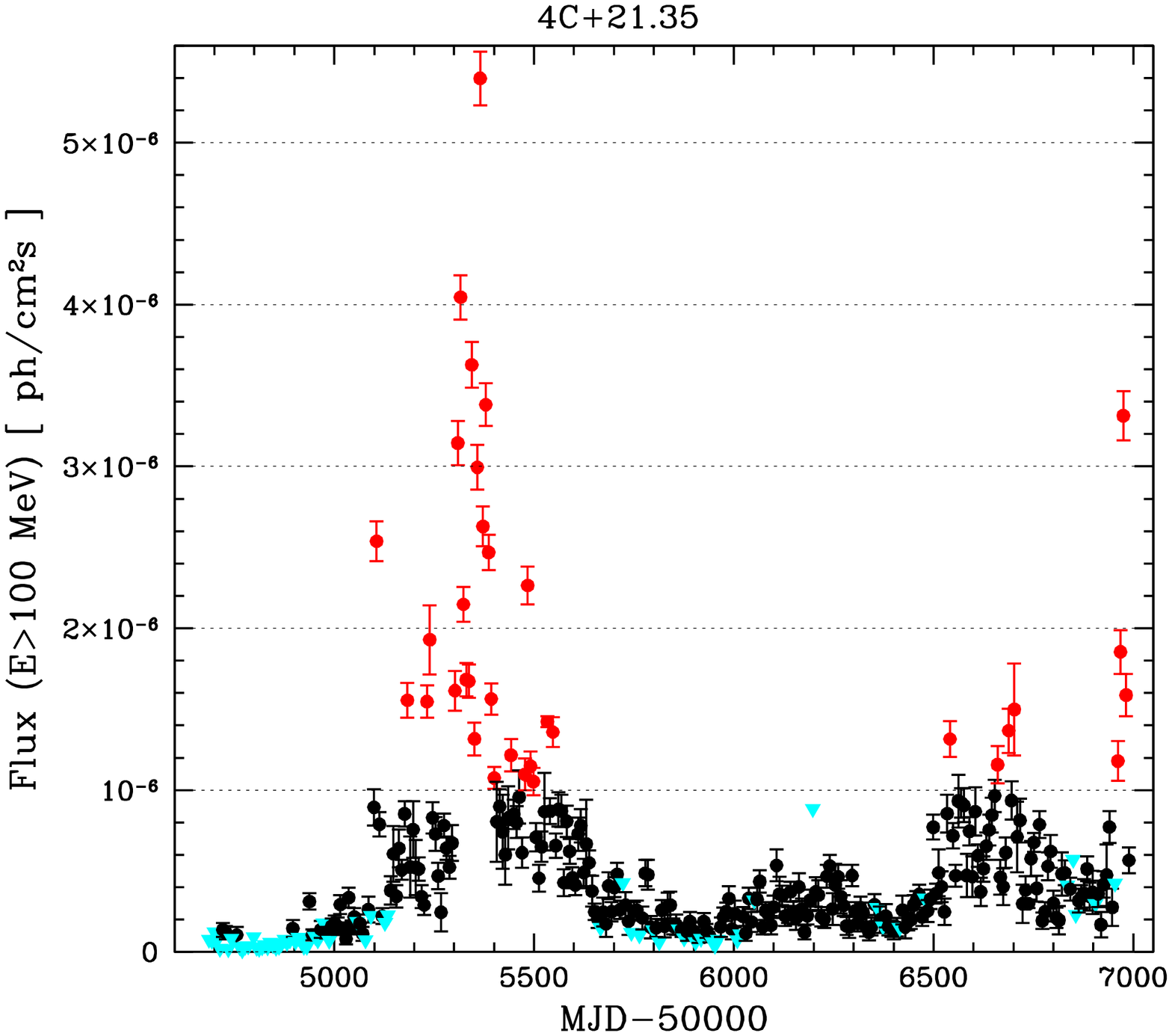}
\includegraphics[width=7.5cm]{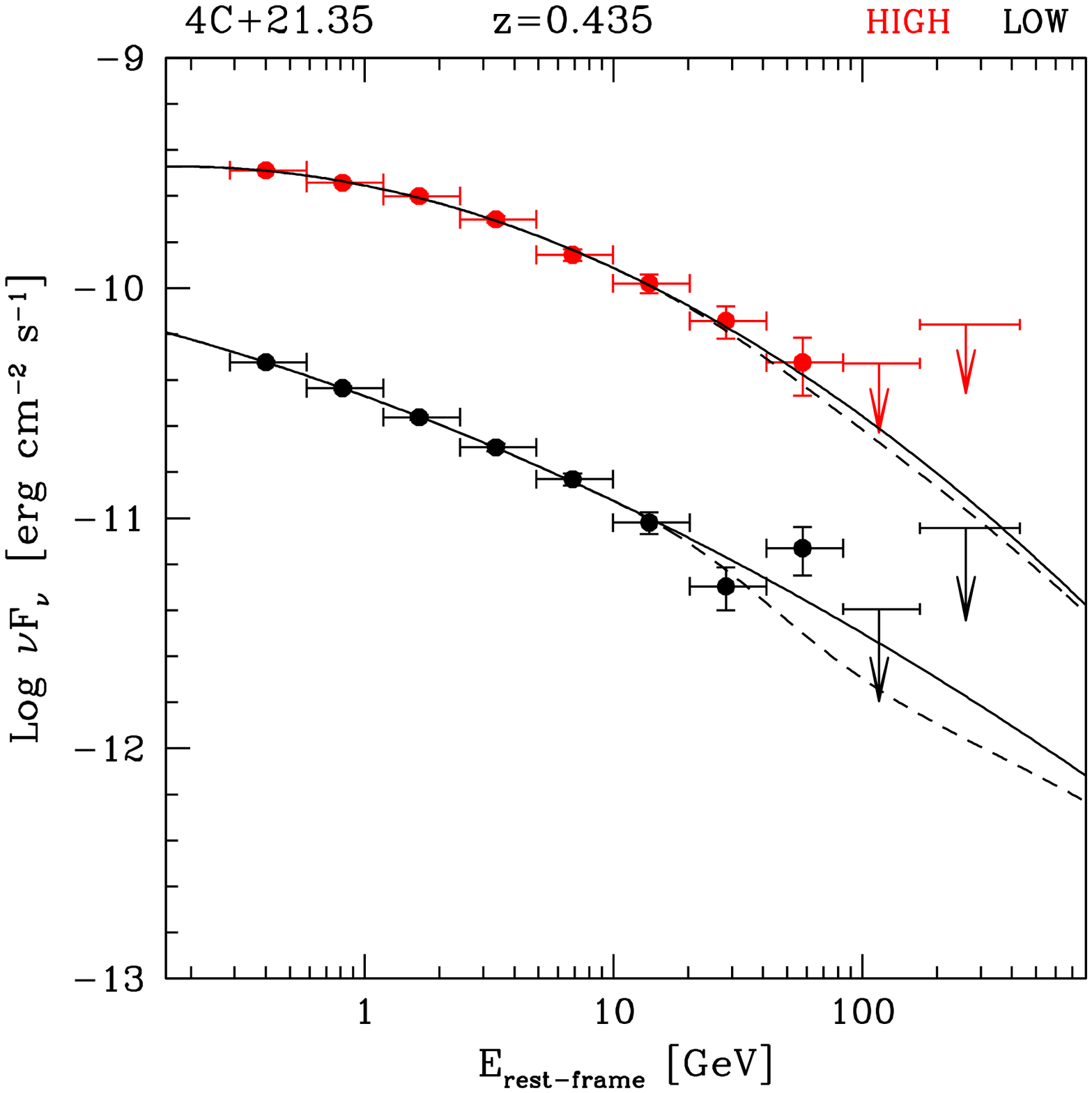} \\
\includegraphics[width=7.5cm]{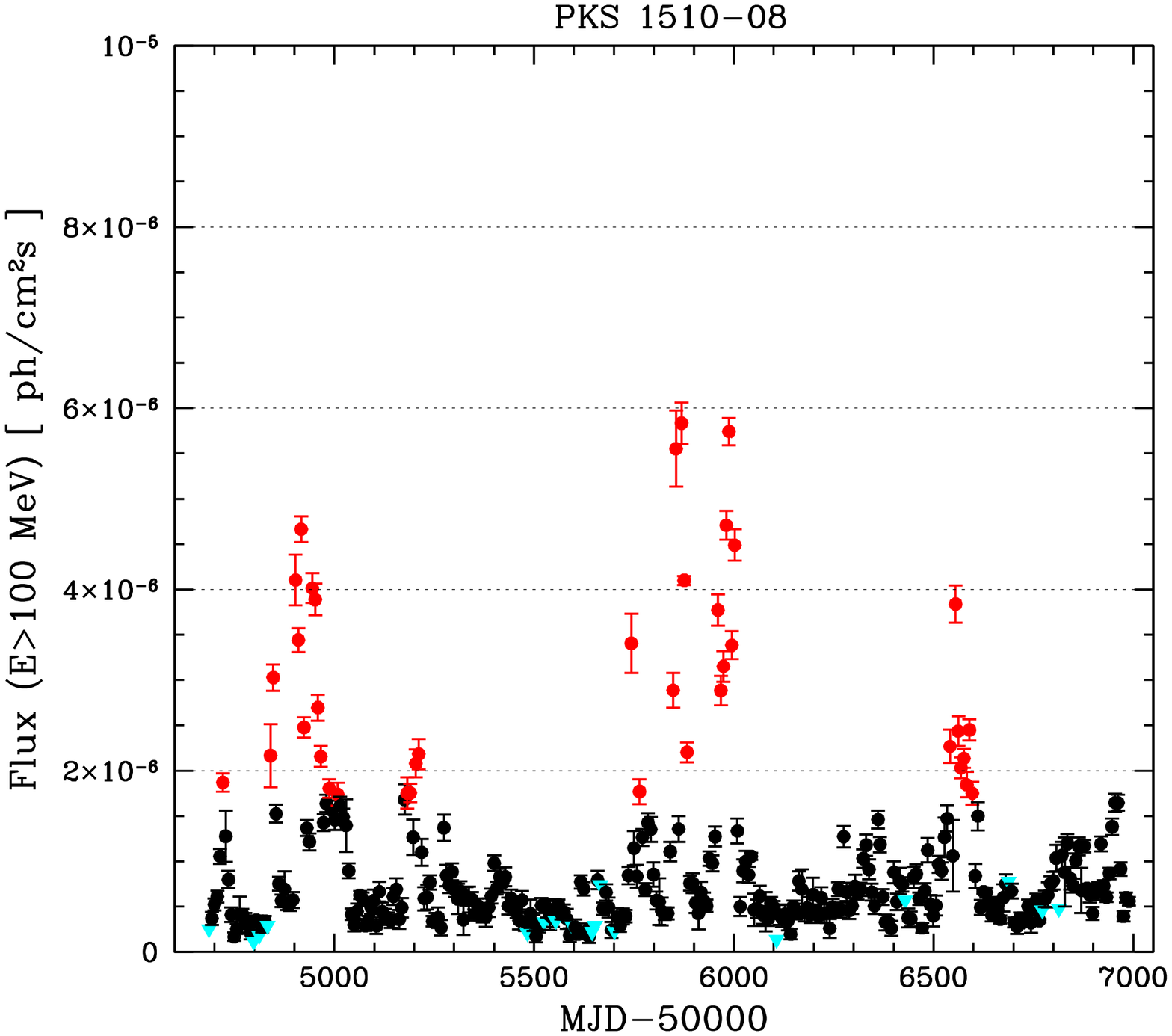}
\includegraphics[width=7.5cm]{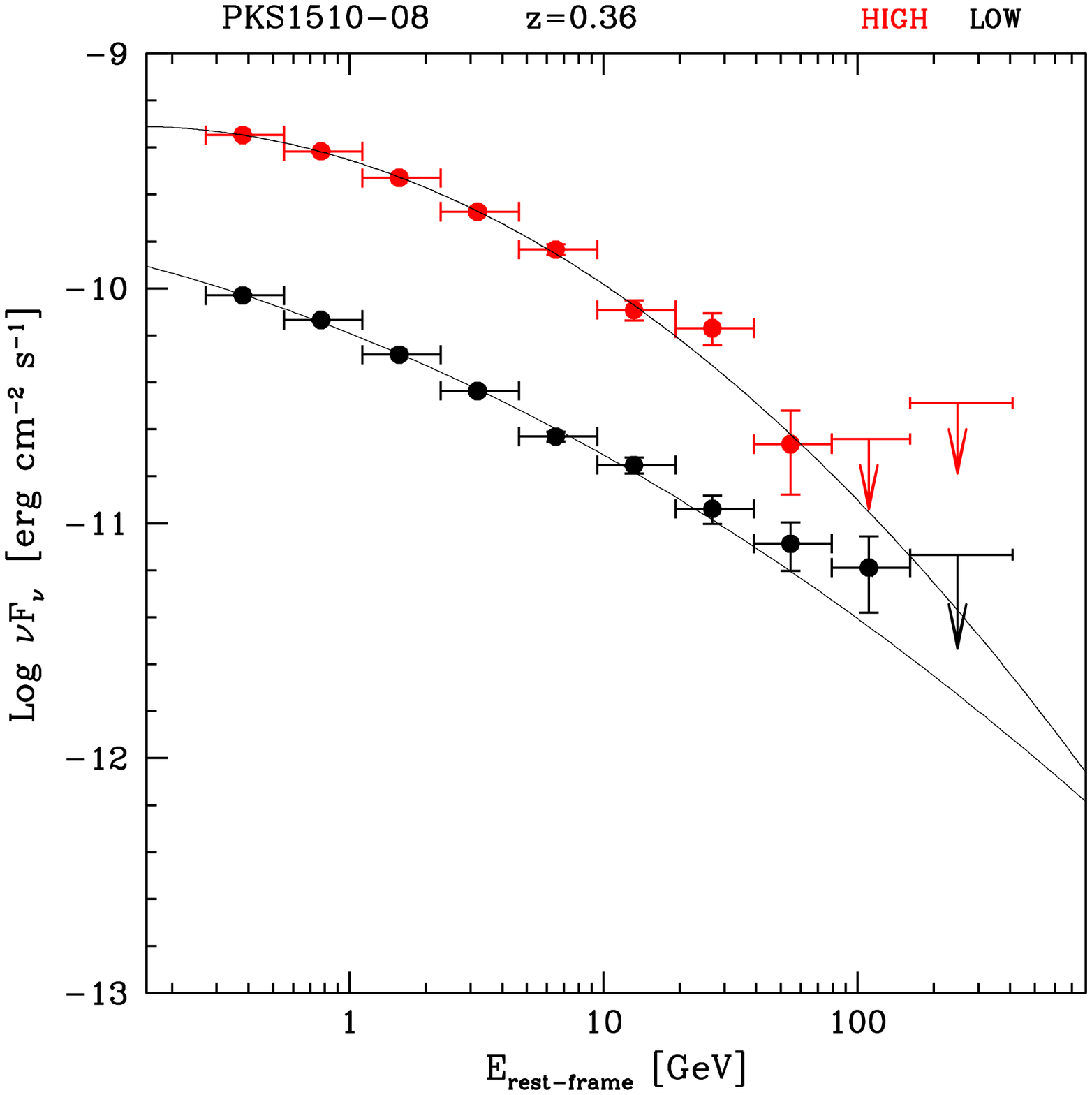}
\caption{Lightcurves and spectra of the TeV-detected FSRQ 4C$+$21.35 and PKS\,1510-08, divided 
in High (red) and Low (black) states. Upper limits are at the $2\sigma$ level (see text).}
\label{tevs}
\end{figure*}

\subsection{VHE-detected FSRQ}
A particularly interesting case is represented by the FSRQ 4C\,+21.35 and PKS\,1510-08.
Their detection at VHE  by MAGIC \citep{magic4c} and H.E.S.S. \citep{hess1510} 
provided evidence for an emitting region beyond the BLR, also in objects with strong BLR and IR emission.
This is confirmed by the LAT data, with spectra in the  high state extending  to $\sim$80 GeV rest-frame
with no signs of cut-off (see Fig. \ref{tevs}).

Remarkably, we find that also during the low states  the LAT spectrum 
extends to very high energies -- almost $\sim$100 GeV -- with no sign of BLR absorption.  
This means that for both 4C\,+21.35 and PKS\,1510-08,
the dominant gamma-ray emitting region is {\it always} located beyond the BLR, and not only 
during the high, VHE-bright and fast-flaring episodes.  
The lack of any indication of high-energy cut-off in the spectrum does not leave much room 
for \gg absorption by other radiation fields as well, like soft photons from other parts of the jet, 
at least in the UV-optical range. 
Adopting an EC scenario, the LAT data indicate threrefore that the seed photons 
should be provided by the infrared torus and not by the BLR.

However, even in this case the EC scenario might face some problems. 
IR photons are targets for \gg collisions as well, this time with TeV photons.  
The expected  optical depth from the torus radiation is again large, and should start to be important 
already around $\sim$500 GeV rest-frame, reaching the maximum from 1 TeV onward \citep[e.g.][]{protheroe97,donea03}. 
Given the strong thermal IR emission in these two sources \citep{malmrose11},
the EC(IR) scenario predicts a strong cut-off in the VHE spectrum above 500 GeV rest-frame,
which is already reached by the MAGIC spectrum in the case of  4C\,+21.35.

If future observations were to detect emission well beyond 500 GeV with no cut-off after correction for
EBL absorption, it will mean that the gamma-ray region is located even beyond the torus inner radius, 
at tens of parsecs from the central engine, where the VLBA core is often located \citep[e.g.][]{pushkarev12}
and where the jet becomes transparent to millimeter wavelengths \citep{marscher,sikora2008}.

It remains a puzzle to explain very fast gamma-ray variability (implying very compact regions) at pc distances from the BH, 
which has been observed also in PKS\,1510-08 at GeV energies \citep[see e.g.][]{lat1510tram,saito13,brown13}, 
with flux-doubling timescales as short as $\sim20$ min \citep{foschini13}. 

\begin{figure*}
\includegraphics[width=5.7cm]{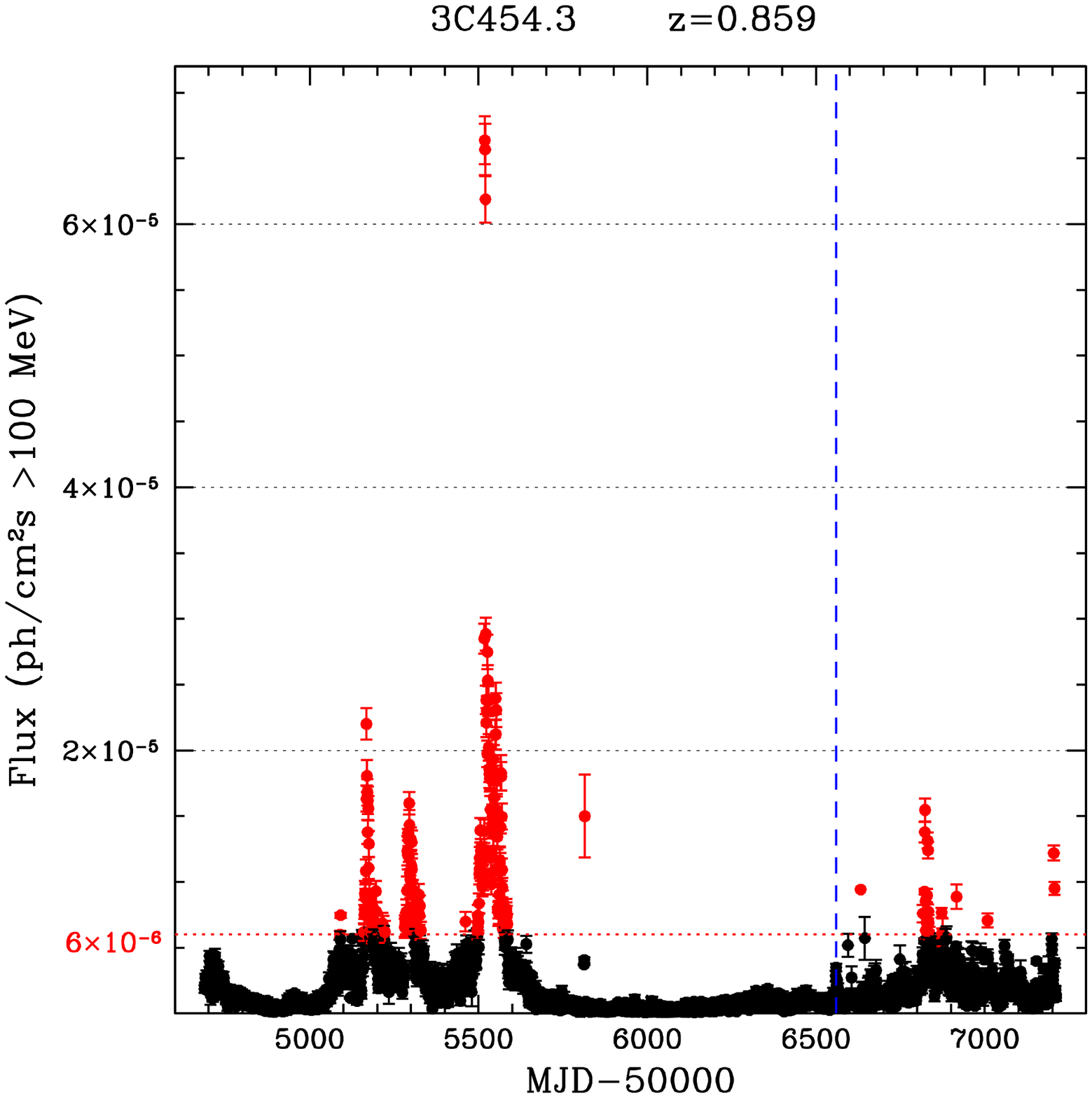}
\includegraphics[width=5.7cm]{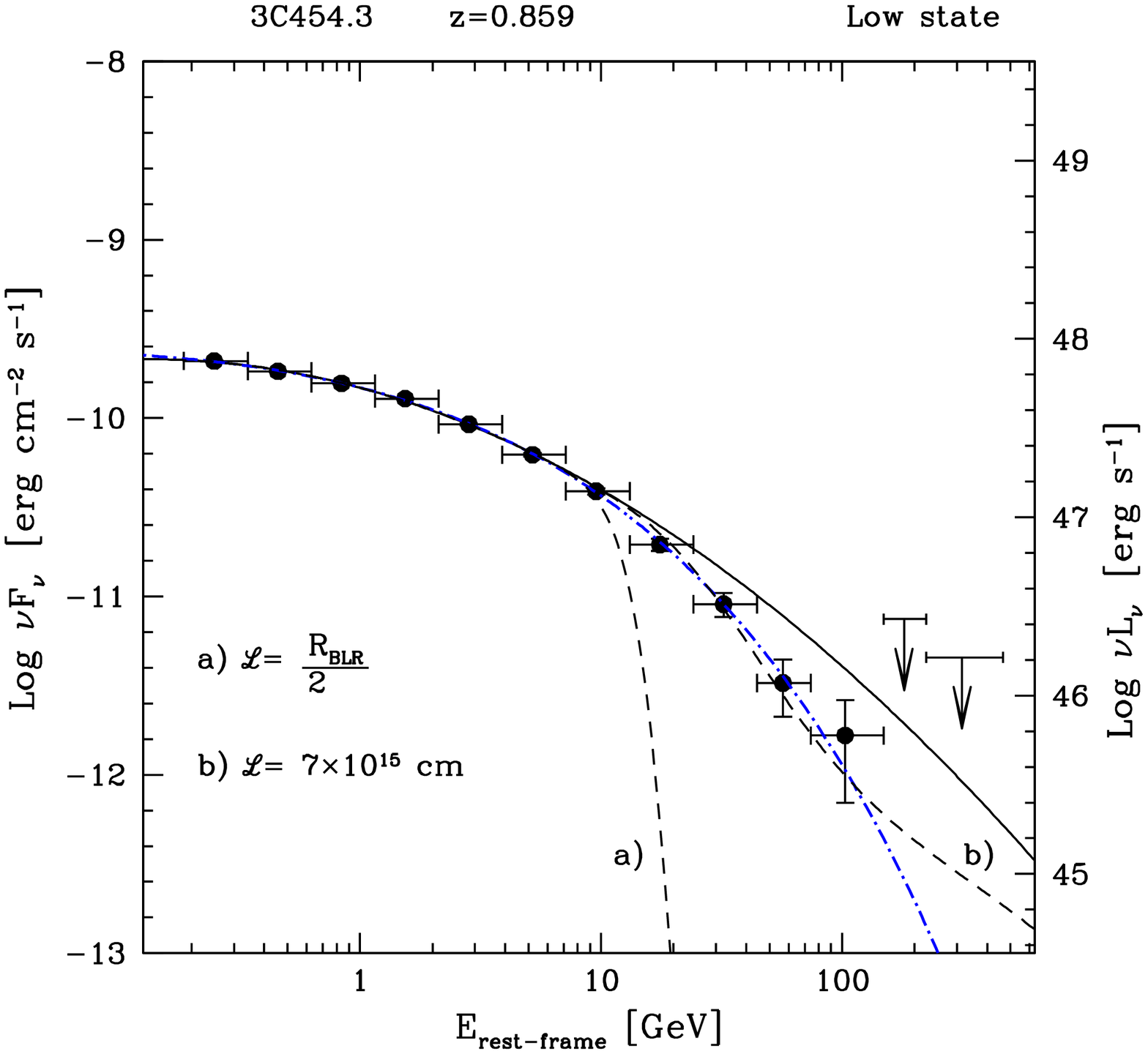}
\includegraphics[width=5.7cm]{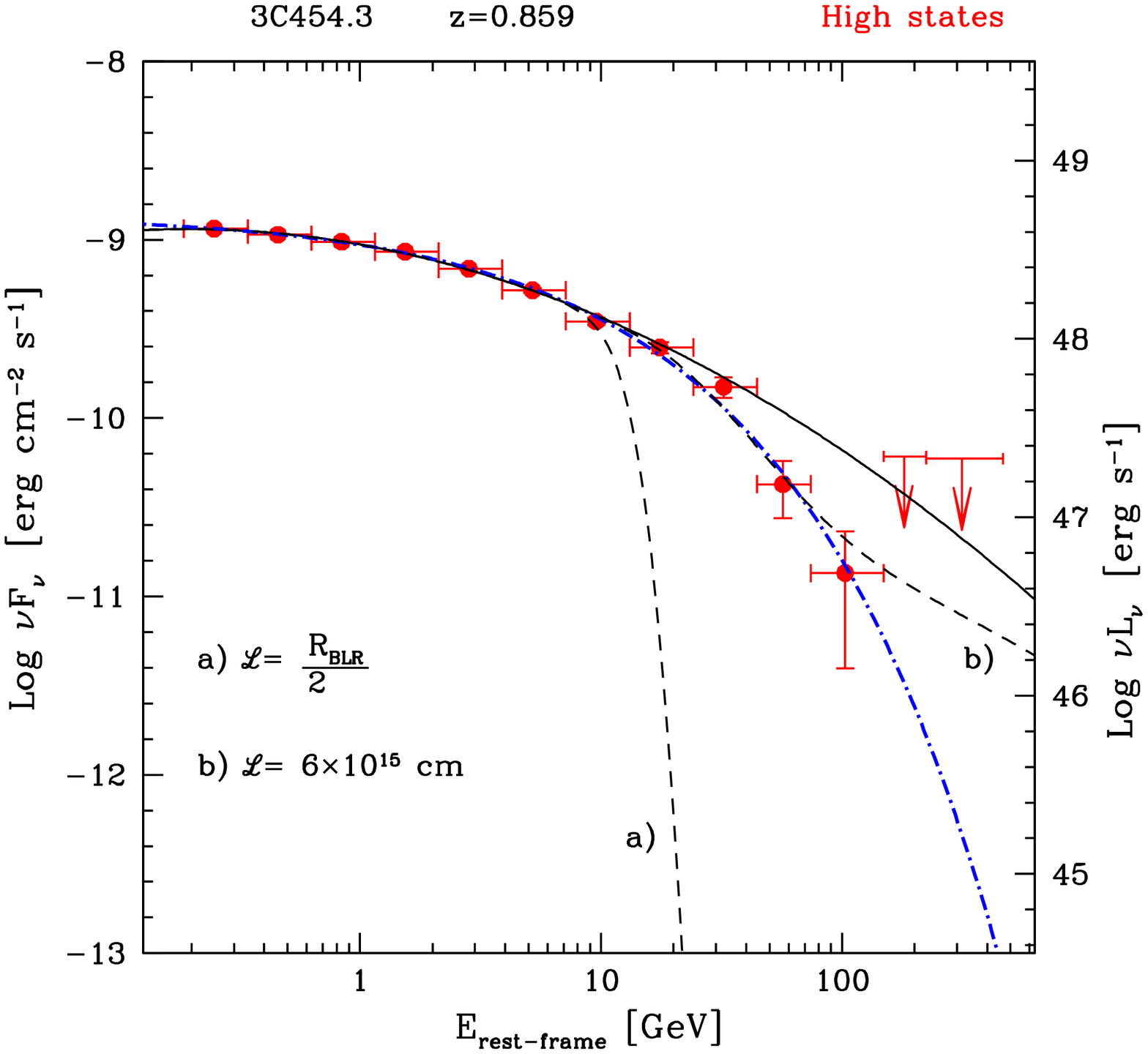} 
\caption{Left: 
lightcurve of 3C\,454.3 in 1-day bins, from the automatic quick-look analysis of the LAT monitored bright source list.
This lightcurve is used only to show the chosen cuts,  with different colors if the bin belongs to the
``{\it High}'' (in red) or  ``{\it Low}'' (black) flux state. 
The dividing line  chosen  for the extraction of the spectra is $6\times10^{-6}$ \pcms.  
The blue dashed line marks the epoch of the $\sim13$-hr flare  reported in \citet{pacciani14} as ``Period A'', 
whose spectrum is shown in Fig. \ref{pacciani}.  Center and Right: 
SED of the {\it Low} and {\it High} states, at rest-frame energies. In both cases the spectrum extends up to $\sim$100 GeV,
with no sign of strong BLR cut-off.  Full lines correspond to the best fit of the spectrum below 13 GeV.
Dashed lines correspond to the BLR-absorbed spectrum, with two different path lengths inside the BLR:
a) assuming the emitting region is deep within the BLR (at $R_{\rm BLR}/2$), 
and b) at the value from the best fit of the spectrum  ($7$ and $6$ $\times10^{15}$ cm). 
The blue dot-dashed line shows how a power-law model with under-exponential high-energy cut-off 
($\beta_{c}=1/3$), corresponding to an intrinsic cut-off in the particle distribution with emission
in the Thomson regime \citep{lefa12}, can provide an excellent fit, reproducing all the properties of
the LAT spectra (see text).}
\label{spec454}
\end{figure*}

\subsection{The case of 3C\,454.3}
Since the launch of {\it AGILE} and {\it Fermi},  3C\,454.3  has been  one of the brightest sources in the gamma-ray sky \citep{vercellone,lat3c454_09}.
It is also one of the most violently variable blazars, with flux variations of several order of magnitude and
observed down to shortest timescales allowed by photon statistics \citep[sub-hour, see][]{lat3c454_11}.
With strong and variable  broad emission lines \citep{isler13,leontavares13},
for a BLR luminosity $L_{\rm BLR}\sim2-5\times10^{45}$ \ergs and $R_{\rm BLR}\sim6\times10^{17}$ cm,
3C\,454.3 is one of the prototypical FSRQ-type blazars.

Its SED has been commonly modeled with EC on BLR photons \citep{ggcanonical,lat3c454_09,bonnoli454},
but specific outbursts have been explained in several alternative ways:
with EC on IR photons from the torus  (for example the huge flare of year 2005, though no gamma-ray telescope was in operation at that epoch, \citealt{sikora2008})
or with IC on radiation from clouds beyond the BLR reflecting the jet emission itself
\citep[``mirror models'', see][and refs therein]{vittorini}, 
with SSC emission outside the BLR \citep{pian2006}  and also as the result of star-jet interaction \citep{mitya3c454}.
The fact that the dissipation zone can move outside the BLR during single flares has been generally supported  
by the detection of several photons above 20 GeV  \citep{lat3c454_11,britto16}
and by the hard spectrum with no steepening \citep{pacciani14}.

Our analysis of the average gamma-ray spectrum  
confirms an emitting region beyond the BLR in high state.
Remarkably, it reveals that the spectrum extends with a smooth shape up to $\sim$100 GeV rest-frame
also in the low state, with a slight steepening beyond 30 GeV  which leaves room only for a small
amount of possible \gg absorption (see Fig. \ref{spec454}). 
The allowed photon path inside the BLR is only $\ell= (0.74\pm0.28)$ and $(0.55\pm0.17)$ $\times10^{16}$ cm, for low and high states respectively, 
corresponding to a $\tau_{\rm max}=1.4$ and $1.1$. 
These values are incompatible with a dissipation region well within the BLR.

The shape of the spectrum, however, is better reproduced by a power-law model with under-exponential cut-off 
rather than the log-parabolic model with free BLR absorption (see Fig. \ref{spec454}, $\redchi\simeq1.3$ vs $2.4-4.0$).
With $\beta_c\equiv 1/3$ (see Eq. \ref{cut-off}), the model fits the gamma-ray spectrum very well
with a photon index $\Gamma=1.83\pm0.02$ and $E_\text{cut-off}=0.14\pm0.02$ GeV 
for the low state, and $\Gamma=1.82\pm0.02$ with $E_\text{cut-off}=0.30\pm0.05$ GeV  for the high state.
Together, these two results (spectrum extending to 100 GeV and better fitted without BLR absorption)
indicate that the steepening seen in 3C\,454.3 is most likely intrinsic, 
related to the end of the emitting particle distribution rather than due to \gg interactions with BLR photons.

\begin{figure}
\centering
\includegraphics[width=8.2cm]{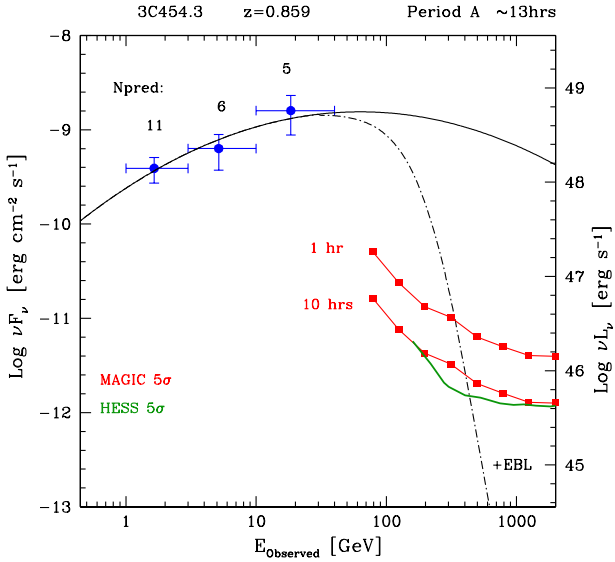}
\caption{SED of 3C\,454.3 during the flare defined ``Period A'' in \citet{pacciani14}, re-analyzed with Pass 8 data.
The number of predicted source photons in each bin is also shown. 
We note that here the spectrum is plotted in the {\it observed energy scale} (not at rest-frame), for easier comparison with sensitivity curves.
The spectrum is hard ($\Gamma=1.8\pm0.1$), as typical in {\it Fermi}-detected HBLs but not FSRQ.
This locates the gamma-ray peak in the SED at or above 100 GeV, with consequently copious emission at VHE.
As example, we show an hypothetical log-parabolic spectrum (with curvature parameter $\beta=0.25$) 
normalized to the LAT flux, assuming that the SED peak is around 100 GeV (full line).
The $5\sigma$ sensitivity curves for 10 and 1 hour of observations with MAGIC in stereo mode \citep{magic16}
are also plotted, together with the 10-hour curve for the present H.E.S.S. performance for the 4-telescope 
array \citep{holler15}.  After accounting for EBL absorption using \citealt{dominguez11} (dot dashed line),  
the sensitivity curves show that 3C\,454.3 can be easily detected up to $\sim$3-400 GeV during such flares.
}
\label{pacciani}
\end{figure}

If this scenario is correct, we can expect that under the right acceleration conditions  
3C\,454.3 could become a strong VHE emitter and detectable by present air-Cherenkov telescopes.
Indeed this seems to have been the case during a relatively low-flux flare of 3C\,454.3 in 2009, 
reported by \citet{pacciani14}.
During a 13-hours timespan (``period A'' in \citealt{pacciani14}, marked by the dashed line in Fig. \ref{spec454}), 
the gamma-ray spectrum was hard (power-law photon index $\Gamma<2$), namely rising with energy in the SED.
We have re-analyzed this flare with Pass 8 data (using in this case an unbinned maximum likelihood fit to the LAT event data, 
given the low counts)  and can confirm the result, obtaining $\Gamma=1.8\pm0.1$. 
This means that 3C\,454.3 in that circumstance was behaving like an high-energy-peaked BL Lac object (HBL),
with the gamma-ray emission in the SED peaking around or above 100 GeV (Fig. \ref{pacciani}).

Given the flux measured by {\it Fermi}-LAT and the emission beyond the BLR, 
Fig. \ref{pacciani} shows that 3C\,454.3 can be easily detected at VHE in less than an hour
with the present generation of air-Cherenkov telescopes, and up to 300-400 GeV despite the large redshift and EBL attenuation.
This makes 3C\,454.3 an excellent and important target for VHE observations during such flares,
which can reveal the spectral evolution of the freshly accelerated particles 
near the cut-off region \citep{lefa12,romoli16}.

\section{Discussion}
The {\it Fermi}-LAT spectra indicate that for 2/3 of our FSRQ sample there is no evidence of BLR absorption ($\tau_{\rm max}<1$),
while for the remaining 1/3 of the objects the possible optical depths are a factor 30-100$\times$ lower than expected 
in EC(BLR) models.    

To keep a low optical depth consistent with a gamma-ray emitting region well inside the BLR, 
one can envisage two possibilities: 1) to decrease the photon densities by enlarging the size of the BLR, 
or 2) to shift the \gg threshold at higher energies by  selecting preferred angles of interaction. 

In the first case, since the optical depth $\tau\propto 1/R_{\rm BLR}$, a factor 100$\times$ less in $\tau$ means that
the size of the BLR should be 100$\times$ larger than it appears from reverberation mapping, in fact close to the torus location.
This is much larger than the scatter on the reverberation mapping relation \citep[2 vs 0.13 dex, e.g.][]{Bentz2013}.
However, this size would imply that the energy density $U_{\rm BLR}$ available for the external Compton process is 
lower as well, and by a factor $10^{-4}$ given that  $U_{\rm BLR} \propto L_{\rm d} R_{\rm BLR}^{-2}$.  
Such low values would make the EC process highly inefficient, requiring a much larger jet power to compensate,
and would put the energy density of BLR photons well below the other local energy densities 
in the jet comoving frame \citep{ggcanonical,sikora2009}.
These arguments hold also considering a wide stratification of the BLR \citep[e.g.][]{abolmasov17},
in the only scenario that could reduce the opacity with respect to our estimates. Namely, 
if the bulk of the BLR emissivity were produced much further out with respect to the radius
from reverberation mapping or photoionization methods,
which should correspond to the virial radius of the BLR clouds \citep[see e.g.][]{negrete13}.
Stratification due to higher-ionization lines instead develops towards inner radii, and thus 
does not impact our results (see Sect. 2.1).

In the second case, one can assume that the BLR is actually flattened onto the accretion disc, such that
the interaction of the jet gamma rays with BLR photons occurs preferentially at low angles \citep{LeiWang14,abolmasov17}.
To push the BLR absorption feature --starting at $\sim$20 GeV in the isotropic case--  outside the 
observed data, for example at 100 GeV rest-frame,  one needs an average shift of $\sim5\times$ of the energy threshold. 
Since the latter is given by  
\begin{equation}
E_{\rm thr} = \frac{2 m_e^2 c^4}{(1-\mu) \epsilon}
\label{threshold}
\end{equation}
where $\epsilon$ is the energy of the target photon, $\mu = \cos \theta$ and $\theta $ is the collision angle,
a back-of-the-envelope estimate shows that the typical angle between gamma rays and target photons 
should be less than 30 deg. 
This would reduce again the efficiency of the EC mechanism, due to the similar kinematics of the interaction.
In any case, it would locate the emitting region again very far from the BH and beyond the BLR radius 
(typically by more than a factor $\sqrt{3}$ larger), in order to achieve such small interaction angles.

Both alternative scenarios do not seem able to keep the external Compton on BLR UV photons 
as a viable mechanism to produce the gamma-ray emission in FSRQ.
When the LAT data allow for possible BLR absorption (in only 1/3 of the objects),
the  low optical depths indicate that, if indeed inside the BLR, the location of the dissipation region must be 
close to the BLR radius given by reverberation mapping ($R_{\rm diss}\approx R_{\rm BLR}$), 
 probably between the inner and outer borders of a more realistic representation of the BLR \citep[e.g.][]{boettcher16}.
In these conditions,  also the isotropic assumption for $\gamma-\gamma$ absorption 
is not valid anymore, and the actual values of $\tau$ will strongly depend on the details of the  
(yet uncertain) geometry of the BLR in relation to the size of the dissipation region \citep[see e.g.][]{finke16}.
In any case, the low  optical depths imply that also the EC cooling on BLR photons is depressed, under the same conditions.

However,  the same data are equally and often better reproduced by a spectrum with intrinsic curvature and no BLR absorption.
In addition, with these fits there is no relevant difference in properties and parameters values  
among objects with (1/3) or without (2/3) possible mild absorption.
The inside-BLR scenario would instead require a dichotomy in jet parameters, 
given the very different energy of the main seed photons for IC, in the two cases (UV or IR).
While not definitive, therefore,  the simplest explanation is that also in these objects the emission occurs mostly outside the BLR.

This leads to the conclusion that for the large majority of FSRQ (94\%) 
and/or for most of the time, the gamma-ray emission is produced outside the BLR, 
and  not by IC scattering of BLR photons.

This conclusion has two important consequences.
First, it means that the steepening often seen in the LAT gamma-ray spectra at higher energies is 
intrinsic,  and not due to interaction with the environment.  
The shape of the cut-off depends on both the electron distribution and the target photon field, 
as well as the regime of the interaction (Thomson or Klein-Nishina).
This dependence has been studied by \citet{lefa12} and 
provides a powerful diagnostic tool for the acceleration and cooling mechanisms in the jet, especially
if coupled with the knowledge of the synchrotron cut-off \citep[see also][]{romoli16}.
Indeed the gradual steepening seen in the majority of our sample seems to favour
a SSC scenario or an EC on IR photons, both in the Thomson regime, with the parameter $\beta_e$ regulating the cut-off
of the electron distribution between 2 and 1, depending on the importance of the synchrotron losses in the acceleration scenario.

The second consequence is an apparent discrepancy: 
if  there is no interaction of the jet's
electrons and photons with the BLR, why the observed strong correlation between the gamma-ray luminosity 
and  the broad line emission \citep{sbarrato12}  ?
Such correlation suggests a tight jet-disc connection, between jet power (traced by gamma-ray luminosity)
and accretion rate and disc luminosity (traced by the BLR luminosity with BH mass estimates), 
with the transition between radiatively efficient and inefficient accretion discs occurring at $L_d/L_{\rm Edd}\sim 10^{-2}$
\citep{sbarrato14}.
Our result 
suggests that the BLR acts as a proxy for the accretion regime and the disc luminosity,
but does not affect directly the gamma-ray luminosity through more efficient cooling.
In other words, jets seem born that way: the division between FSRQ and BLLacs seems determined by the conditions
of accretion (if radiatively efficent or inefficient, and by the accretion rate) 
rather than the environment and location of the dissipation region, inside or outside the BLR.
Alternatively, the role of the BLR is simply taken by the IR emission from the torus.
In such case the BLR would act as a proxy for the torus emission, via the disc luminosity.

We note that our values for the possible optical depths are consistent with those  
reported in \citet{stern14}, 
after accounting for the different scaling (since they scale the optical depth to the Thomson cross section,
their values must be divided by $\approx$5 to be compared with ours). 
However we arrive to the opposite interpretation considering the much larger sample,
with the majority of FSRQ without BLR absorption,  the extension of the spectra at high energies without clear breaks, 
the very low optical depths inconsistent with EC models and the better fits with intrinsic models.

\begin{figure}
\centering
\includegraphics[width=9cm]{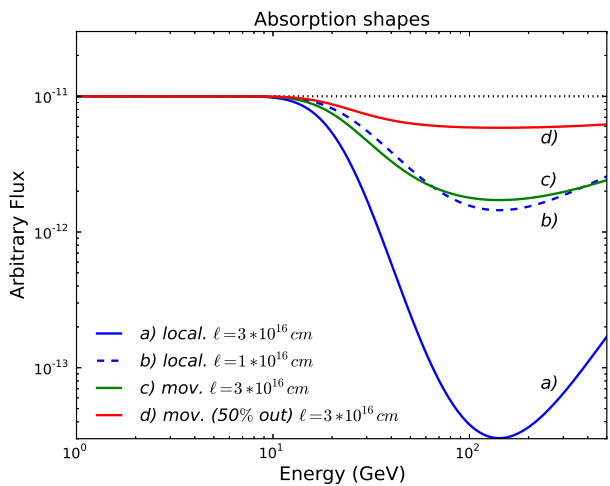}
\caption{
Absorption shapes for different emission scenarios:  
{\it a)} localized emission, with the same path $\ell$ for all photons ($e^{-\tau}$ absorption);
{\it b)} same as a) but closer to $R_{\rm BLR}$ (path $\ell= 1\times 10^{16}$ cm vs $3\times 10^{16}$ cm); 
{\it c)} constant emission from moving blob fully inside the BLR, up to $R_{\rm BLR}$ (Eq. \ref{shallow});
{\it d)} constant emission from moving blob, with 50\% of the flux coming from outside the BLR. 
Scenarios {\it c)} and {\it d)} present shallower attenuation features than scenario {\it a)}, for the same overall path $\ell$ inside the BLR.
This translates to the possibility of slightly longer paths inside the BLR for approximately the same absorption shape 
(compare lines {\it b)} with {\it c)} ). 
}
\label{blob}
\end{figure}

\subsection{Caveats}
Our result comes with two important caveats.
One concerns the long integration time of the spectra, needed by the extremely low countrates at high energy.
In principle, there is the possibility that a series of short hard flares occurring beyond the BLR might provide all 
the detected high-energy photons,
while for most of the time the emission is actually inside the BLR. This could skew the fit, simulating the effects 
of a lower absorption in the averaging process. It seems unlikely because flares should be frequent, short and not too strong 
in order to remain below the flux threshold of the high/low-state separation, for nearly all sources in our sample.
Unfortunately the limited collection area of the LAT detector ($\sim1$ m$^2$) 
cannot provide a definitive answer to this issue. 
Such answer could be provided by future imaging air-Cherenkov arrays like CTA \citep{cta1,cta2}, 
thanks to their huge collection areas and lower energy threshold.
The optimum would be represented by the proposed  ``5@5'' array \citep{5at5}, with a threshold as low as 5 GeV.
A low threshold below 10 GeV is necessary to get a handle also on the unabsorbed part of the gamma-ray spectrum.

The second caveat concerns the kinematics of the emission.  
Our assumption is that the bulk of the emission originates on average at roughly the same location inside the BLR. 
For example, from a preferred dissipation region within the jet (e.g. a standing shock) at a given distance from the 
central engine. 
In this case the produced gamma rays travel the same path $\ell$ inside the BLR,  
and thus have all the same probability of survival $e^{-\tau(E,\ell)}$.
The time-integrated  spectrum  is therefore attenuated by $e^{-\tau(E,\ell)}$.
This is the absorption factor commonly used in the literature for BLR absorption.

Instead, if we assume that the observed flux is dominated by the emission of a single relativistic blob over a relatively long time,
the actual distance traveled by the blob in that timeframe can be quite large, 
because of the shortening of the arrival times due to the Doppler effect.
Gamma rays originate from different locations inside the BLR, and the net absorption effect can be different.   
The travel $\Delta R$ of a blob with velocity $\beta c$ in an observed time interval $\Delta t_{\rm obs}$
is $\Delta R \simeq \Delta t_{\rm obs}*\beta c/(1-\beta \cos(\theta))$.
For a blazar jet seen at an angle $\theta \simeq 1/\Gamma$, for which $\cos(\theta)=\beta$, 
$\Delta R \simeq \Delta t_{\rm obs}*\beta c*\Gamma^2$.
With bulk Lorentz factors $\Gamma\geq 10$, an observed emission time of $\geq 10^5$ s can correspond 
to a travel of the emitting region  $\Delta R \geq 10^{17}$ cm, i.e. from deep within the BLR to its borders or beyond.  
In such a case, the last part of the emission is less affected by BLR attenuation than the first part,
since the actual photon path inside the BLR becomes shorter and shorter.
The resulting attenuation shape on the total integrated spectrum becomes shallower than exponential (see Fig. \ref{blob}), 
and this could relax the constraints on the path inside the BLR.   

For example, assuming that all the blob emission takes place inside the BLR 
but up to the BLR border (the most favourable scenario for EC on BLR),
the flux becomes attenuated as
\begin{equation}
N(E) \, \propto \, E^{-\Gamma}\,\,\; {(1 - e^{-\tau(E,\ell)}) \over \tau(E,\ell)} 
\label{shallow}
\end{equation}
which goes as $\tau^{-1}$ for large values of $\tau$. 
Here $\ell$  corresponds to the path inside the BLR since the start of the blob emission.

However, this possibility does not change the main outcome of our study.
Fitting all sources with the shape of Eq. \ref{shallow},
the limits on the path length inside the BLR increase only mildly, by a factor 2--3.
The maximum possible path inside the BLR is still limited to 
small values ($\ell < 3\times10^{16}$) for 86\% of the total sample of 106 objects
(compared to the 94\% of the $e^{-\tau}$ attenuation shape).
We can conclude that also in this scenario the gamma-ray emission seems produced outside the BLR for most of the sources.
This scenario however does allow for significantly larger $\tau$ in about 10\% of the FSRQ, 
for which $\tau_{\rm max}\geq10$ with an average maximum path $\ell_{\rm avrg}\simeq 9\times10^{16}$.  
Future observations with Cherenkov arrays in the 10--300 GeV range should provide more detailed answers,
thanks to the much larger collection area at high energies.
Interestingly, if the blob path during the emission were to continue beyond the BLR, the time-integrated spectral shape
could have a step depending on the percentage of the blob emitting time passed inside or outside the BLR 
(see Fig. \ref{blob}),  and this could become a diagnostic tool with CTA. 

\section{Conclusions}
We have analyzed 7.3 years of LAT data to study the spectra of 106 FSRQ  characterized by the largest gamma-ray photon statistics
and/or the largest BLR luminosities and sizes, looking for evidence of \gg absorption on the Hydrogen-based BLR emission. 

The result is that, for the large majority of {\it Fermi} blazars (9 out of 10),
there is NO evidence of strong interaction of the jet gamma rays with BLR photons.
The emission seems to originate \textit{almost always} outside the BLR, 
both on average and during high/flaring or low-flux states.
To this respect there is no relevant difference between high/flaring states and low states.
The percentage of objects where the LAT spectrum is consistent with a relatively strong BLR absorption 
is limited to only a few \% of the FSRQ population (at maximum 10\%, depending on the emission scenario). \\
The implications are remarkable and far-reaching.
\begin{enumerate}
\item External Compton on BLR photons is disfavoured as main mechanism for the gamma-ray emission in {\it Fermi} FSRQ.
This conclusion holds also in the alternative scenario of a BLR  much larger (100$\times$) than given by reverberation mapping,
requiring  10$\times$ higher bulk motion Lorentz factors of the plasma to reach an equivalent energy density in the jet frame.
A higher  Lorentz factor would increase the energy density of the other external photons as well. 
A viable alternative for the EC mechanism is provided by the IR photons from the dusty torus \citep{sikora2009}.

\item The gamma-ray spectrum measured by {\it Fermi}-LAT  is mostly intrinsic, 
up to energies where \gg absorption by the EBL or IR radiation from the torus kicks in.
The mild, smooth cut-off often seen between few GeV to few tens GeV is most likely due to the end of the accelerated particle distribution.
This opens up the possibility of powerful diagnostics on the accelerated particle spectrum, 
studying the shape of the cut-off and in comparison with the synchrotron spectrum \citep{lefa12,romoli16}. 

\item Without suppression by the BLR, the gamma-ray emission in FSRQ can  reach the VHE band 
with potentially very high  luminosities, as demonstrated by the flares of 3C\,454.3.
Therefore also FSRQ can be good targets for VHE telescopes:   
they can be easily detected by present and future air-Cherenkov arrays even at high redshift (around or beyond $z\sim1$), up to 300-400 GeV,
thanks to the very high photon fluxes and the low intensity of the EBL \citep{nature_ebl,lat_ebl}.
\end{enumerate}

If indeed gamma rays originate mostly outside the BLR, our result indicates that
the VHE sky with CTA should be much richer of FSRQ  than  previously expected.

Given  the strong correlation observed between gamma-ray luminosity
and BLR emission in {\it Fermi} blazars \citep{sbarrato14},
our result suggests that the BLR acts  as a proxy of the accretion properties
but does not affect directly the jet emission through cooling. 
The classical distinction between FSRQ and BL Lacs becomes thus somewhat fuzzier:  
differences like high/low-peaked SEDs and 
Compton dominance cannot  be attributed to the cooling on BLR photons and their high energy density. 
This result may have implications also for the interpretation of the ``blazar sequence"  \citep{gg98,sequence2},
since the jet physical parameters derived from the SED modeling change depending on the 
emission mechanism and dominant seed photons. 
A general re-assessment of the blazar emission parameters 
and re-modeling of their SED  (with external Compton on IR photons, SSC or with 100$\times$ lower $U_{\rm BLR}$)
seems in order.

\section*{Acknowledgements}
We thank the anonymous referee for the helpful comments.
The  {\it Fermi}-LAT Collaboration acknowledges generous ongoing support from a number of agencies and institutes that
have supported both the development and the operation of
the LAT as well as scientific data analysis. These include
the National Aeronautics and Space Administration and
the Department of Energy in the United States, the Commissariat \`a l'Energie Atomique and the Centre National de
la Recherche Scientifique / Institut National de Physique Nucleaire et de Physique des Particules in France, 
the Agenzia Spaziale Italiana and the Istituto Nazionale di Fisica Nucleare in Italy, 
the Ministry of Education, Culture, Sports, Science and Technology (MEXT), High Energy Accelerator
Research Organization (KEK) and Japan Aerospace Exploration Agency (JAXA) in Japan, and the K. A. Wallenberg
Foundation, the Swedish Research Council and the Swedish National Space Board in Sweden.

L. Costamante acknowledges partial support from the Grant 
``Cariplo/Regione Lombardia ID 2014-1980 / RST\_BANDO congiunto Fondazione Cariplo-Regione Lombardia - ERC'',
for the project ``Science and Technology at the frontiers of Gamma-Ray Astronomy with imaging atmospheric Cherenkov telescopes''.

We thank G. Ghisellini for useful discussions, in particular for pointing out 
the change in absorption shape due to the emission from a moving blob.

\bibliographystyle{mnras}
\bibliography{mybibtex}

\bsp	
\label{lastpage}


\appendix
\section{Additional Tables and Figures} 
%
Here we provide the complete version of Table A1 and Figures A1, A2 and A3.
They are available to download as supplementary material on the electronic version of the journal.




\include{tableCa}
\include{tableCb}
\include{tableCc}

\begin{figure*}
\vspace{1cm}
\begin{tabular}{cccc}
 \psfig{file=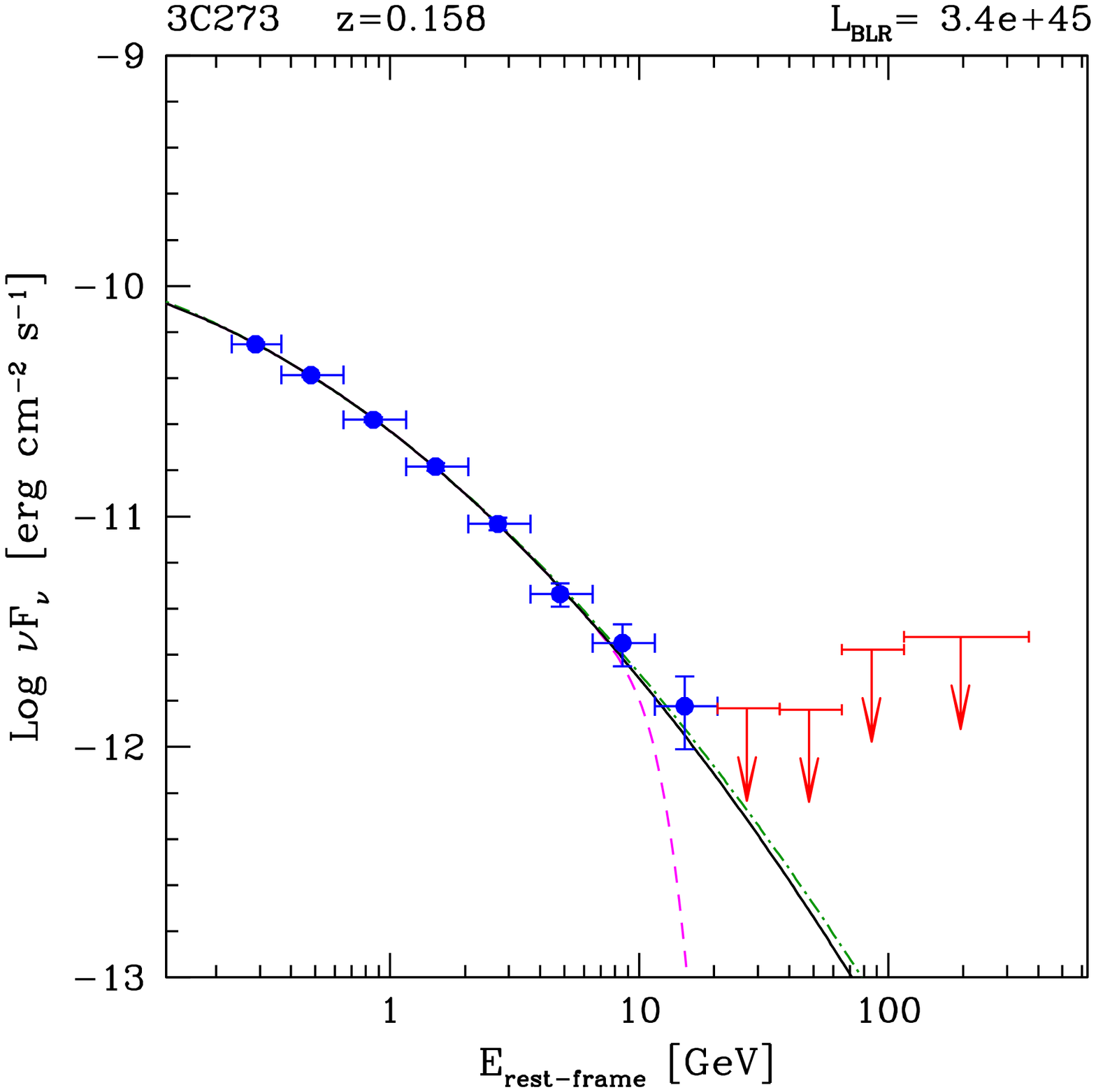,width=4.3cm,height=3.2cm }     
&\psfig{file=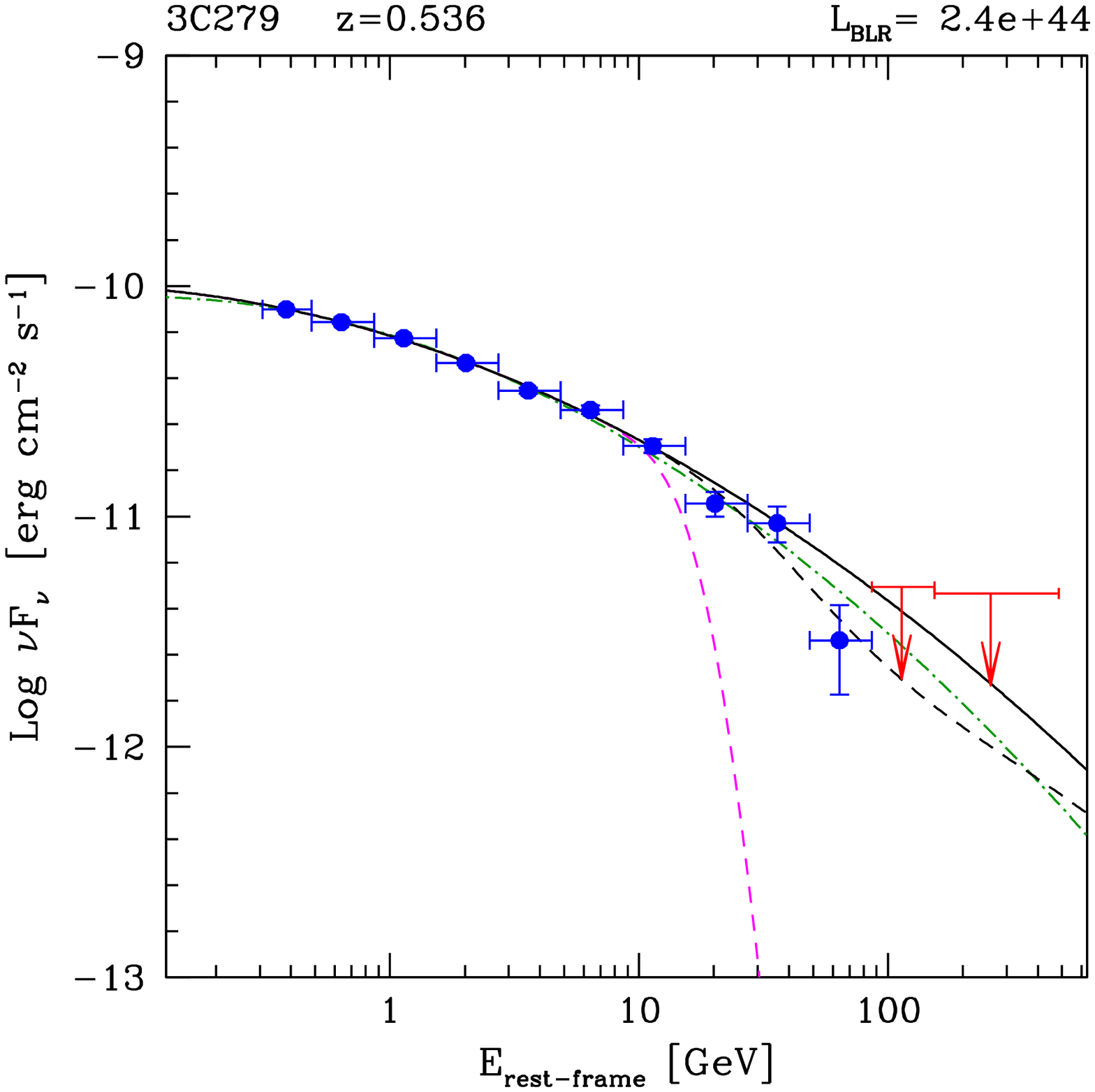,width=4.3cm,height=3.2cm } 
&\psfig{file=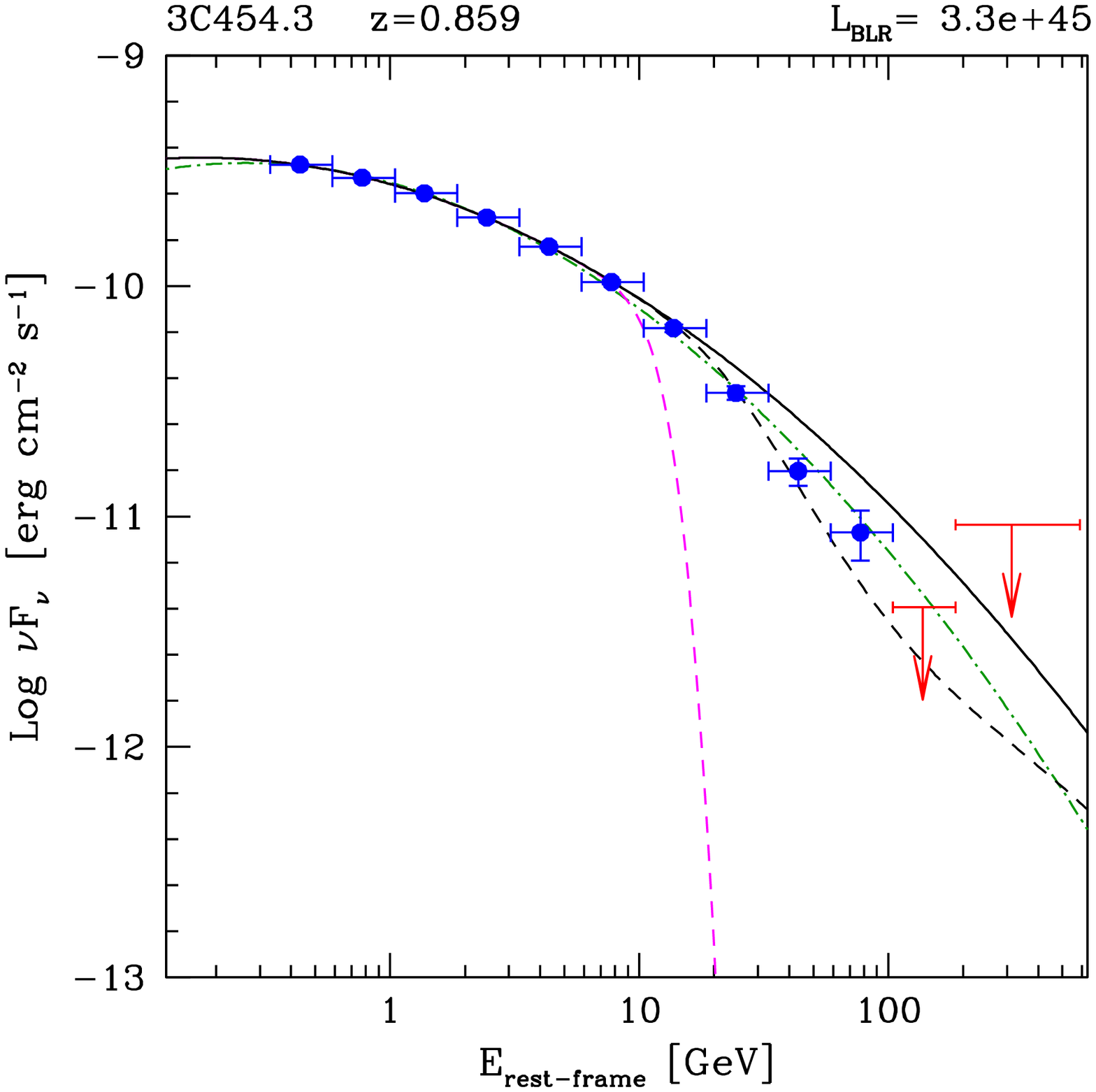,width=4.3cm,height=3.2cm }   
&\psfig{file=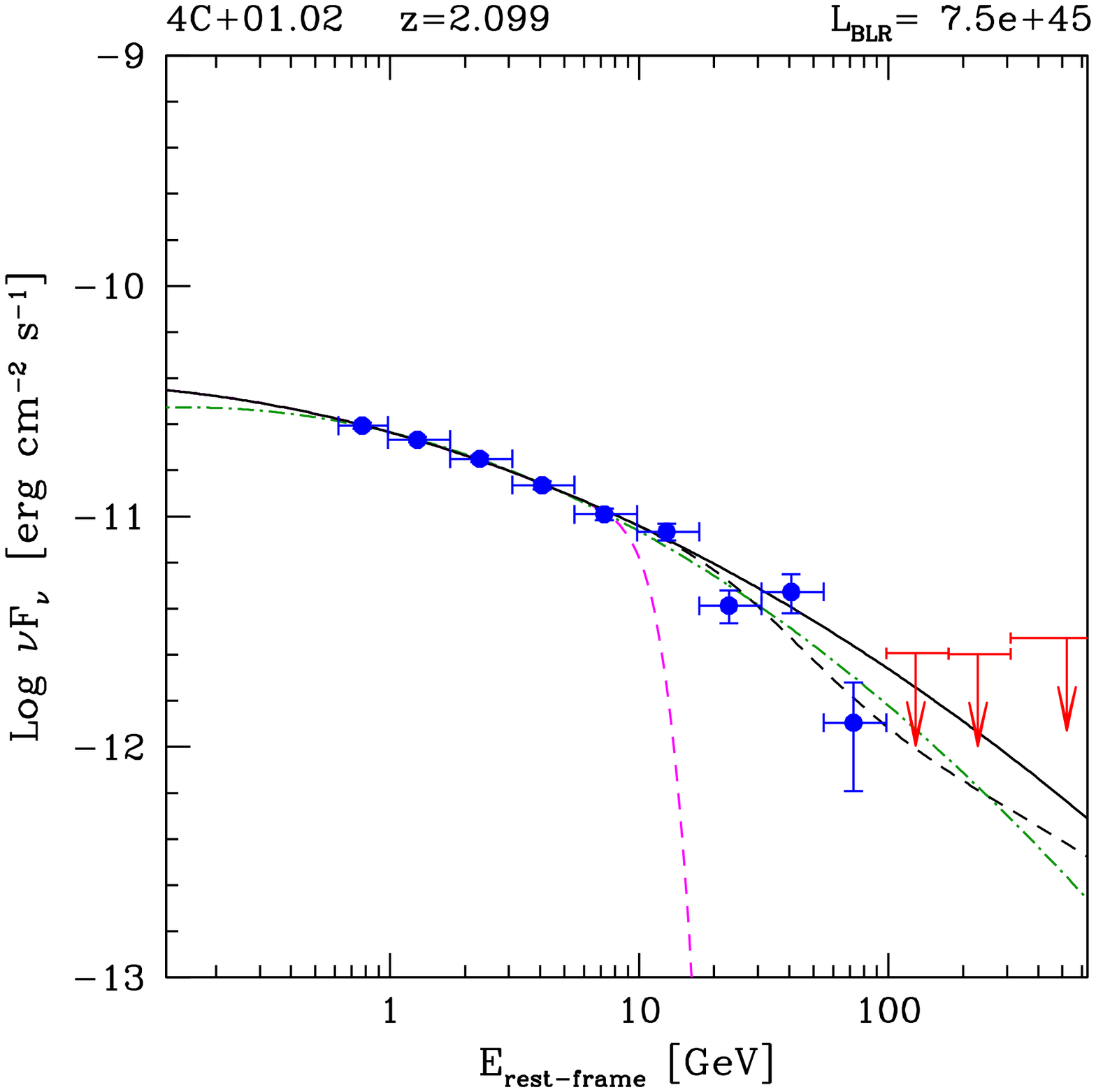,width=4.3cm,height=3.2cm } \vspace{1.2cm}\\
 \psfig{file=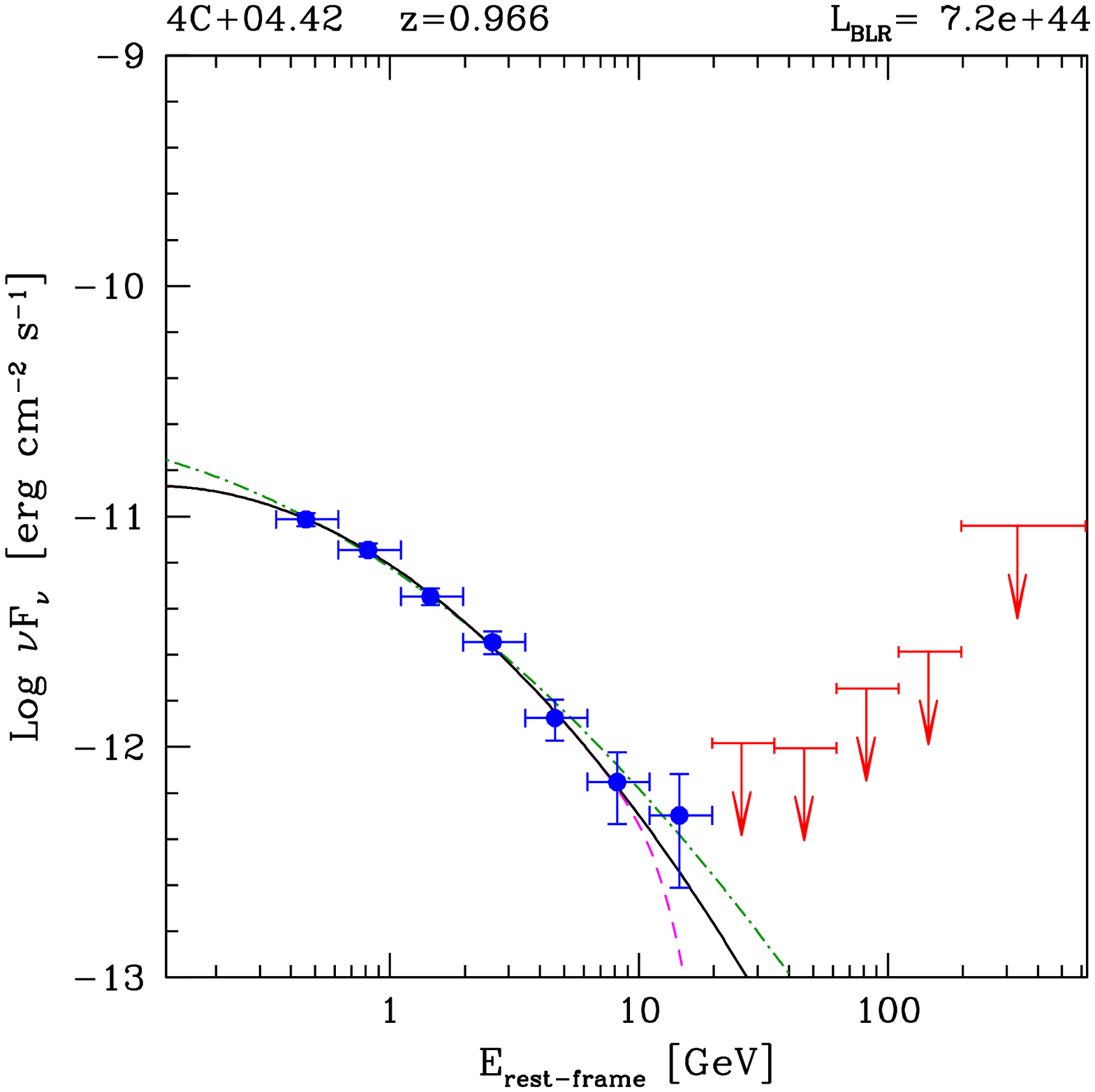,width=4.3cm,height=3.2cm }   
&\psfig{file=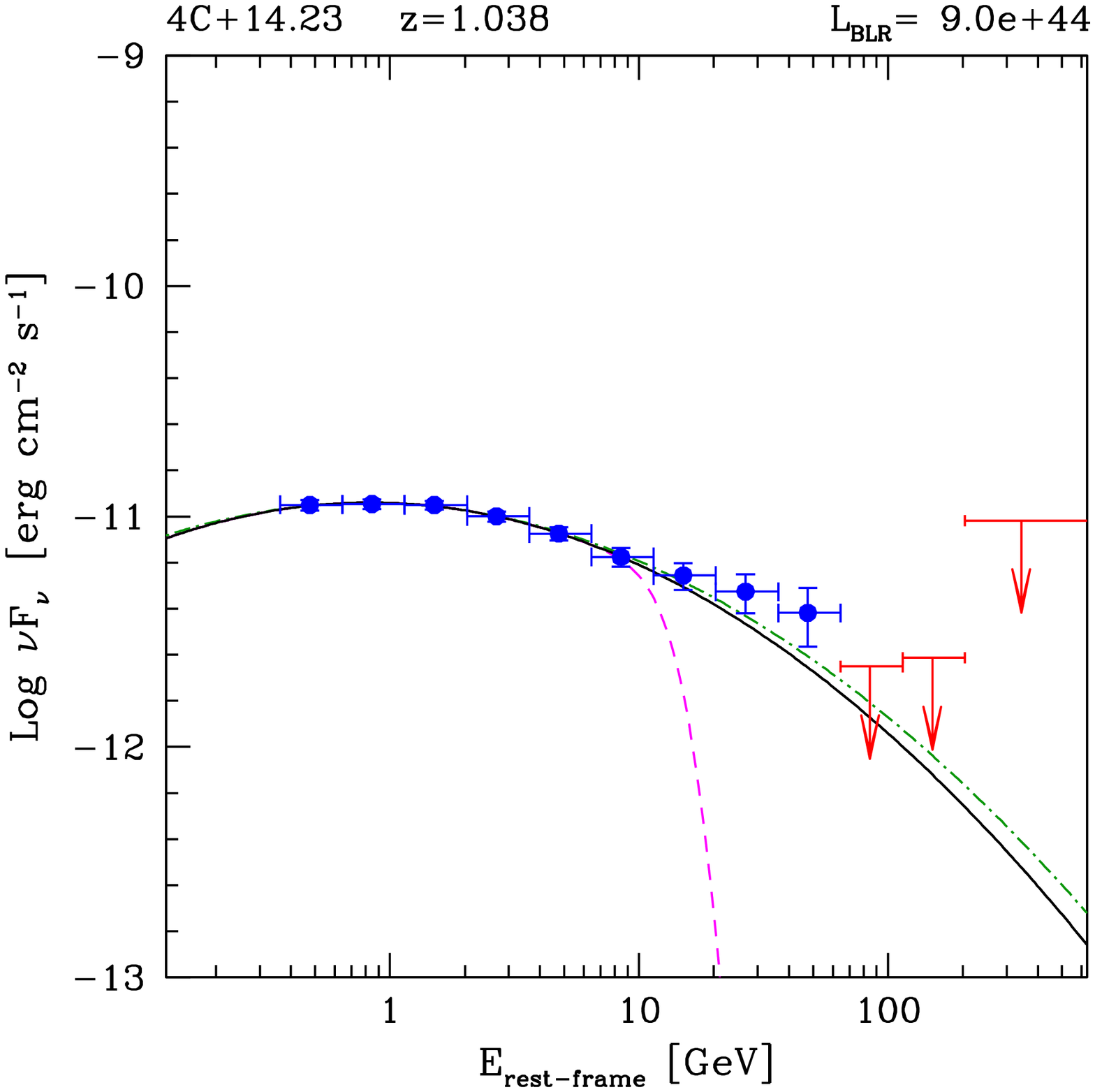,width=4.3cm,height=3.2cm } 
&\psfig{file=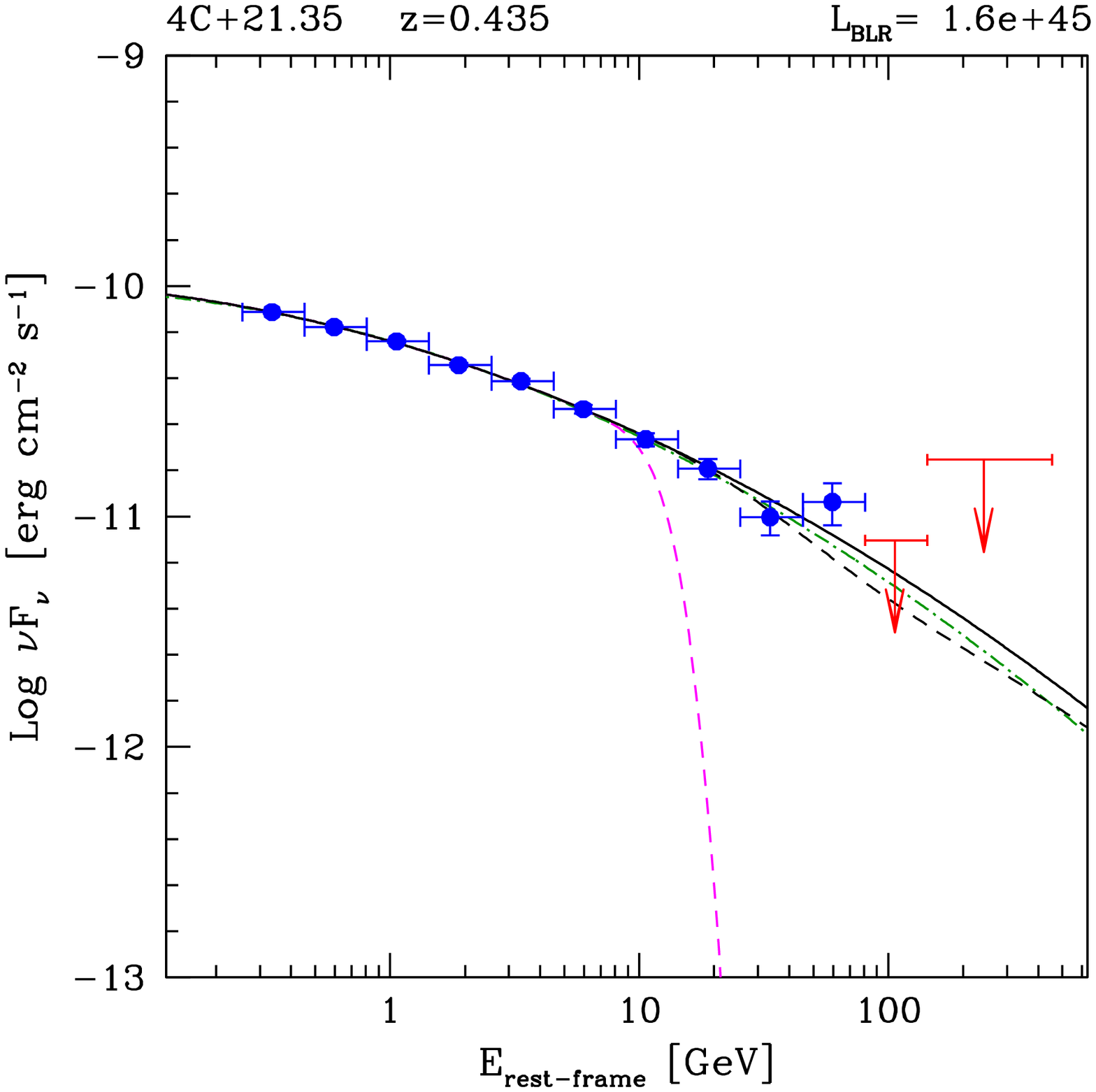,width=4.3cm,height=3.2cm }   
&\psfig{file=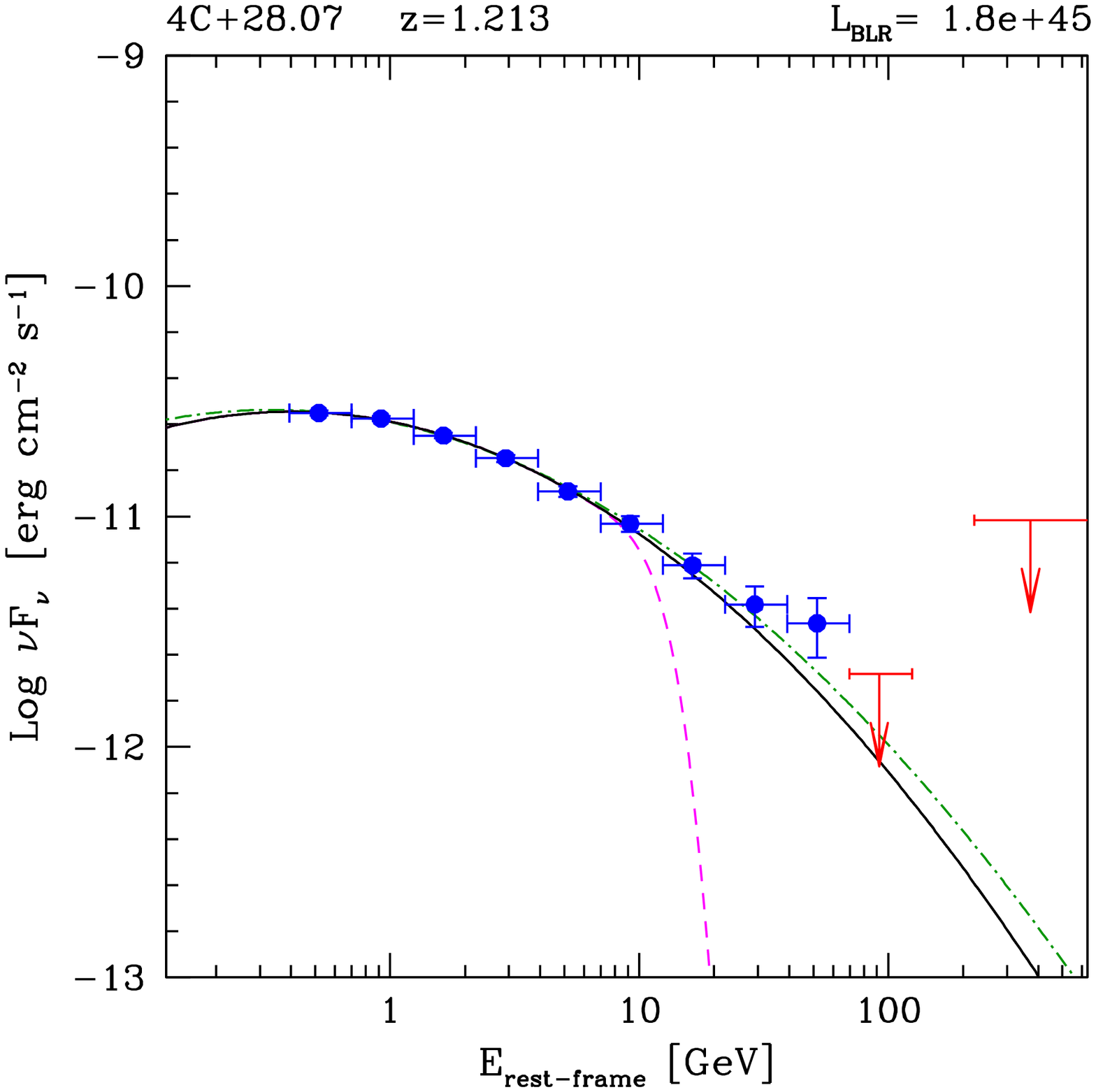,width=4.3cm,height=3.2cm } \vspace{1.2cm}\\
 \psfig{file=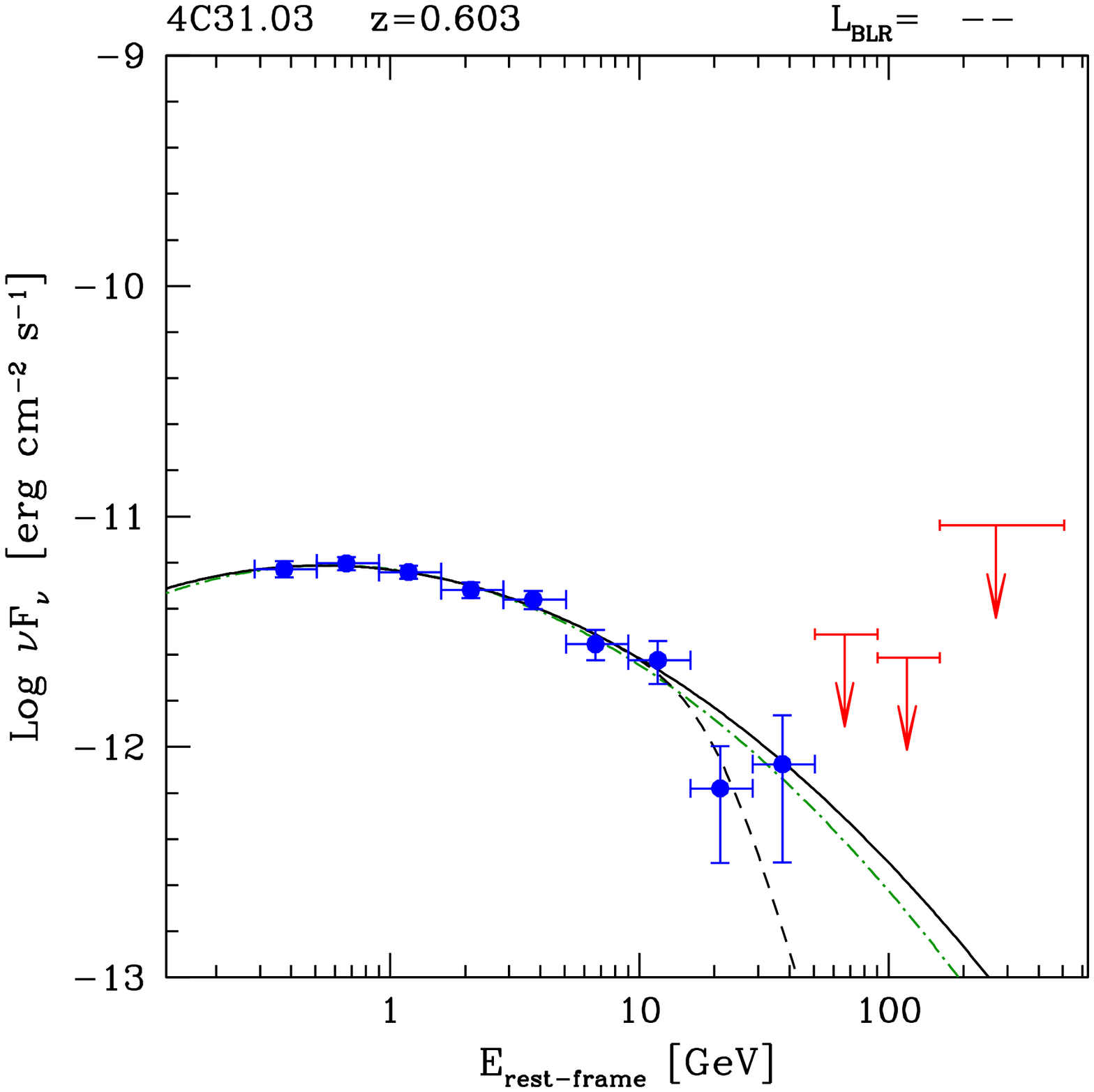,width=4.3cm,height=3.2cm }   
&\psfig{file=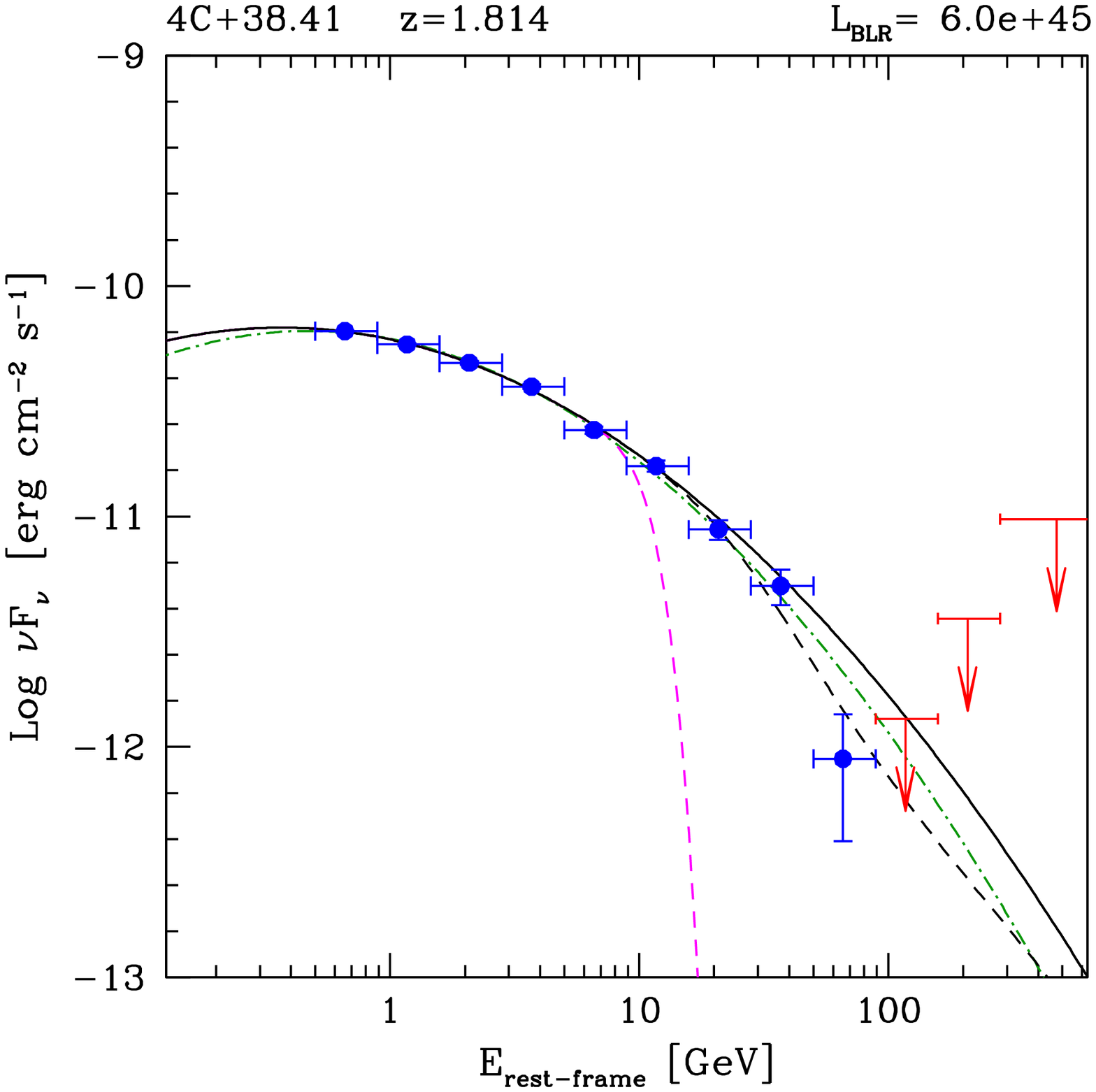,width=4.3cm,height=3.2cm } 
&\psfig{file=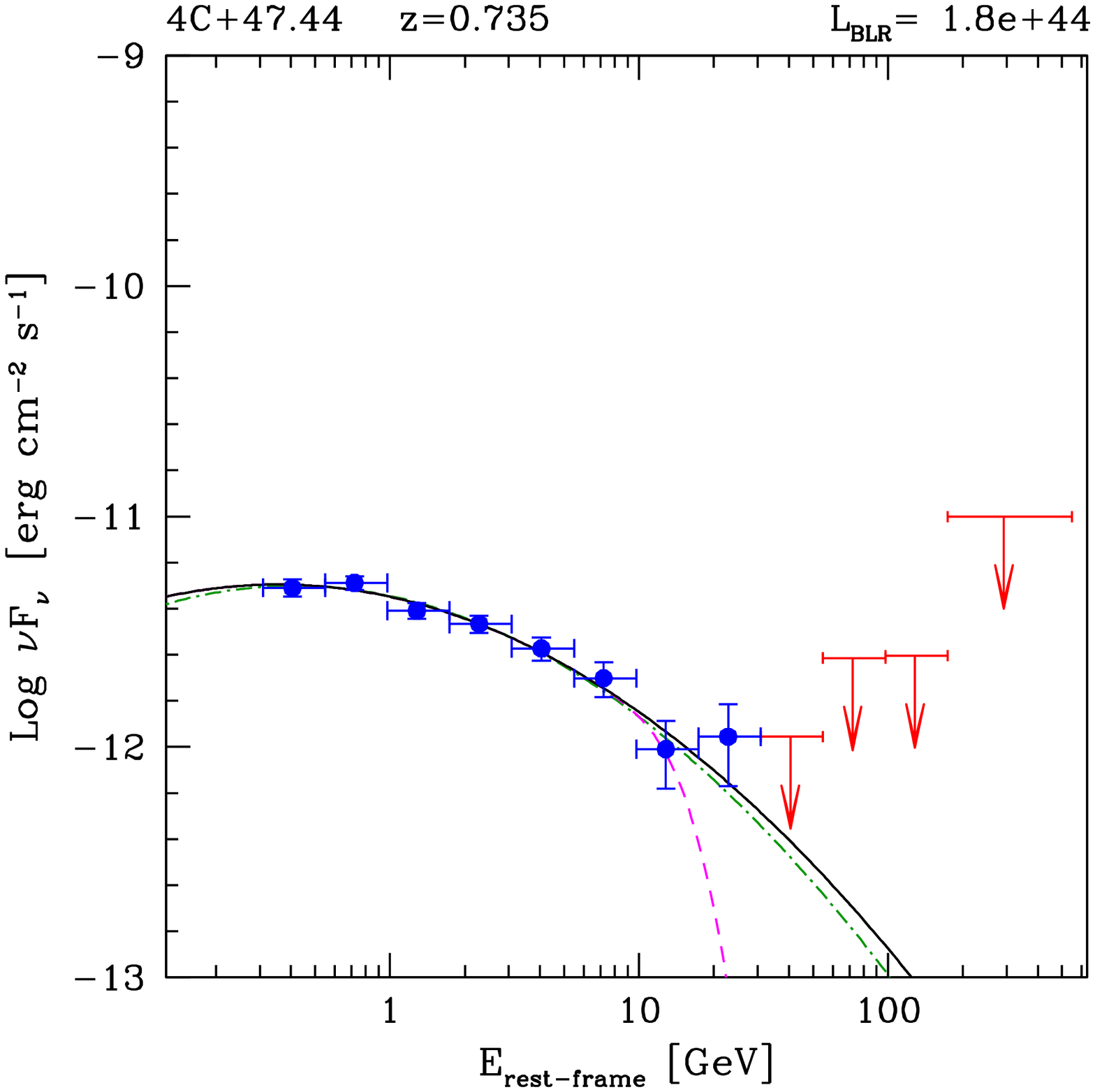,width=4.3cm,height=3.2cm }   
&\psfig{file=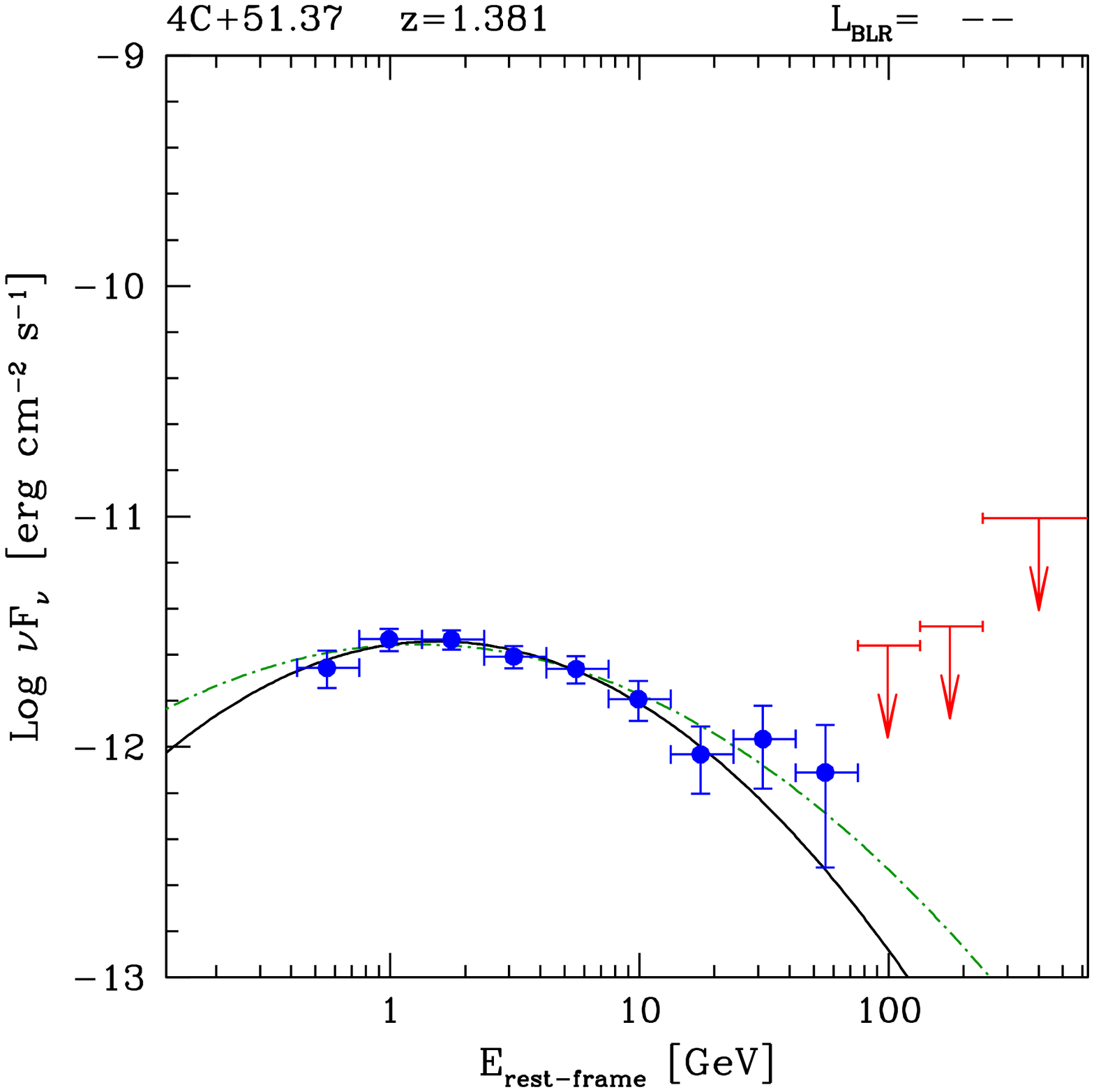,width=4.3cm,height=3.2cm } \vspace{1.2cm}\\
 \psfig{file=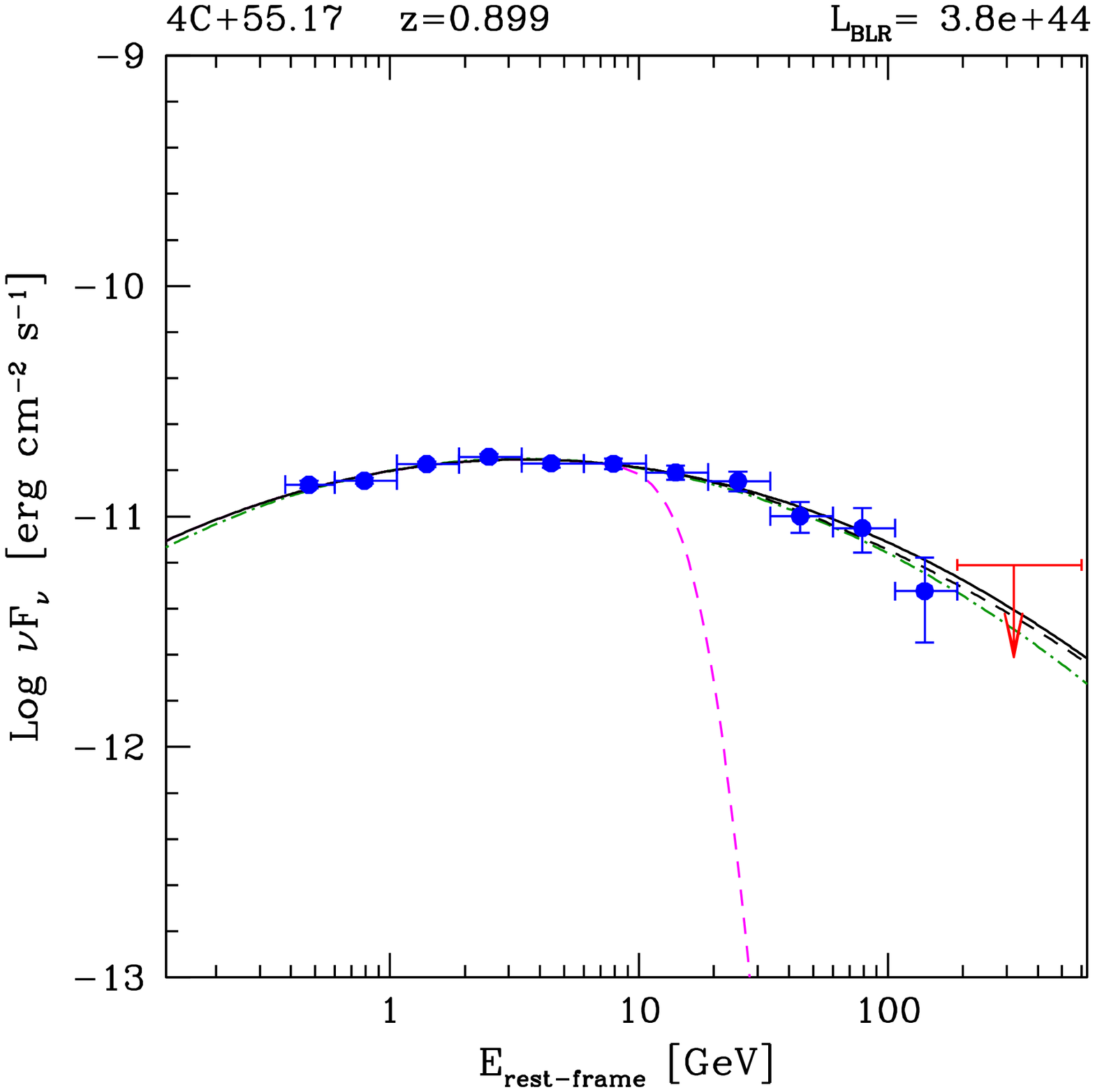,width=4.3cm,height=3.2cm }   
&\psfig{file=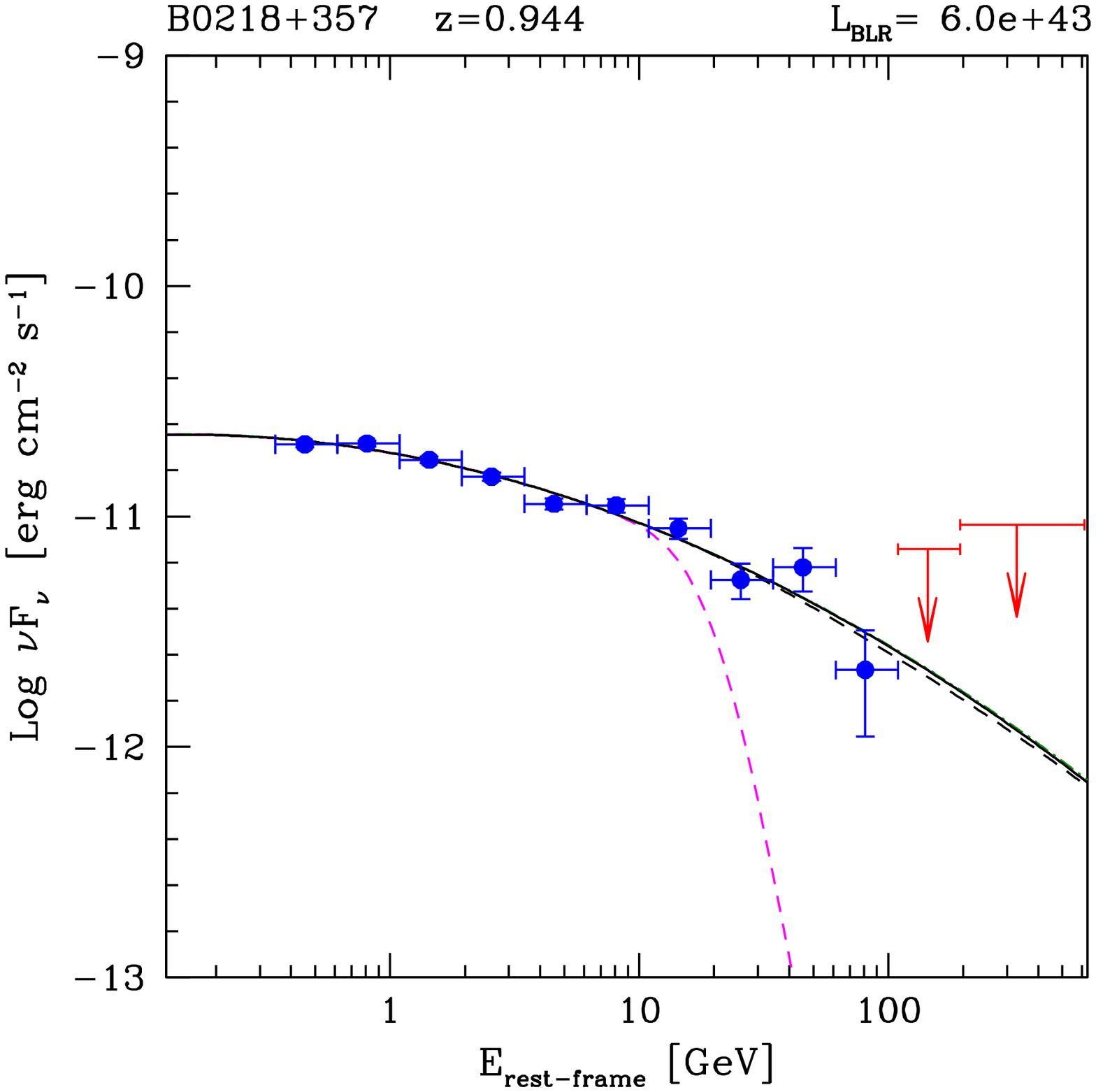,width=4.3cm,height=3.2cm } 
&\psfig{file=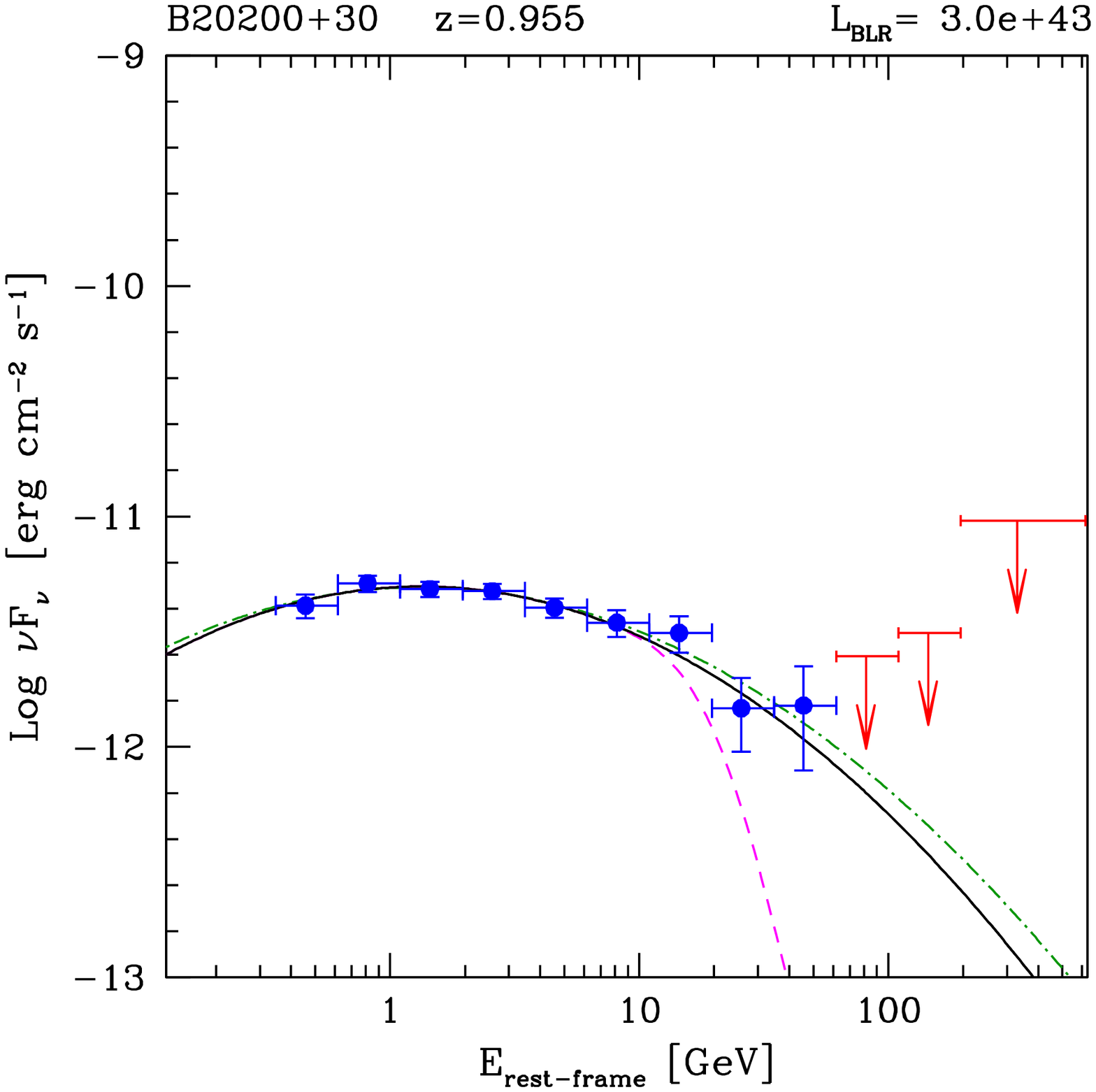,width=4.3cm,height=3.2cm }  
&\psfig{file=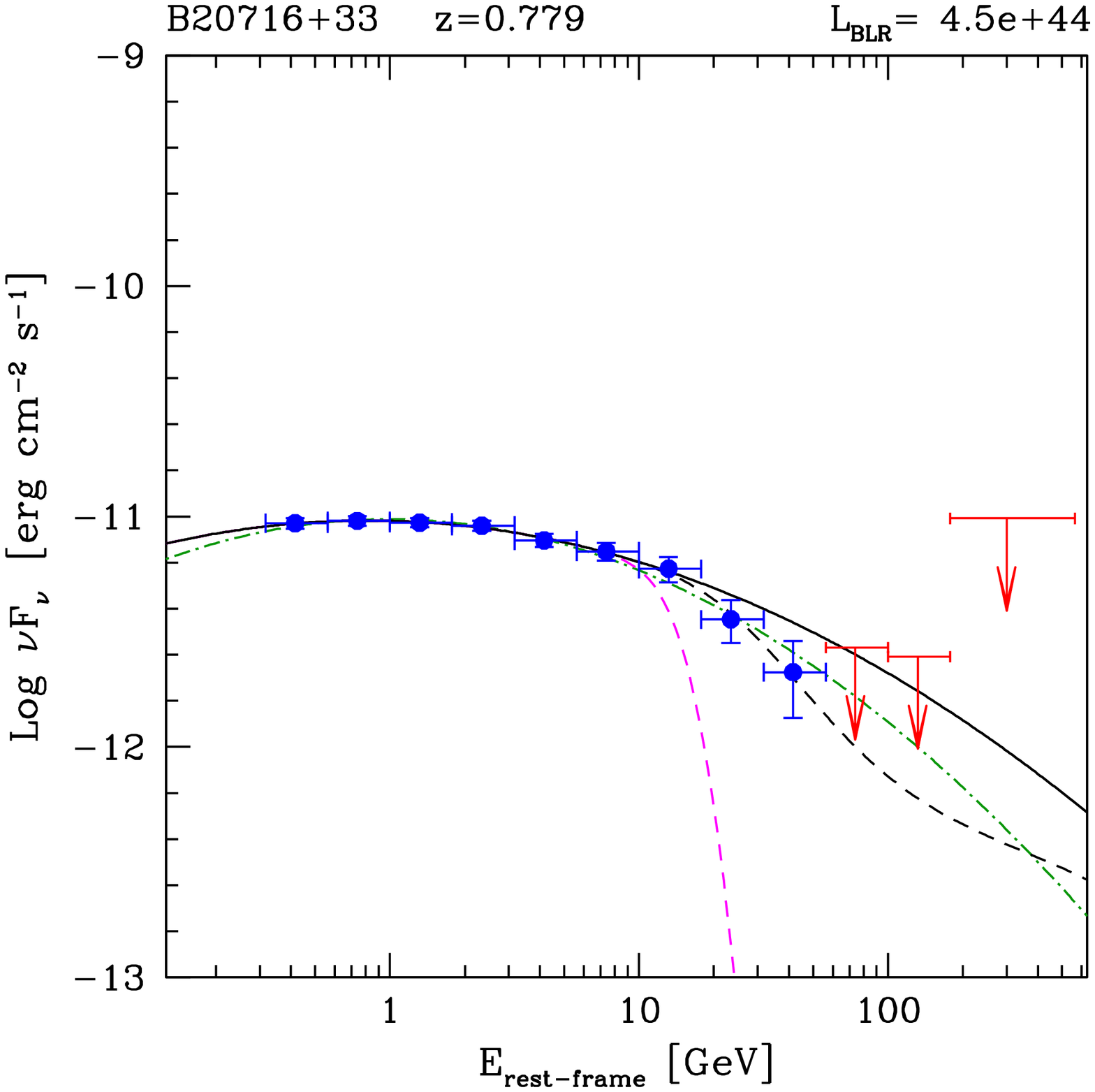,width=4.3cm,height=3.2cm }\vspace{1.2cm} \\
 \psfig{file=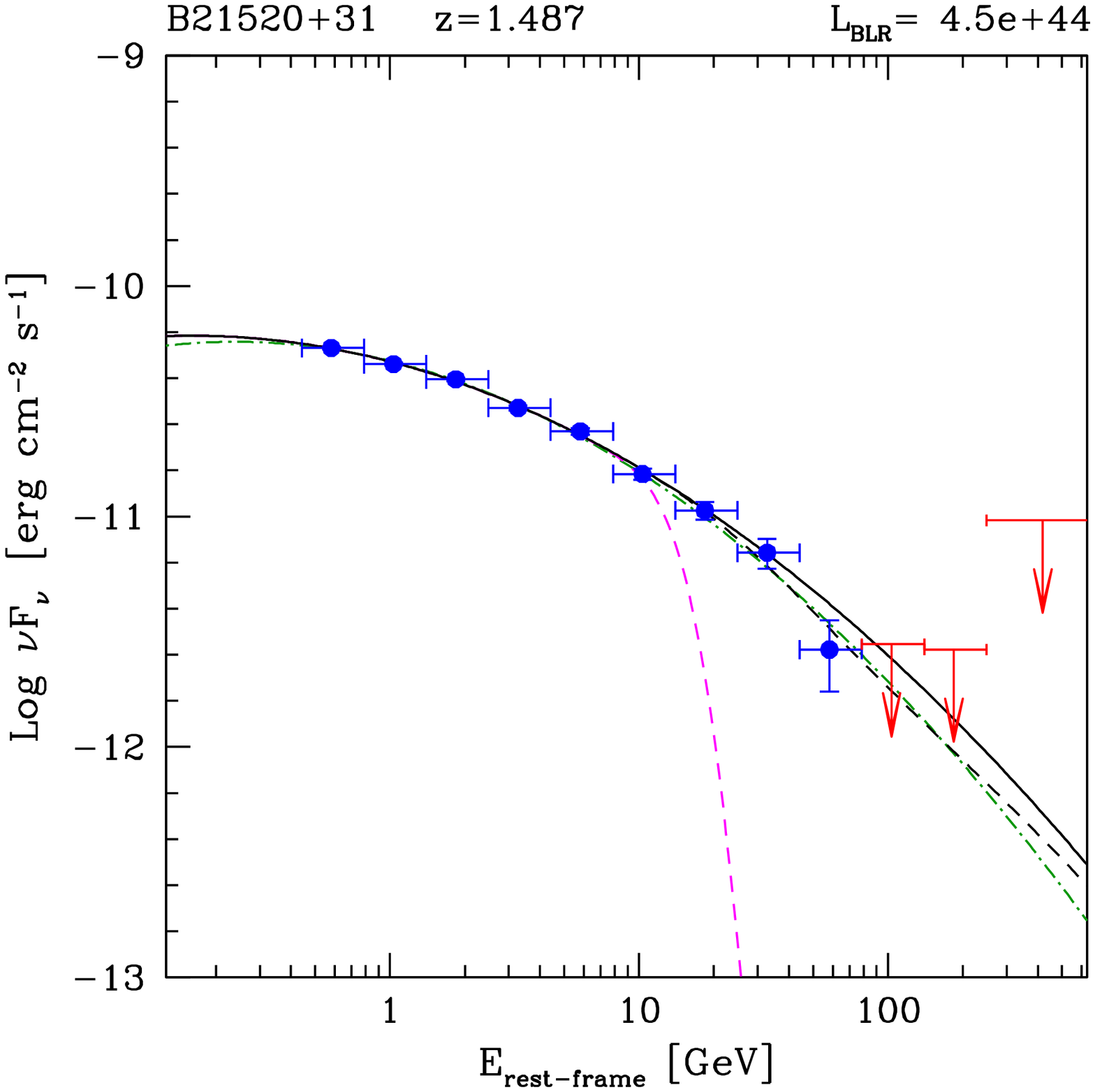,width=4.3cm,height=3.2cm }  
&\psfig{file=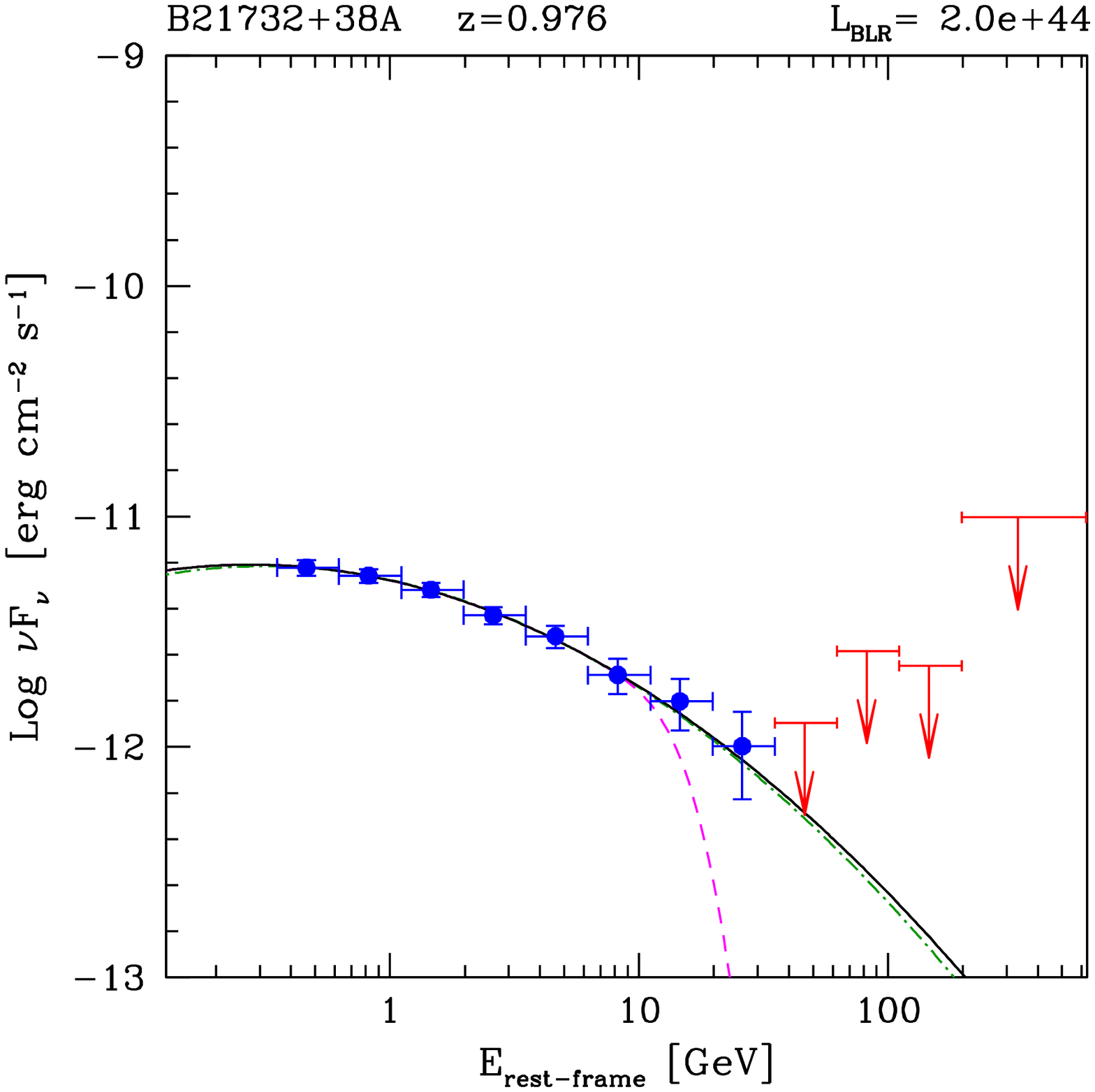,width=4.3cm,height=3.2cm } 
&\psfig{file=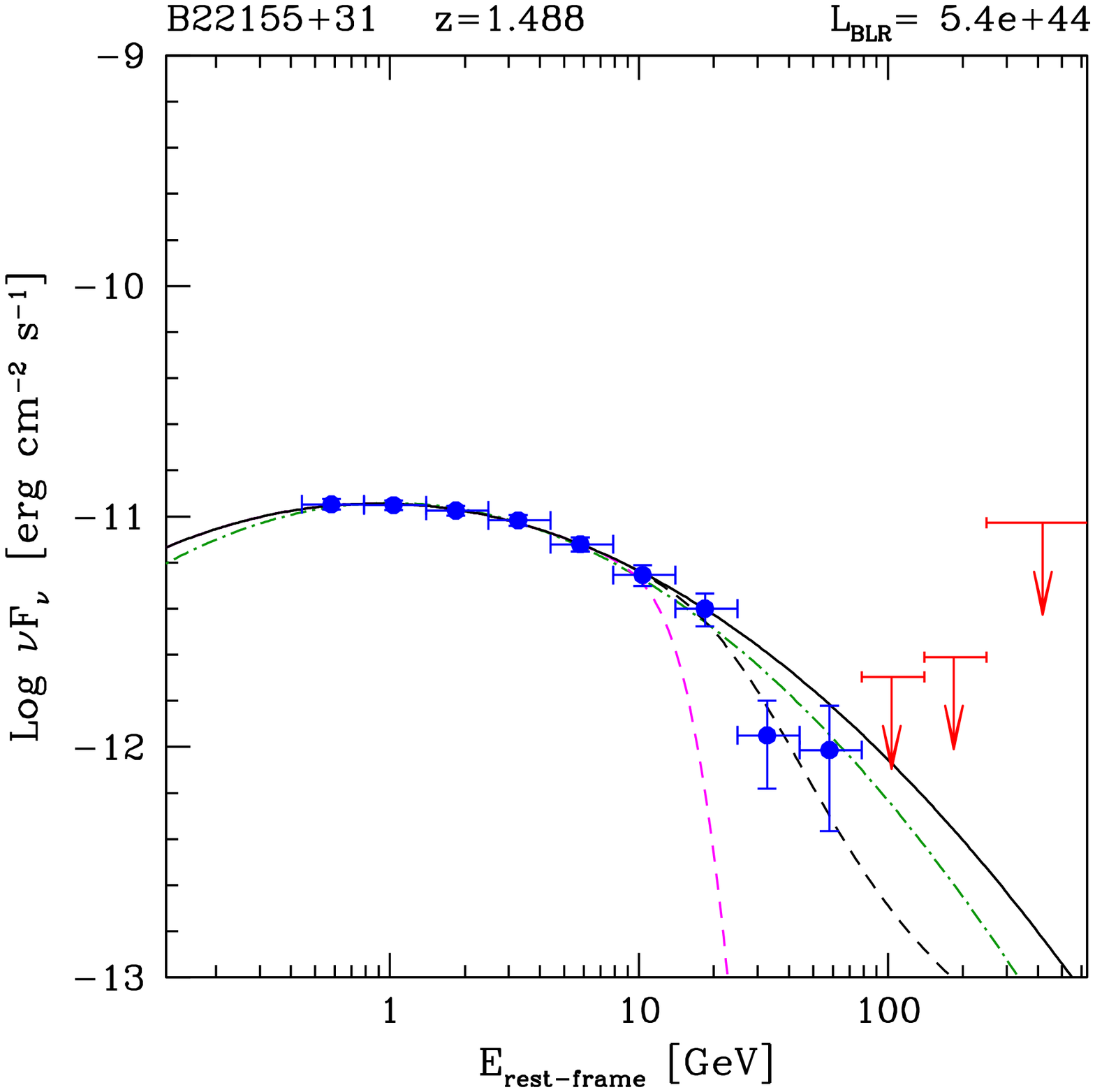,width=4.3cm,height=3.2cm }  
&\psfig{file=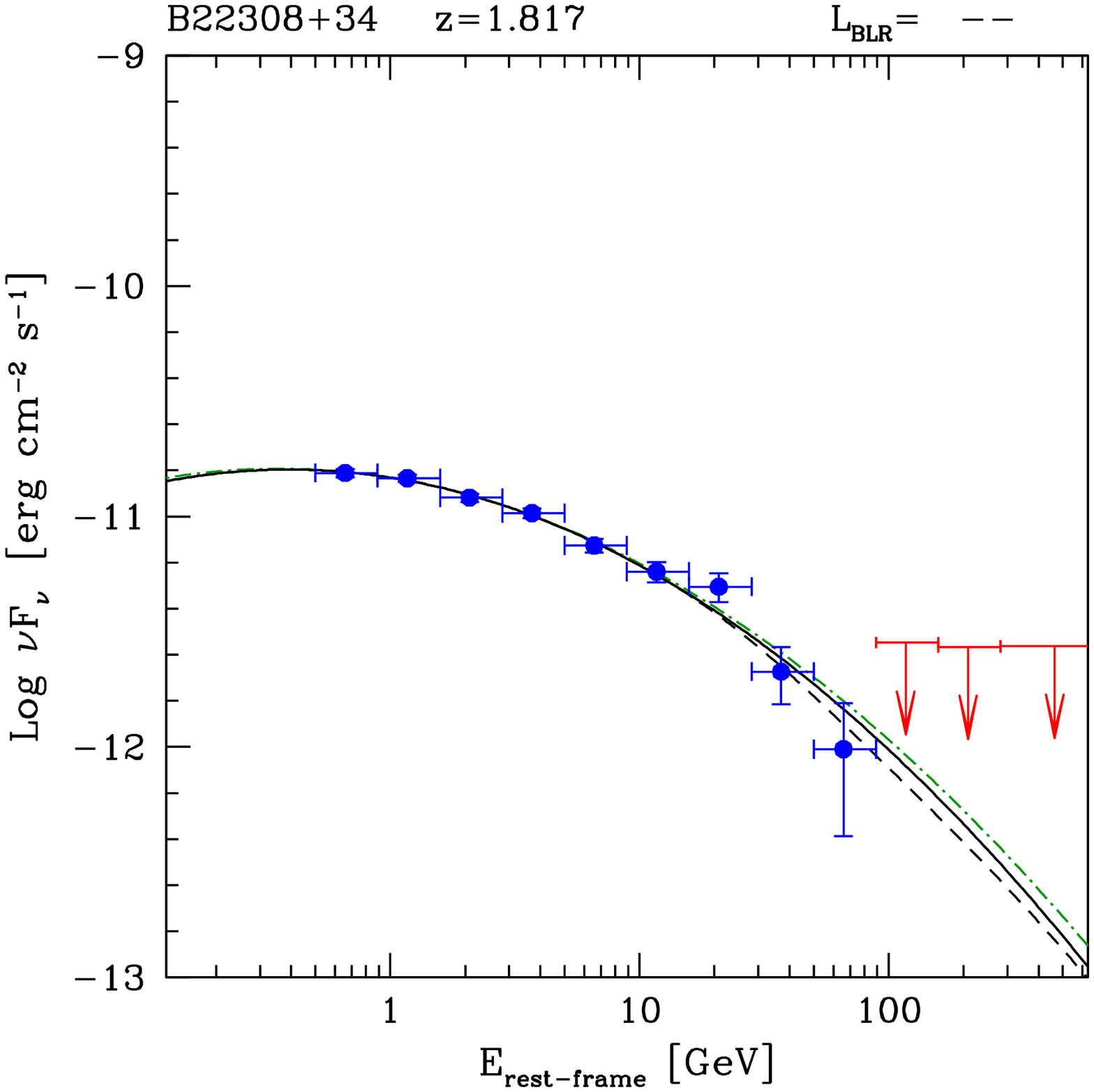,width=4.3cm,height=3.2cm } 
%
\end{tabular}
\caption{
Average {\it Fermi}-LAT spectra (SED) plotted in the source rest frame energy, listed in alphabetical order. 
Upper limits are at the $2\sigma$ level (see text).
Solid lines show the fit of the spectrum  with parameters determined below 13 GeV.
Dashed lines shows the spectrum with BLR absorption fitted to the data, using the solid-line spectrum 
as intrinsic model.  When $L_{\rm BLR}$ is available, the magenta dashed line shows the expected absorbed spectrum 
if the emitting region is located at $R_{\rm diss}=R_{\rm BLR}$/2.  
Dot-dashed (green) lines shows the fit of a pure log-parabolic model to the whole spectrum.}  
\label{spectra1}
\end{figure*} 


\begin{figure*}
\vspace{1cm}
\begin{tabular}{cccc}
 \psfig{file=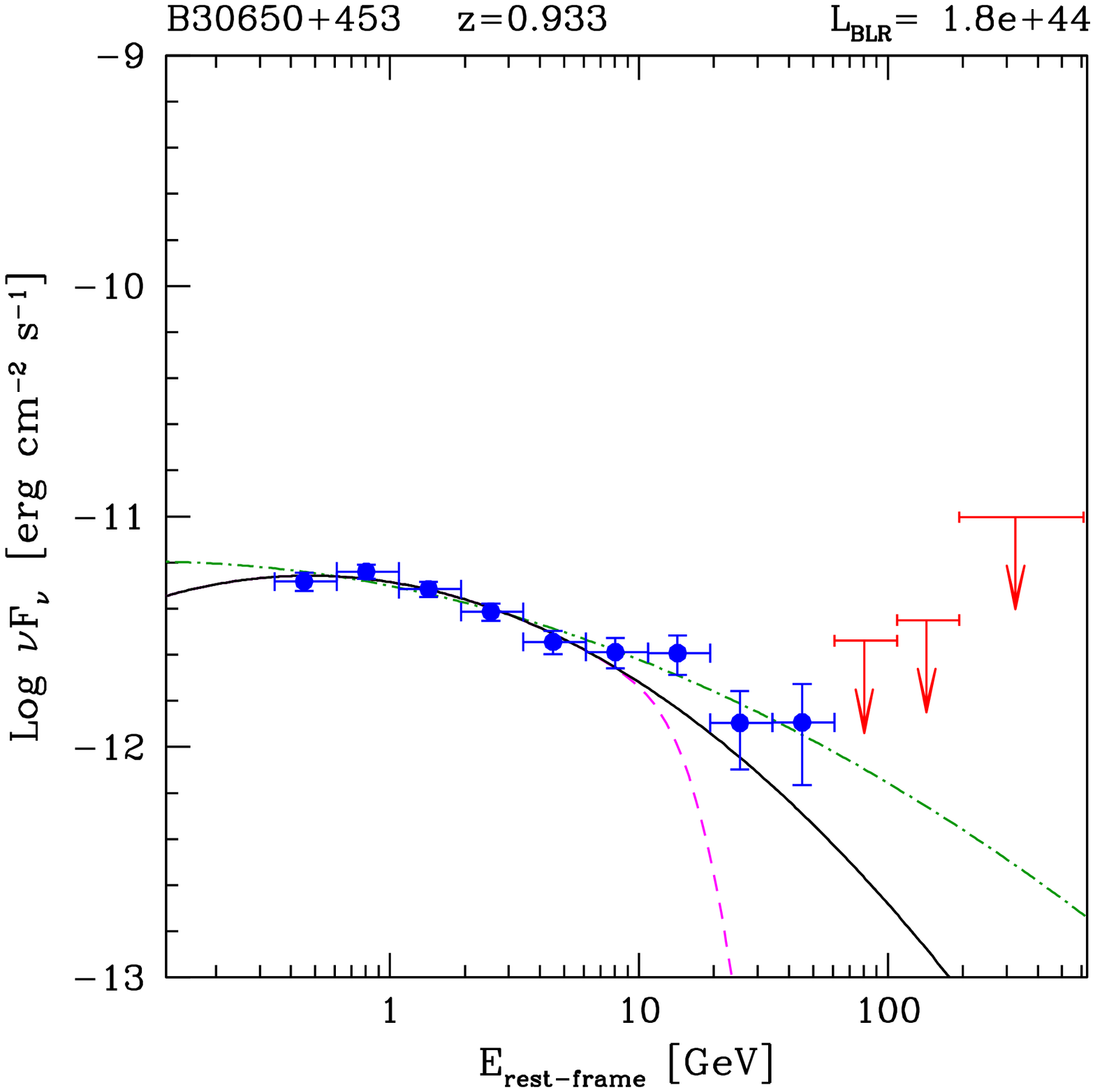,width=4.3cm,height=3.2cm } 
&\psfig{file=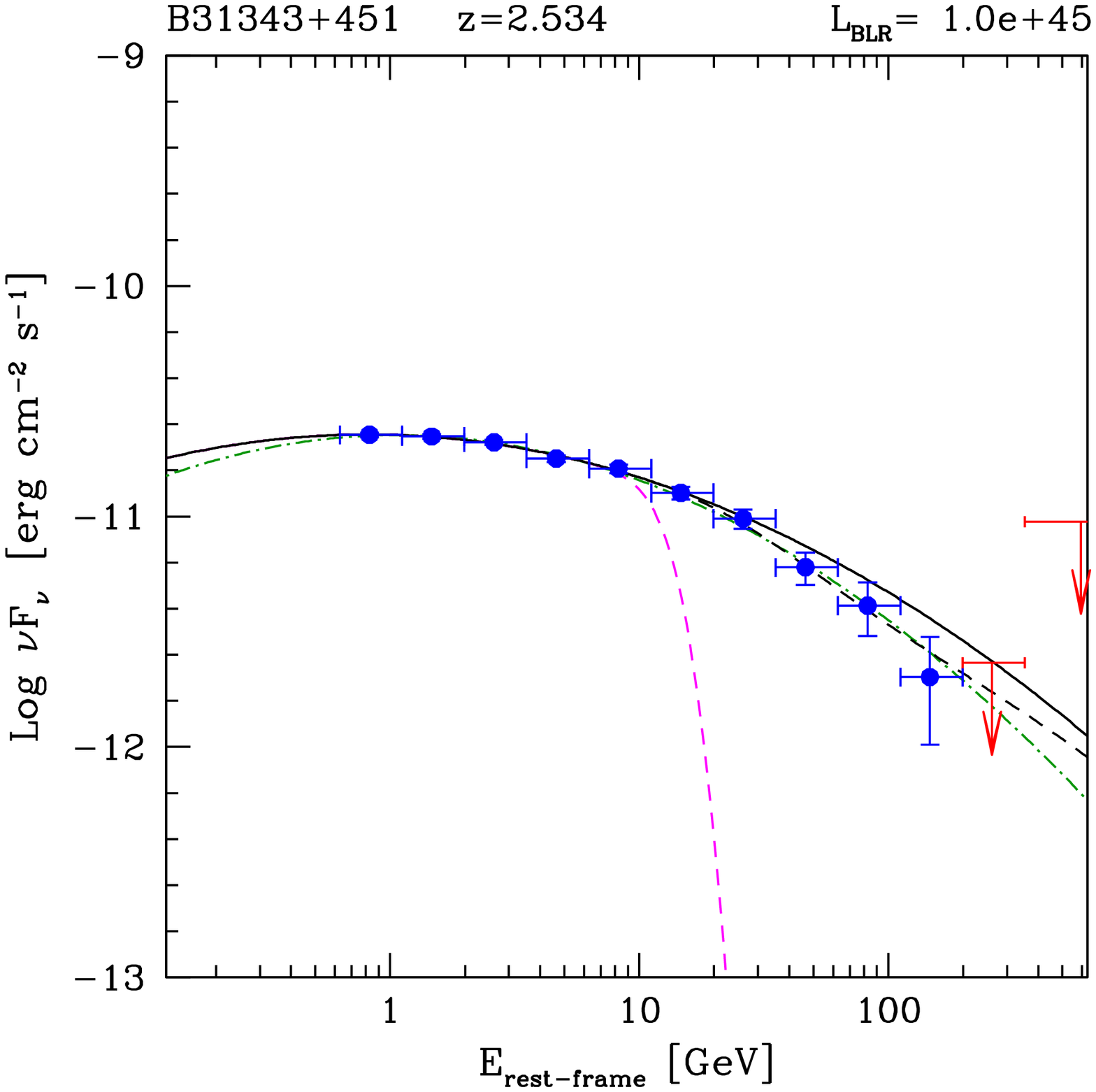,width=4.3cm,height=3.2cm } 
&\psfig{file=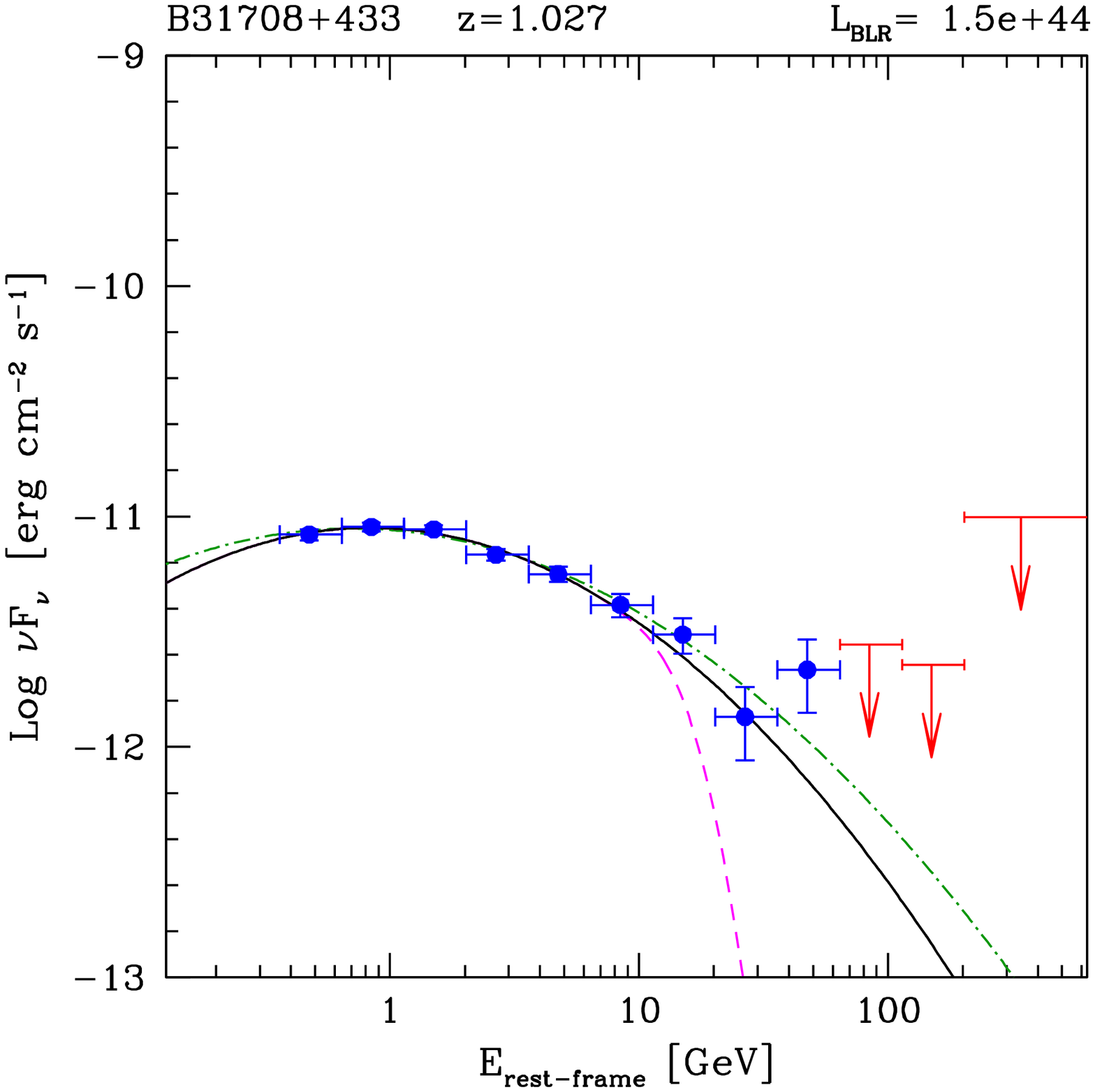,width=4.3cm,height=3.2cm } 
&\psfig{file=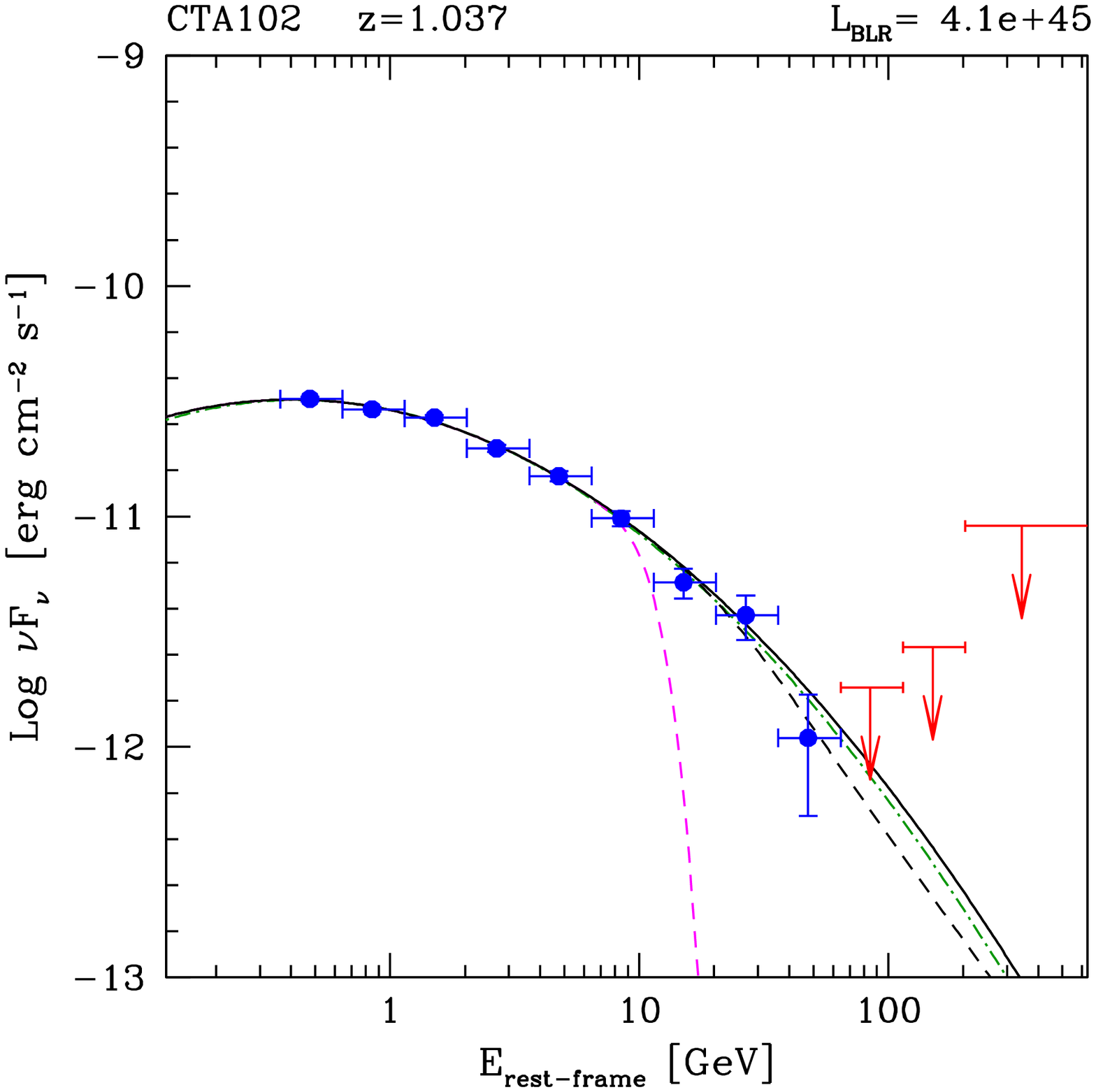,width=4.3cm,height=3.2cm }   \vspace{1.2cm} \\
 \psfig{file=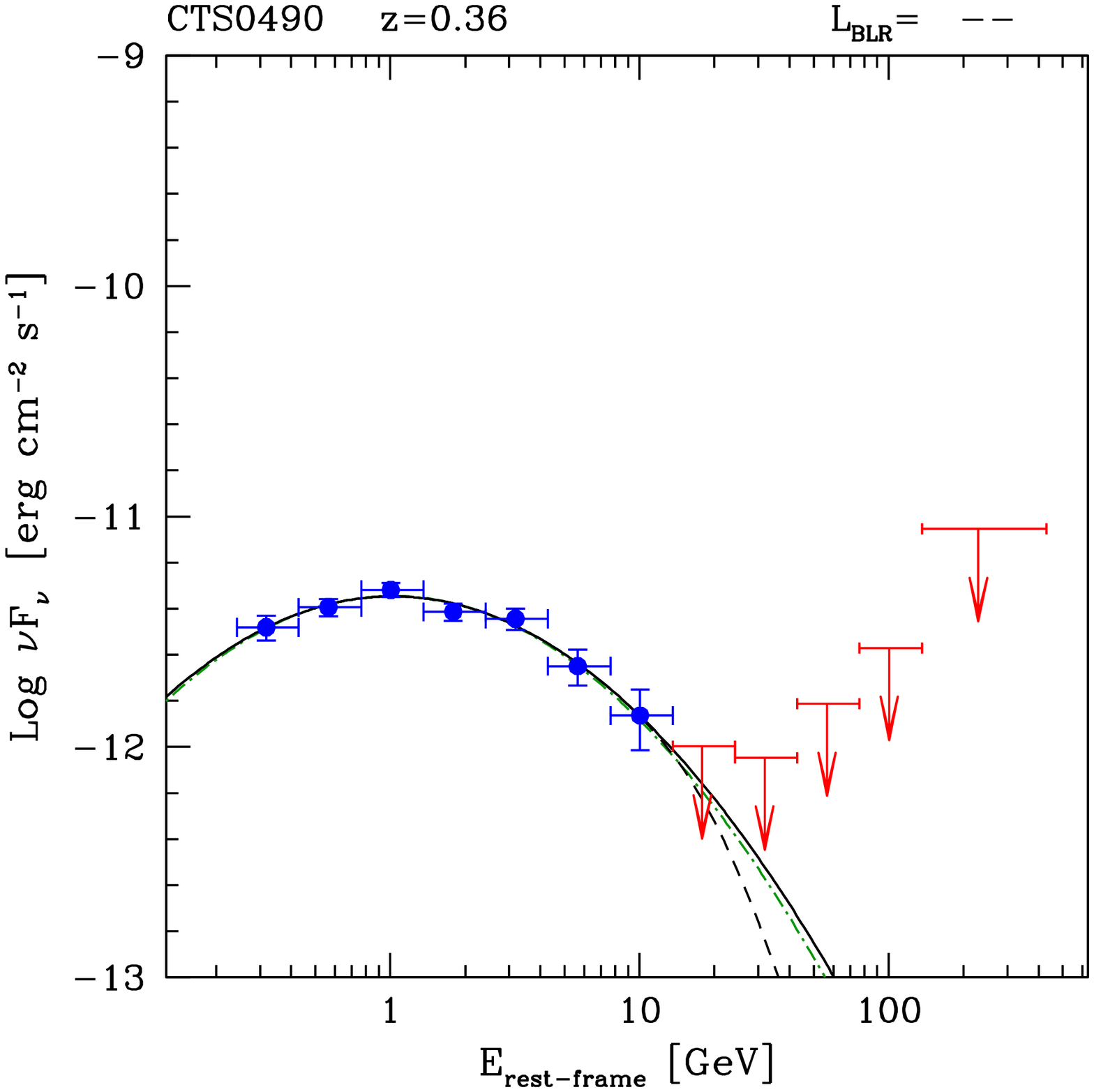,width=4.3cm,height=3.2cm } 
&\psfig{file=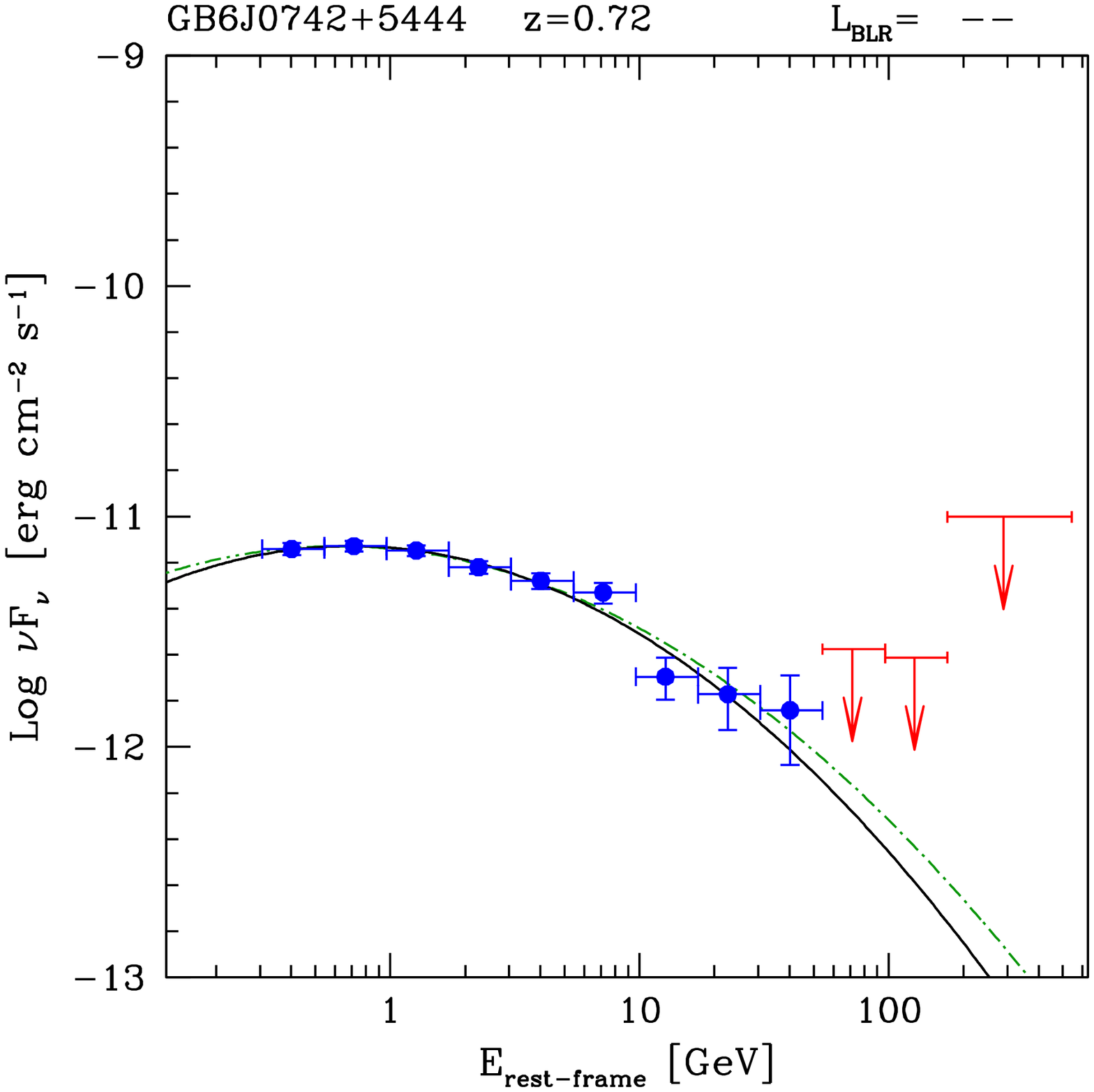,width=4.3cm,height=3.2cm }   
&\psfig{file=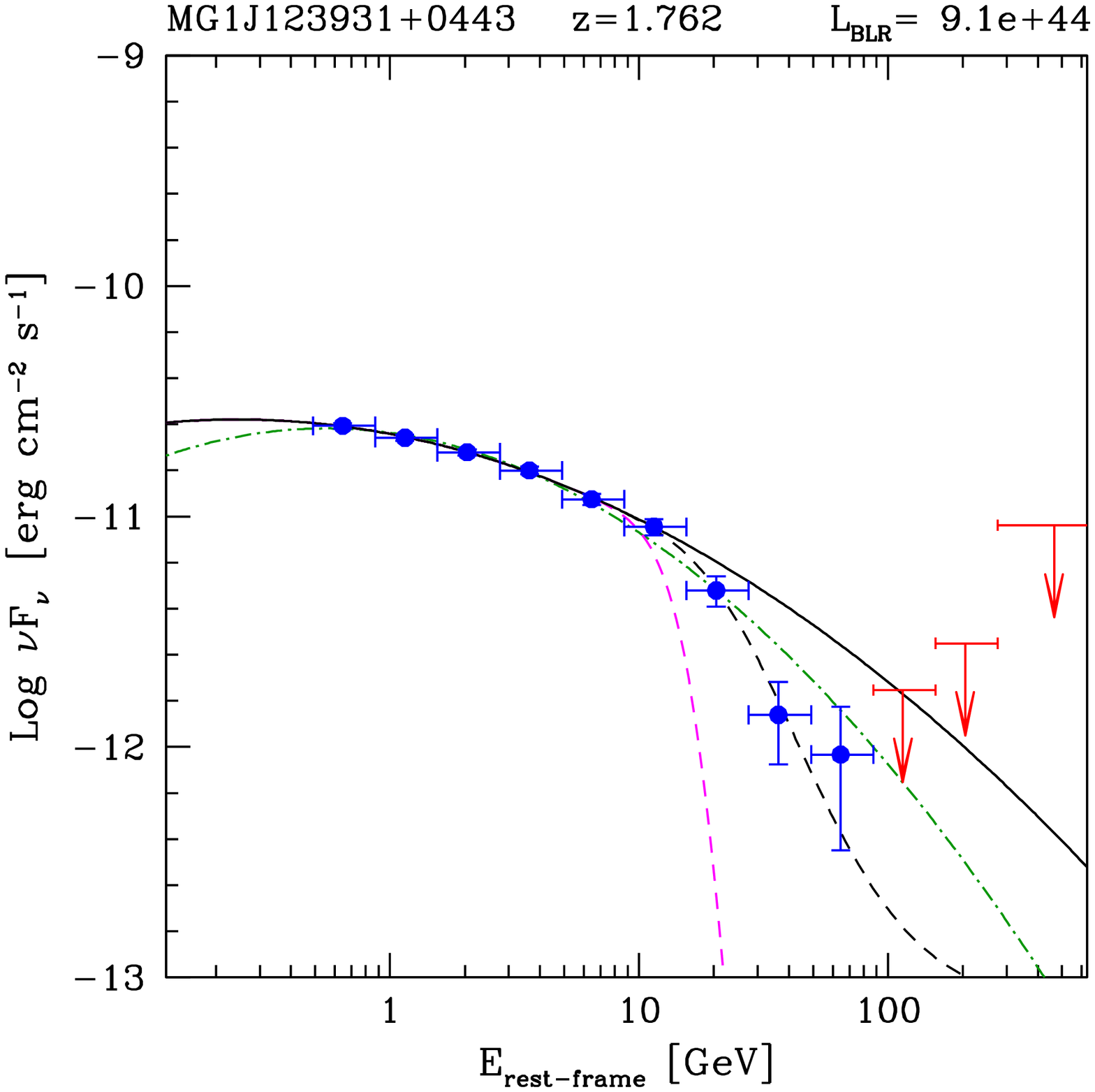,width=4.3cm,height=3.2cm }
&\psfig{file=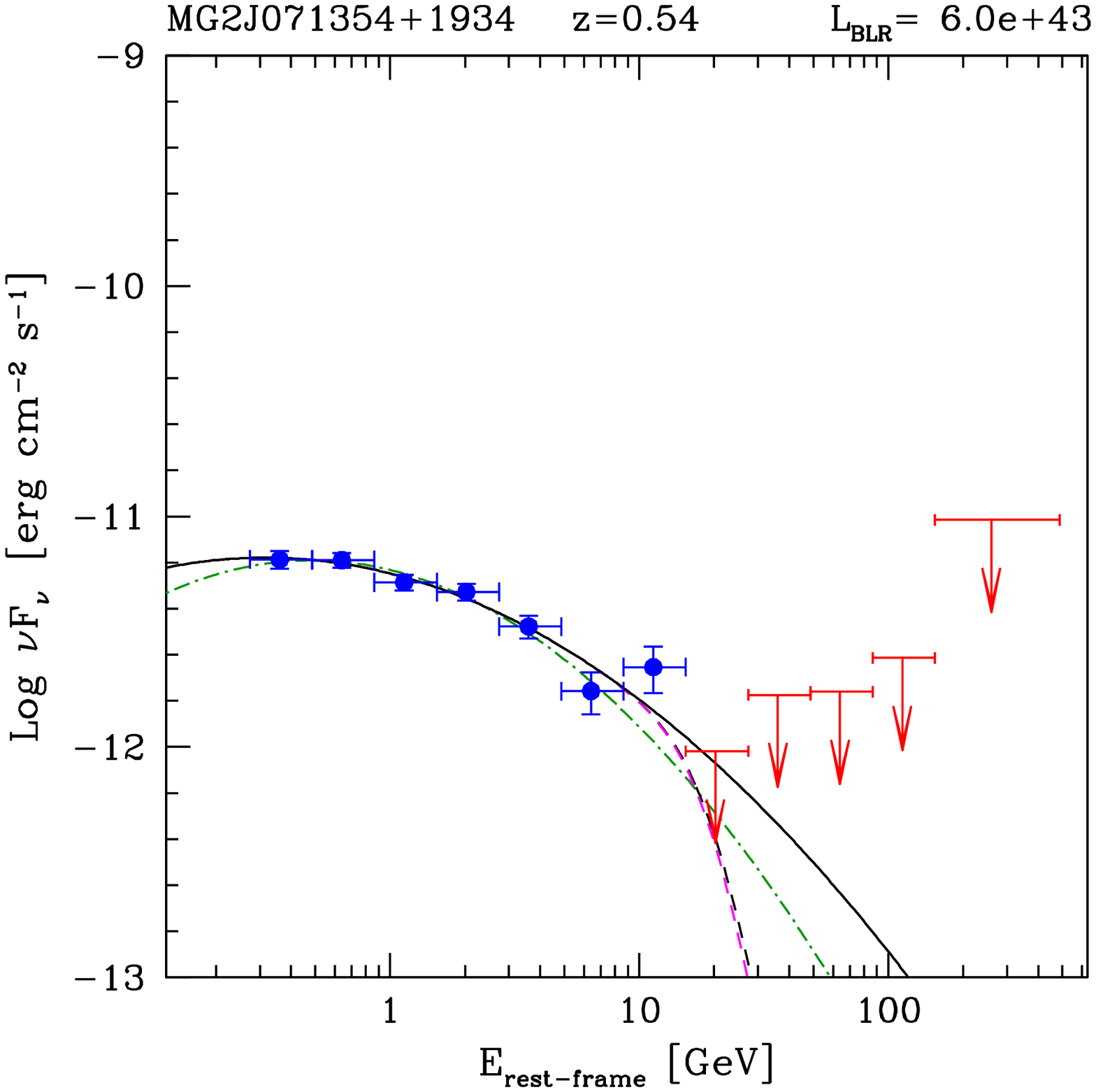,width=4.3cm,height=3.2cm }\vspace{1.2cm} \\
 \psfig{file=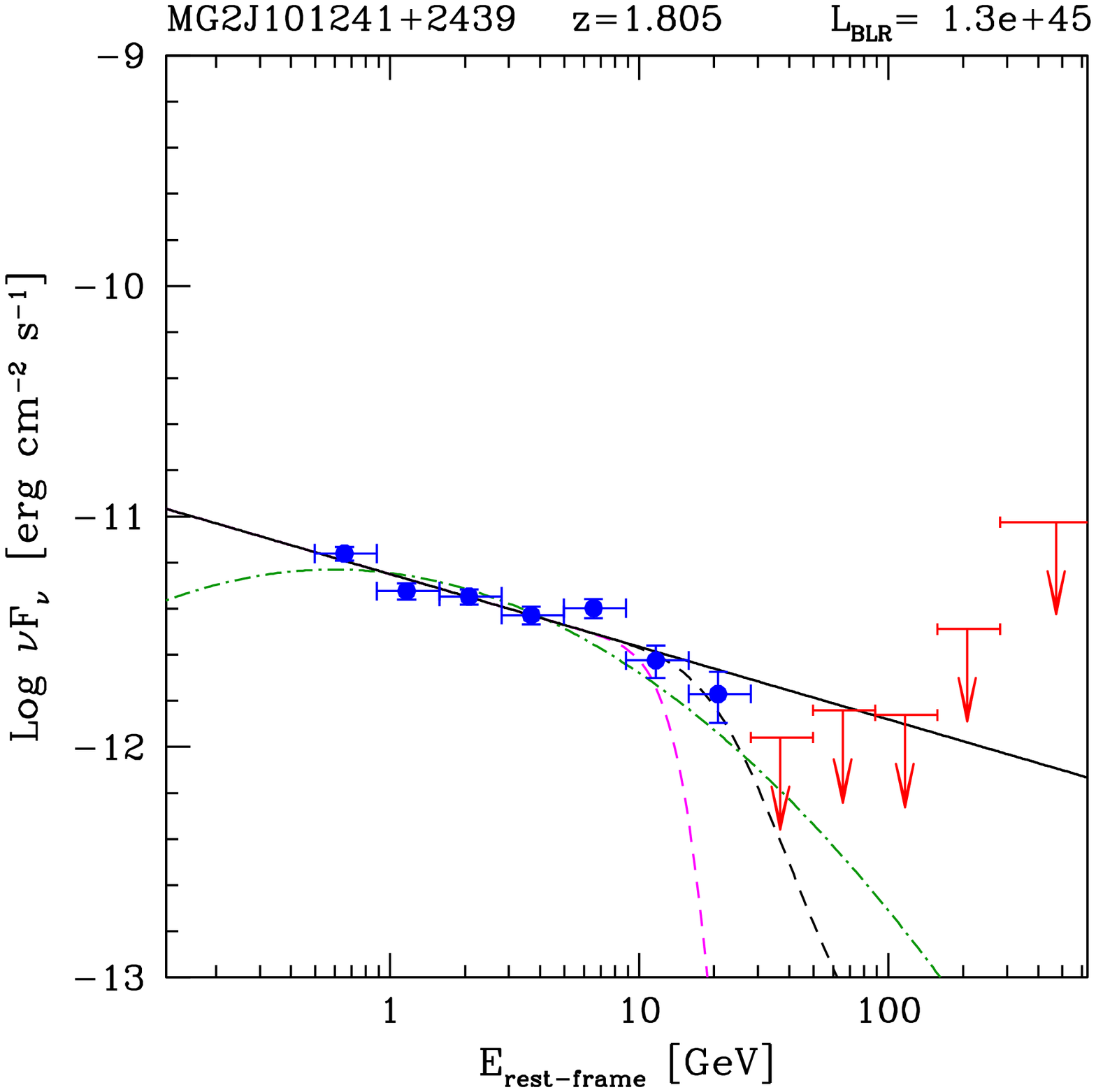,width=4.3cm,height=3.2cm } 
&\psfig{file=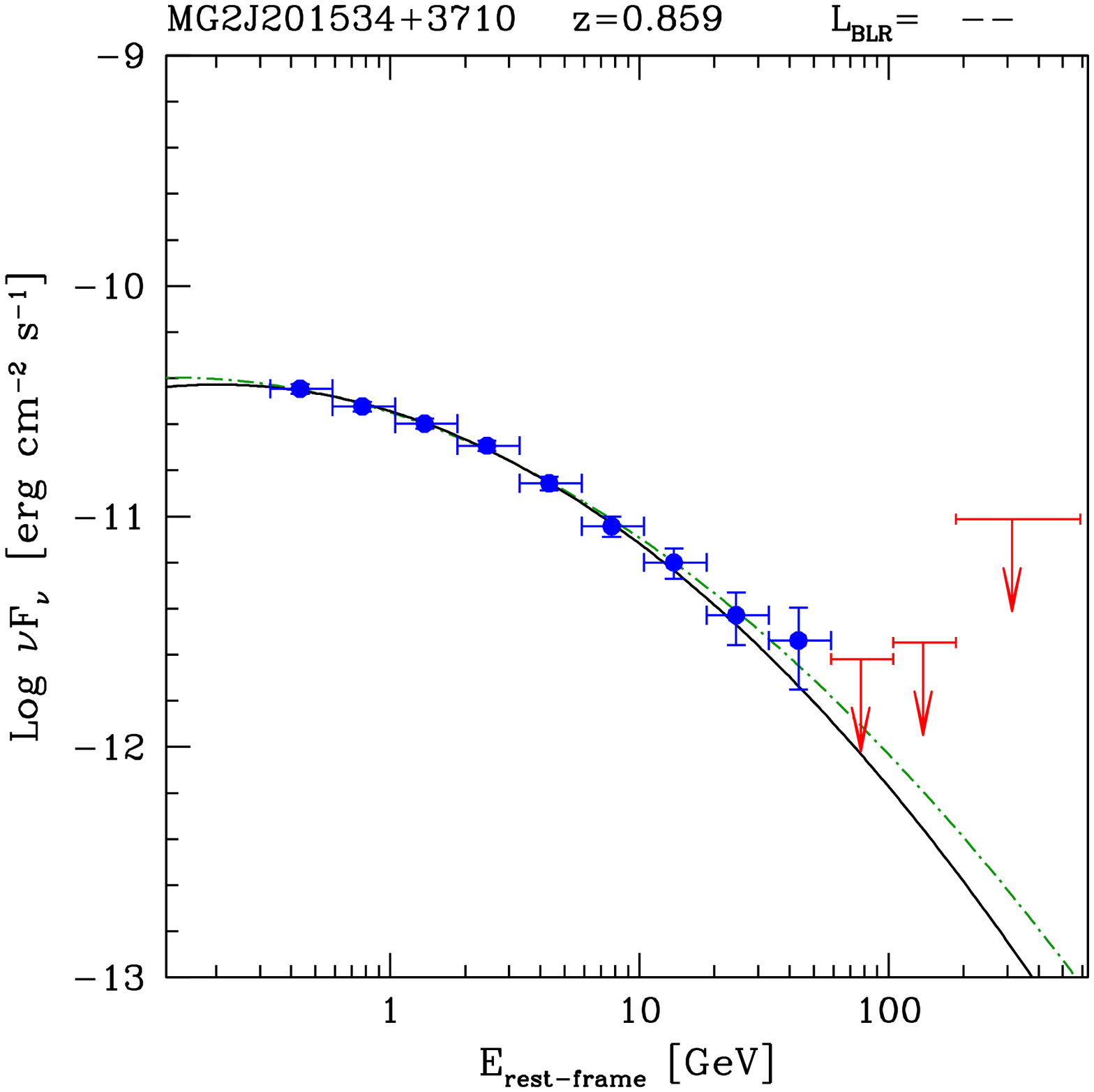,width=4.3cm,height=3.2cm } 
&\psfig{file=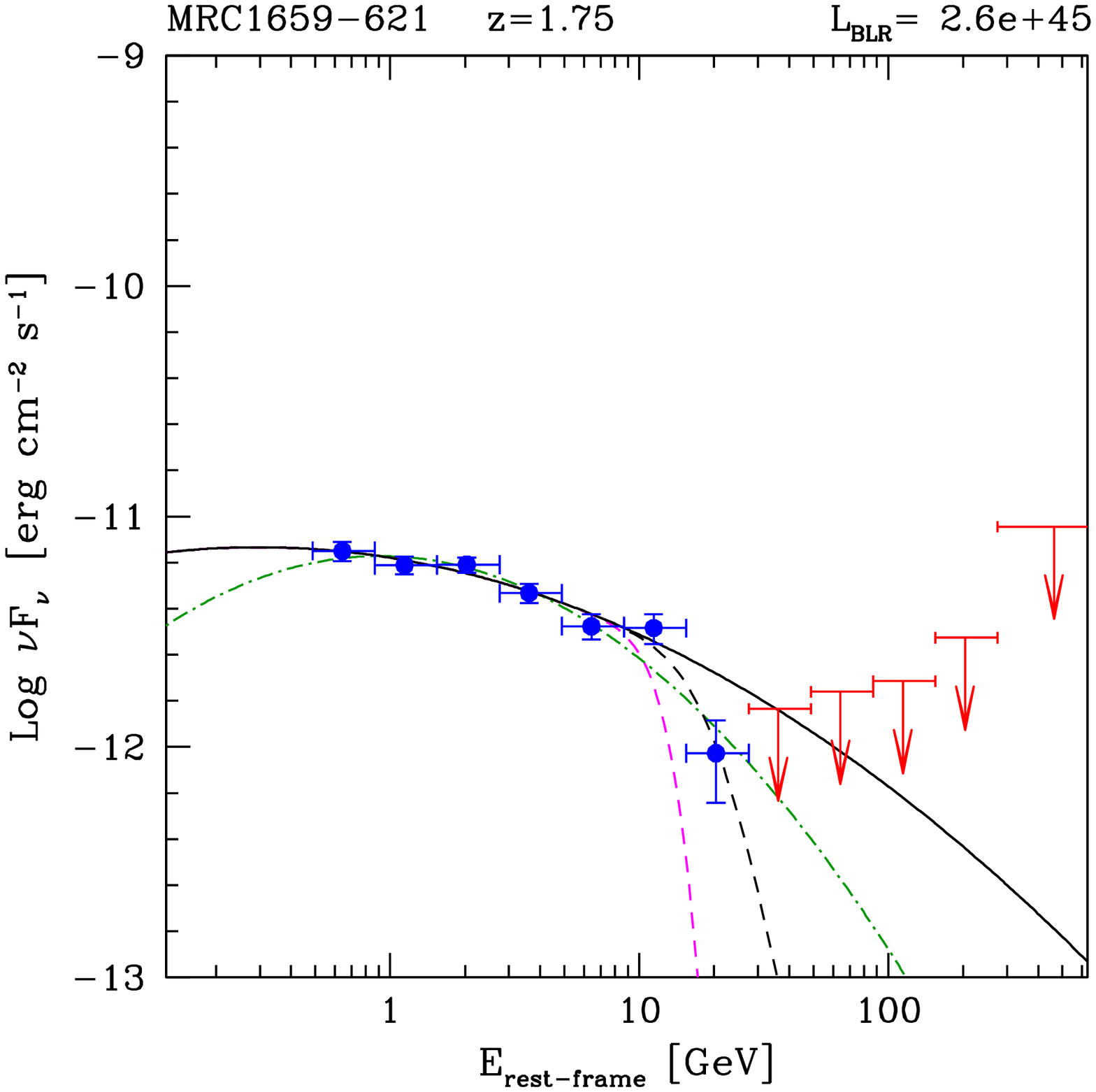,width=4.3cm,height=3.2cm }    
&\psfig{file=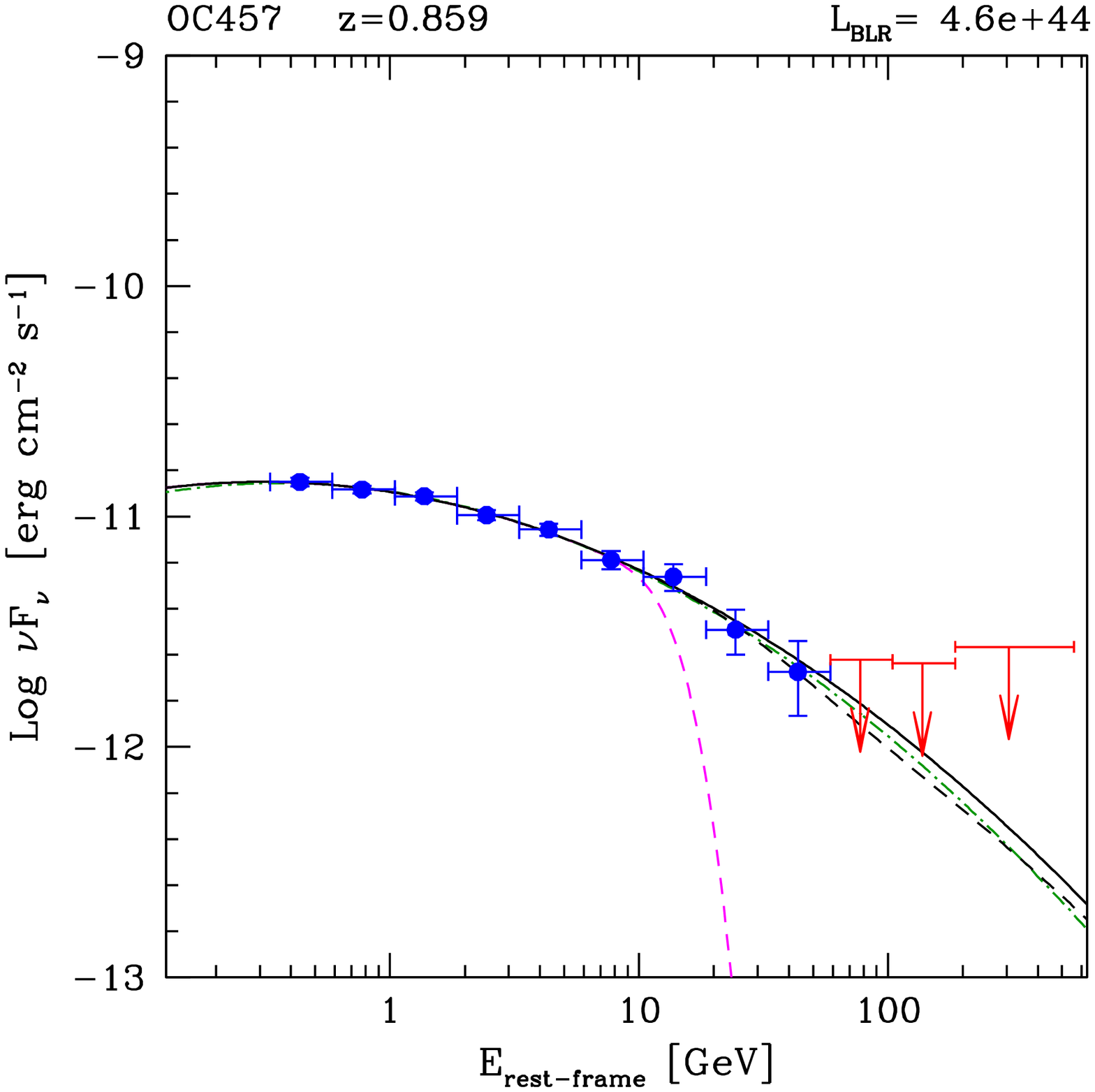,width=4.3cm,height=3.2cm }\vspace{1.2cm} \\
 \psfig{file=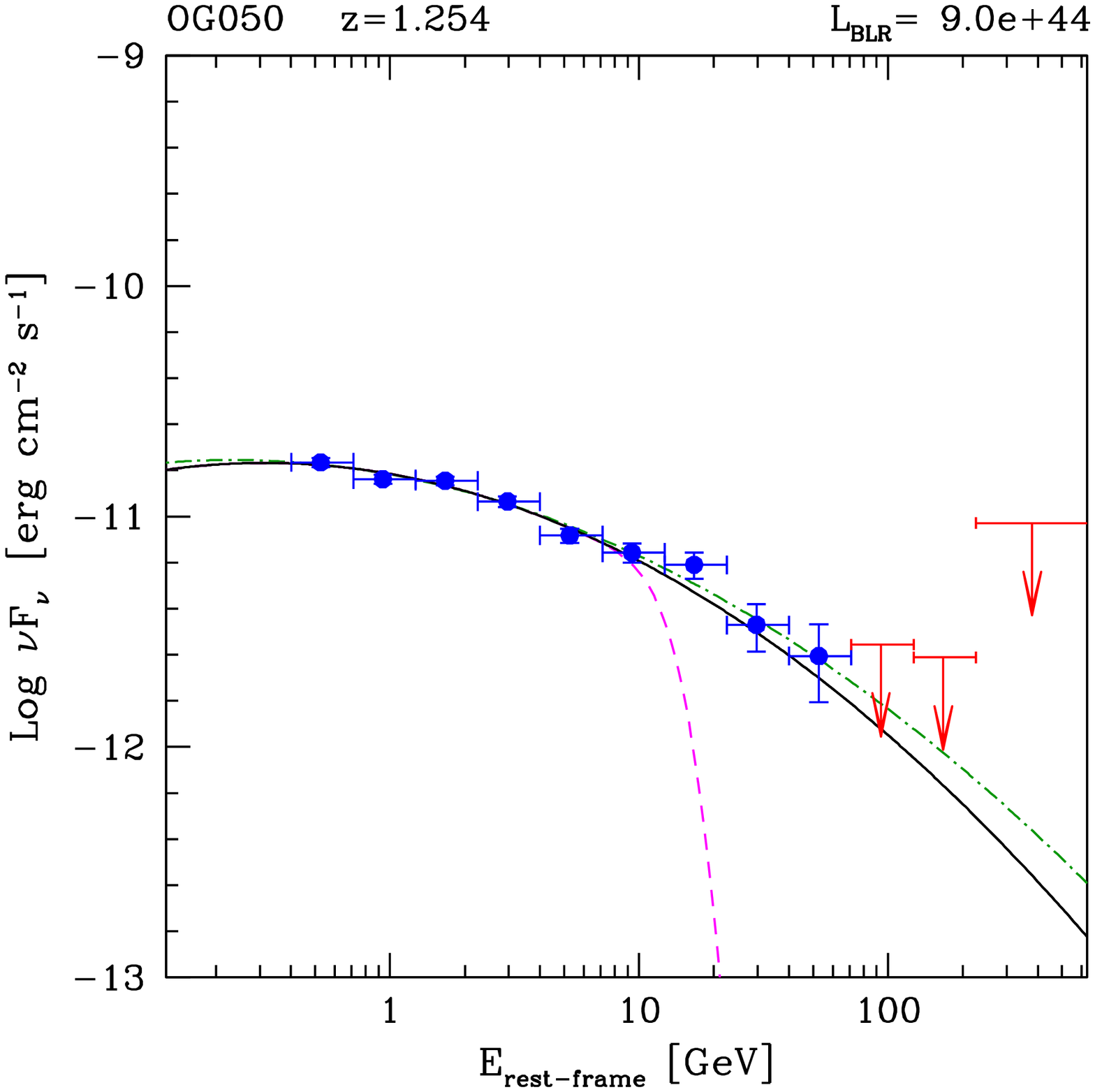,width=4.3cm,height=3.2cm }          
&\psfig{file=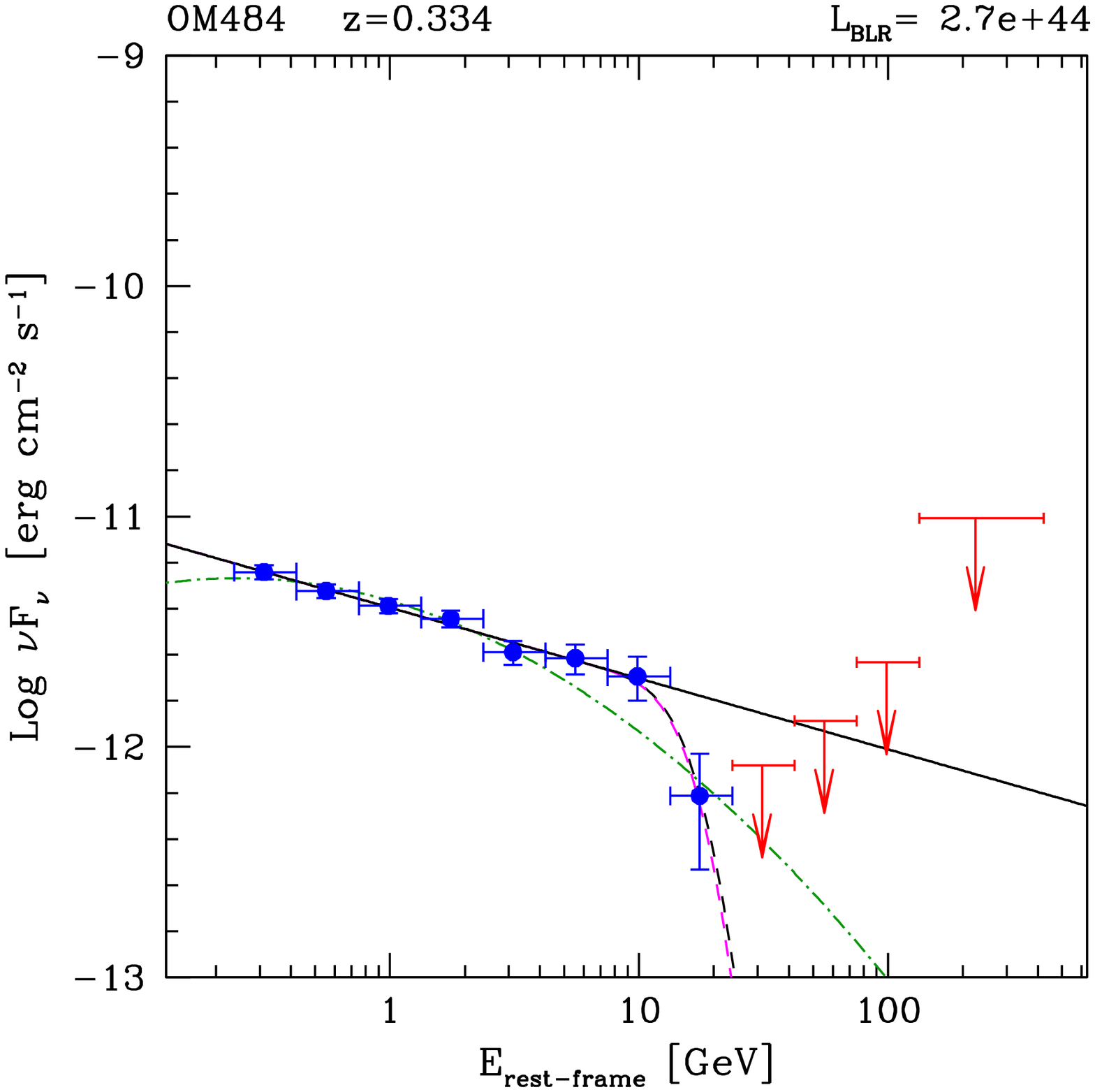,width=4.3cm,height=3.2cm } 
&\psfig{file=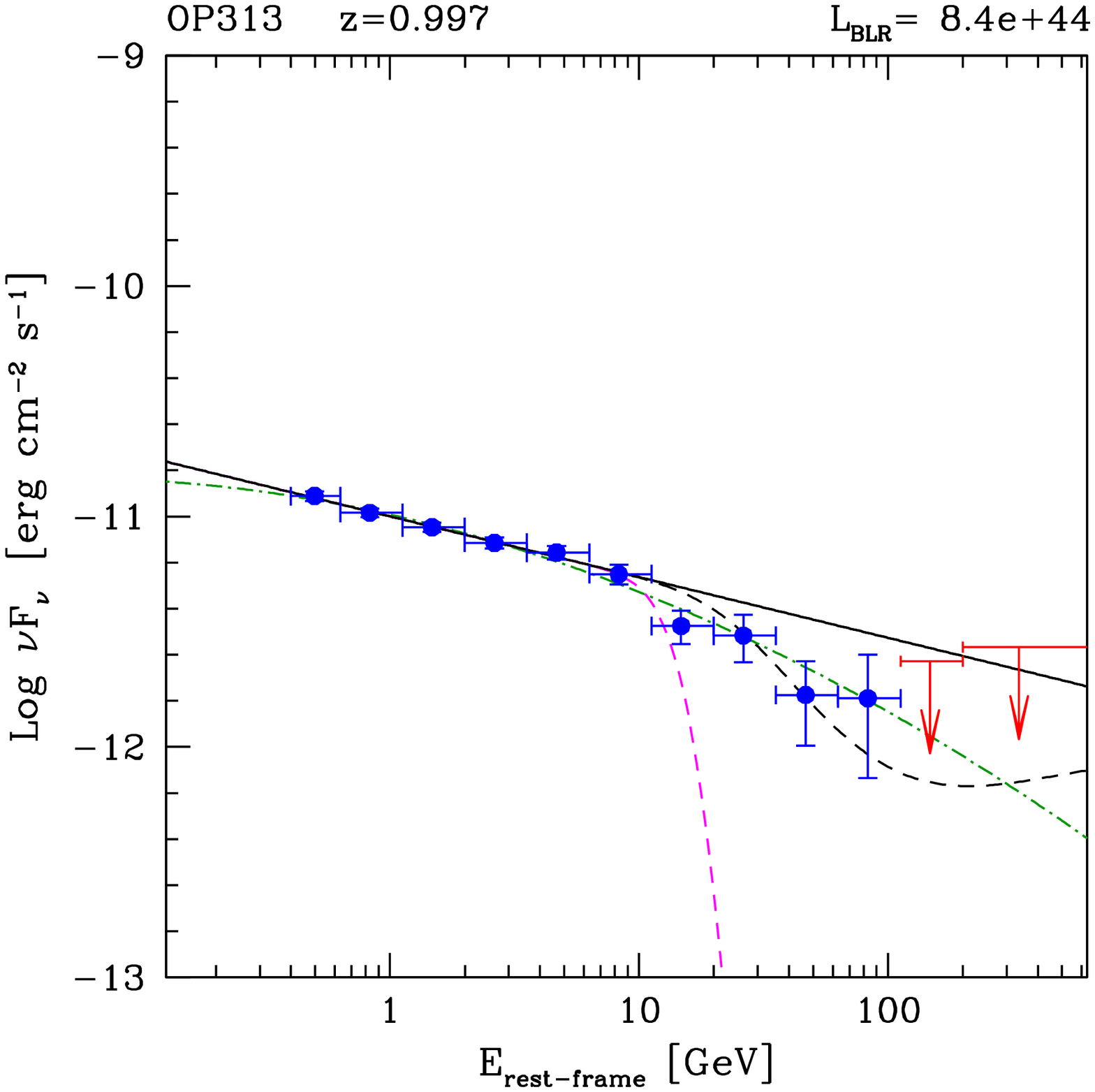,width=4.3cm,height=3.2cm }          
&\psfig{file=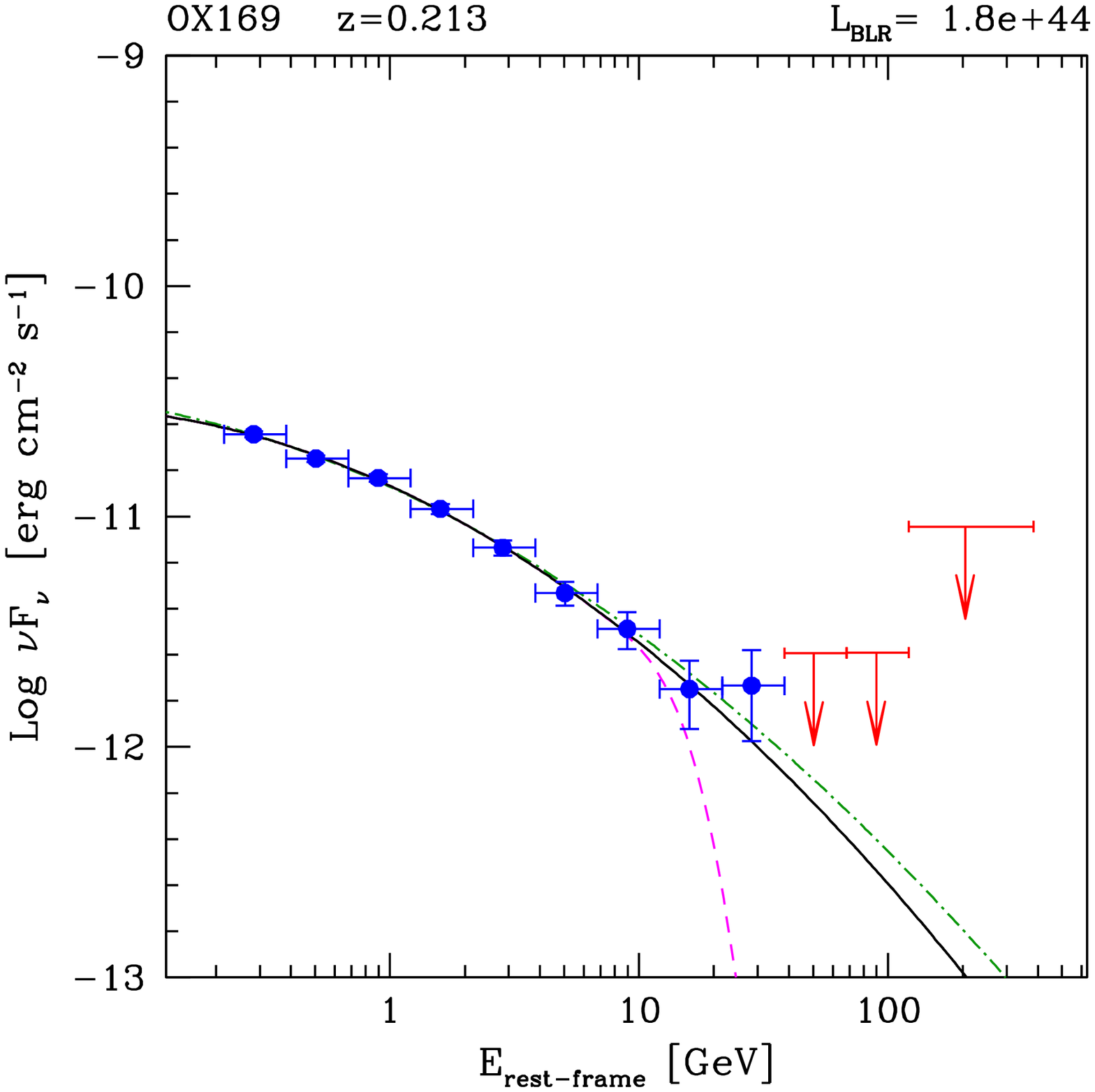,width=4.3cm,height=3.2cm } \vspace{1.2cm}\\
 \psfig{file=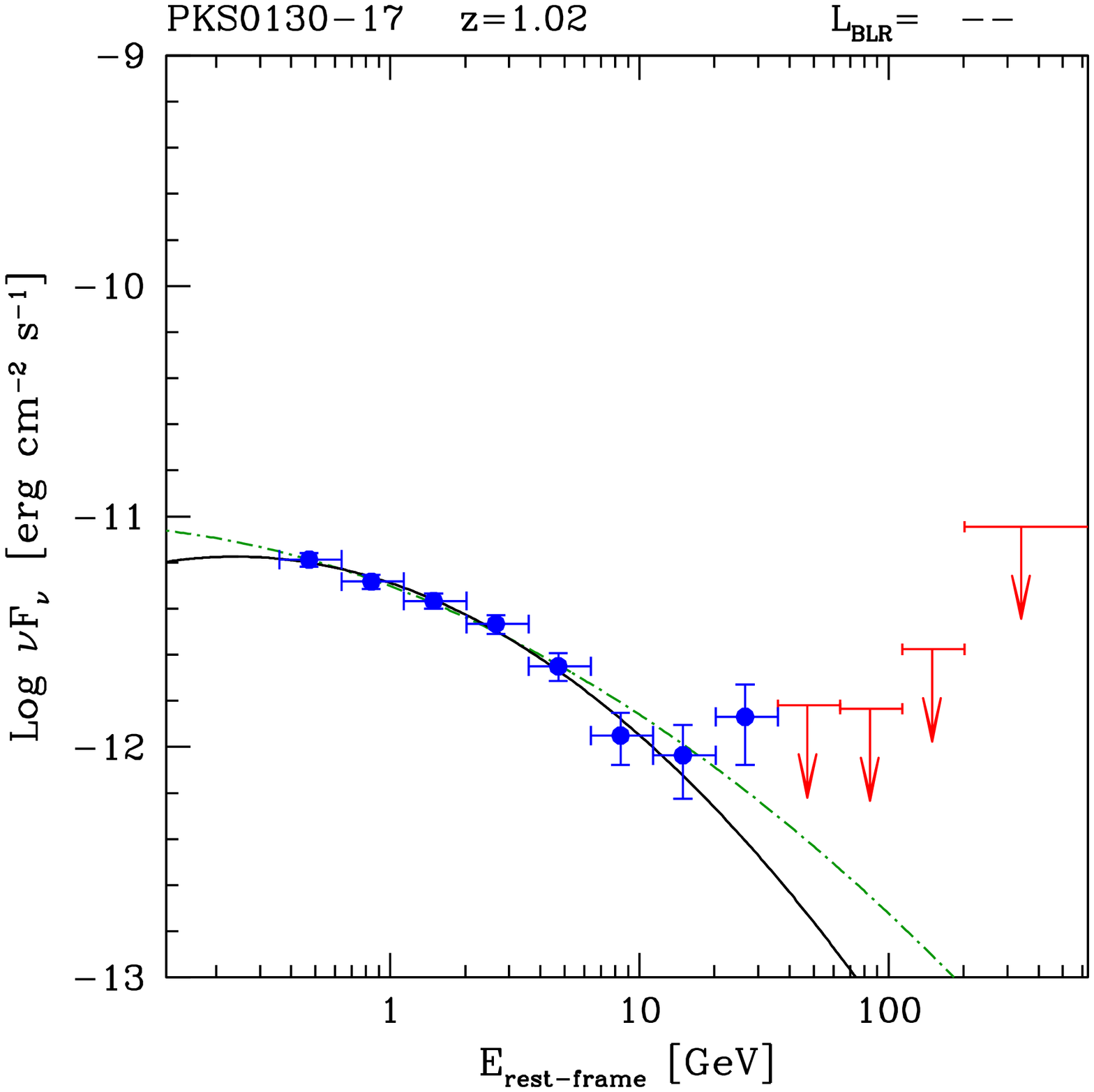,width=4.3cm,height=3.2cm }     
&\psfig{file=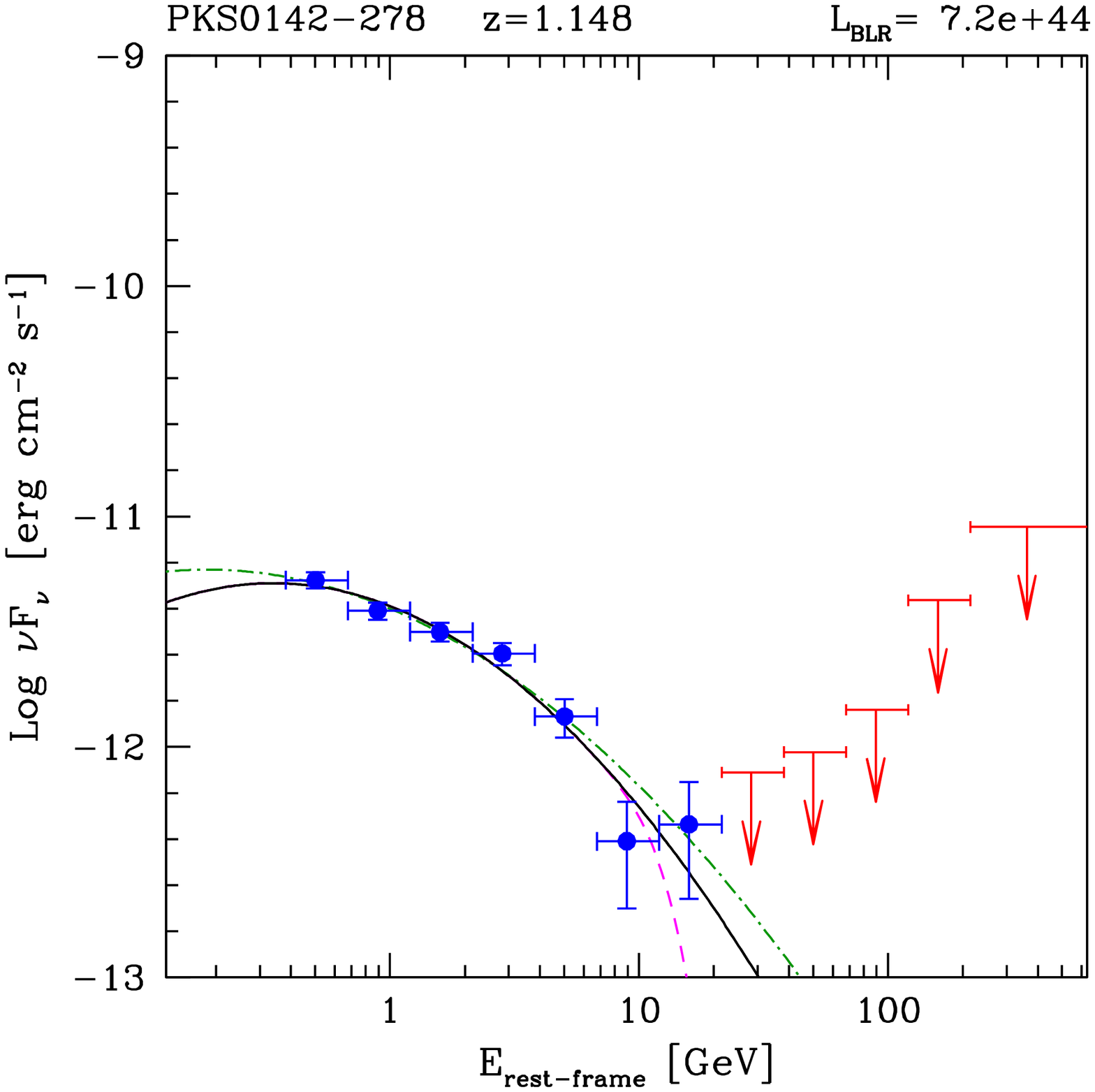,width=4.3cm,height=3.2cm } 
&\psfig{file=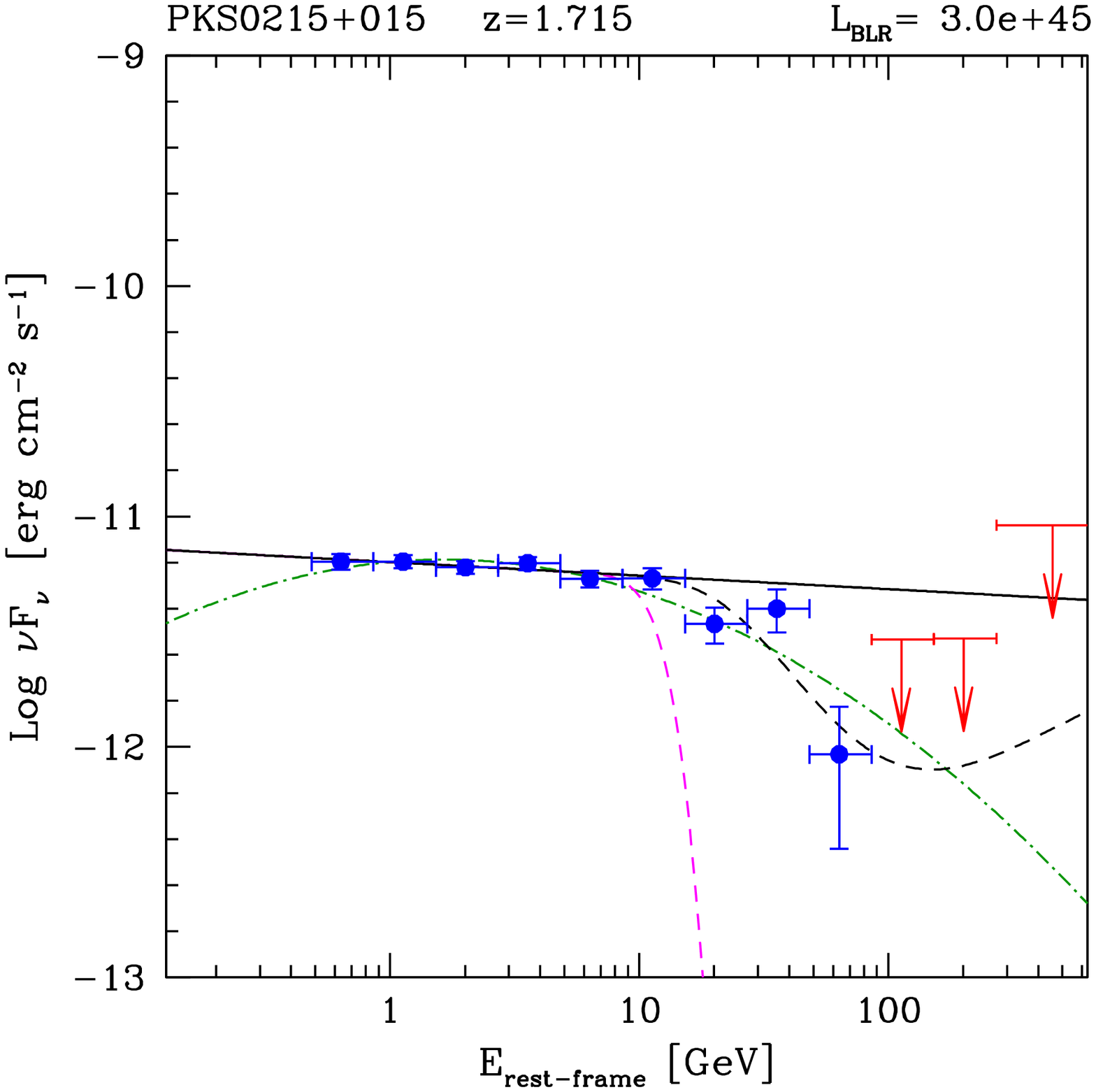,width=4.3cm,height=3.2cm }     
&\psfig{file=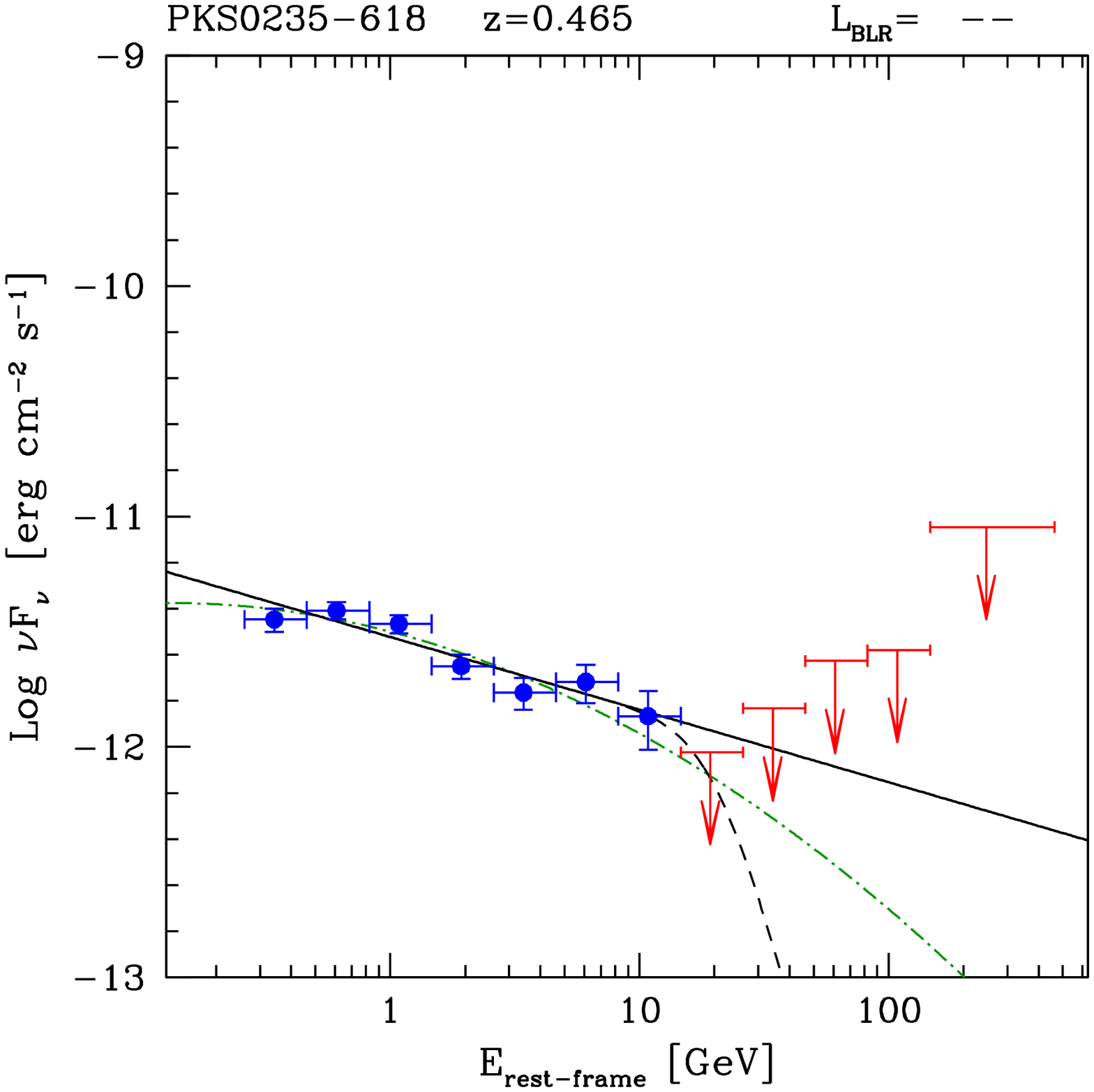,width=4.3cm,height=3.2cm } 
\end{tabular}
\contcaption{}   
\end{figure*}

\begin{figure*}
\vspace{1cm}
\begin{tabular}{cccc}
 \psfig{file=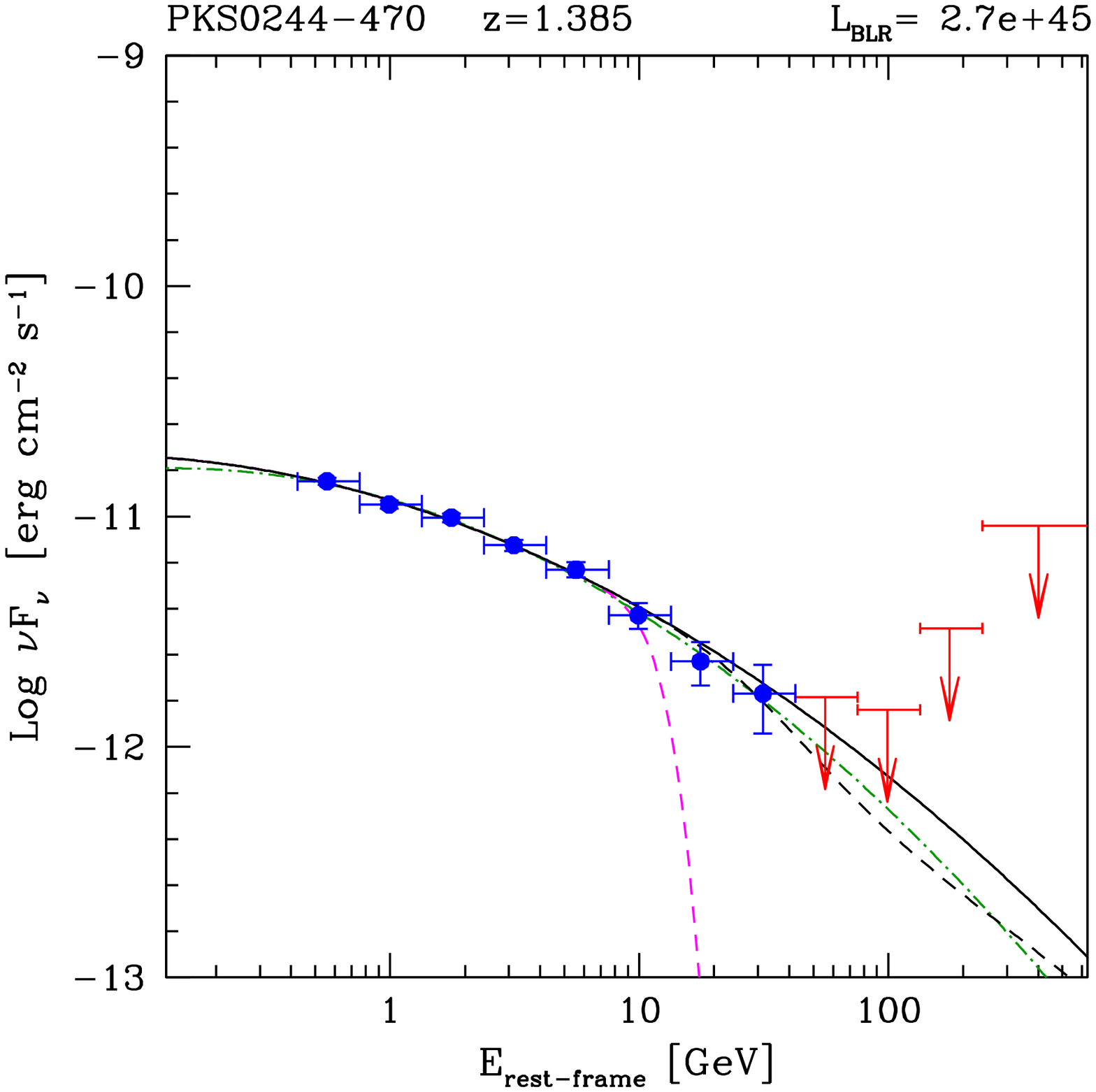,width=4.3cm,height=3.2cm }    
&\psfig{file=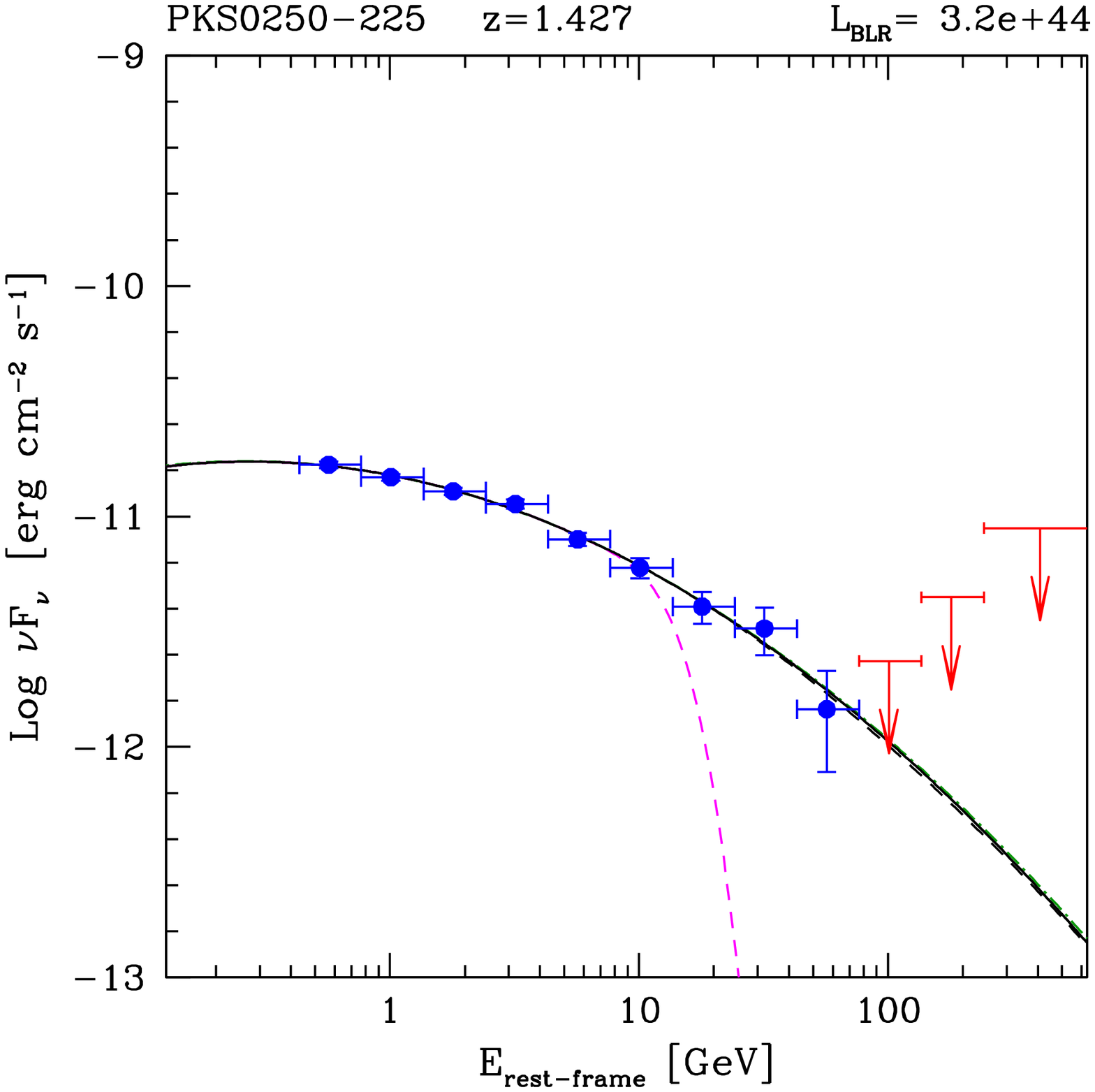,width=4.3cm,height=3.2cm } 
&\psfig{file=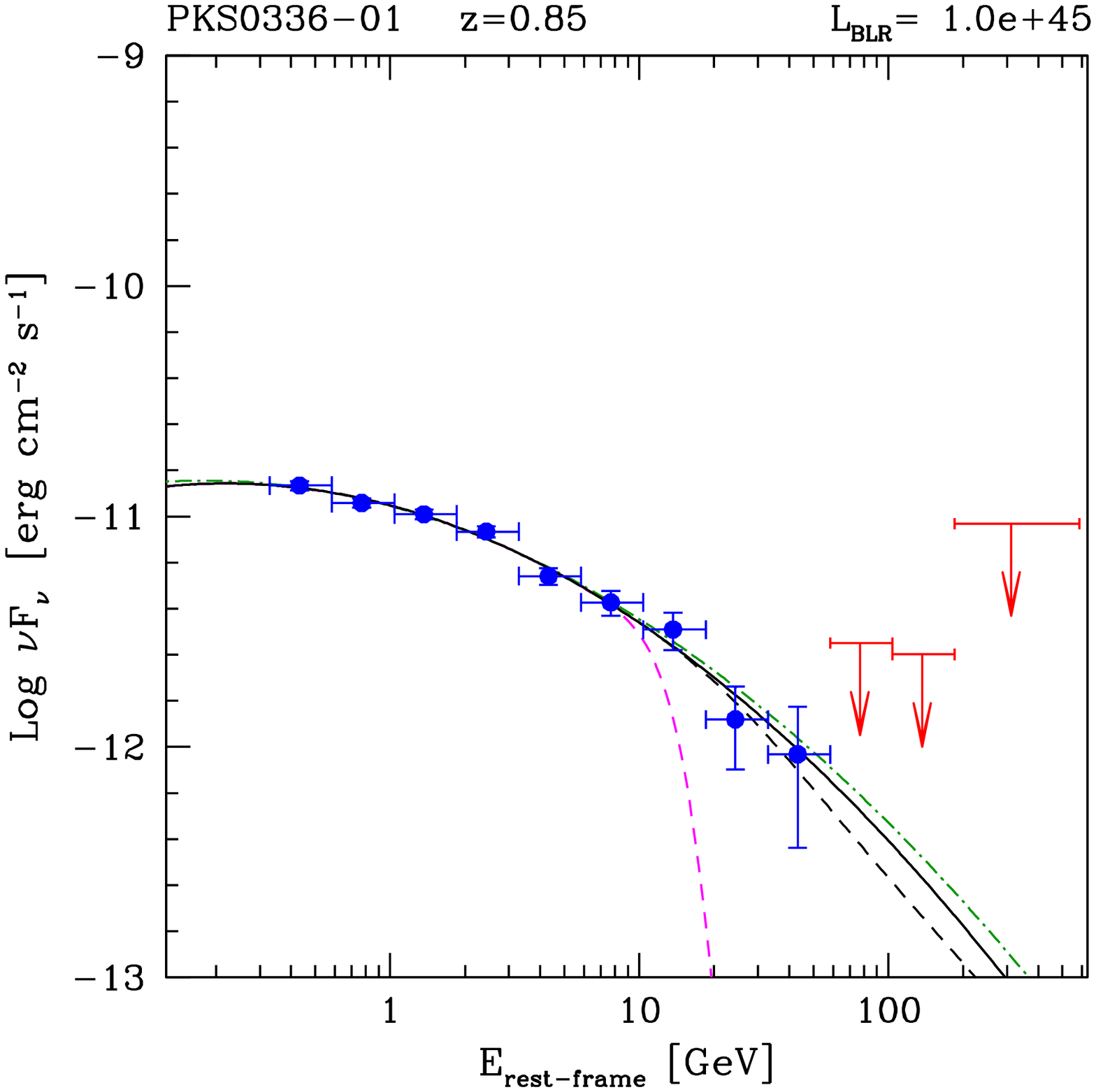,width=4.3cm,height=3.2cm }     
&\psfig{file=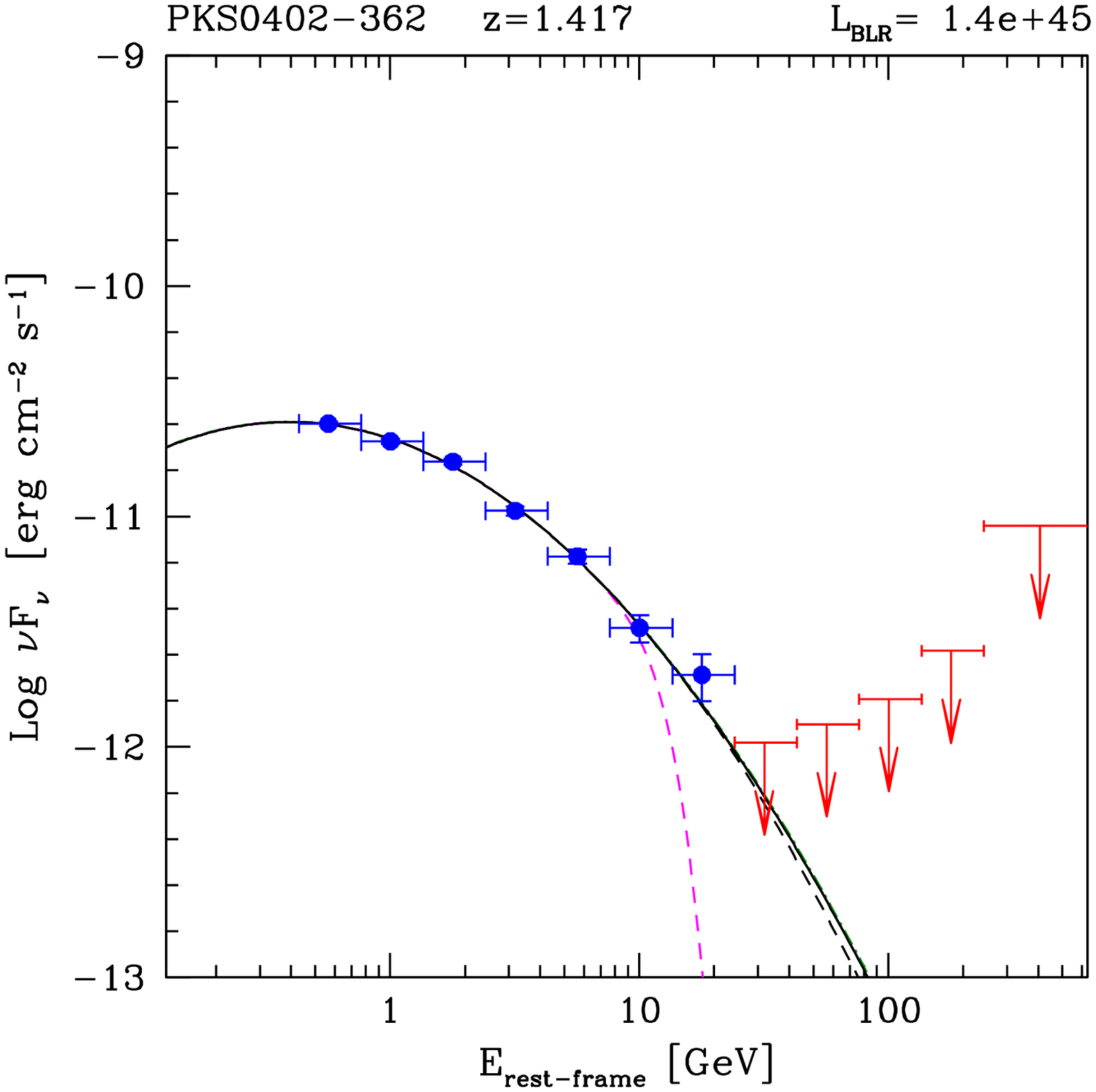,width=4.3cm,height=3.2cm }\vspace{1.2cm} \\
 \psfig{file=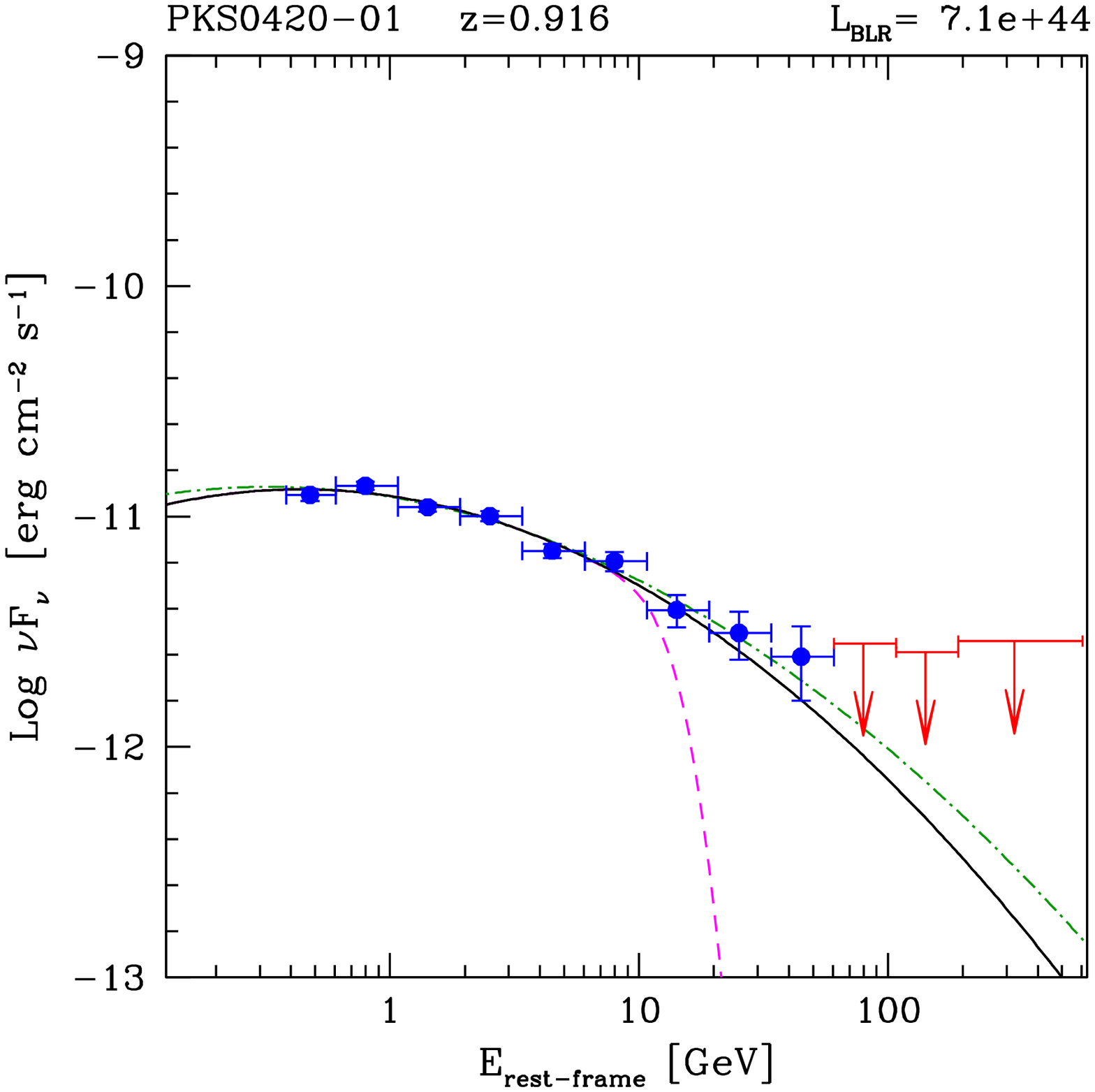,width=4.3cm,height=3.2cm }     
&\psfig{file=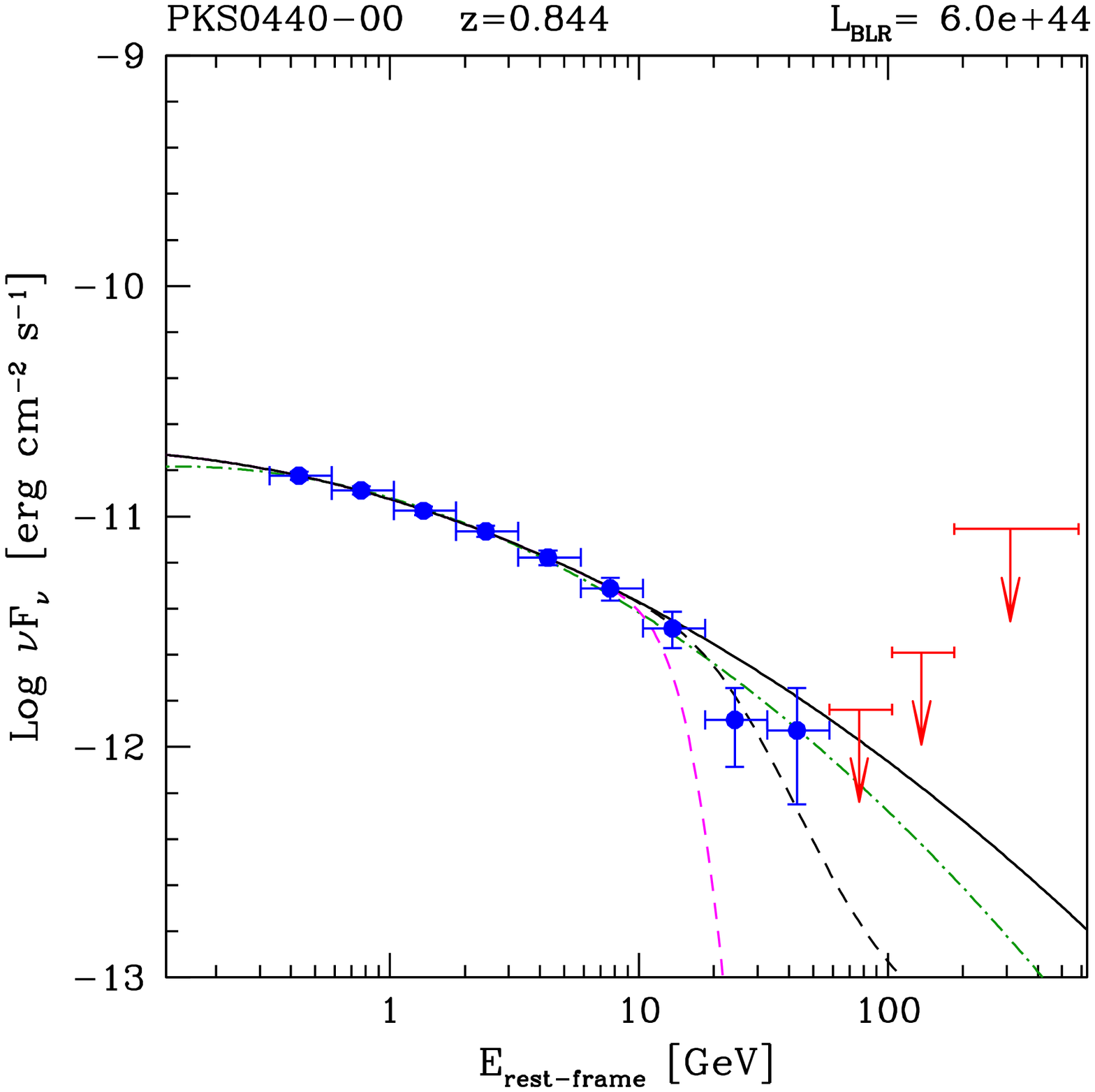,width=4.3cm,height=3.2cm } 
&\psfig{file=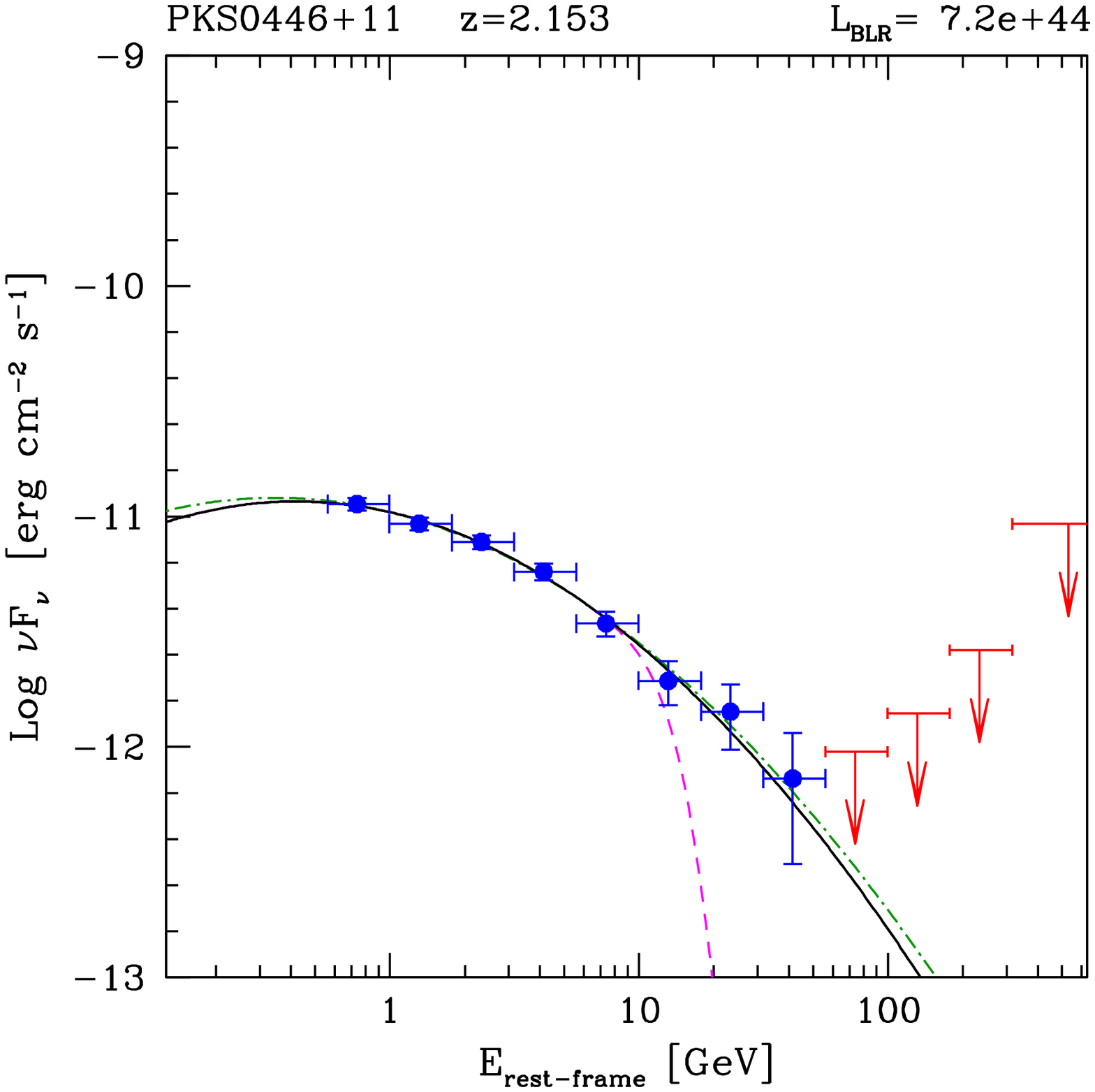,width=4.3cm,height=3.2cm }      
&\psfig{file=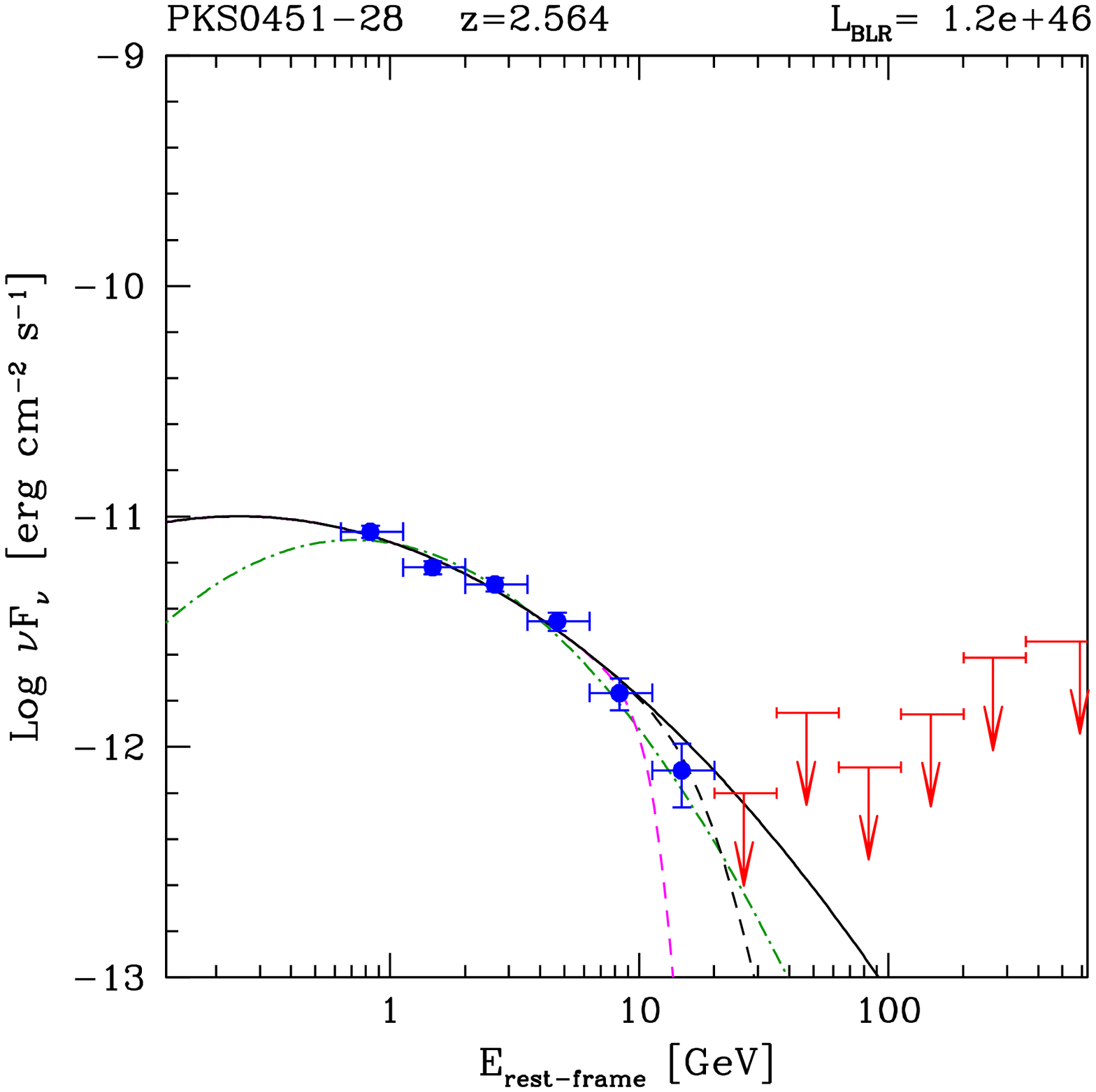,width=4.3cm,height=3.2cm } \vspace{1.2cm}\\
 \psfig{file=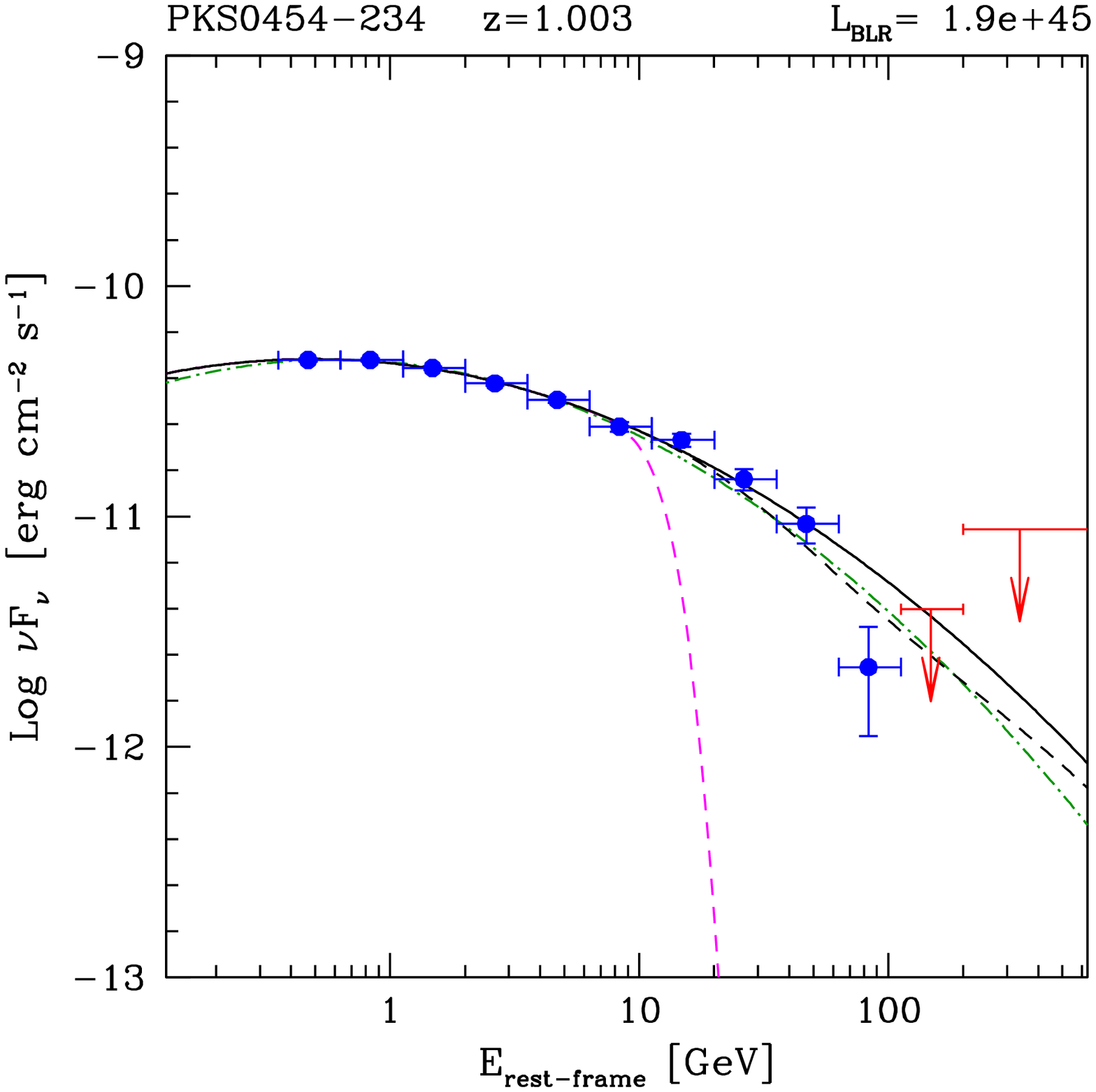,width=4.3cm,height=3.2cm }    
&\psfig{file=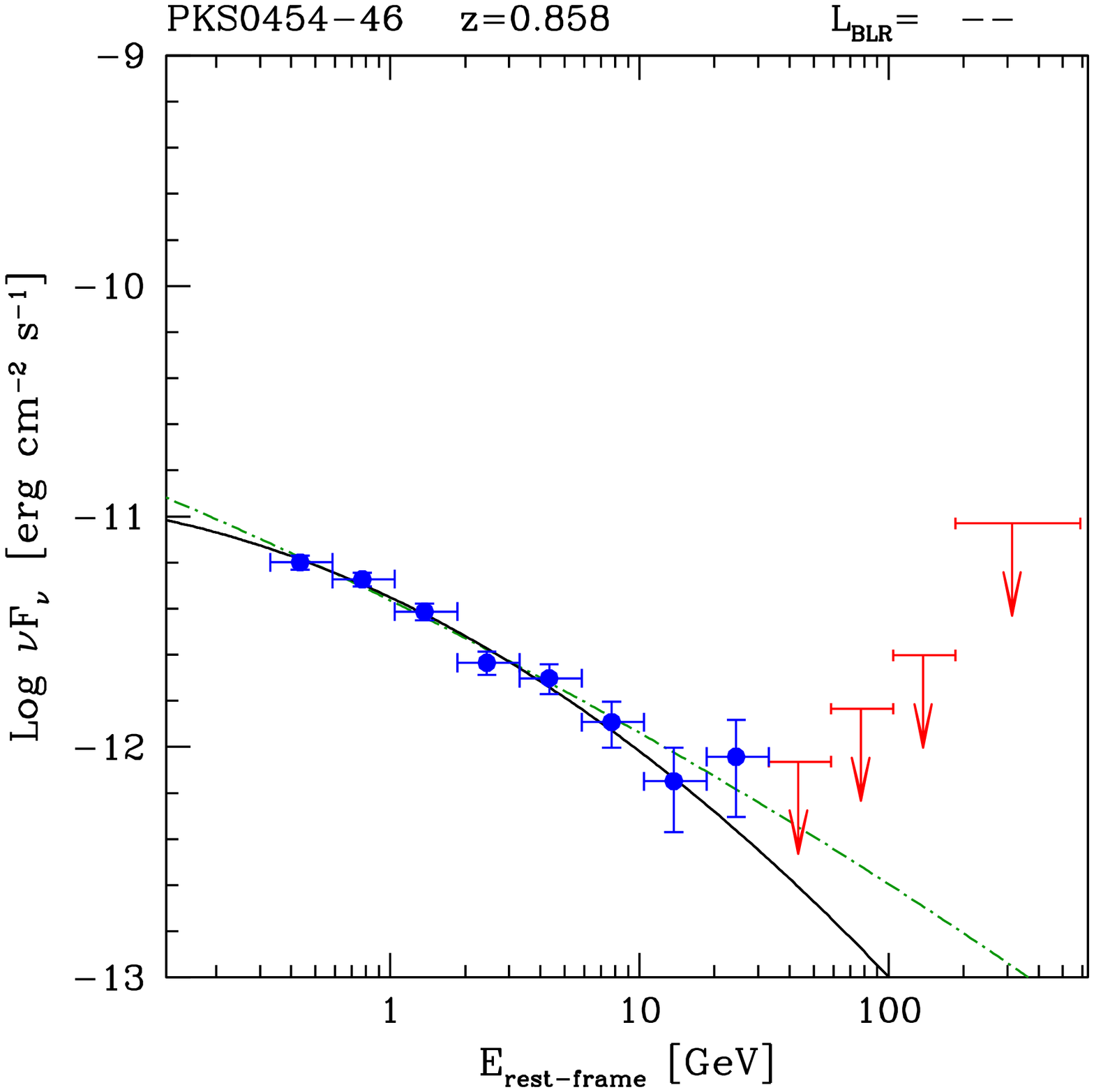,width=4.3cm,height=3.2cm } 
&\psfig{file=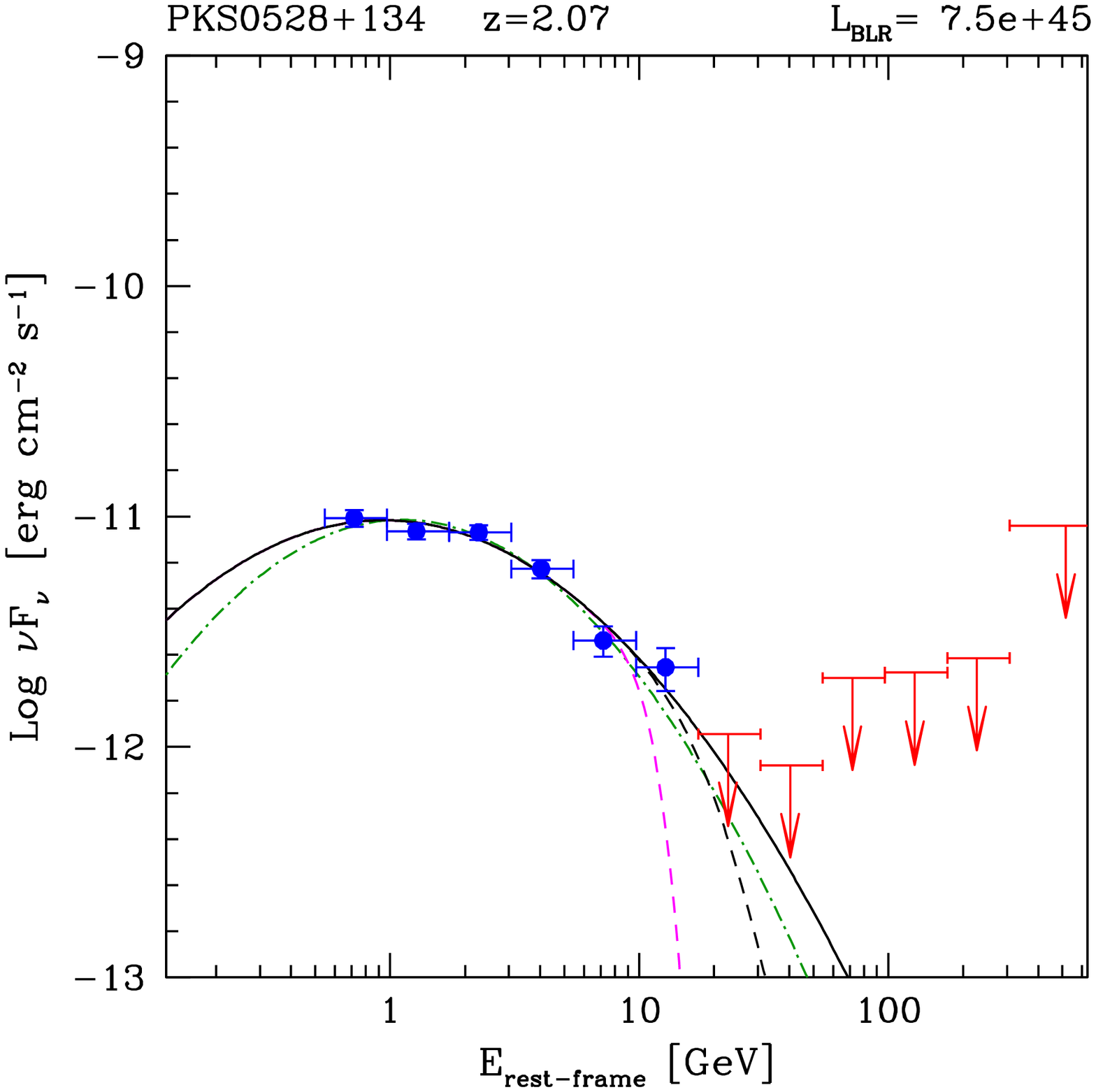,width=4.3cm,height=3.2cm }   
&\psfig{file=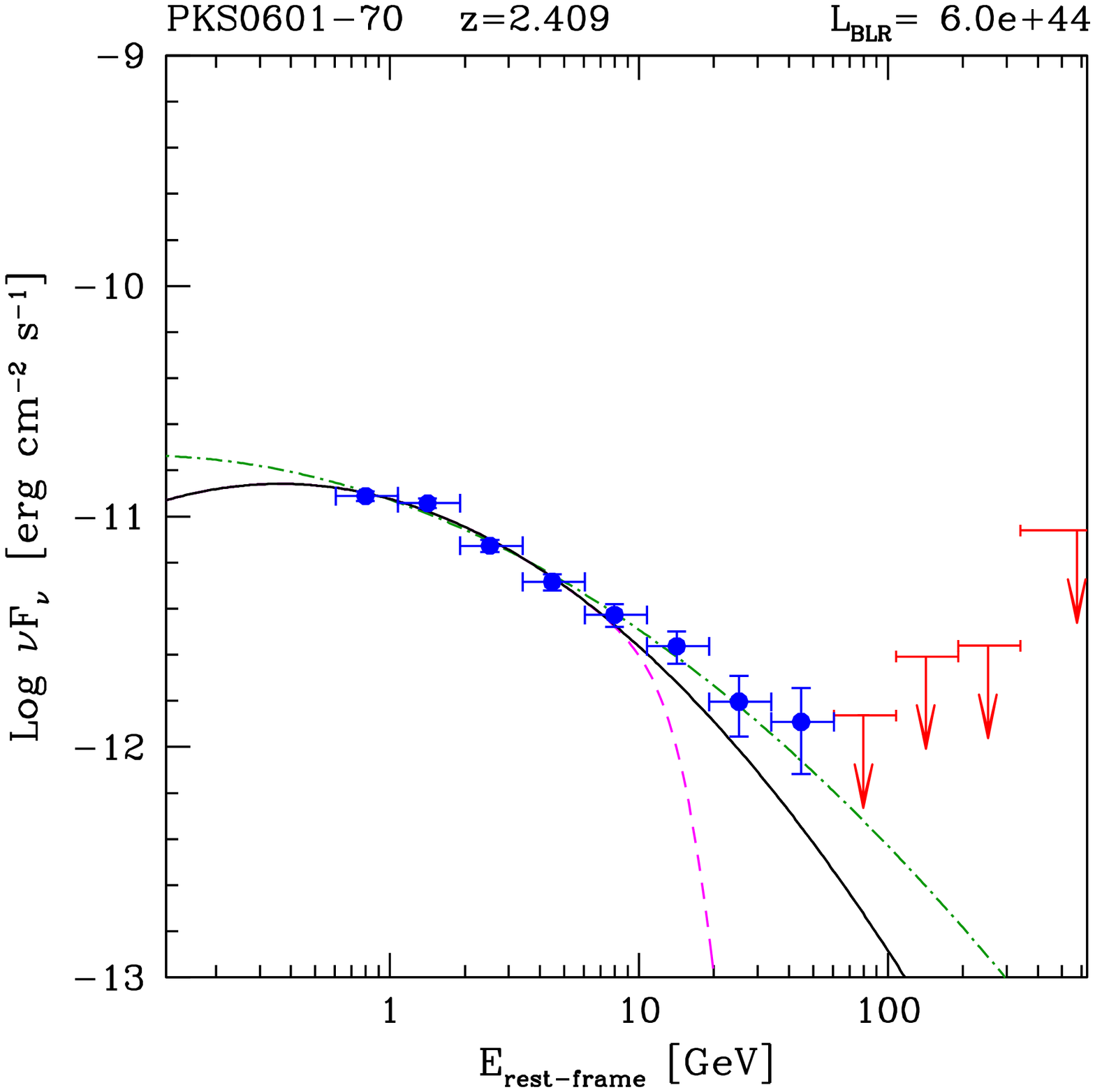,width=4.3cm,height=3.2cm } \vspace{1.2cm} \\ 
\psfig{file=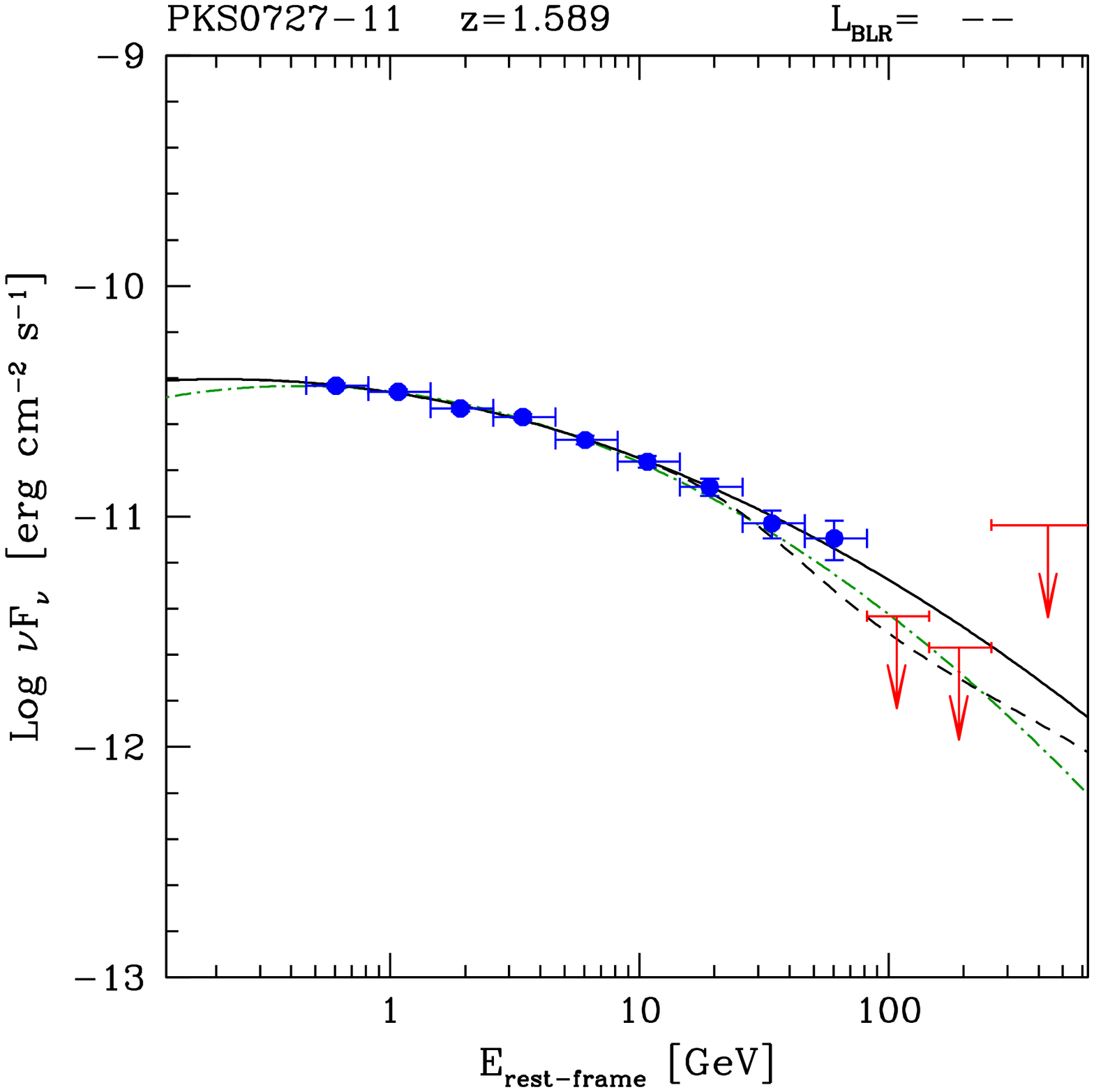,width=4.3cm,height=3.2cm } 
&\psfig{file=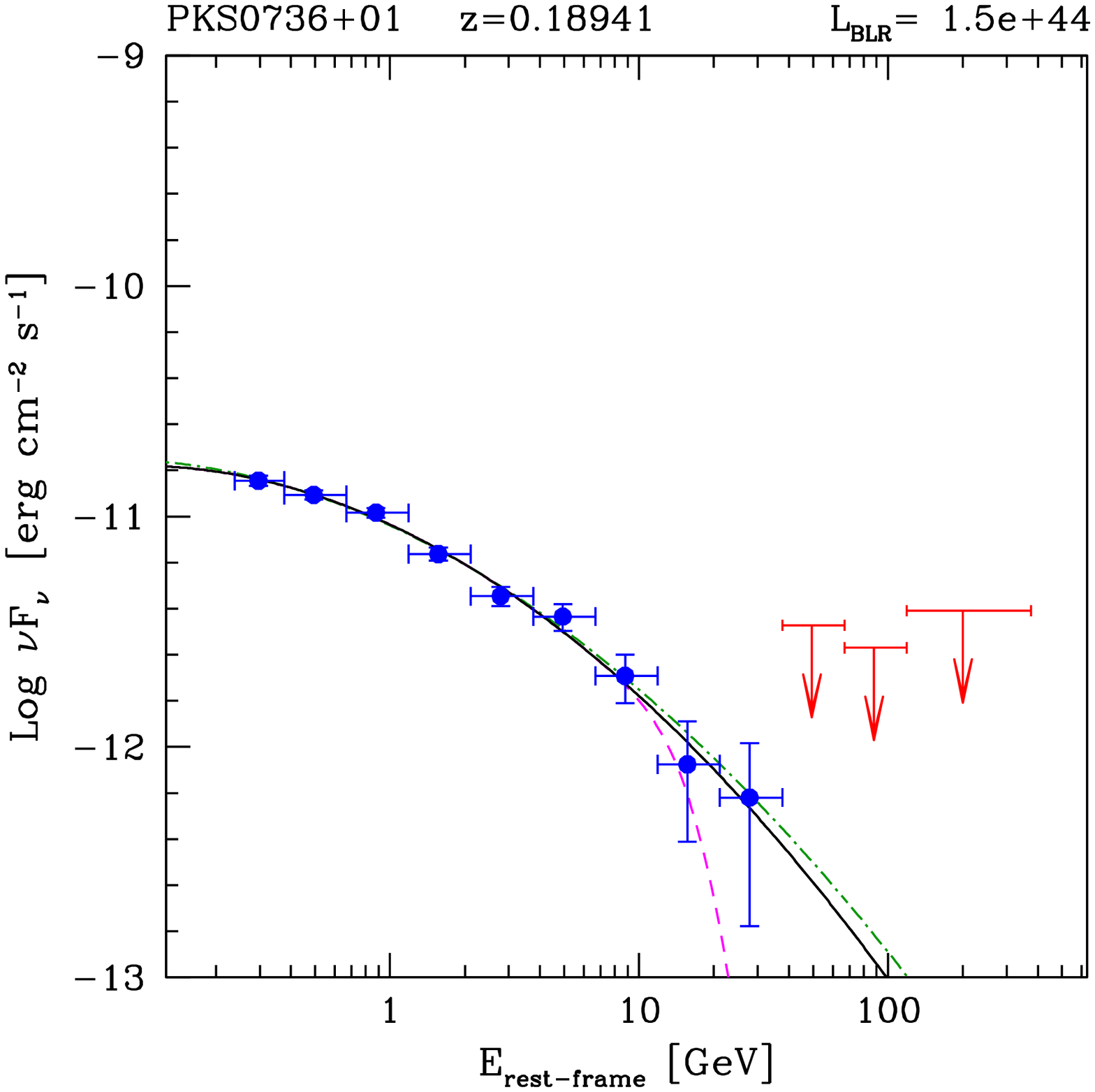,width=4.3cm,height=3.2cm }    
&\psfig{file=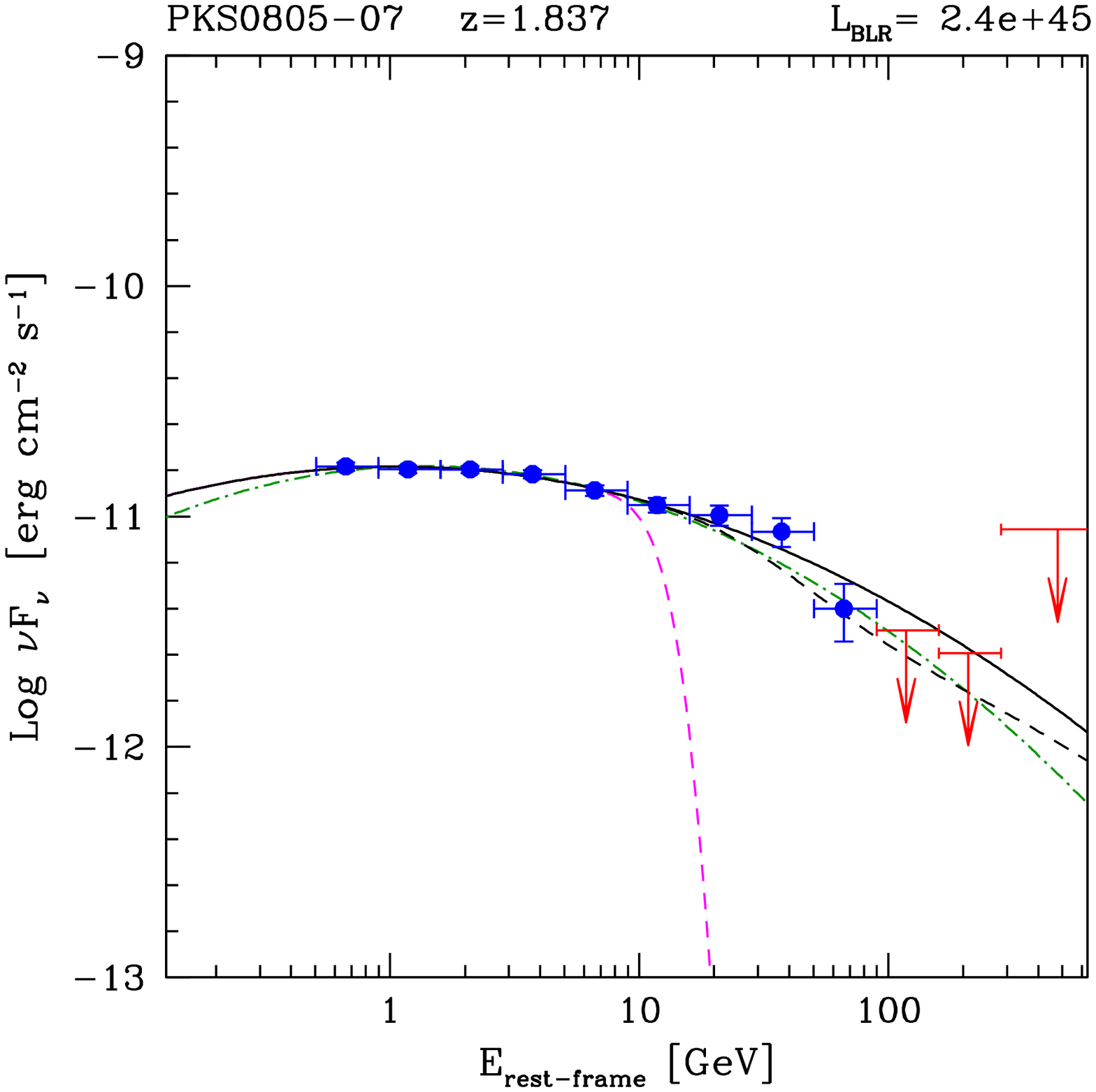,width=4.3cm,height=3.2cm } 
& \psfig{file=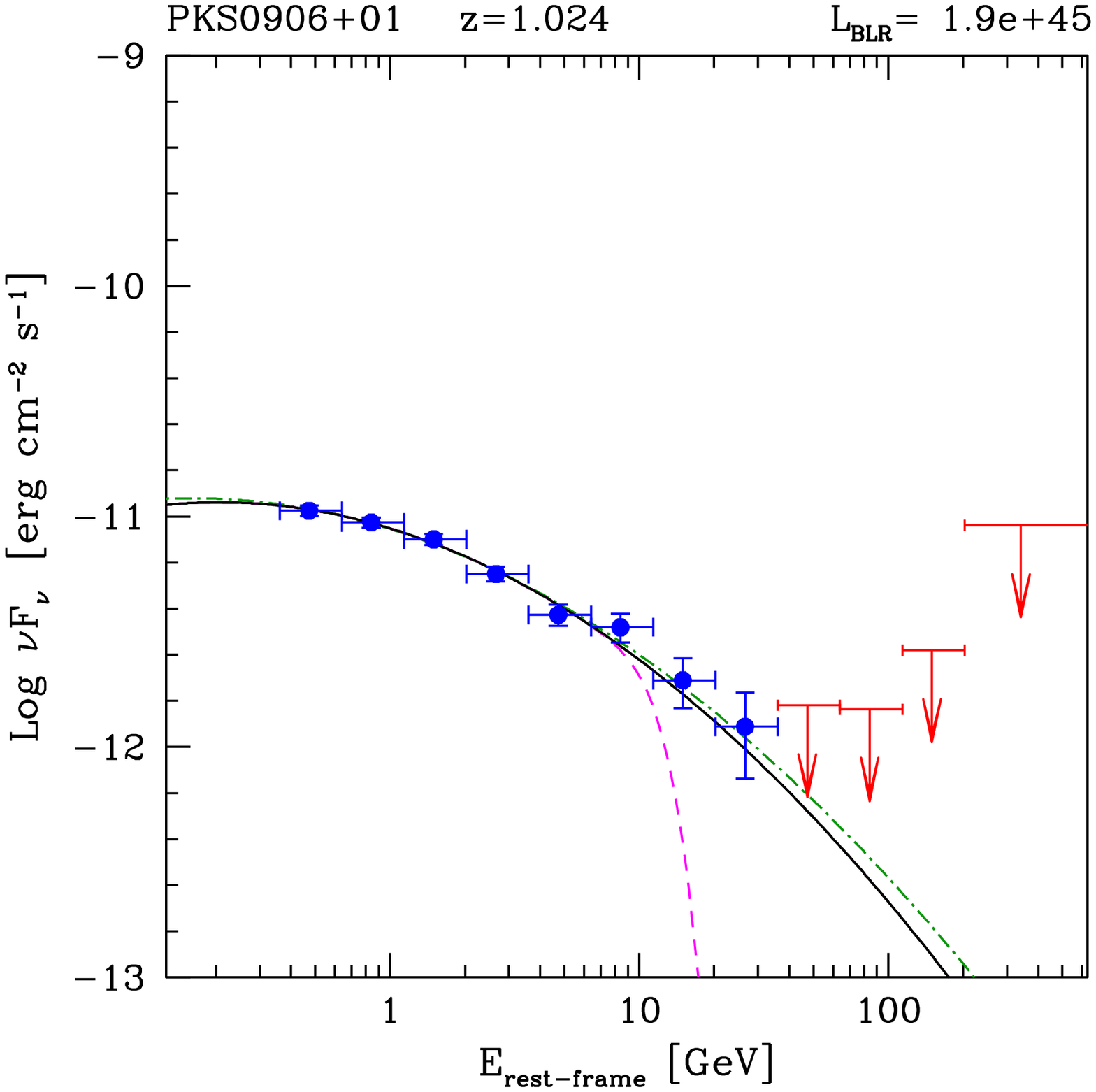,width=4.3cm,height=3.2cm }\vspace{1.2cm} \\   
 \psfig{file=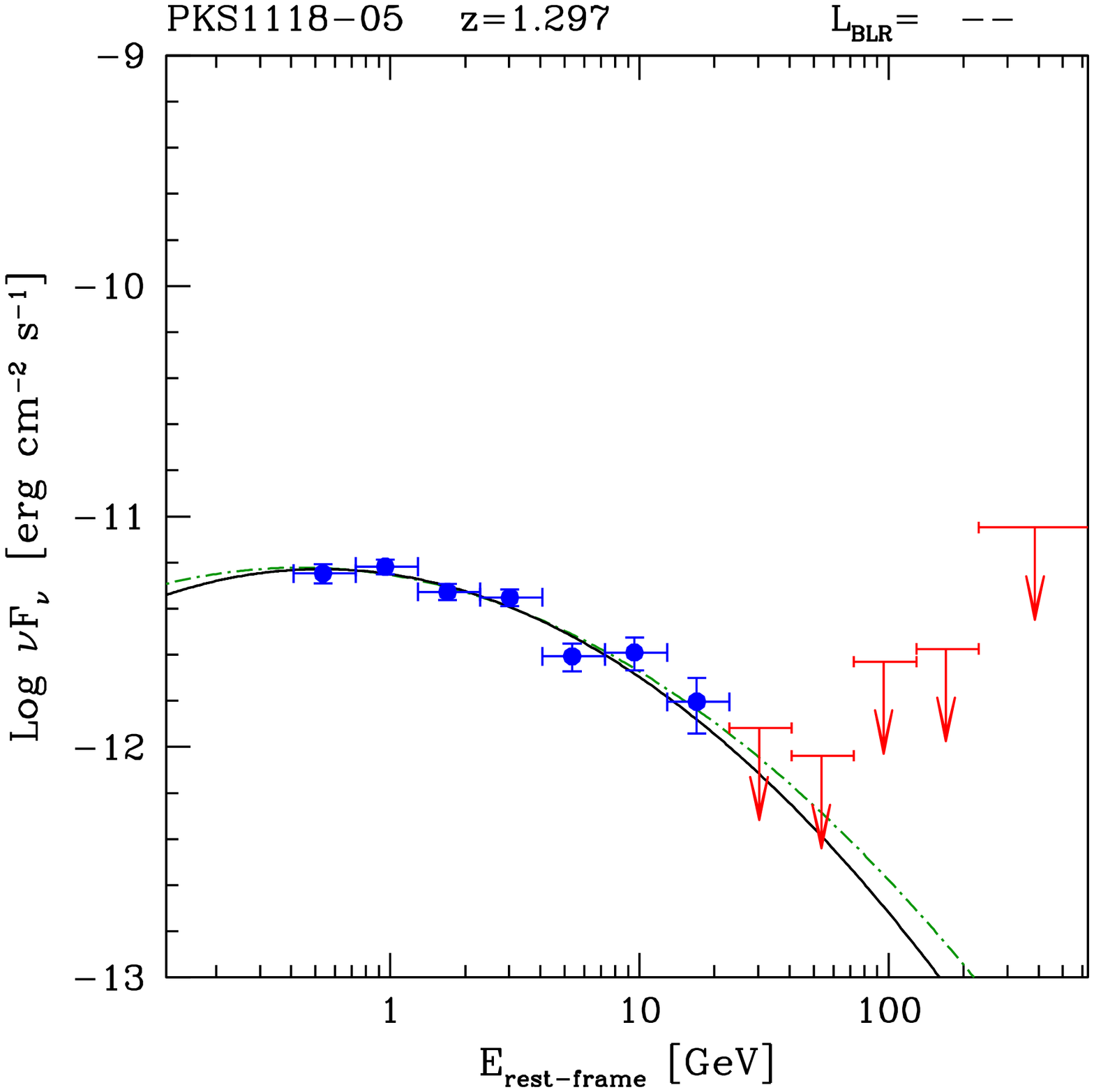,width=4.3cm,height=3.2cm } 
&\psfig{file=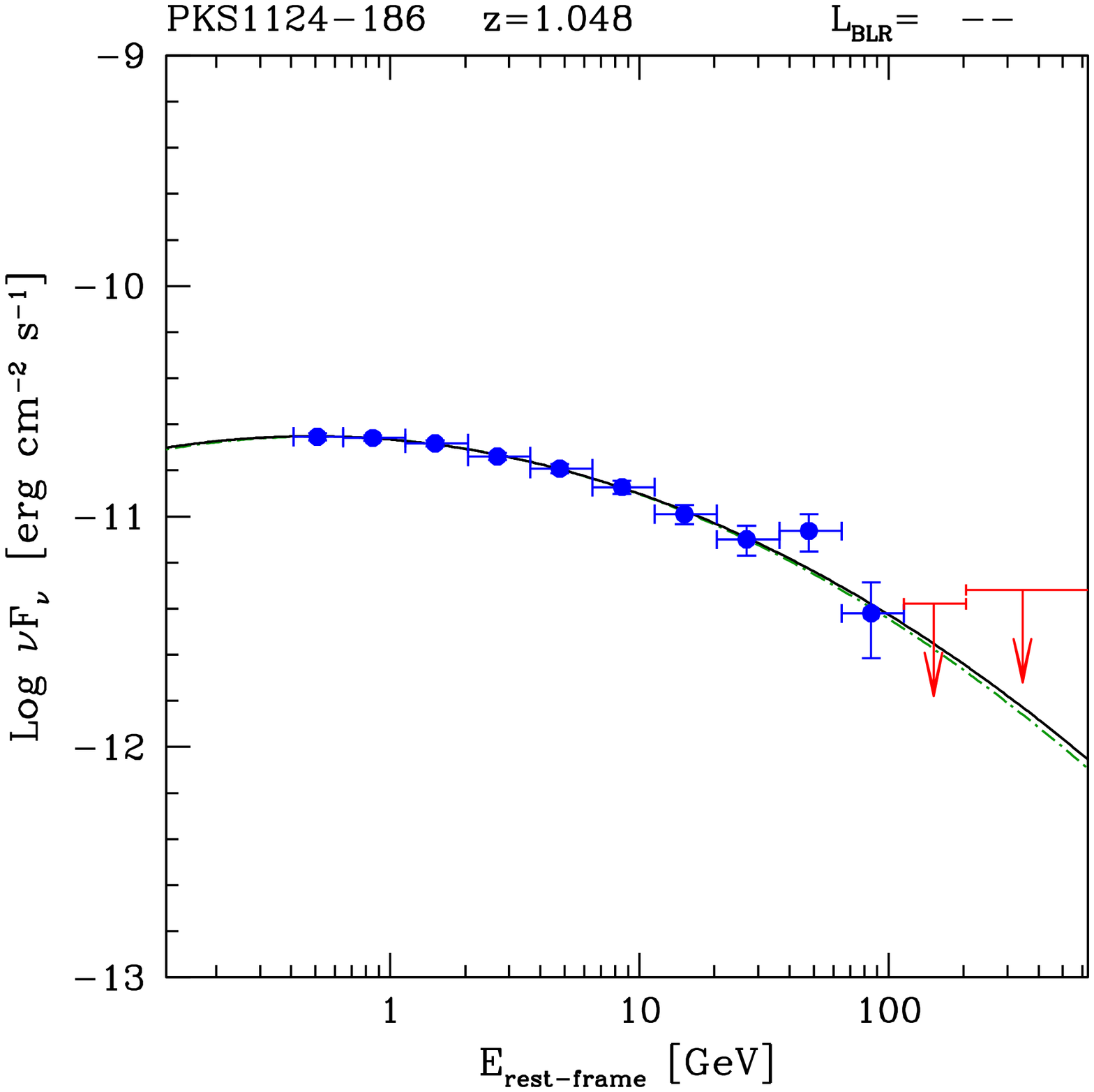,width=4.3cm,height=3.2cm }  
&\psfig{file=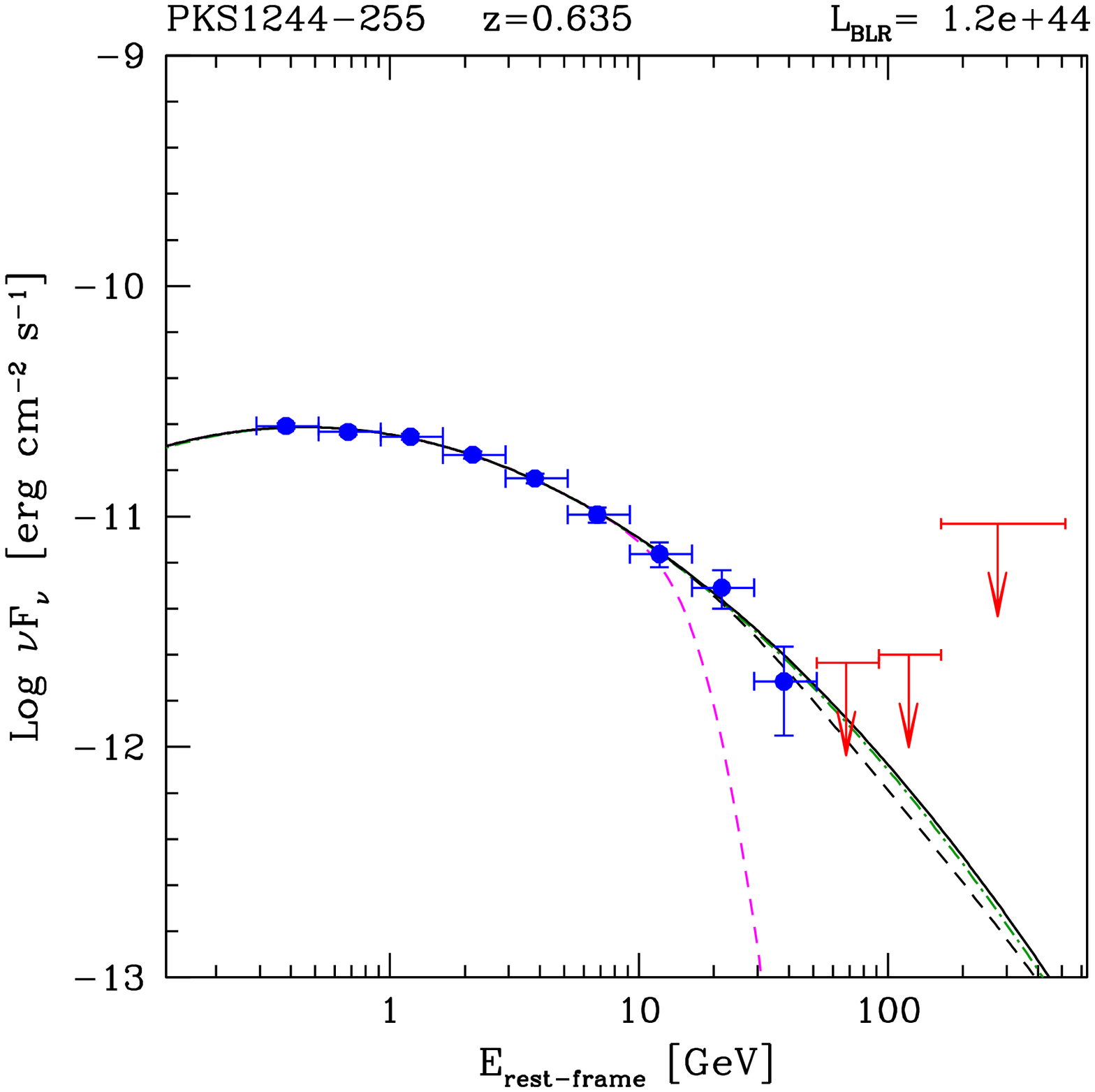,width=4.3cm,height=3.2cm } 
& \psfig{file=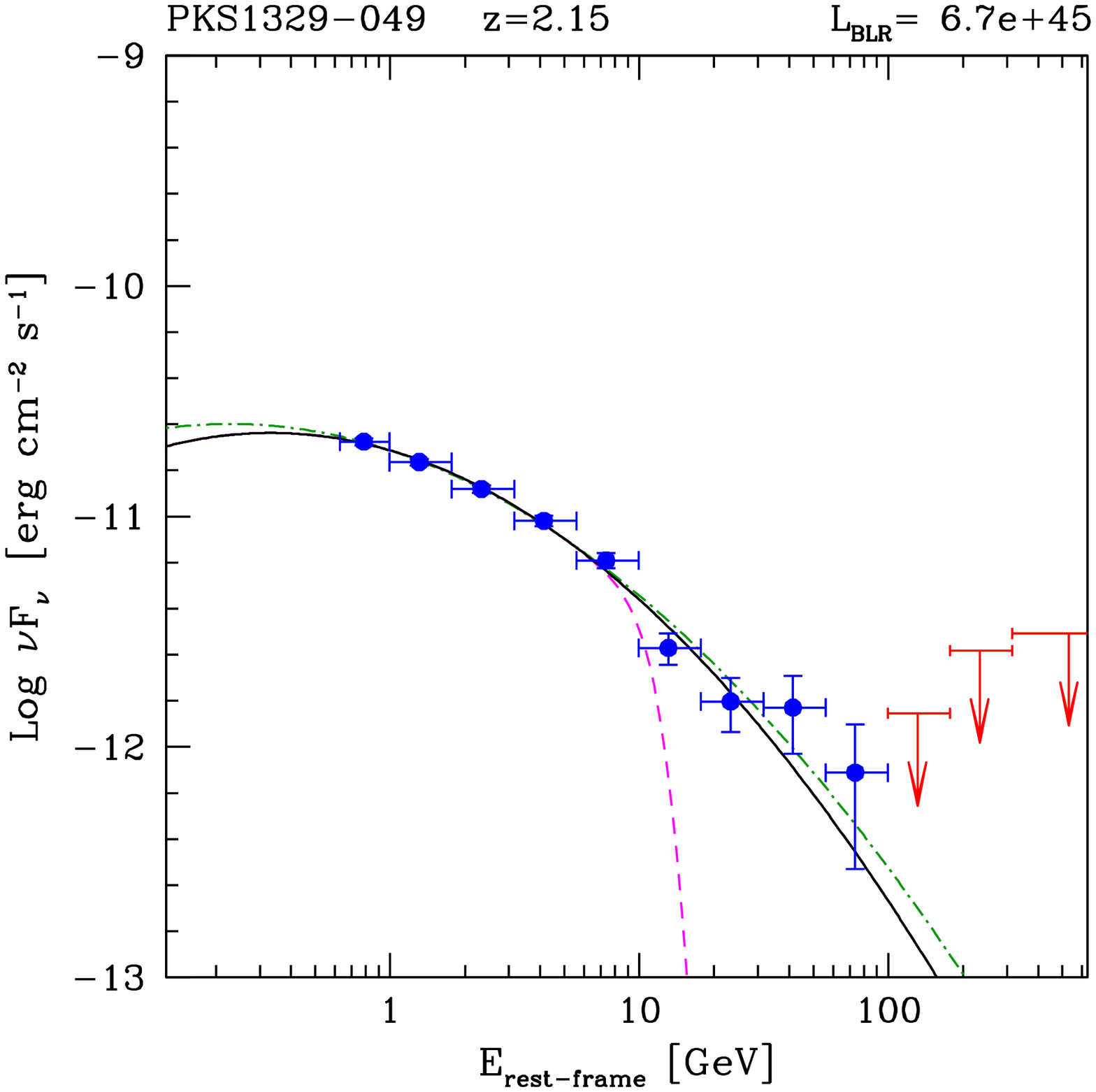,width=4.3cm,height=3.2cm } 
\end{tabular}
\contcaption{}  
\end{figure*}

\begin{figure*}
\vspace{1cm}
\begin{tabular}{cccc}
 \psfig{file=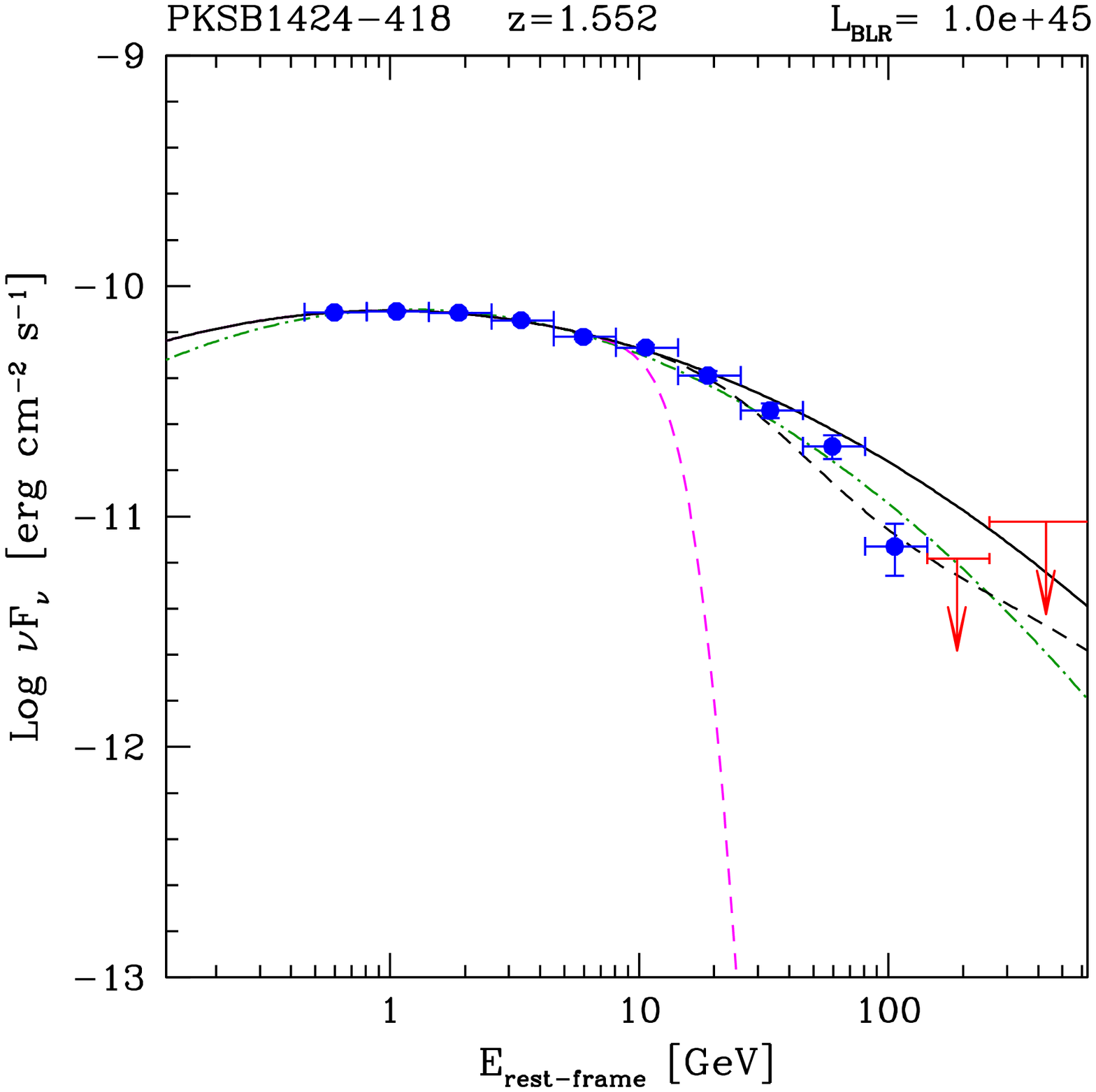,width=4.3cm,height=3.2cm } 
&\psfig{file=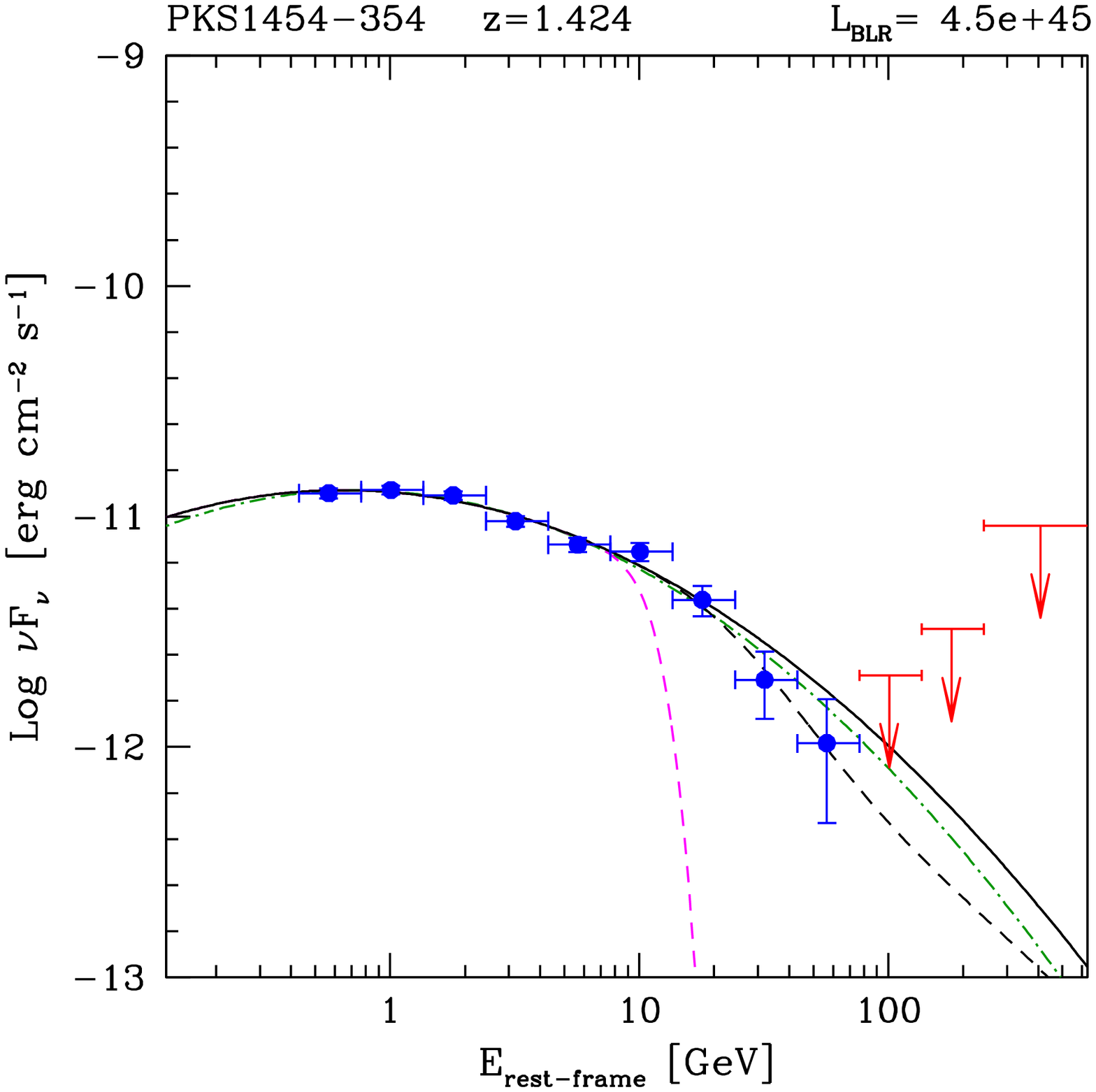,width=4.3cm,height=3.2cm }  
&\psfig{file=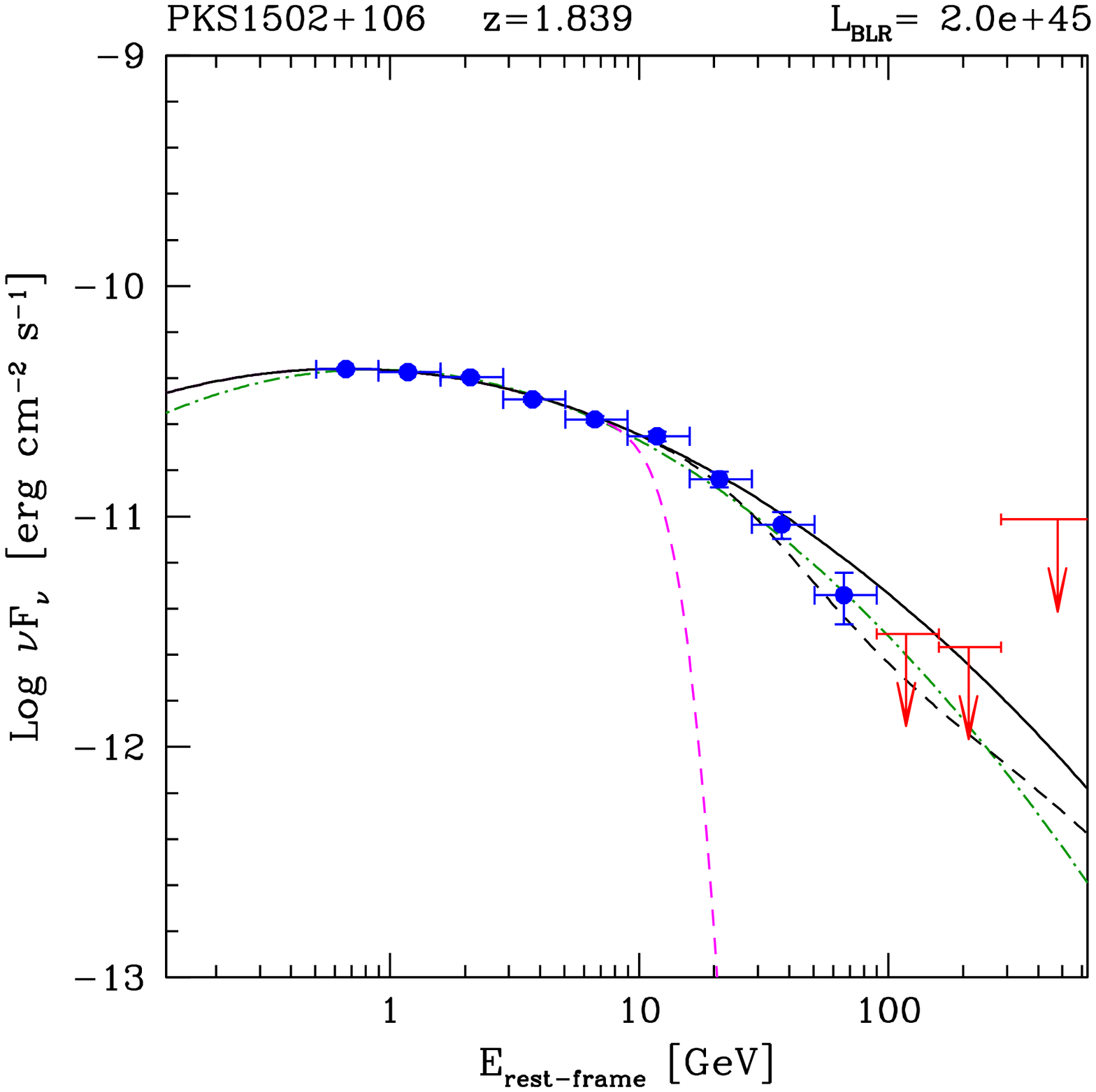,width=4.3cm,height=3.2cm } 
&\psfig{file=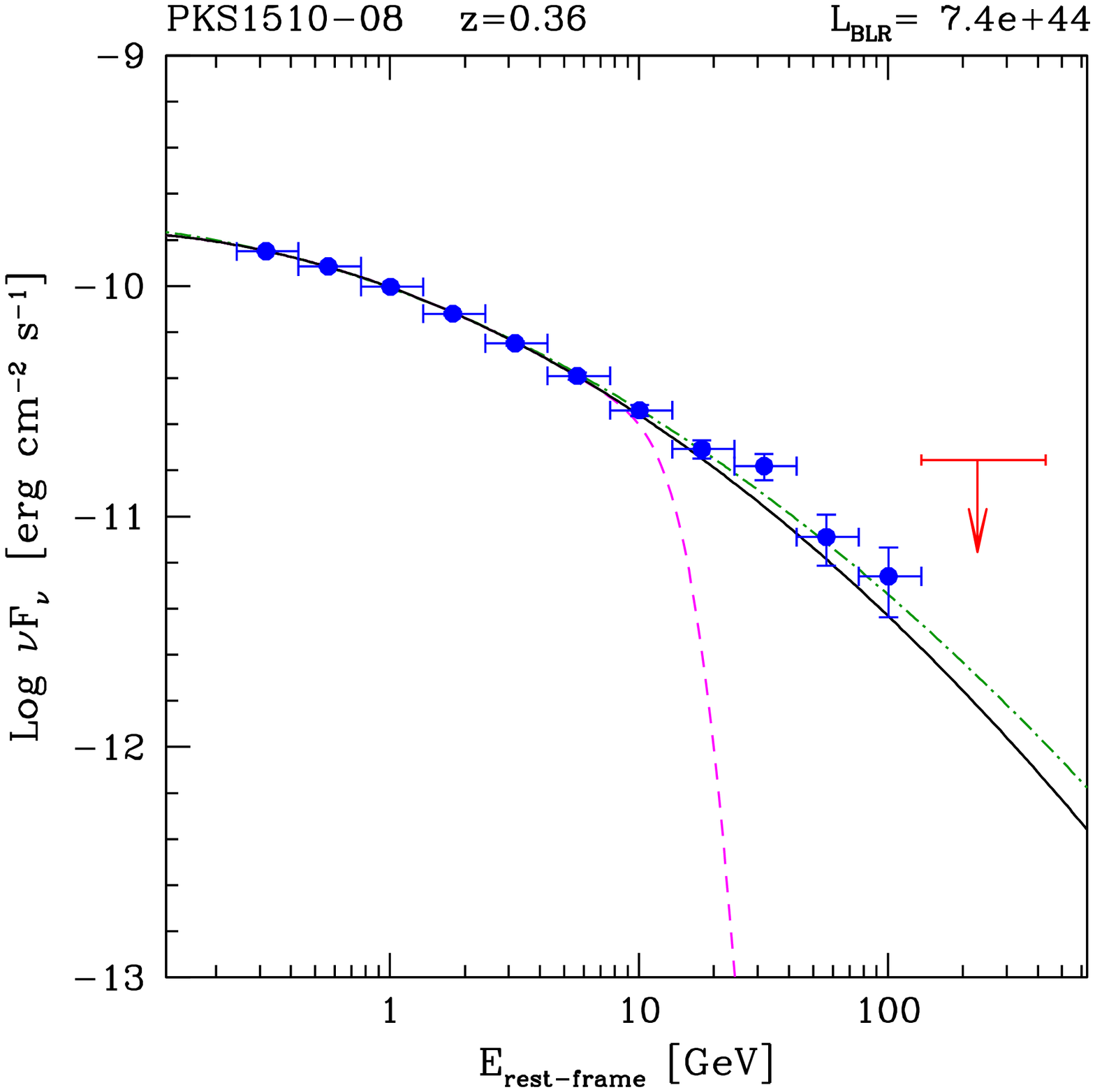,width=4.3cm,height=3.2cm }\vspace{1.2cm} \\    
\psfig{file=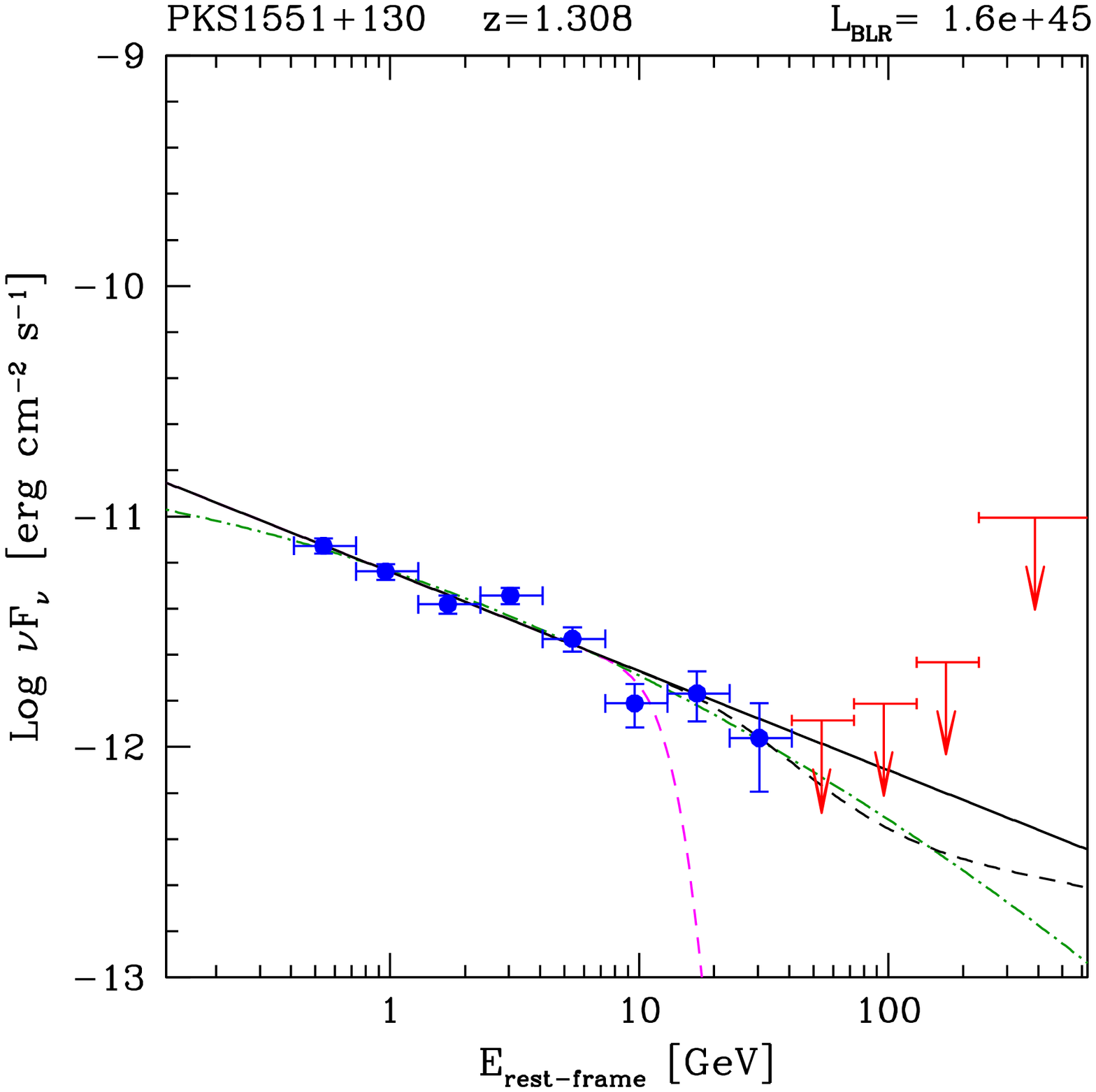,width=4.3cm,height=3.2cm } 
&\psfig{file=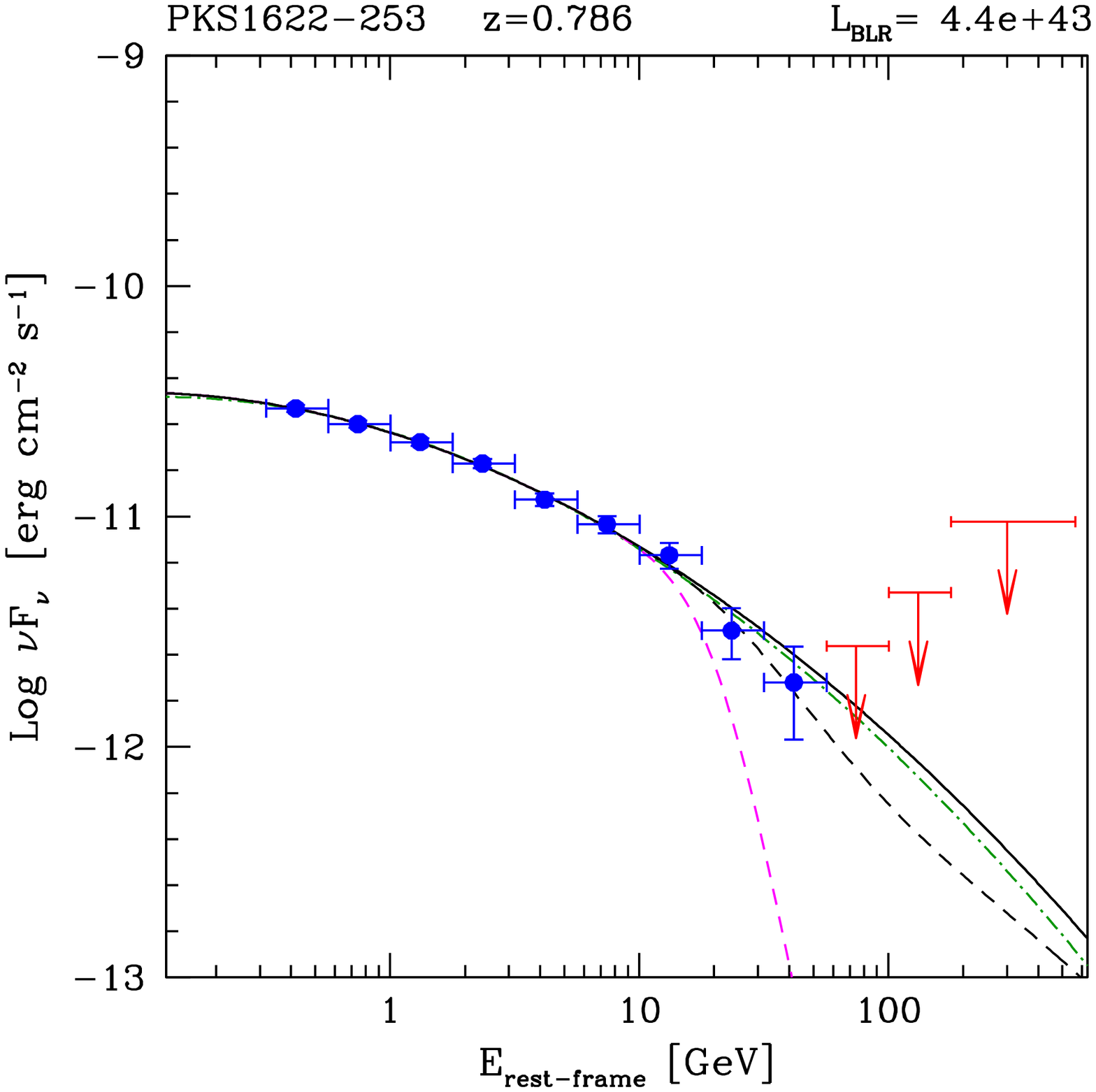,width=4.3cm,height=3.2cm }  
&\psfig{file=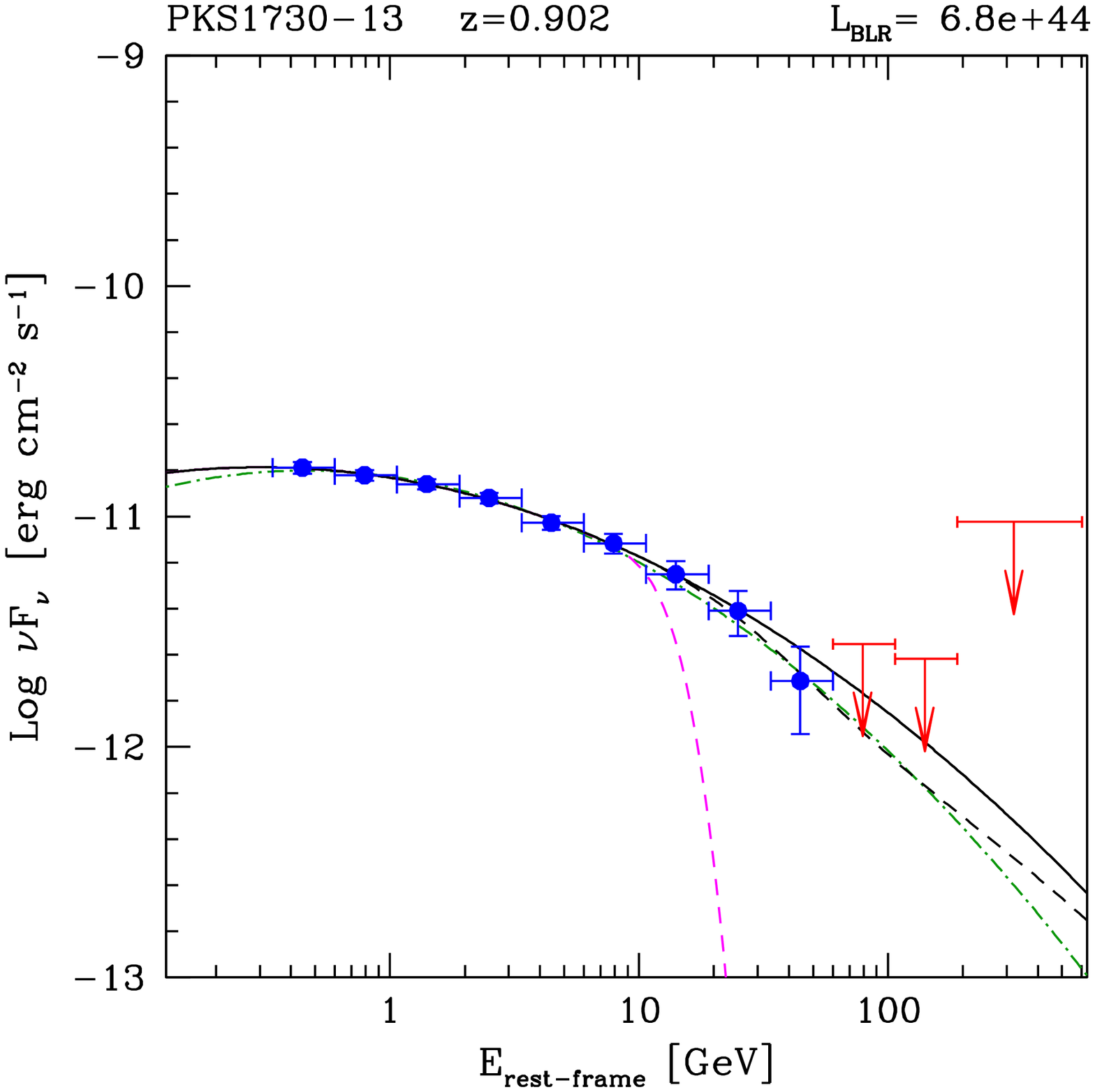,width=4.3cm,height=3.2cm } 
&\psfig{file=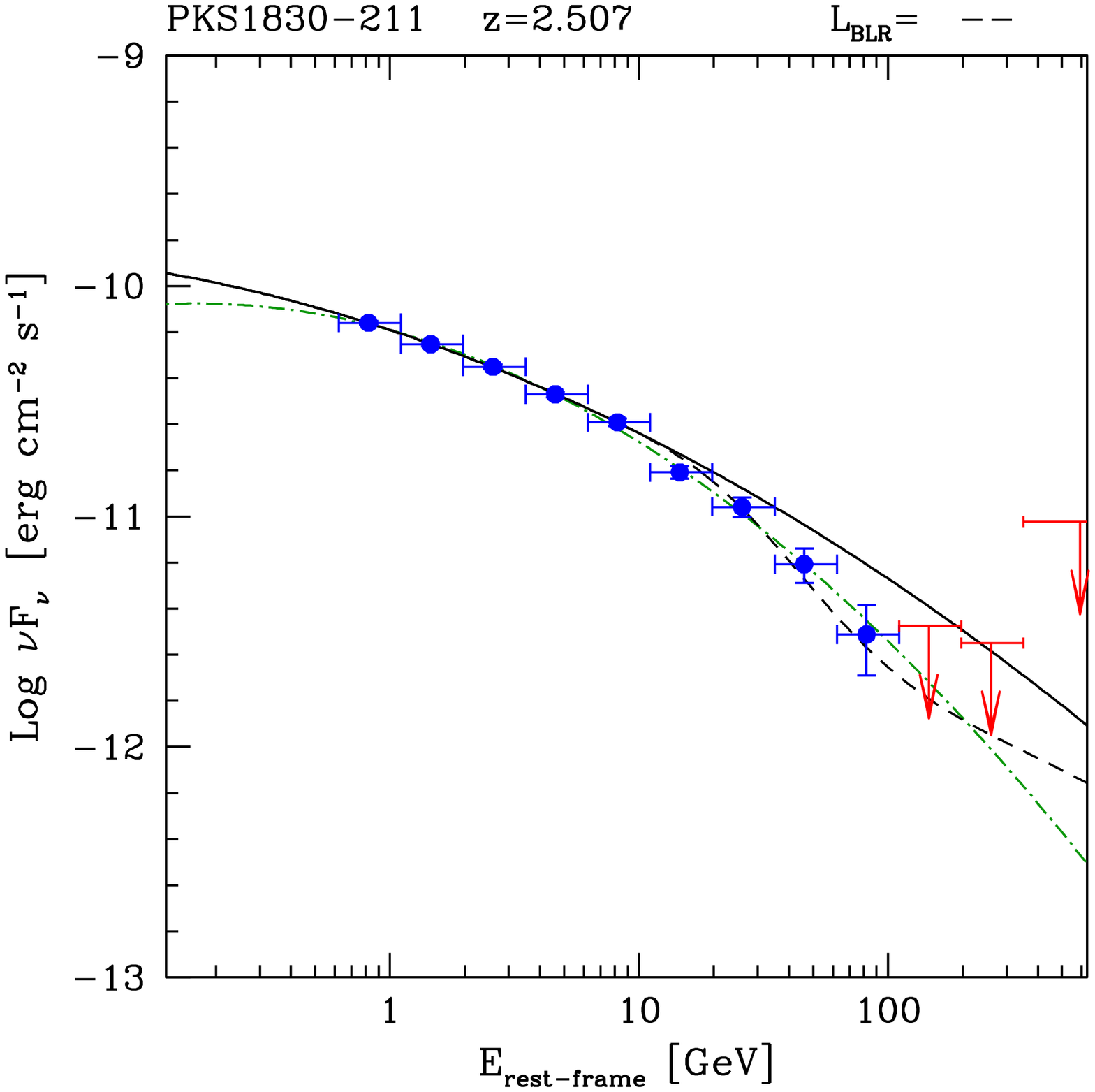,width=4.3cm,height=3.2cm }\vspace{1.2cm} \\ 
\psfig{file=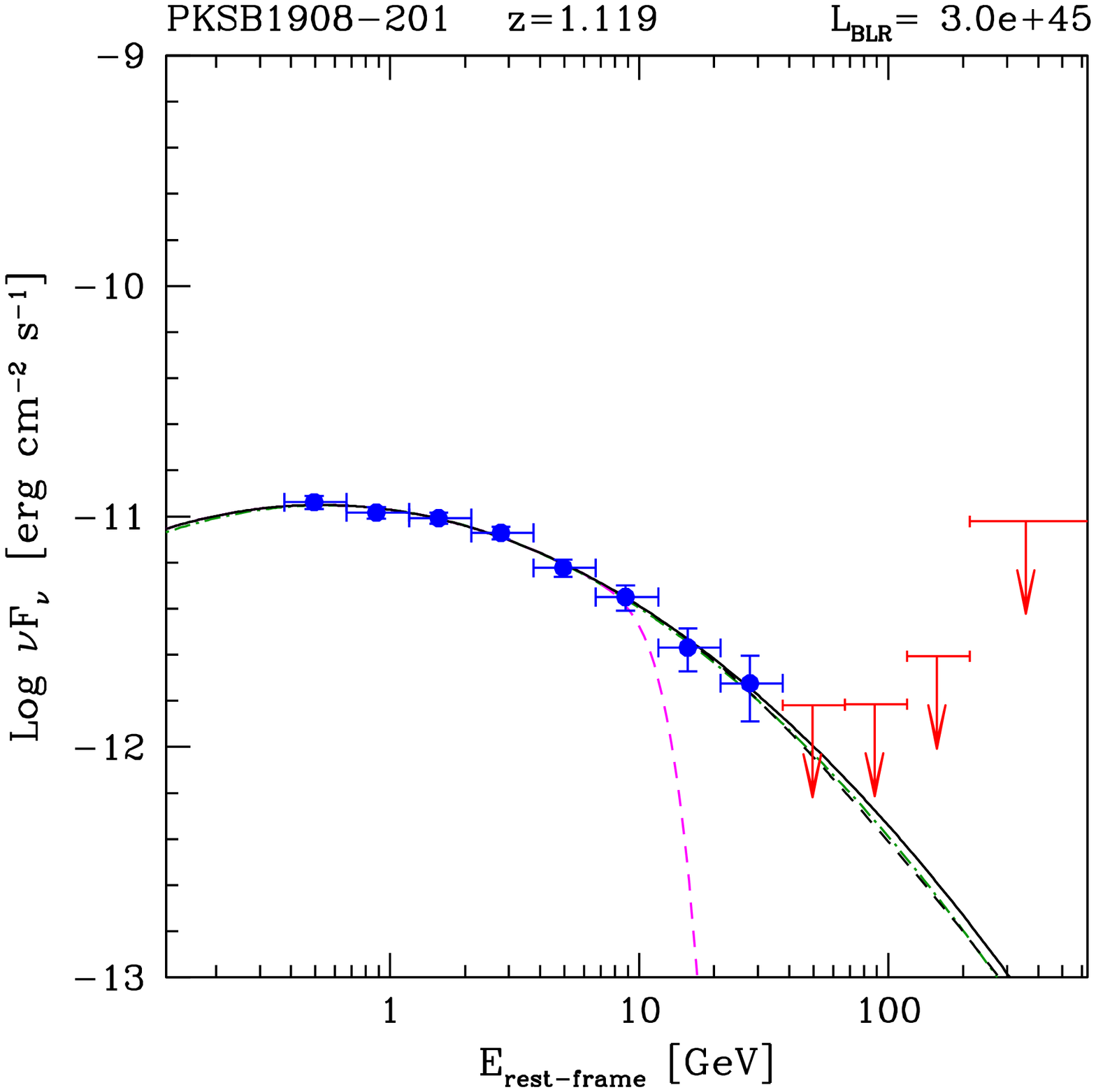,width=4.3cm,height=3.2cm } 
&\psfig{file=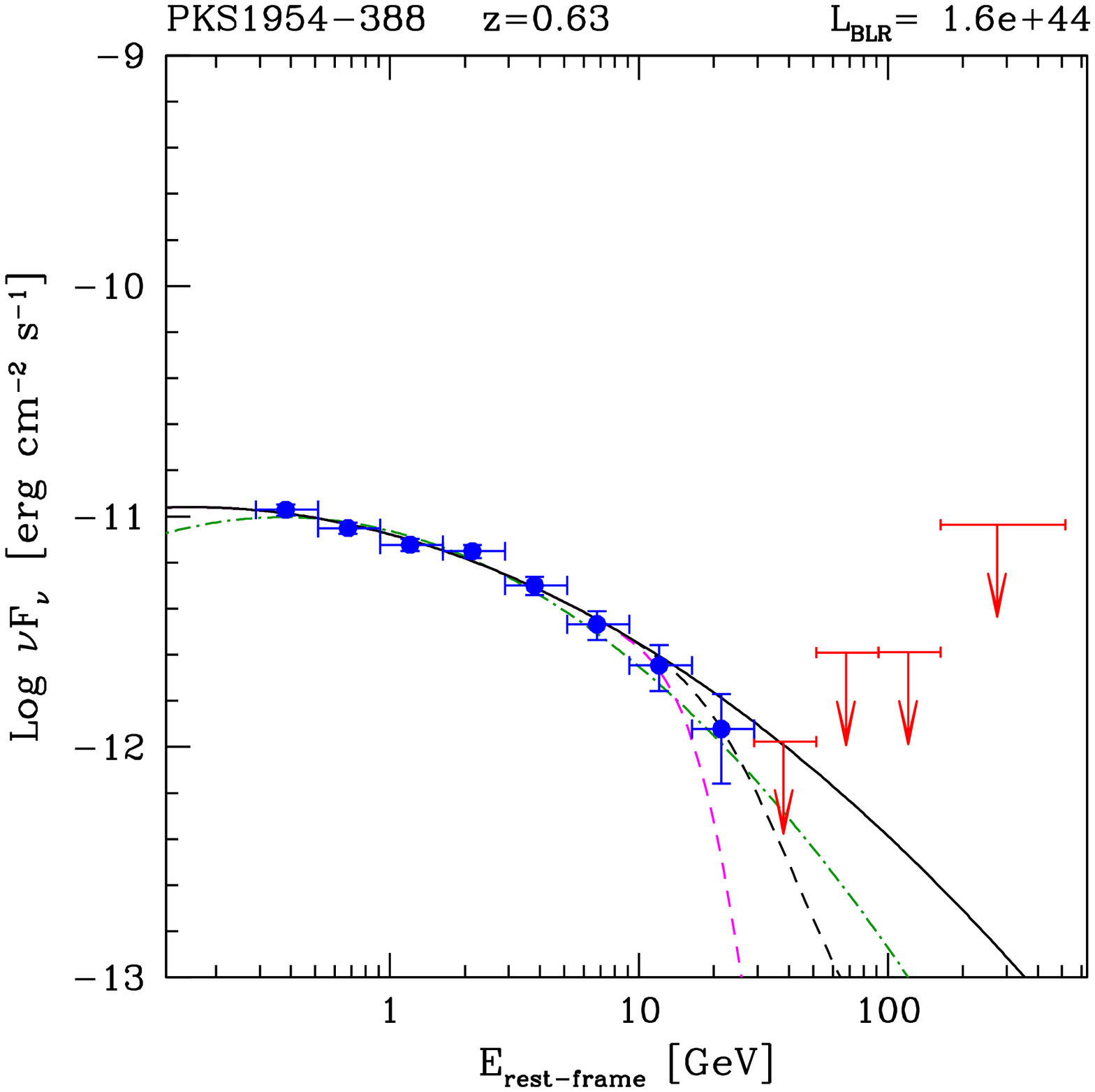,width=4.3cm,height=3.2cm }  
&\psfig{file=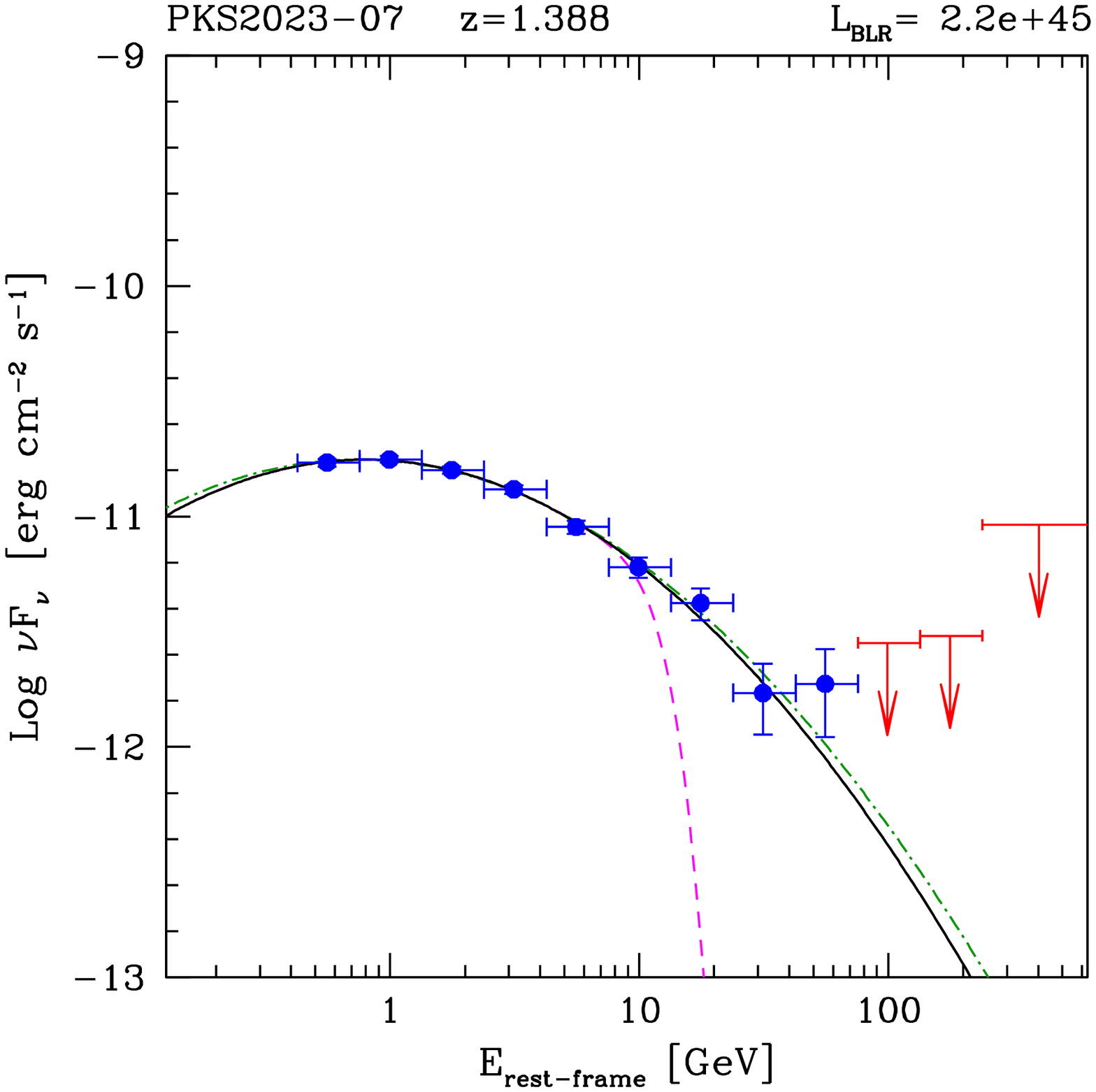,width=4.3cm,height=3.2cm }  
&\psfig{file=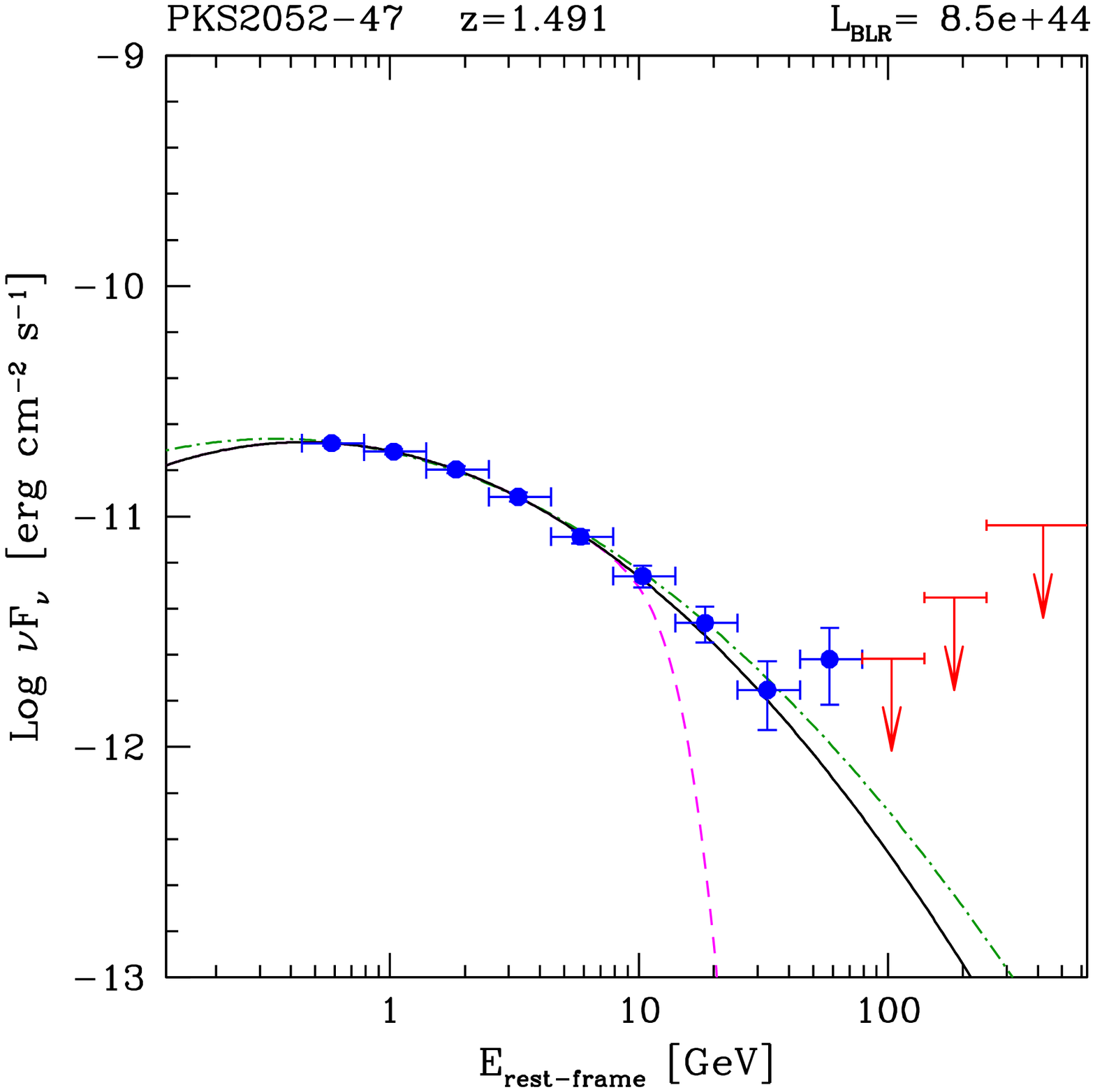,width=4.3cm,height=3.2cm }\vspace{1.2cm} \\
 \psfig{file=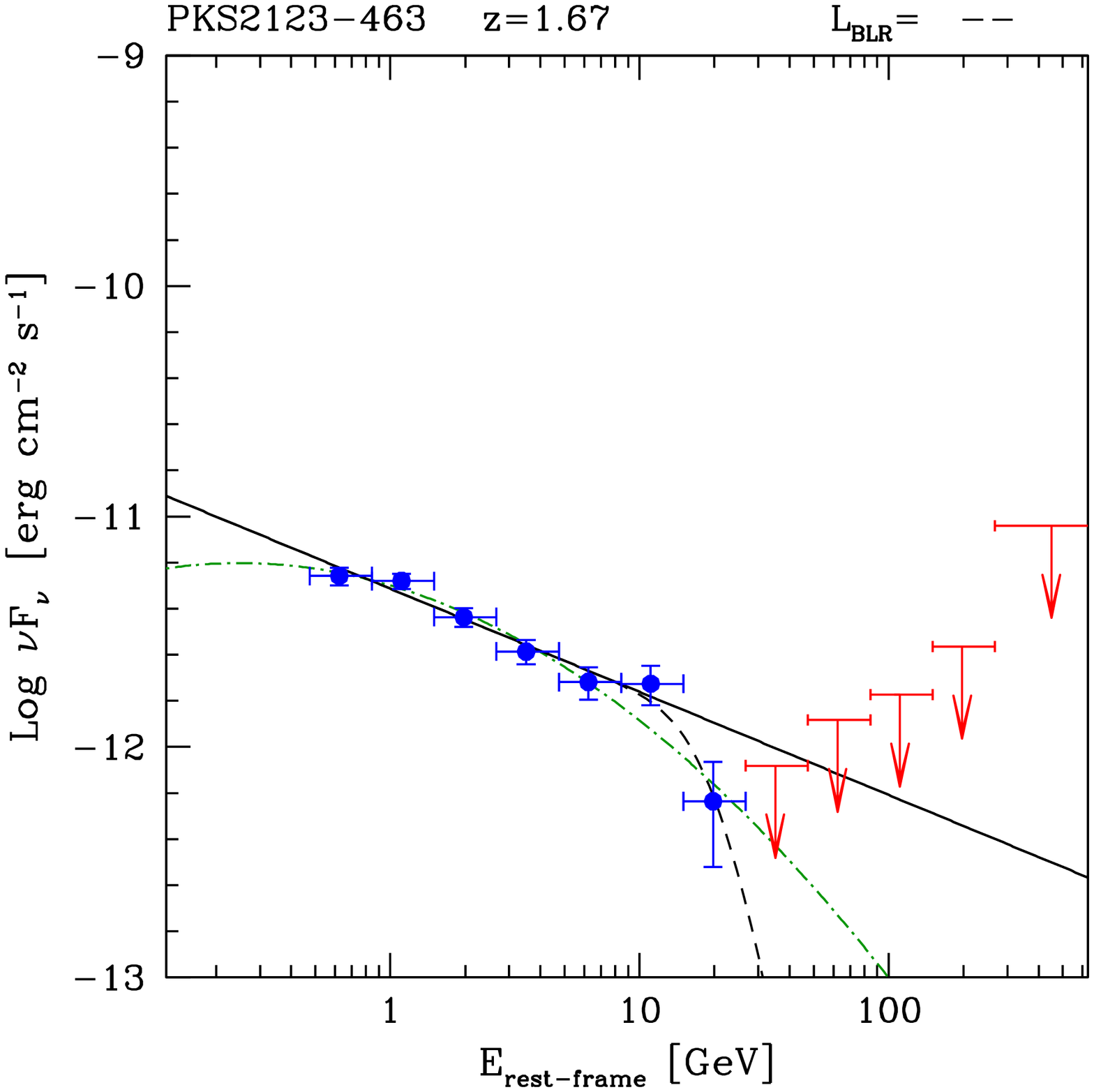,width=4.3cm,height=3.2cm }  
&\psfig{file=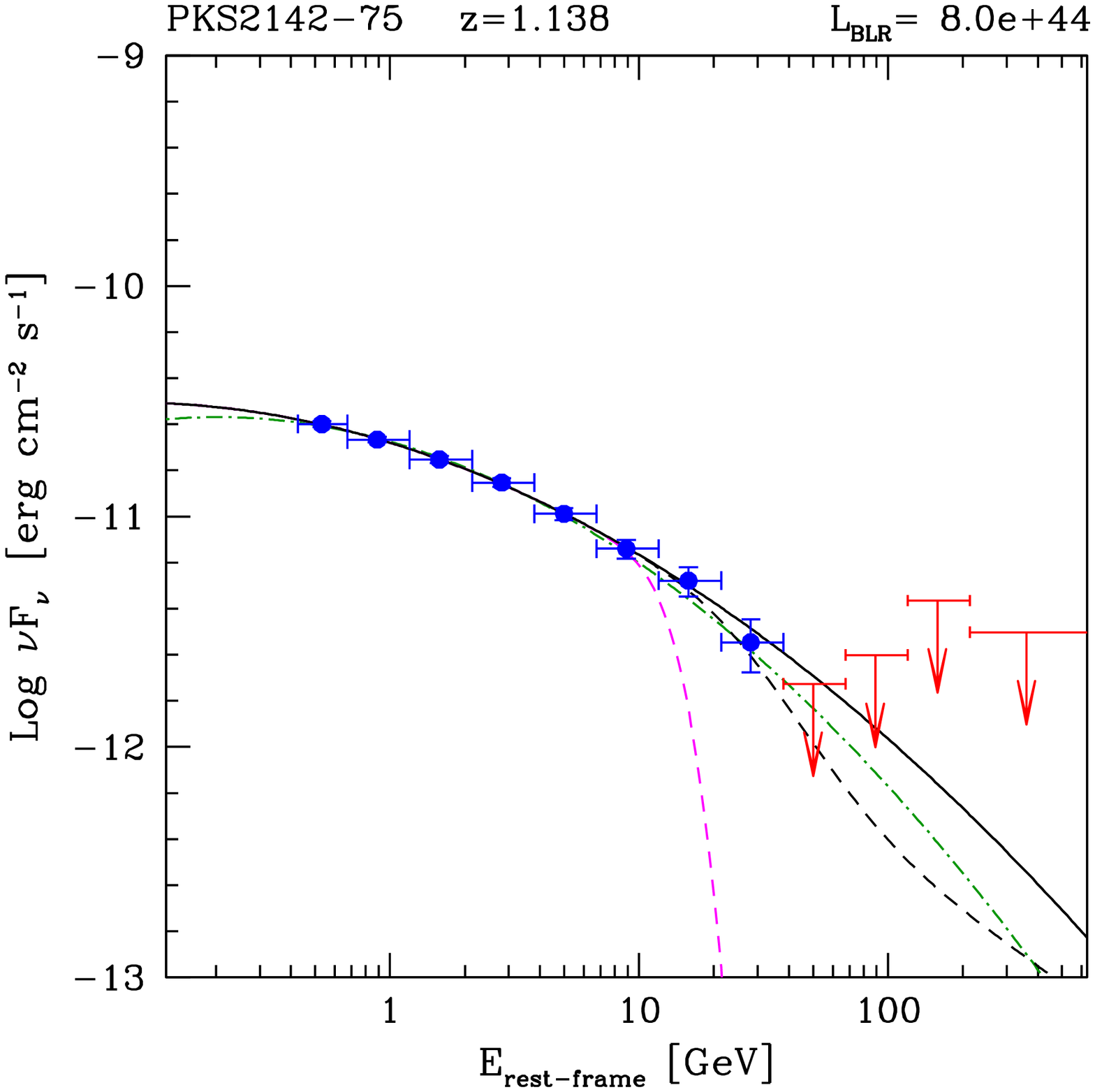,width=4.3cm,height=3.2cm } 
&\psfig{file=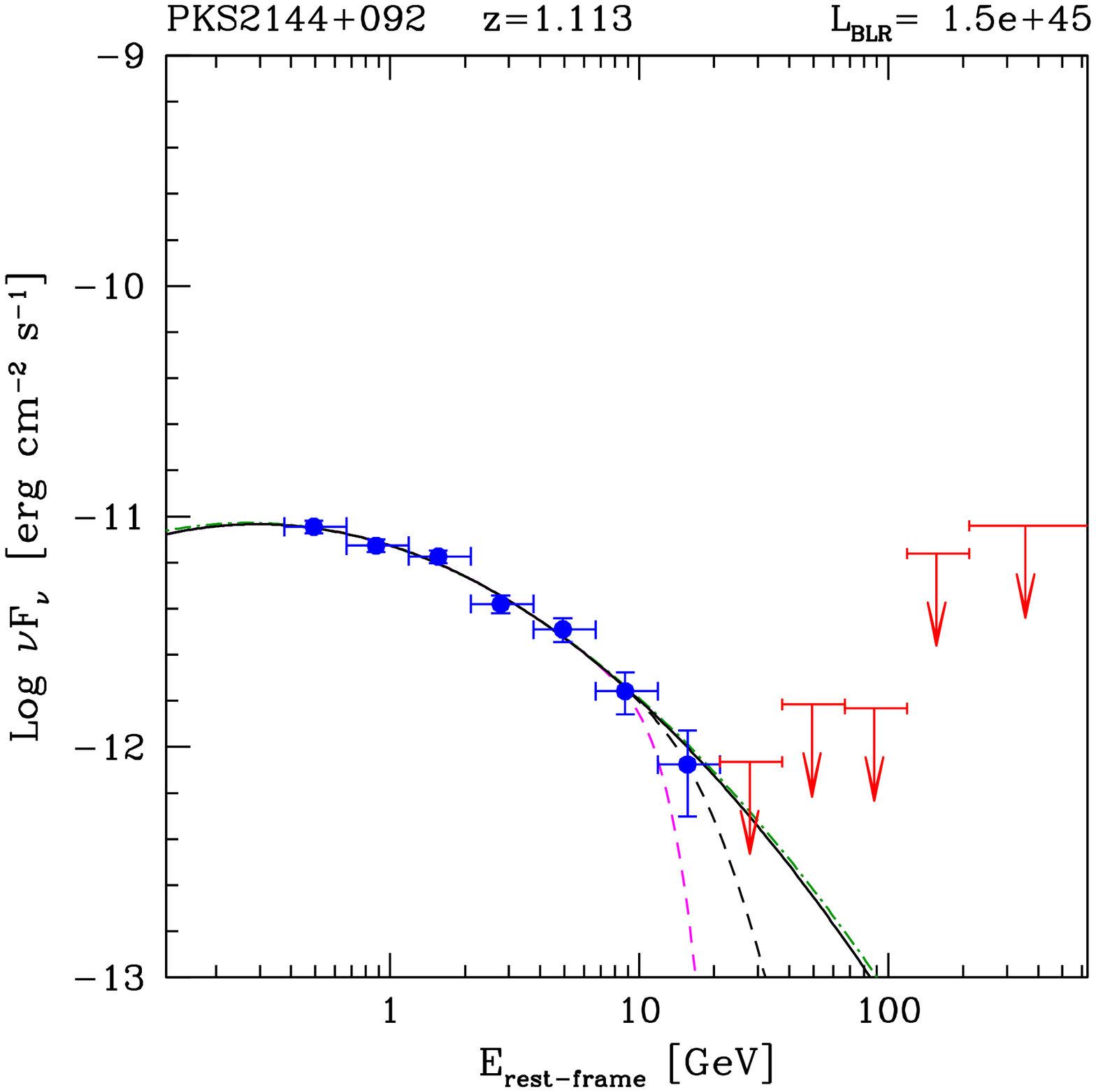,width=4.3cm,height=3.2cm }   
&\psfig{file=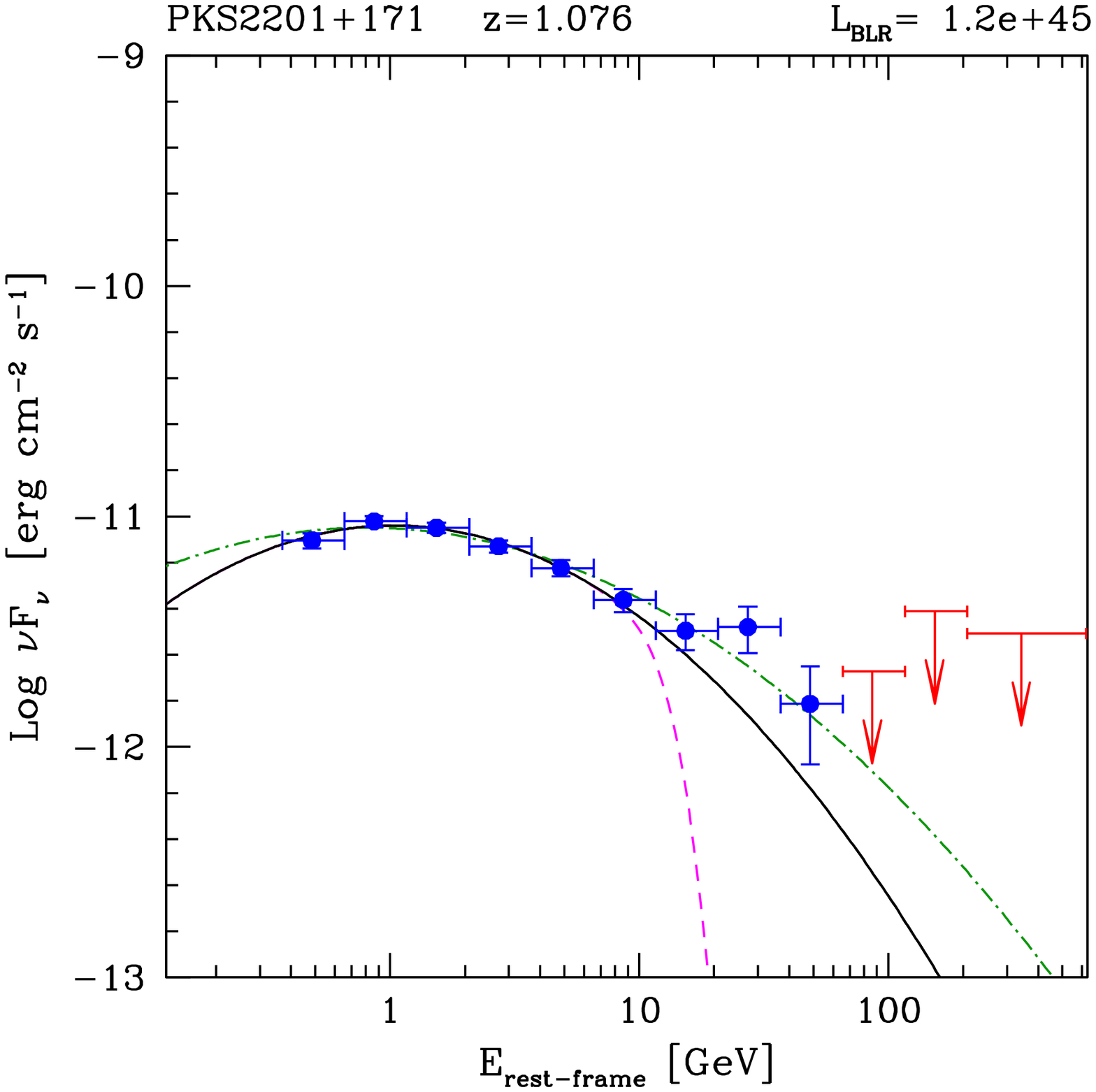,width=4.3cm,height=3.2cm }  \vspace{1.2cm} \\
 \psfig{file=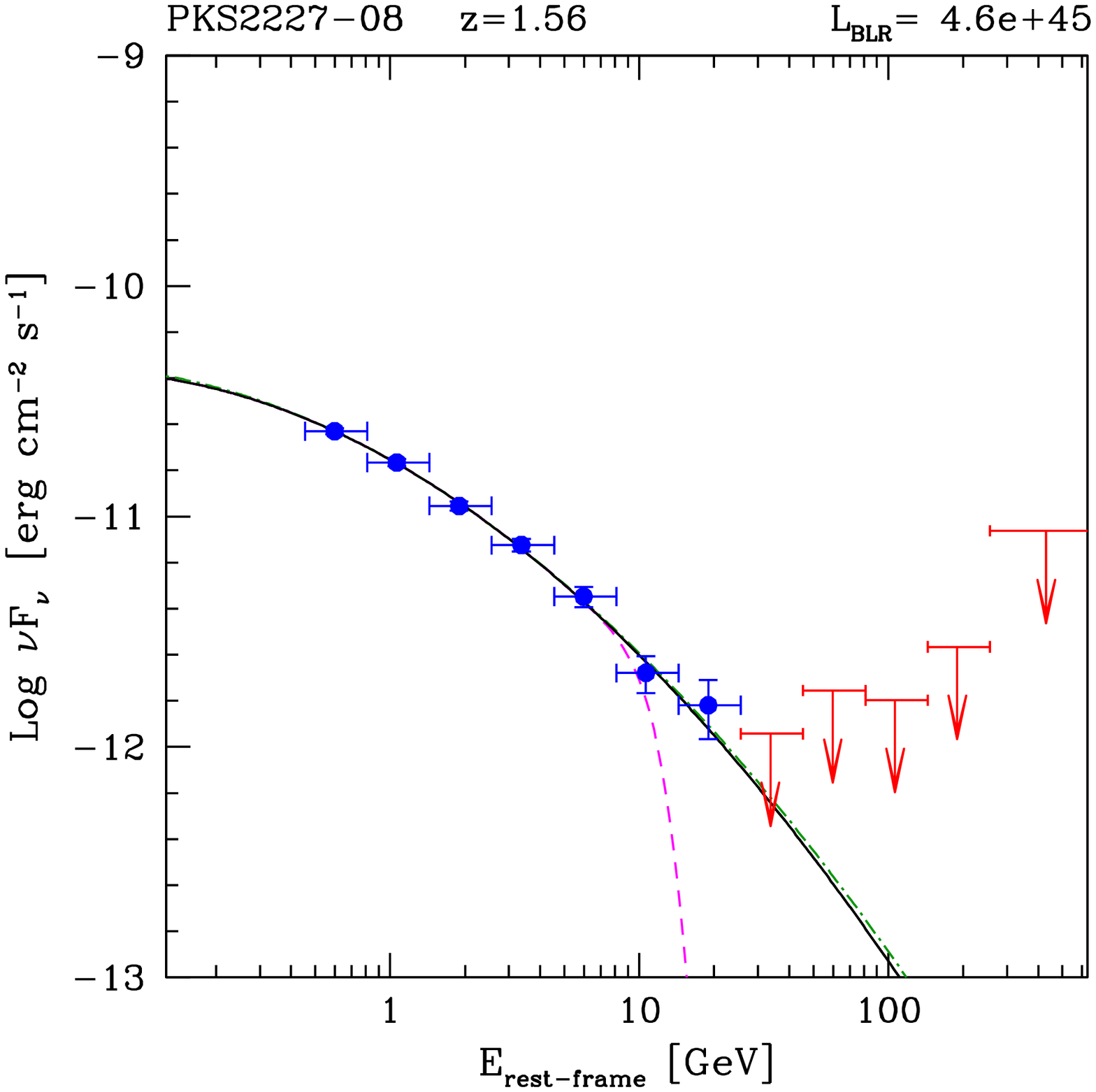,width=4.3cm,height=3.2cm }     
&\psfig{file=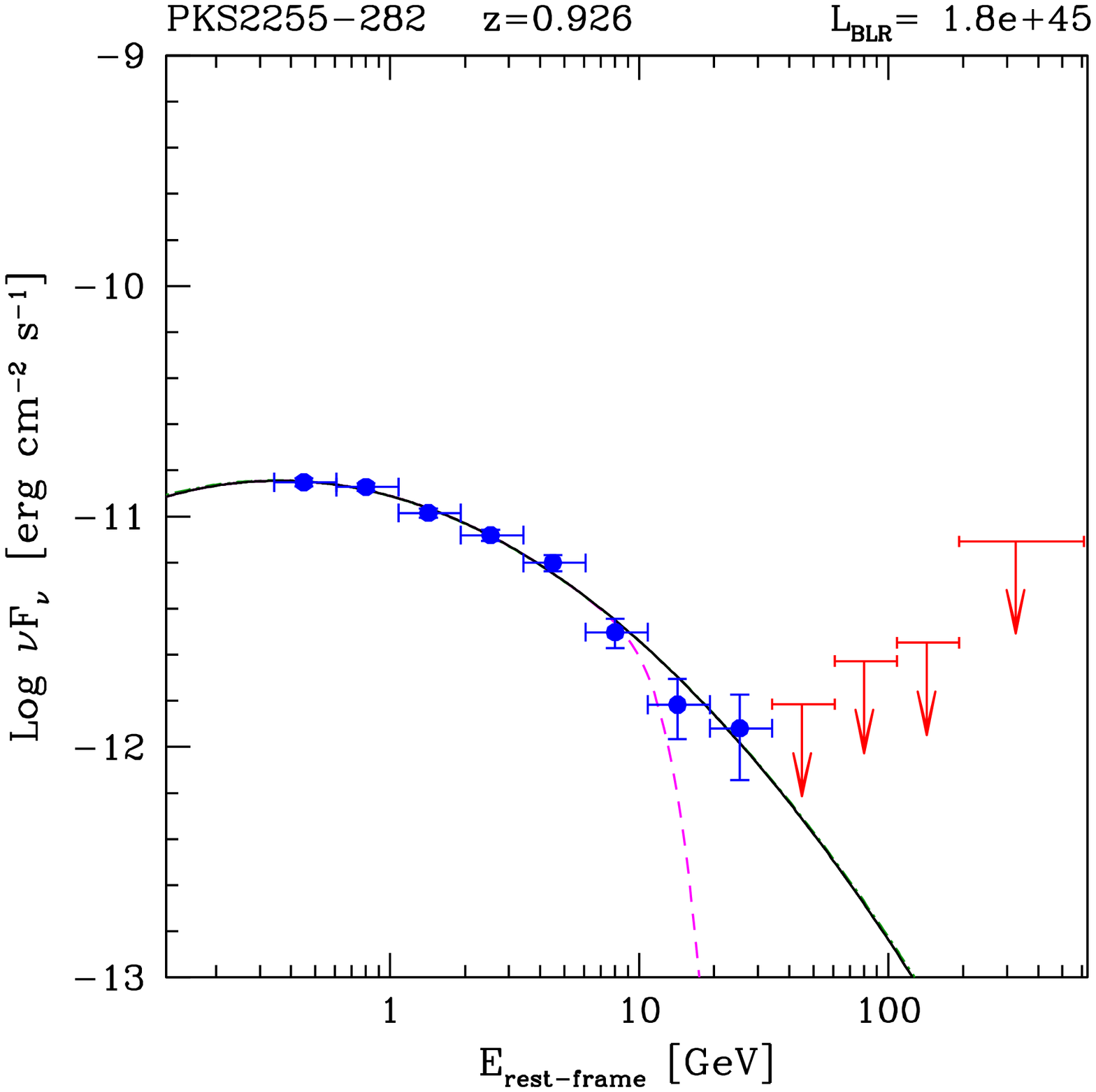,width=4.3cm,height=3.2cm } 
&\psfig{file=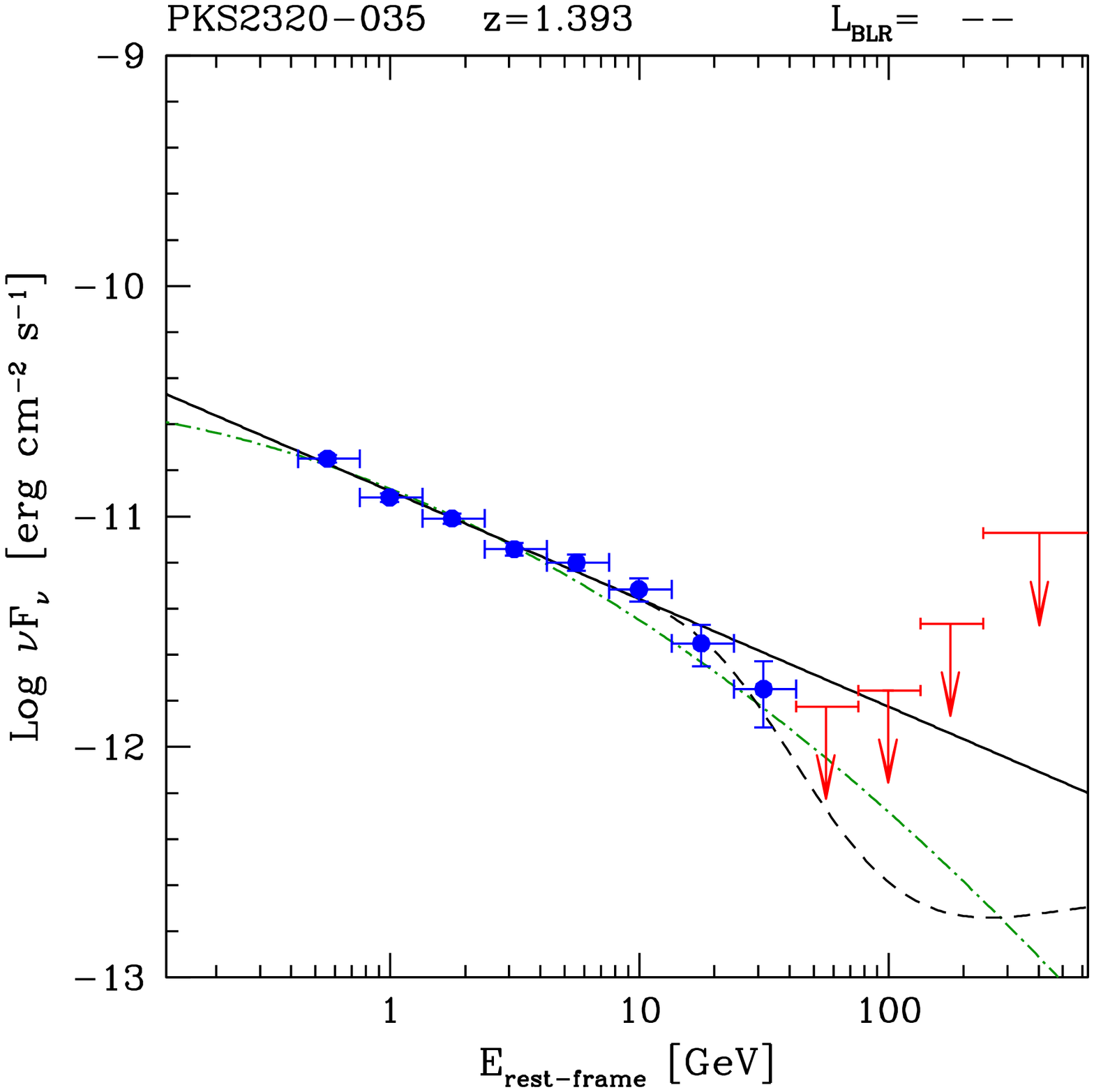,width=4.3cm,height=3.2cm }    
&\psfig{file=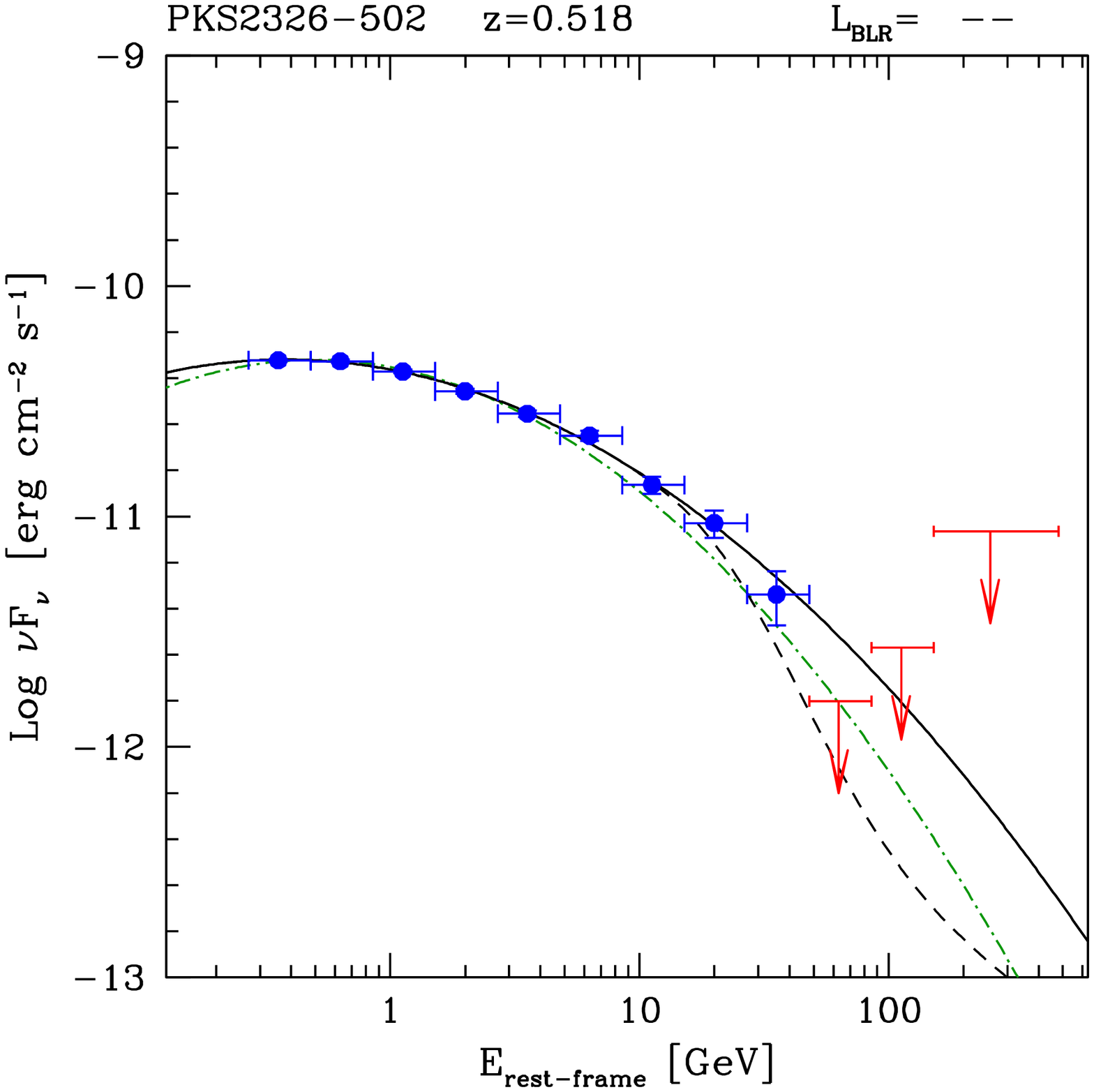,width=4.3cm,height=3.2cm } 
\end{tabular}
\contcaption{}  
\end{figure*}

\begin{figure*}
\vspace{1cm}
\begin{tabular}{cccc}
 \psfig{file=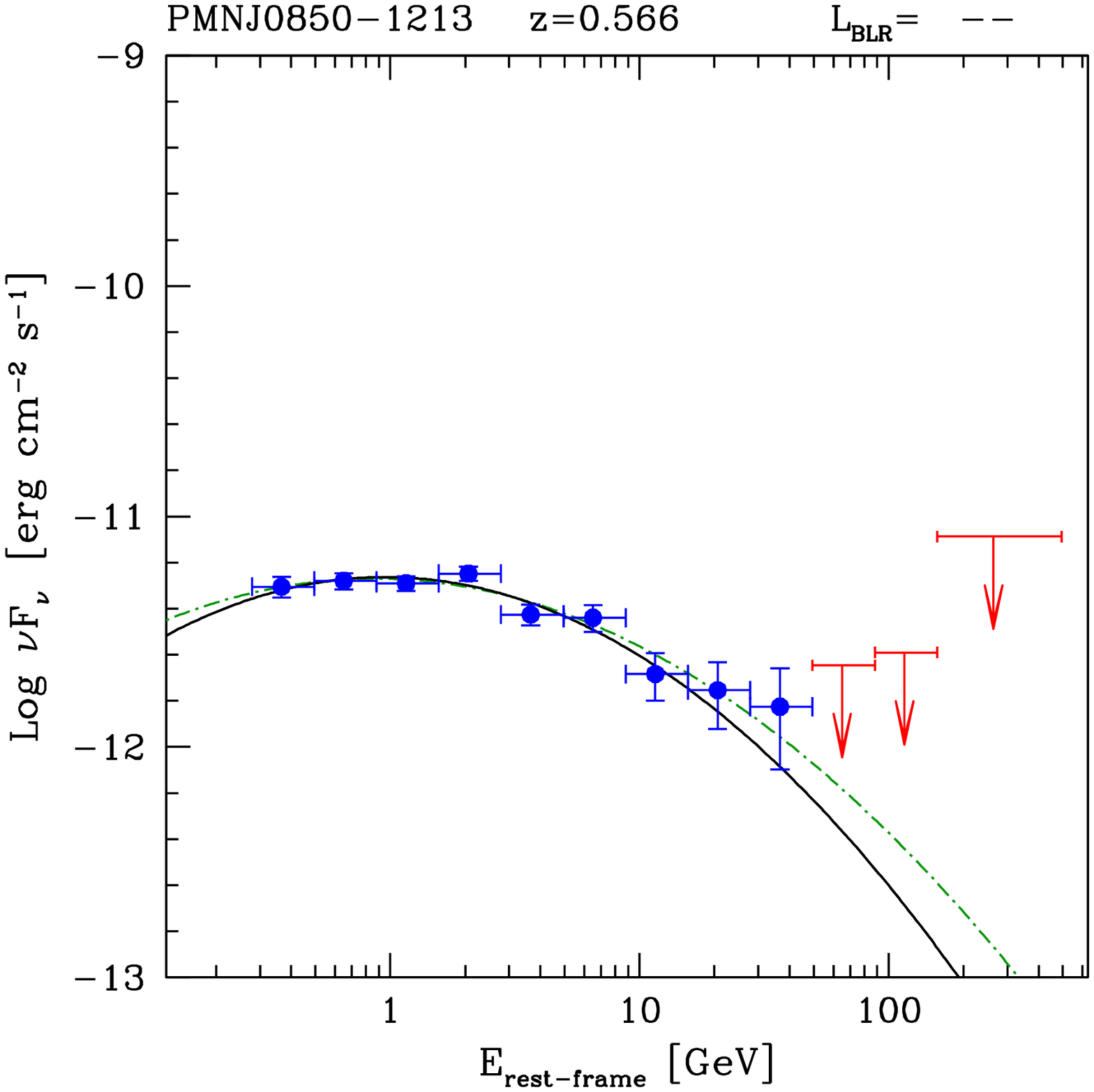,width=4.3cm,height=3.2cm }  
&\psfig{file=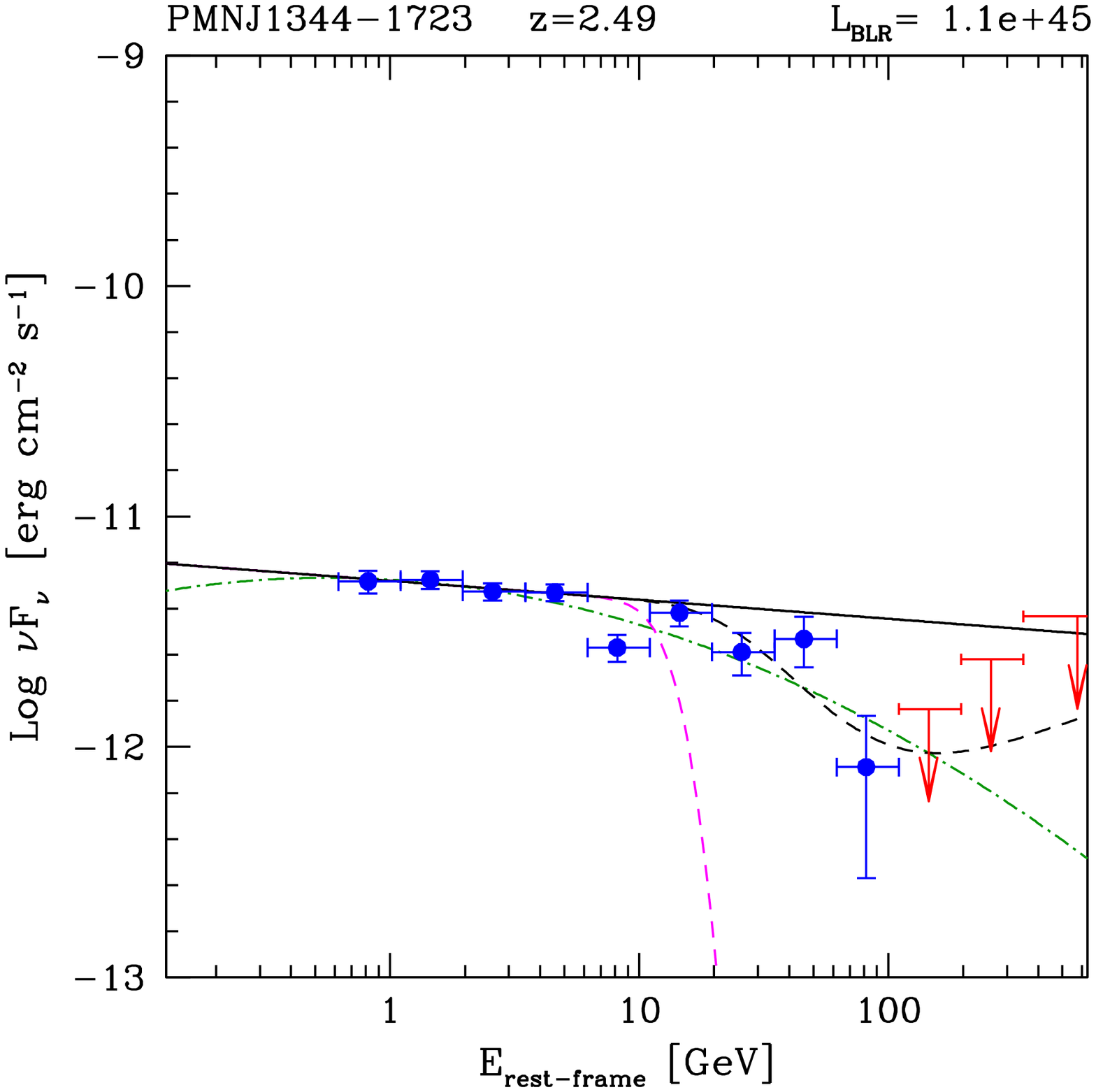,width=4.3cm,height=3.2cm } 
&\psfig{file=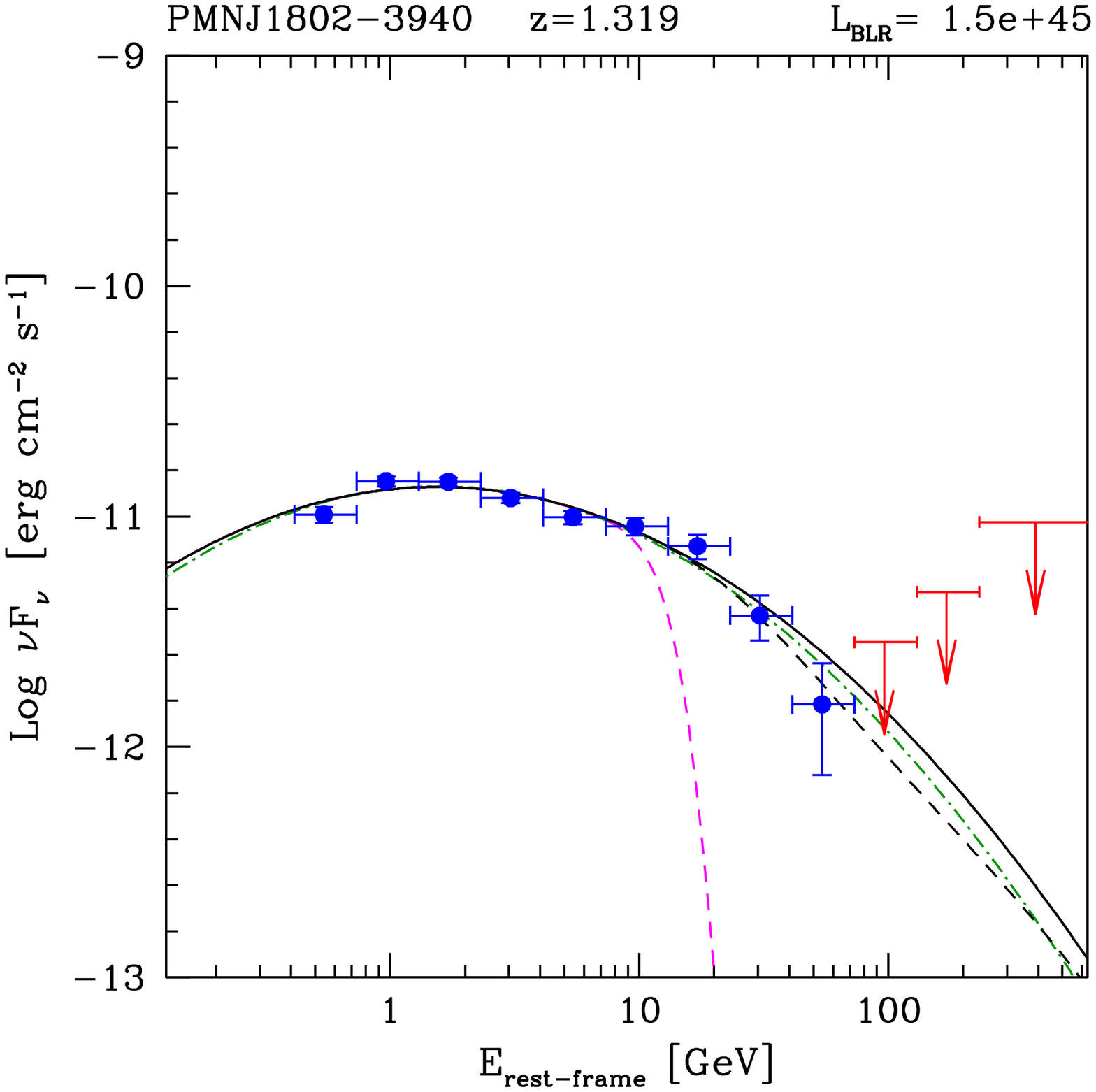,width=4.3cm,height=3.2cm }  
&\psfig{file=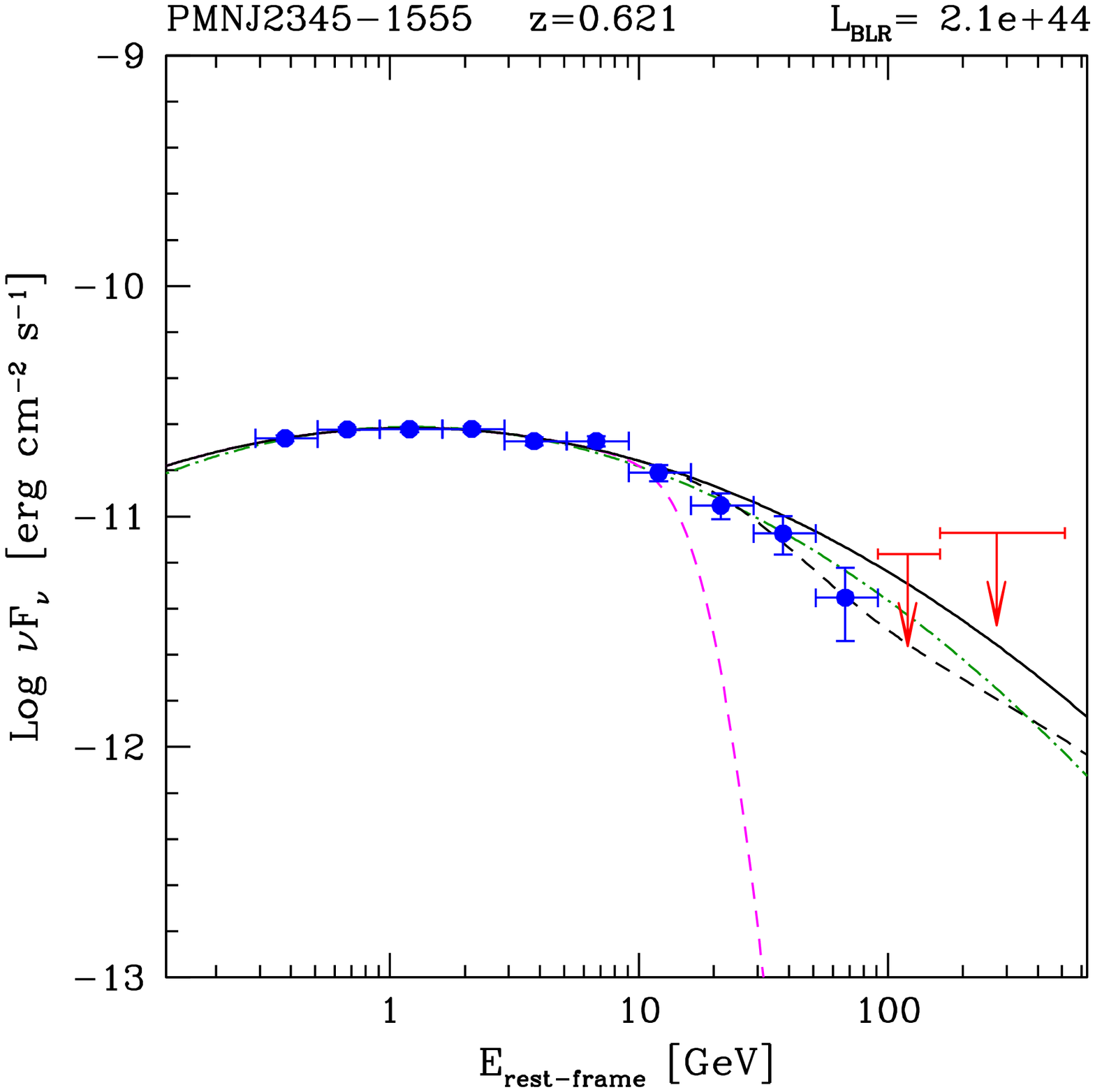,width=4.3cm,height=3.2cm }\vspace{1.2cm} \\
 \psfig{file=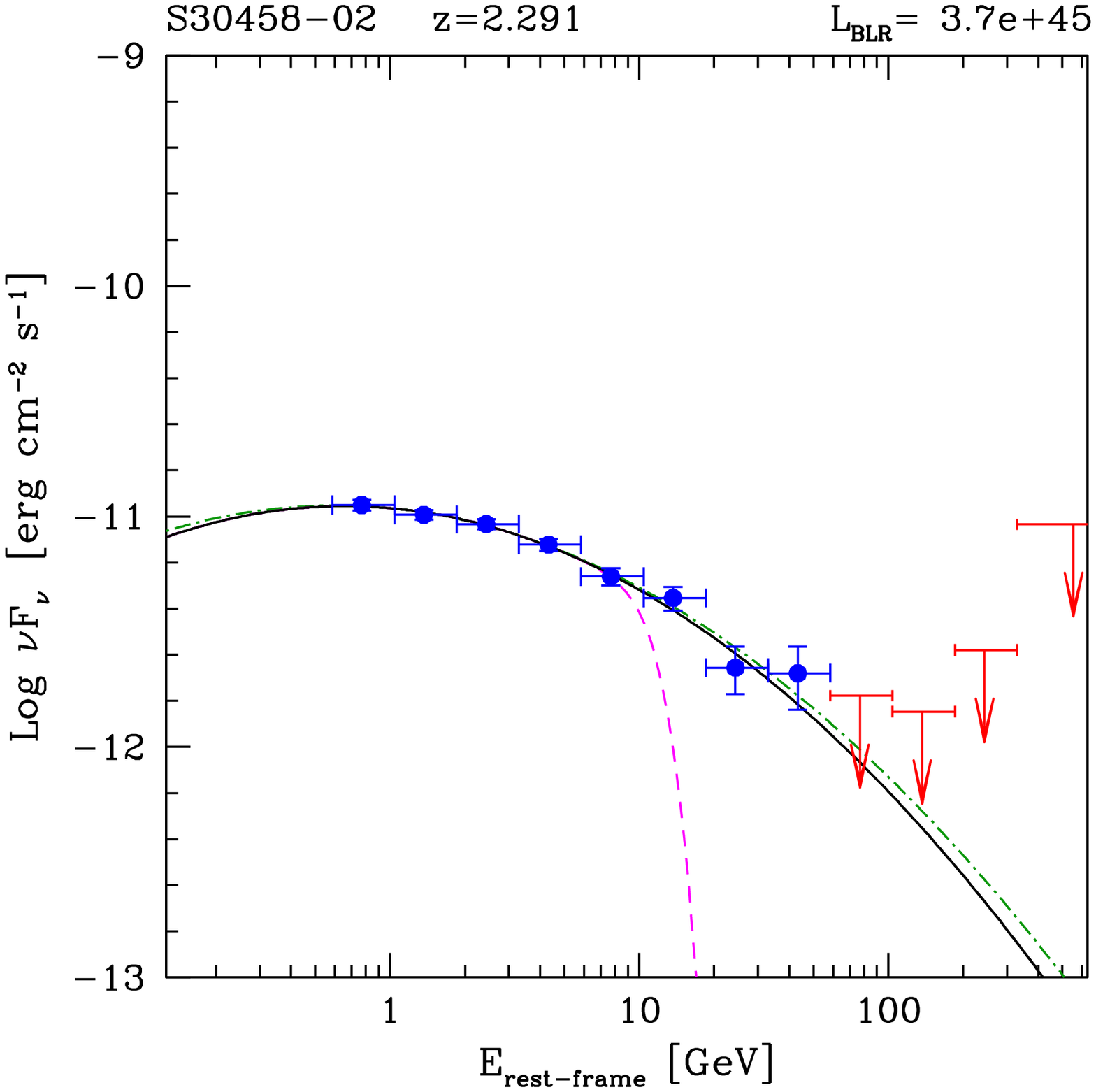,width=4.3cm,height=3.2cm }      
&\psfig{file=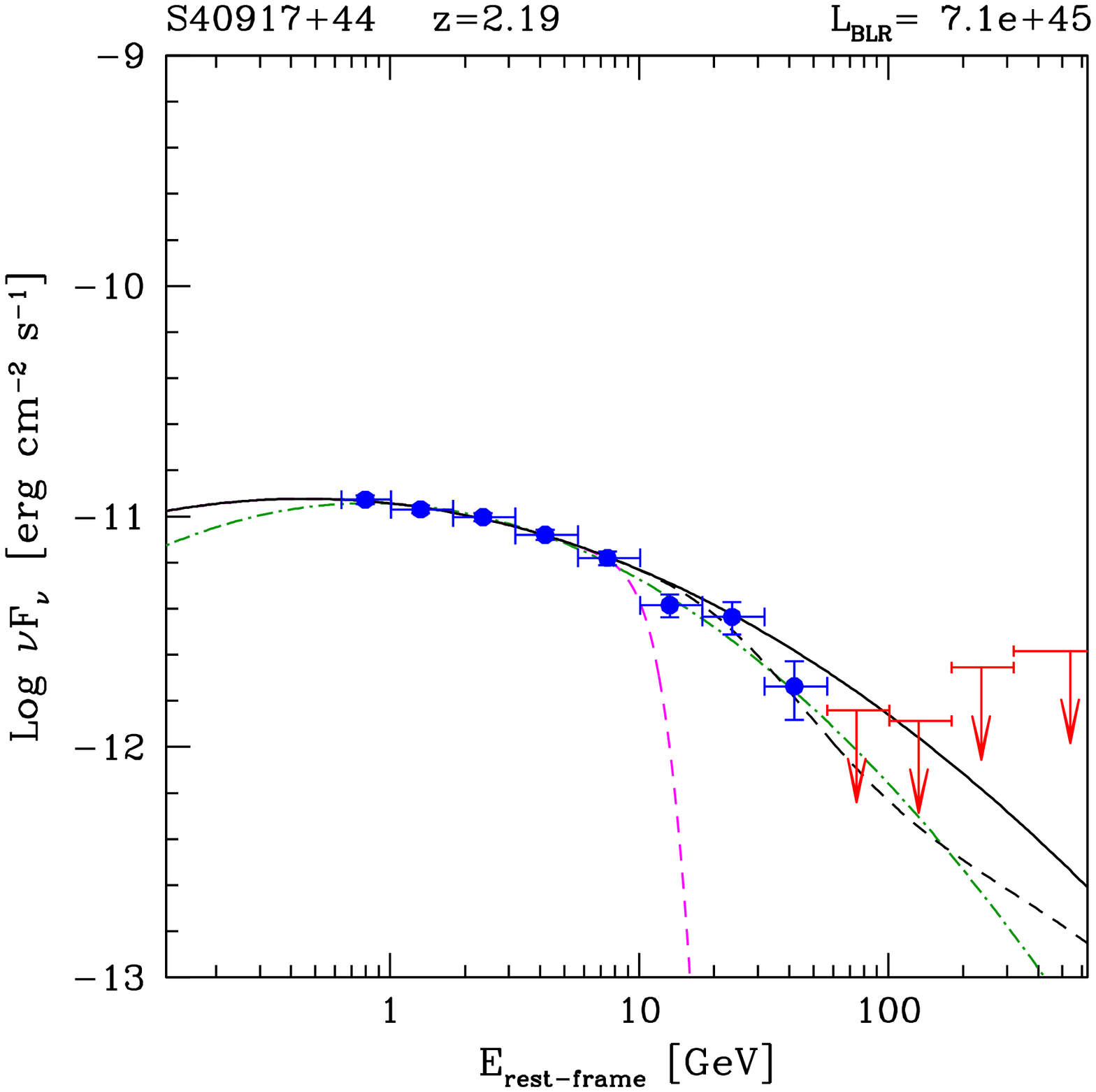,width=4.3cm,height=3.2cm } 
&\psfig{file=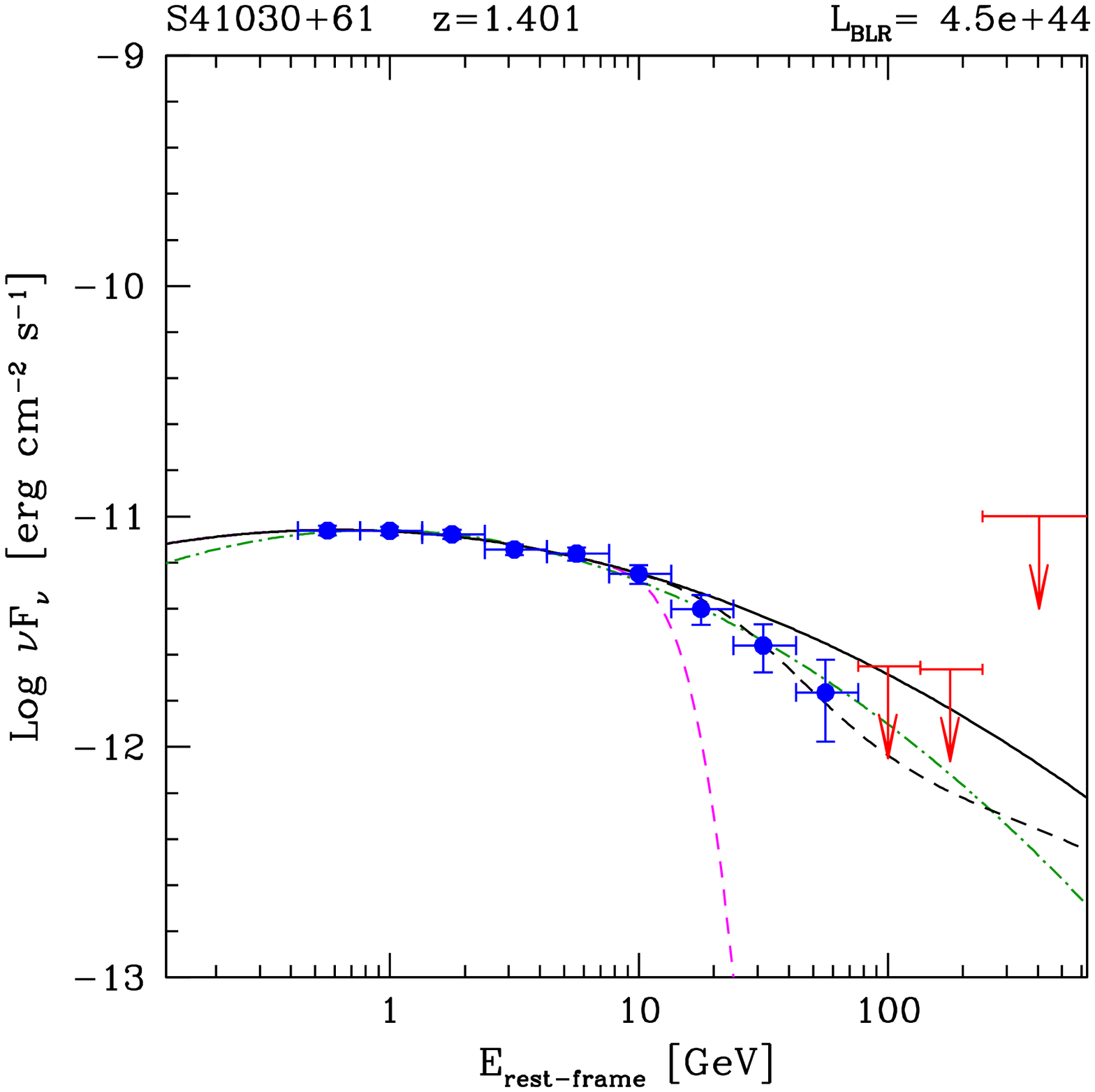,width=4.3cm,height=3.2cm }       
&\psfig{file=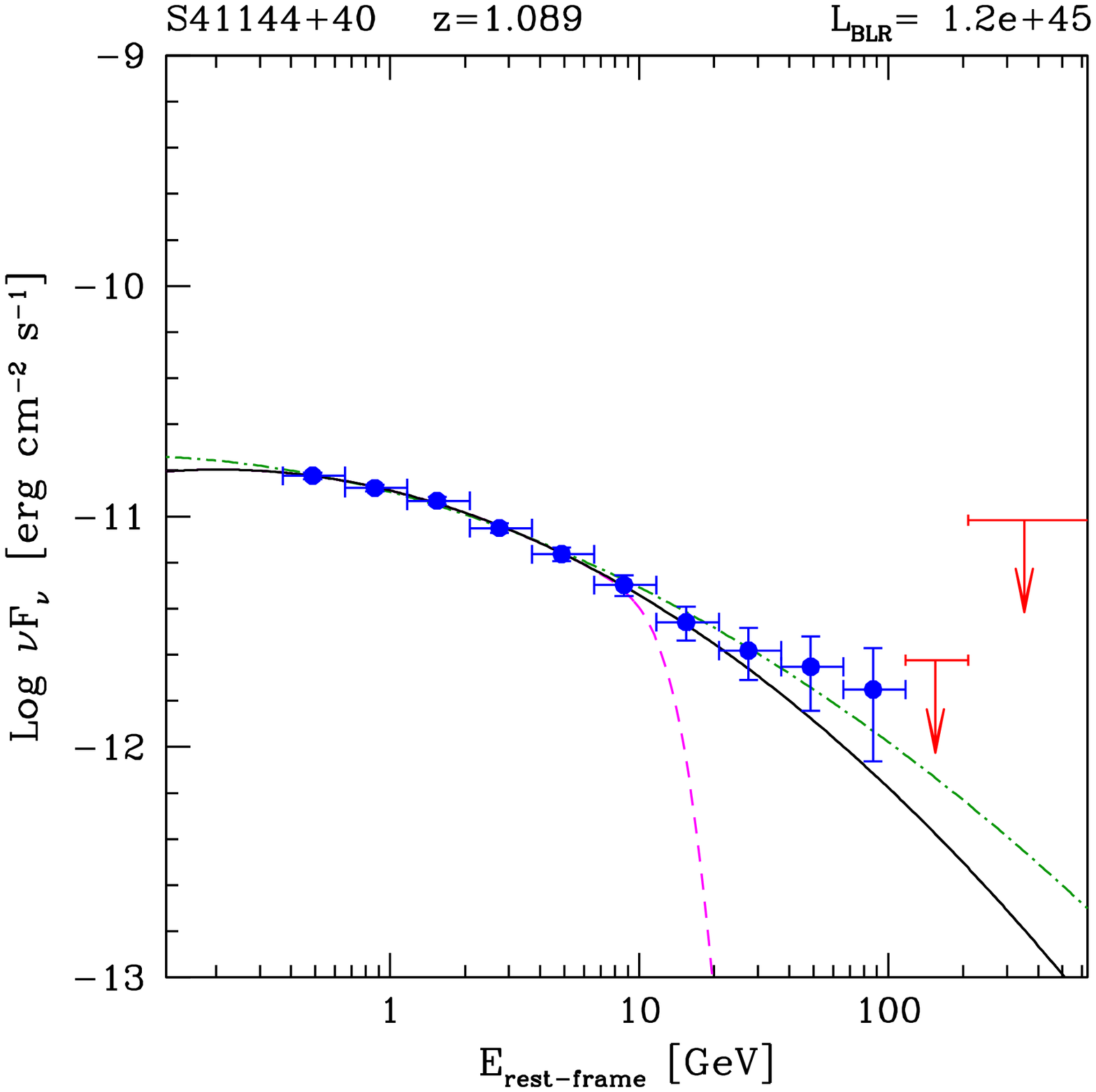,width=4.3cm,height=3.2cm } \vspace{1.2cm}\\
 \psfig{file=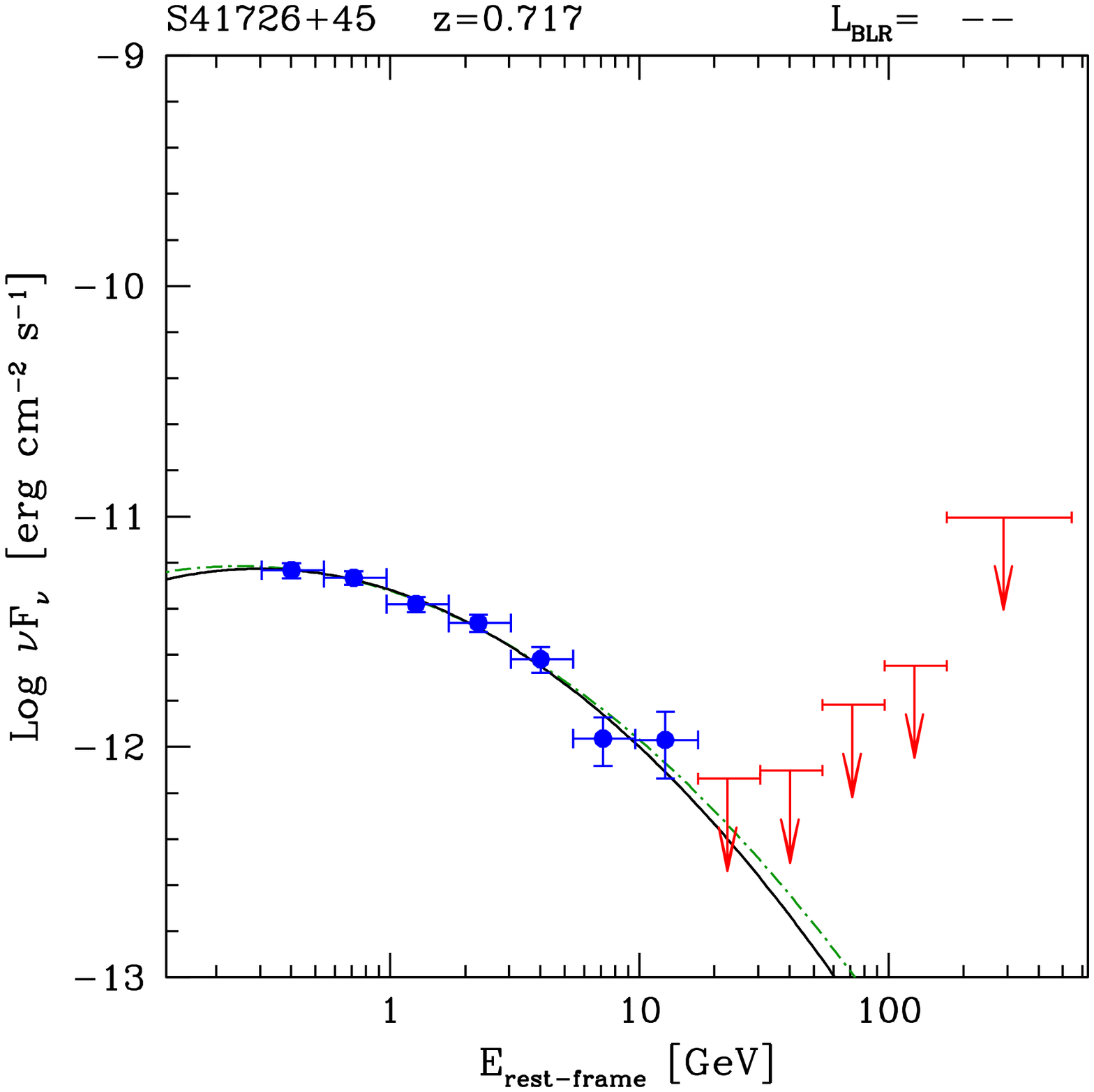,width=4.3cm,height=3.2cm }       
&\psfig{file=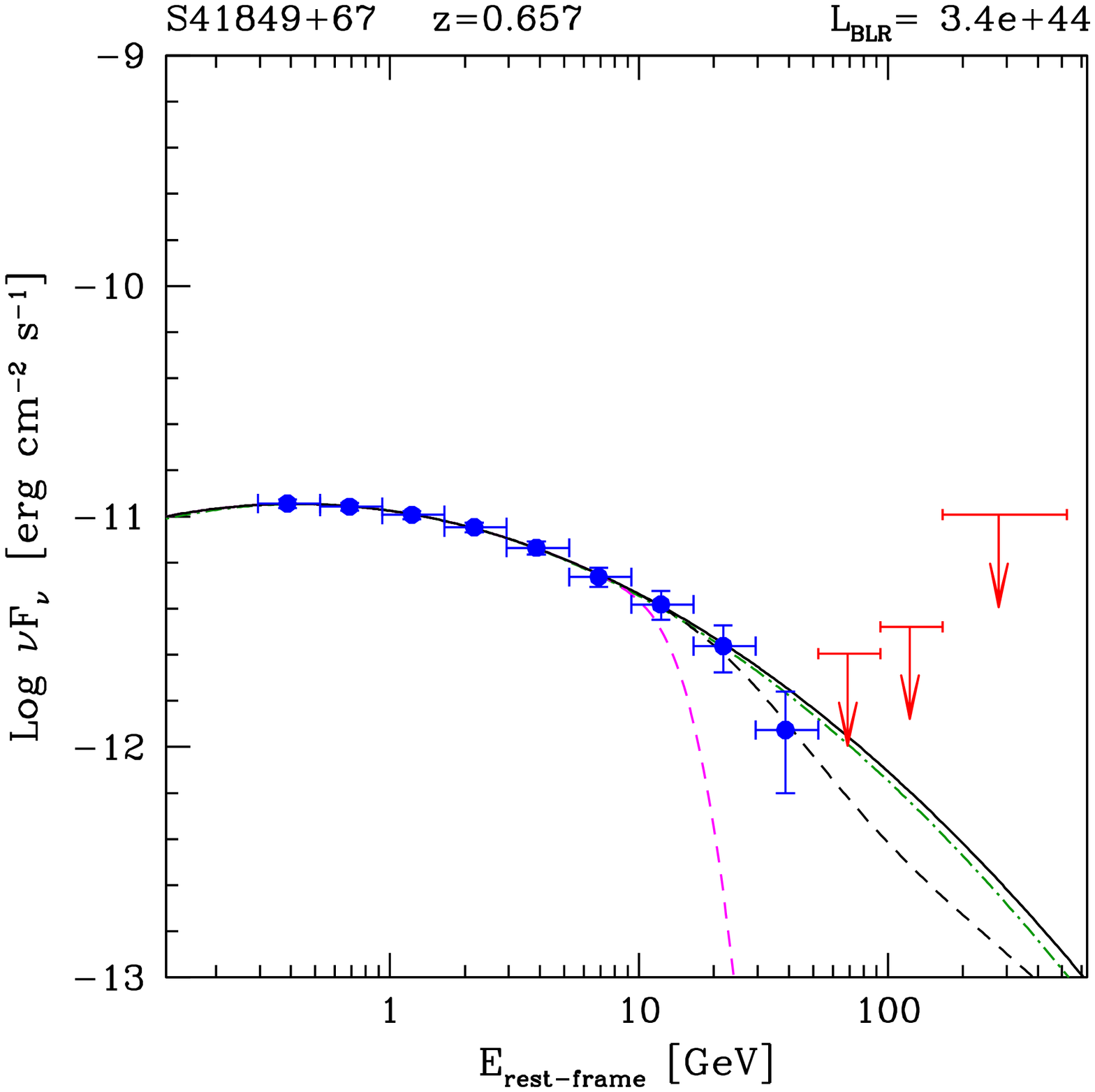,width=4.3cm,height=3.2cm } 
&\psfig{file=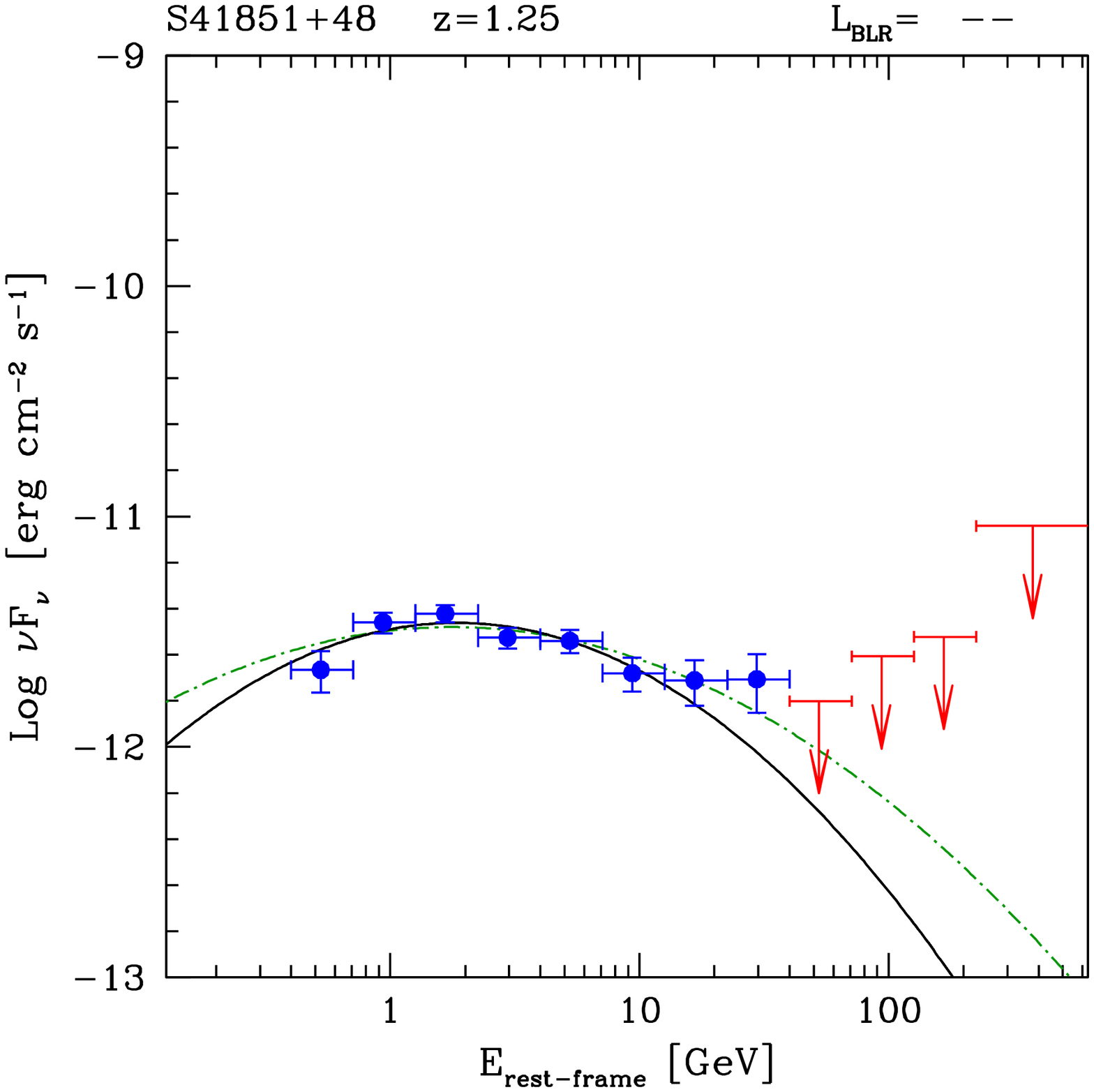,width=4.3cm,height=3.2cm } 
&\psfig{file=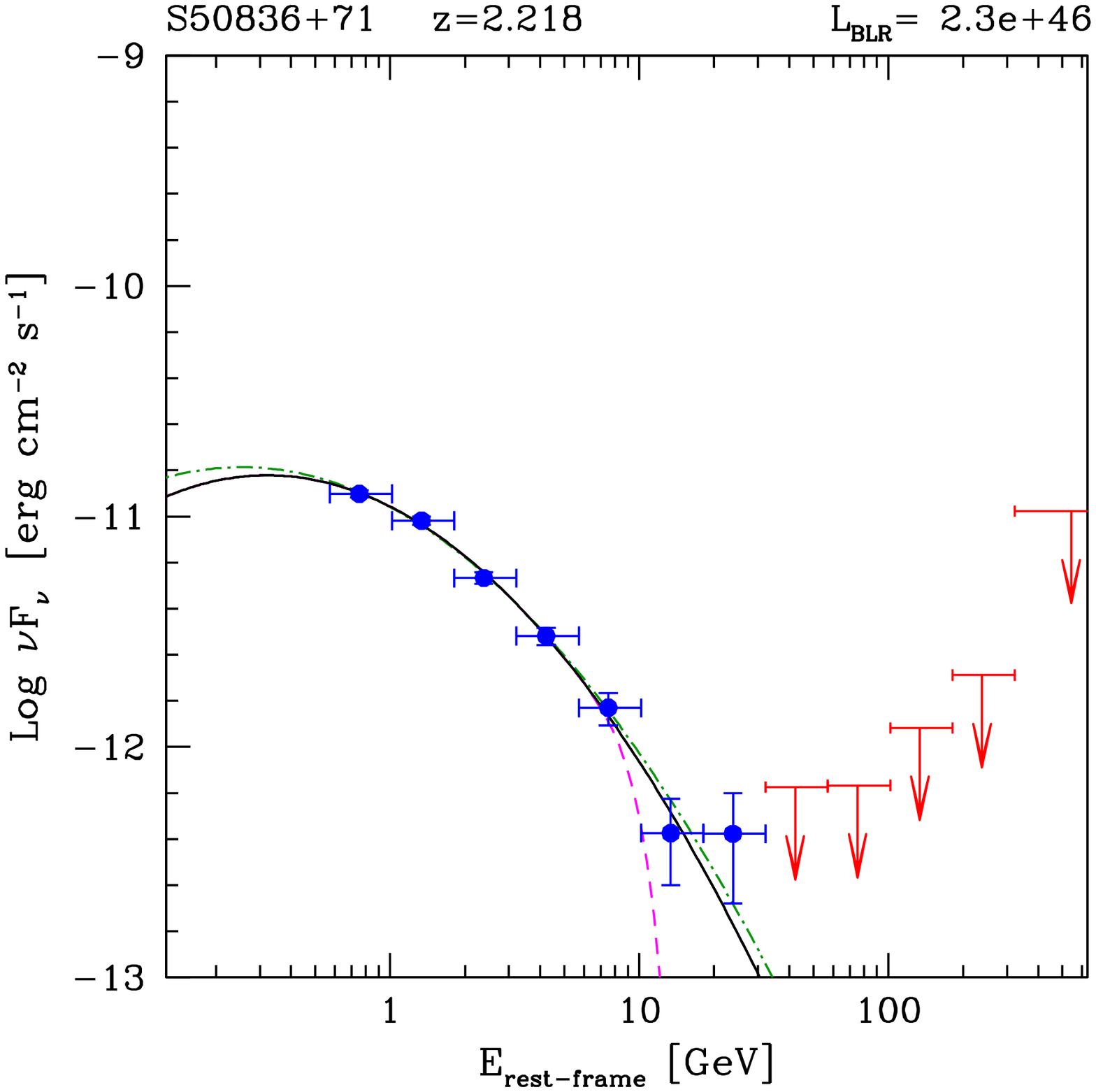,width=4.3cm,height=3.2cm } \vspace{1.2cm}  \\    
 \psfig{file=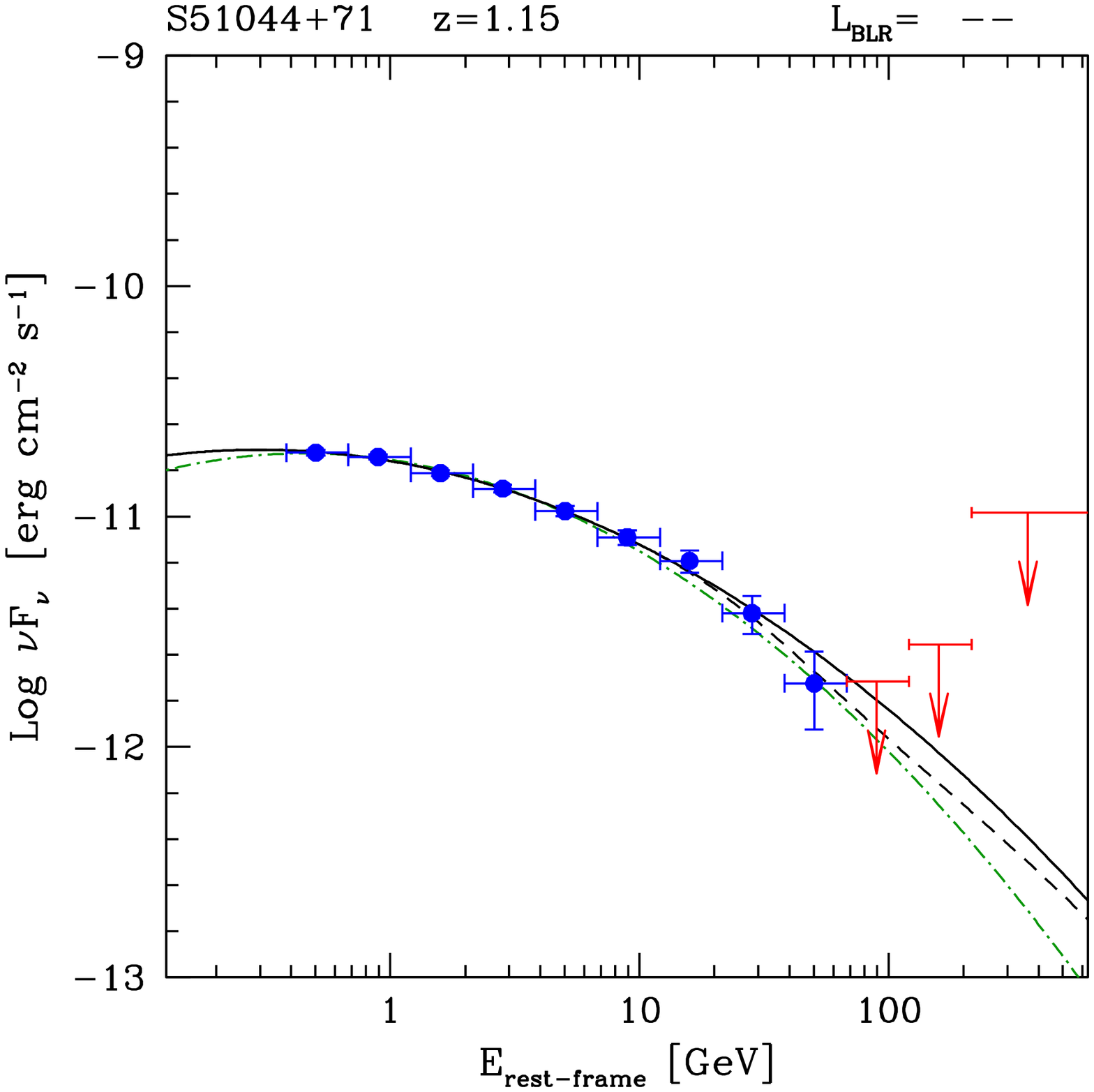,width=4.3cm,height=3.2cm } 
&\psfig{file=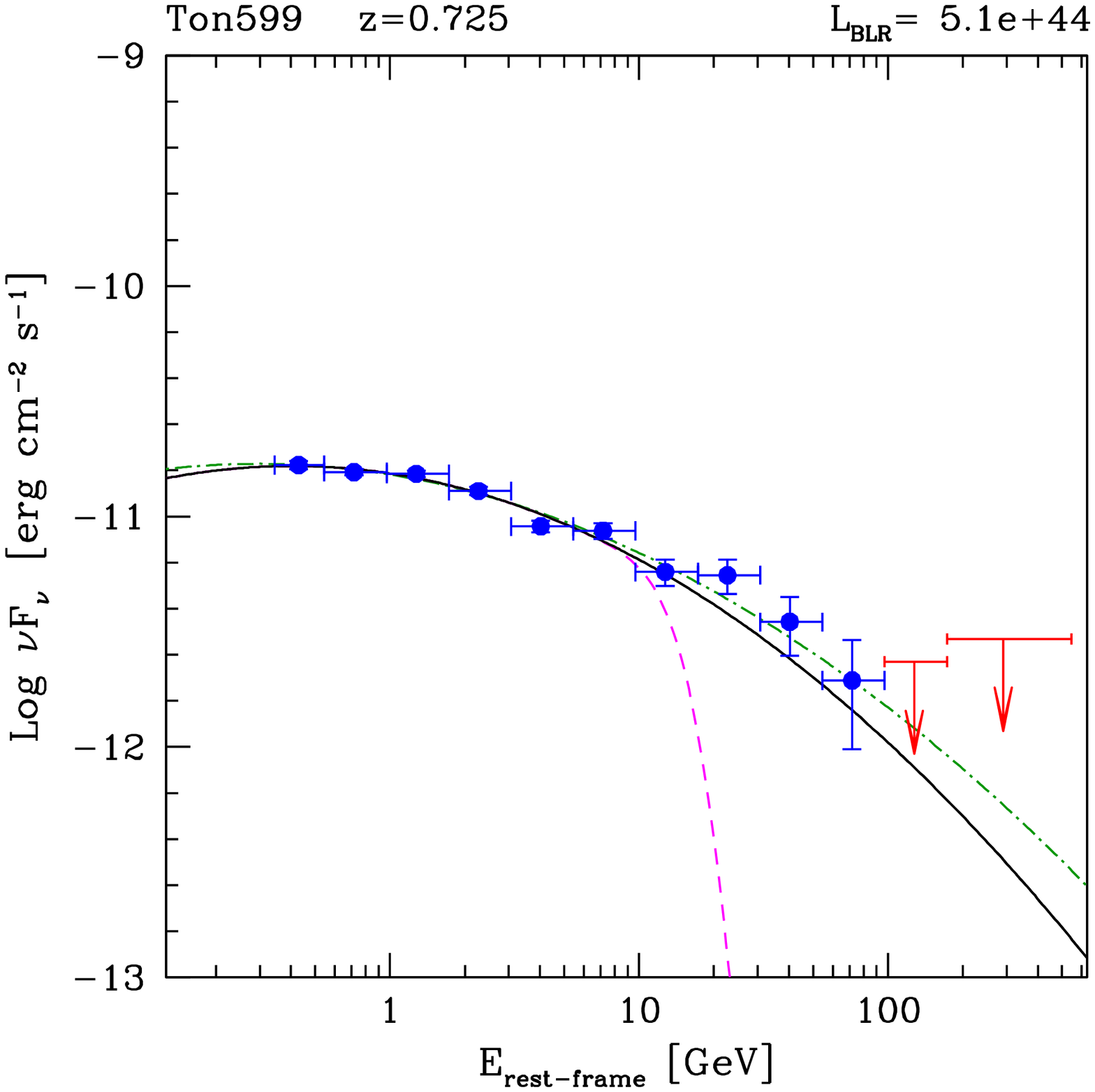,width=4.3cm,height=3.2cm }         
&\psfig{file=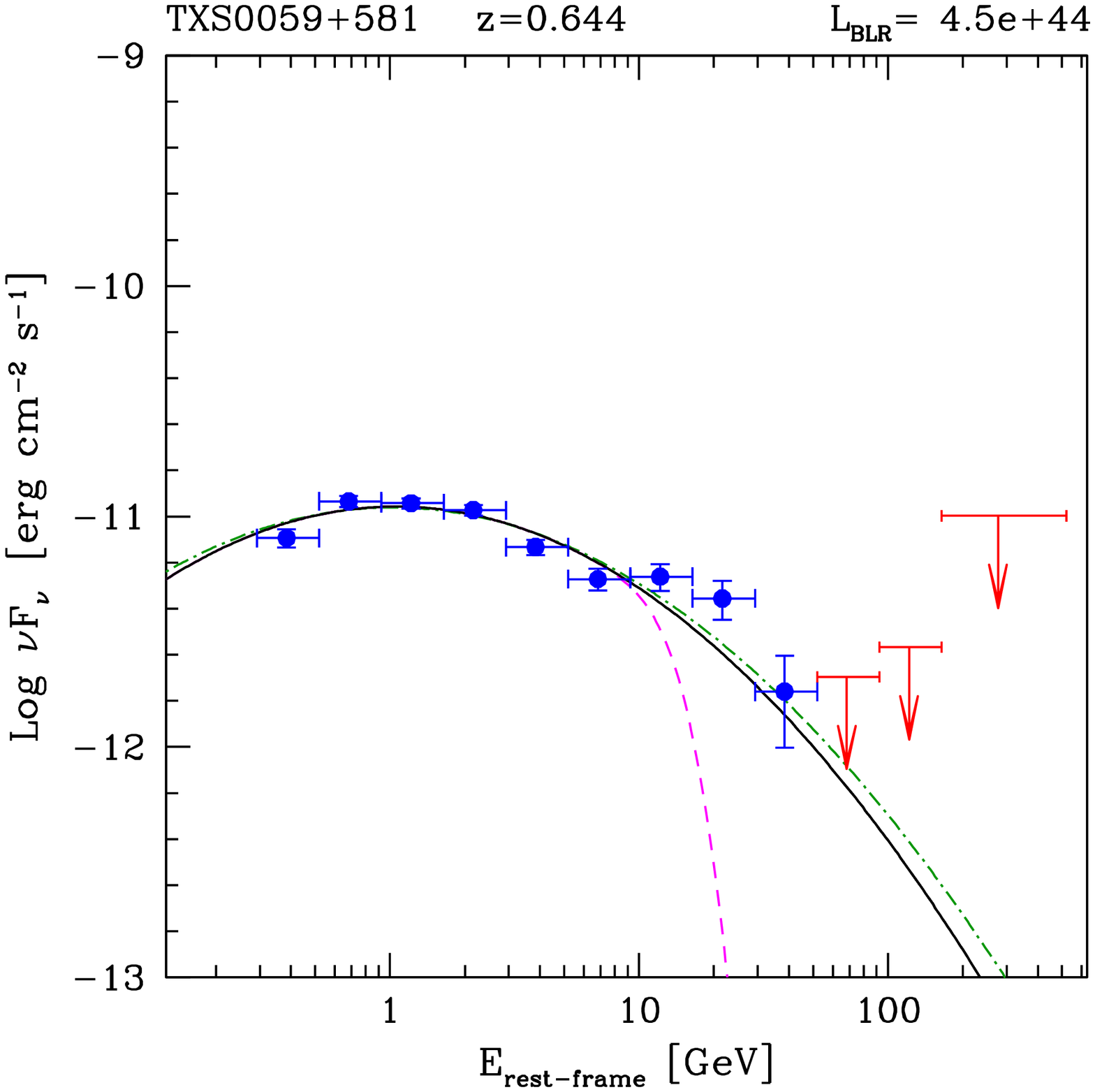,width=4.3cm,height=3.2cm } 
&\psfig{file=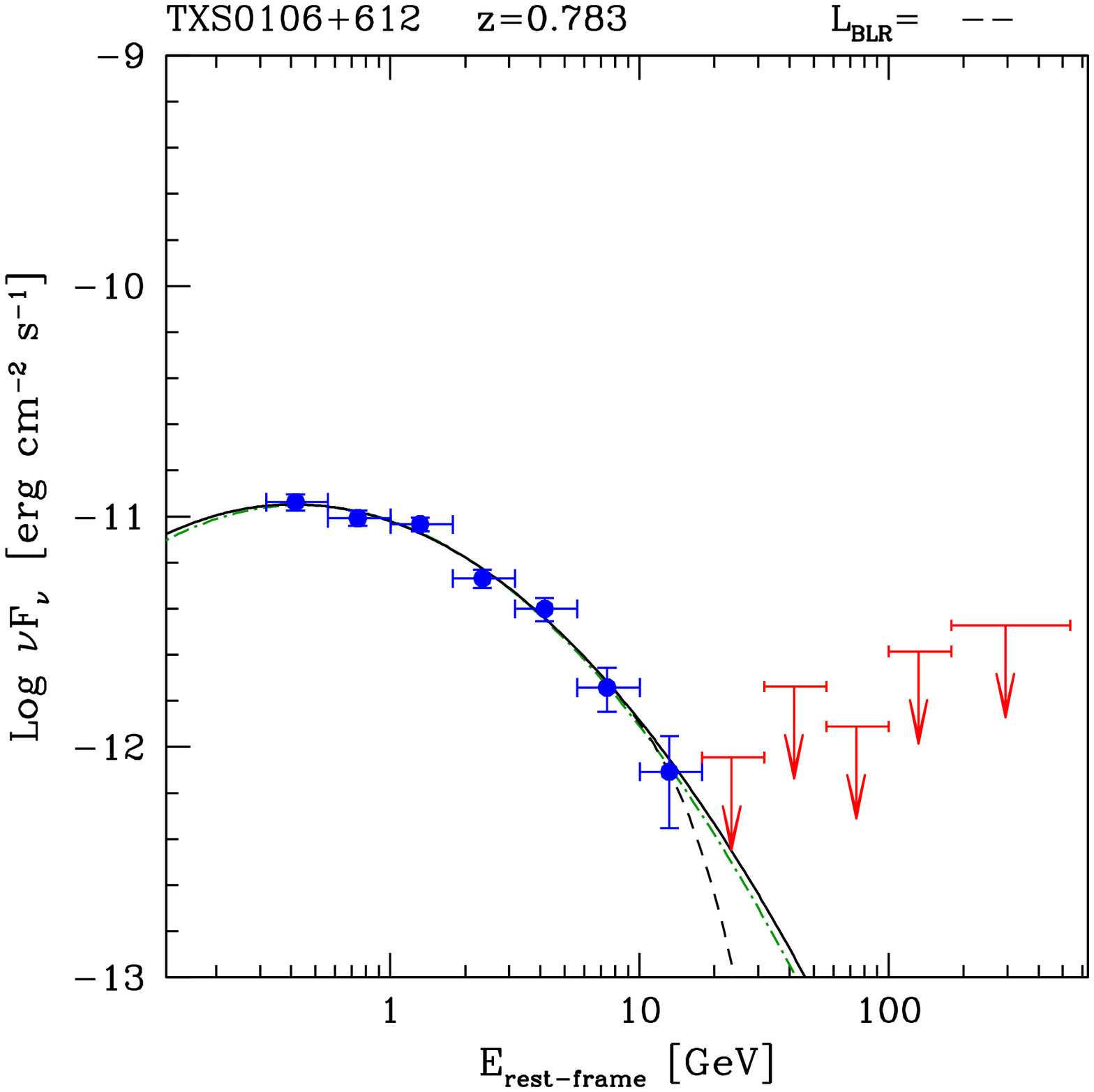,width=4.3cm,height=3.2cm } \vspace{1.2cm} \\ 
 \psfig{file=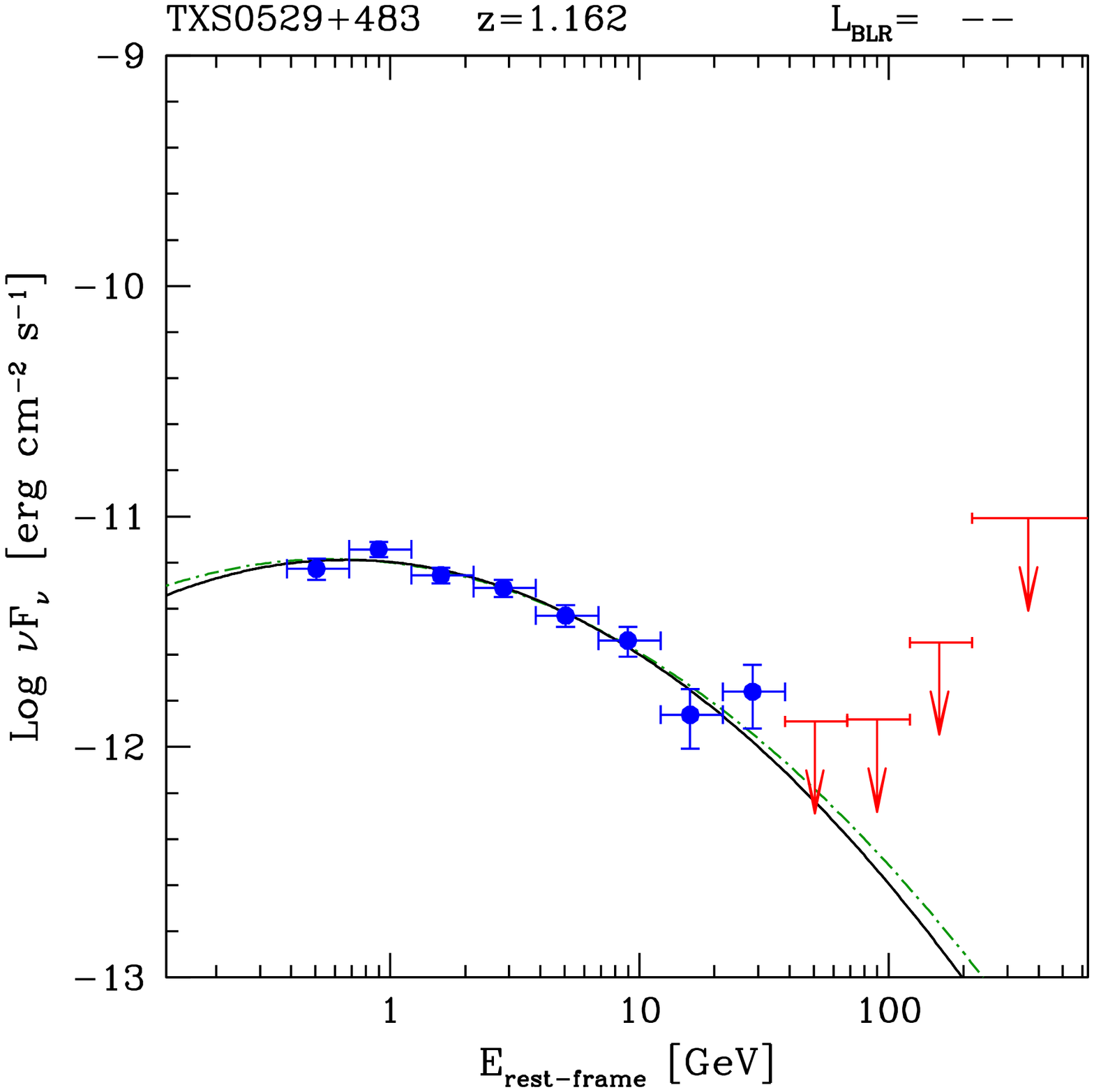,width=4.3cm,height=3.2cm } 
&\psfig{file=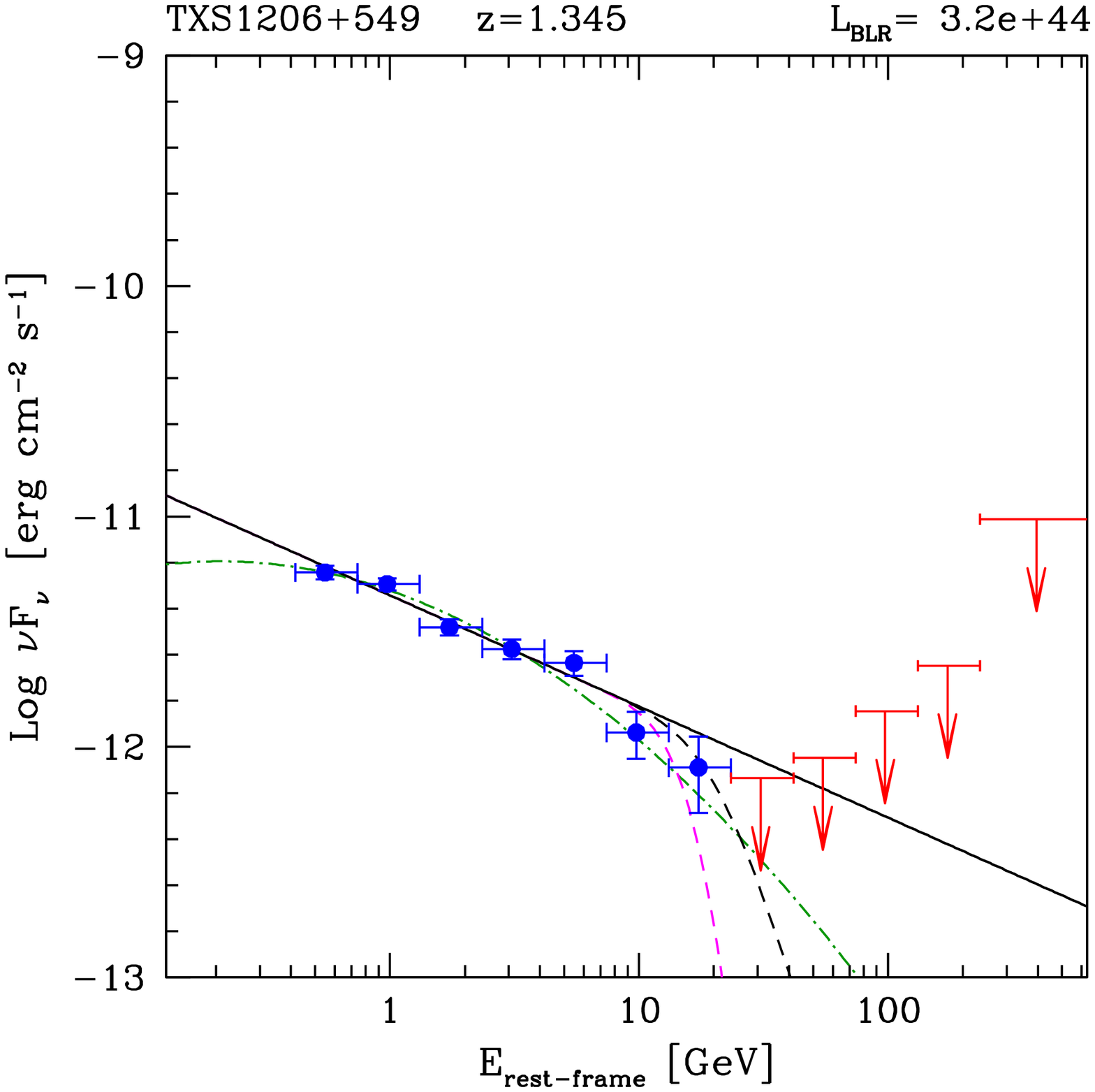,width=4.3cm,height=3.2cm }     
&\psfig{file=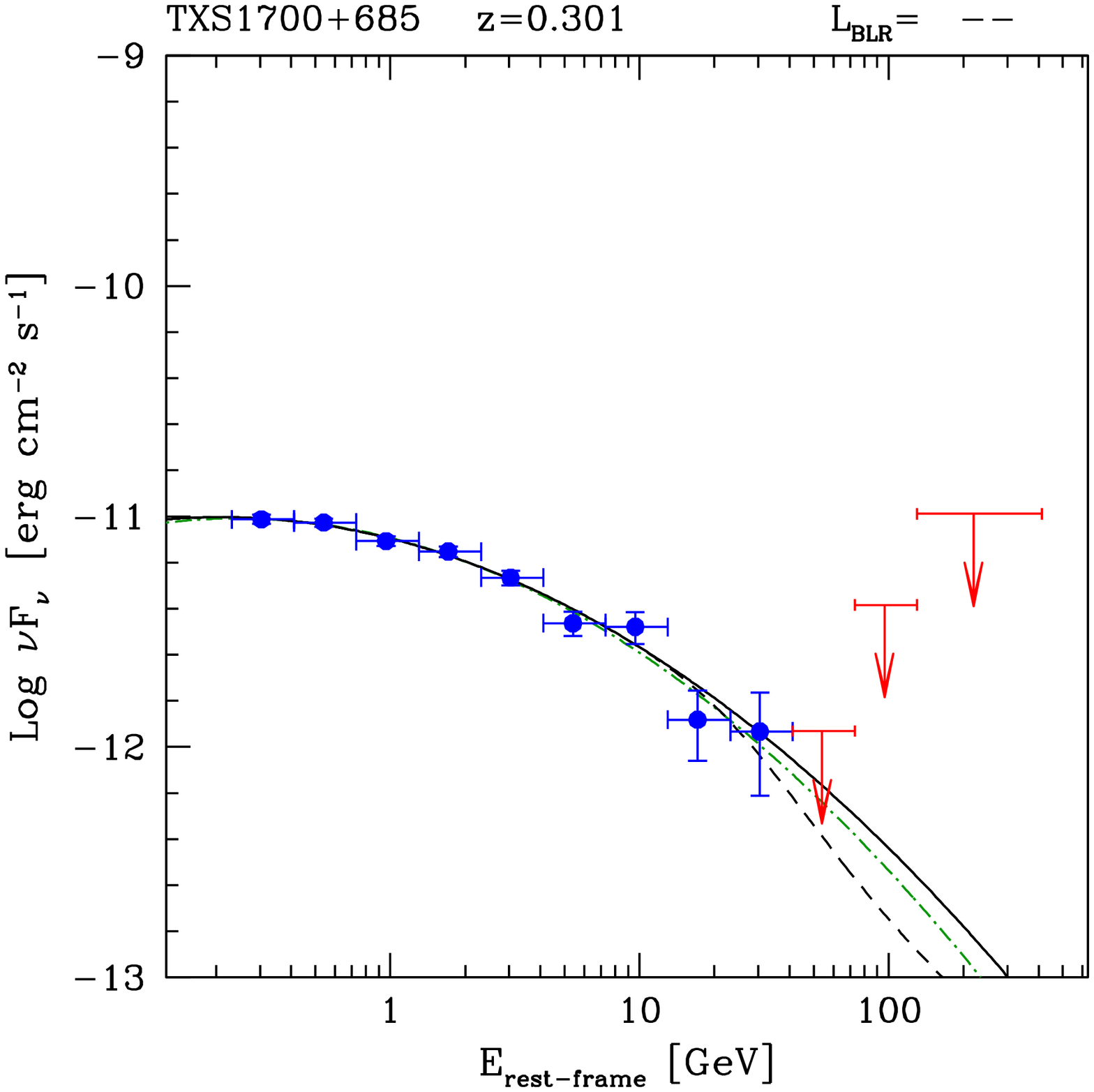,width=4.3cm,height=3.2cm } 
&\psfig{file=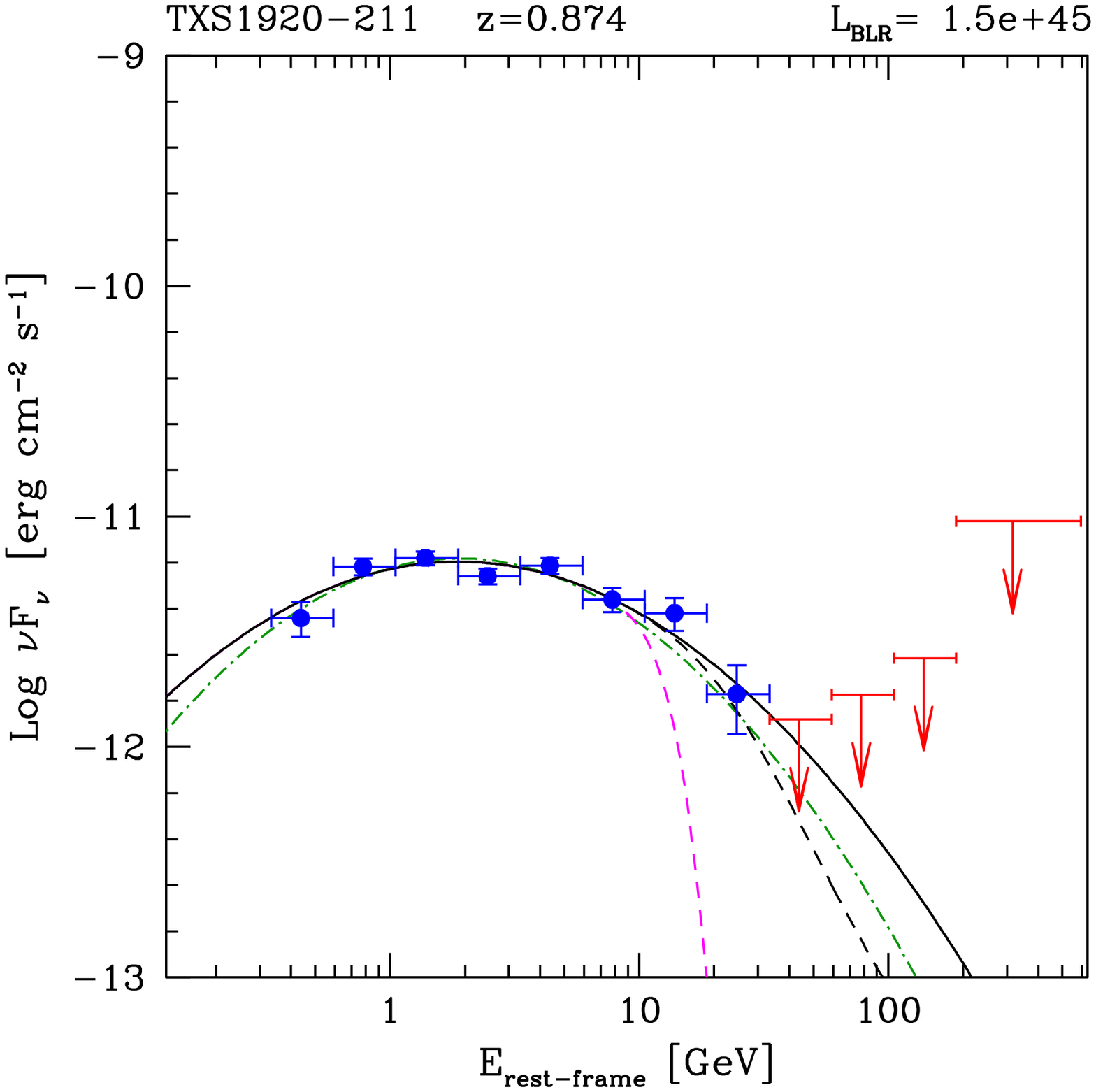,width=4.3cm,height=3.2cm }    
\end{tabular}
\contcaption{}  
\end{figure*}


\begin{figure*}
\includegraphics[width=5.8cm]{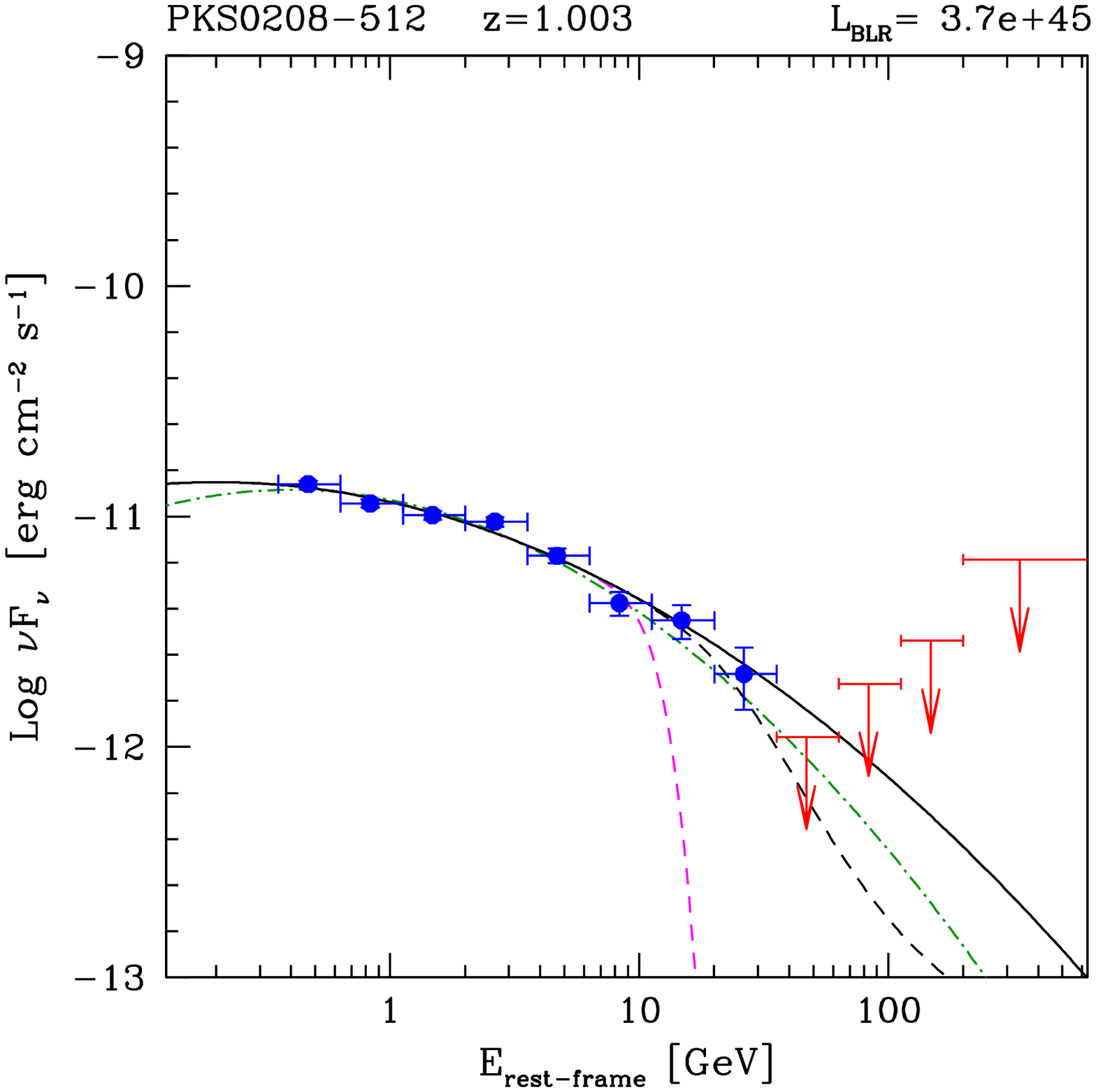}
\includegraphics[width=5.8cm]{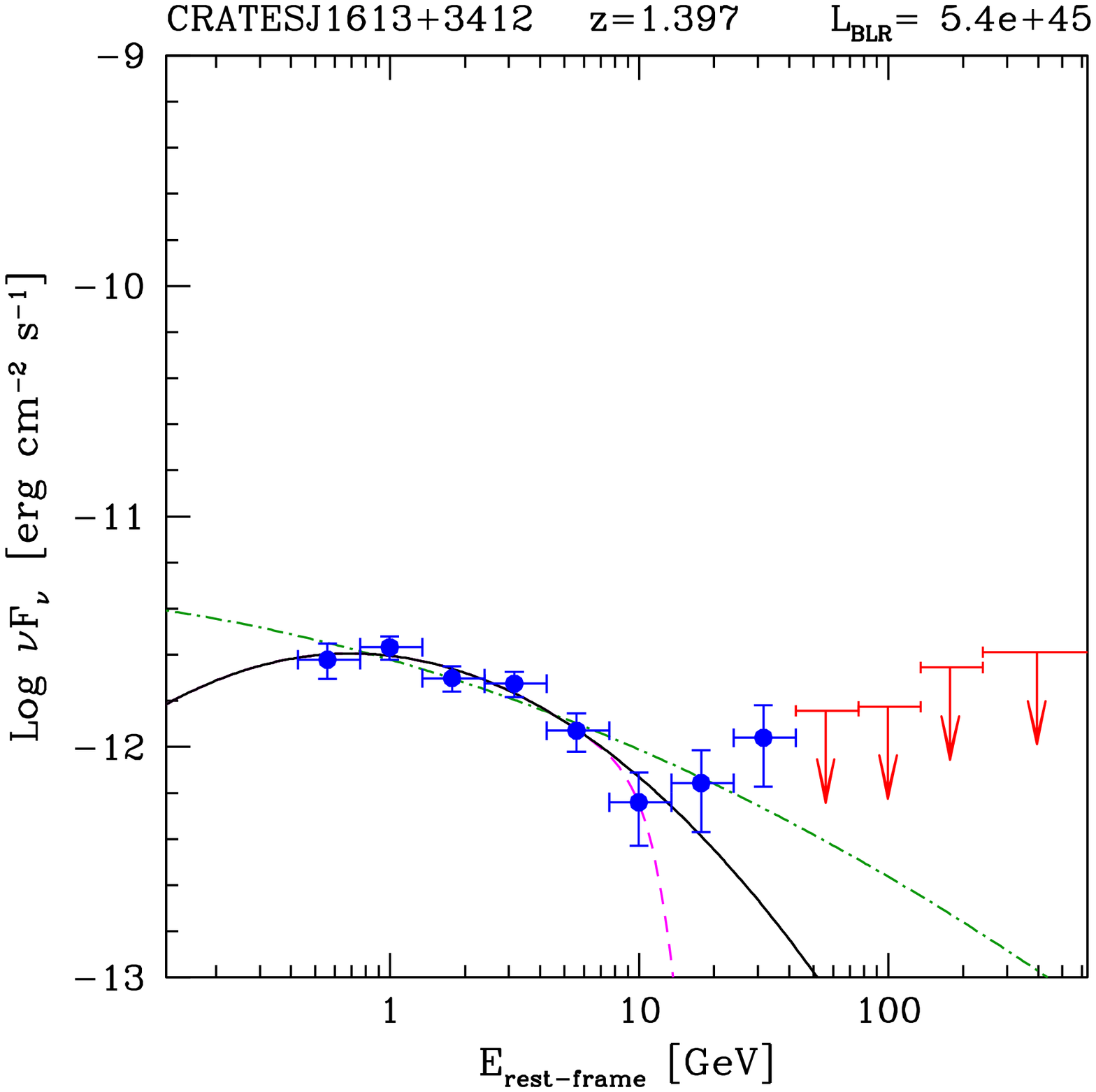}
\includegraphics[width=5.8cm]{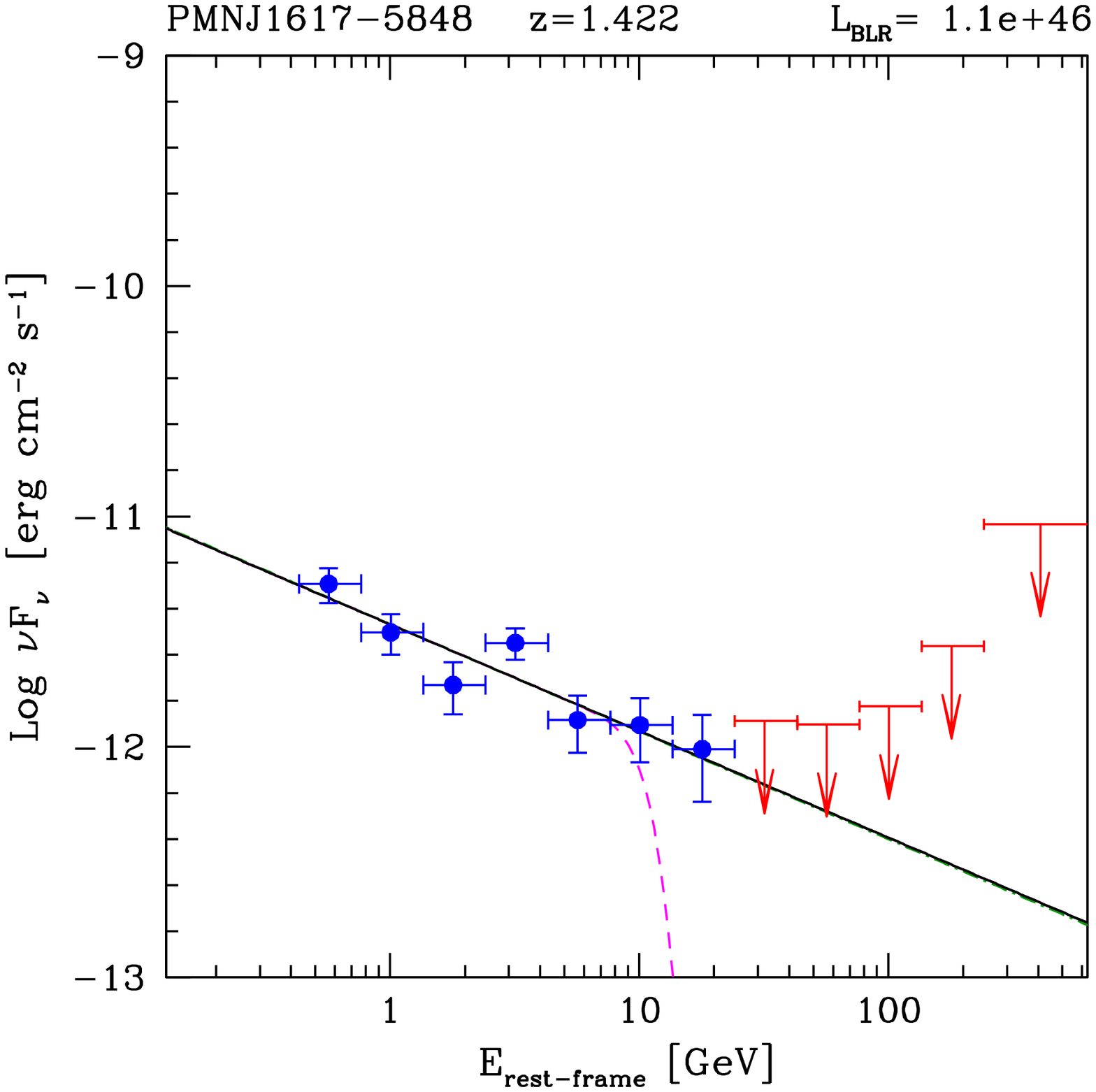} \\
\includegraphics[width=5.8cm]{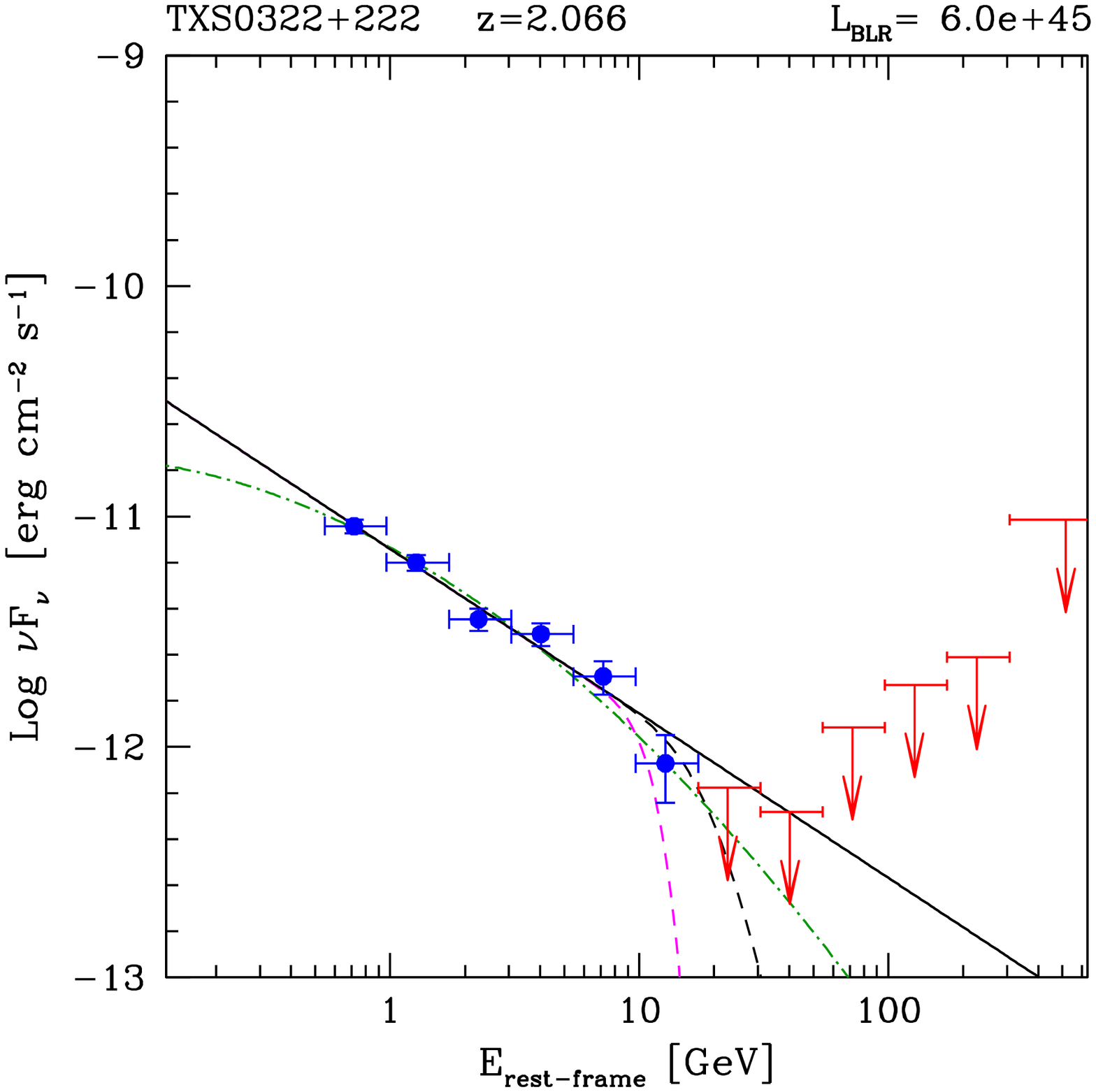}
\includegraphics[width=5.8cm]{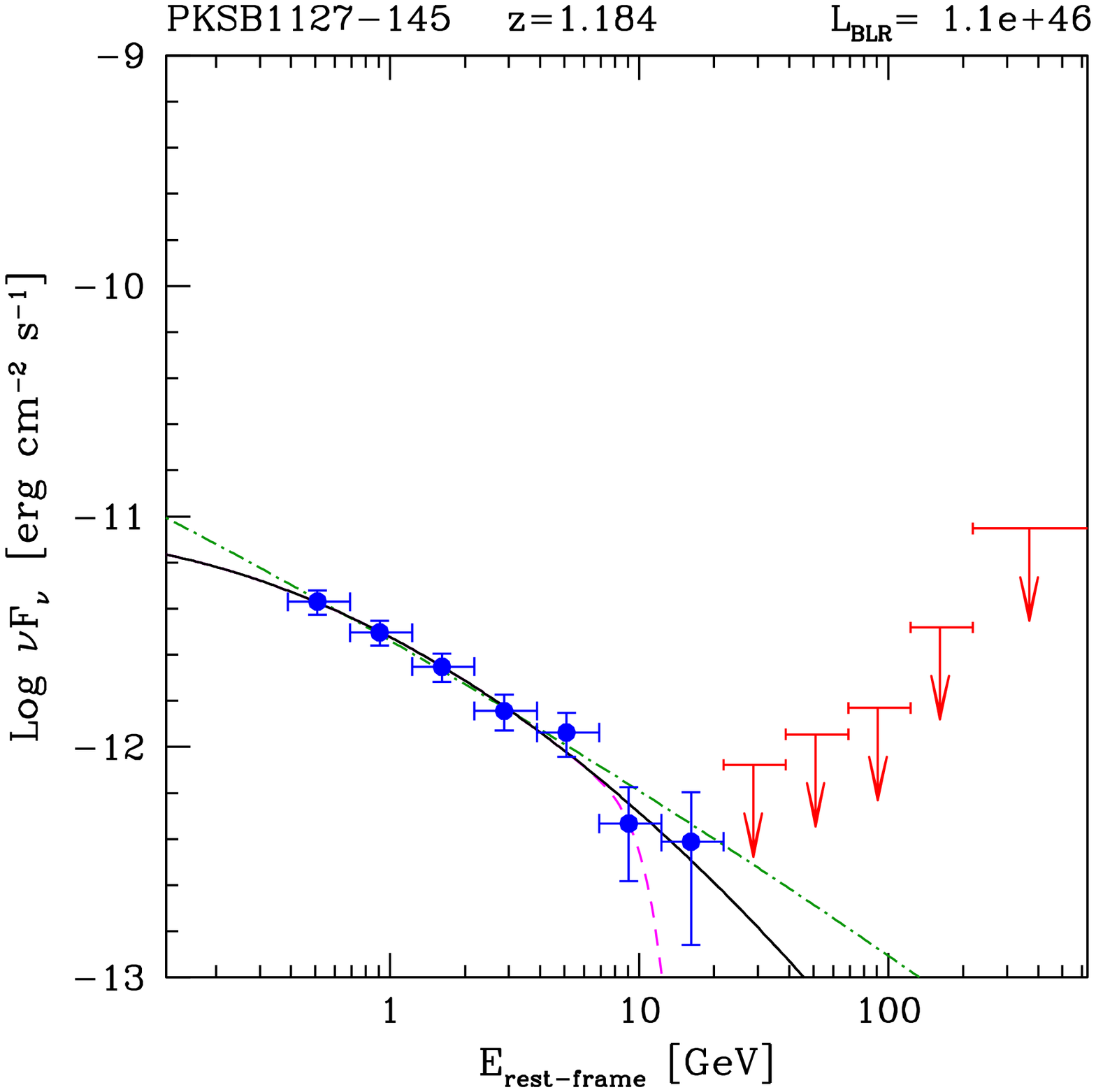}
\includegraphics[width=5.8cm]{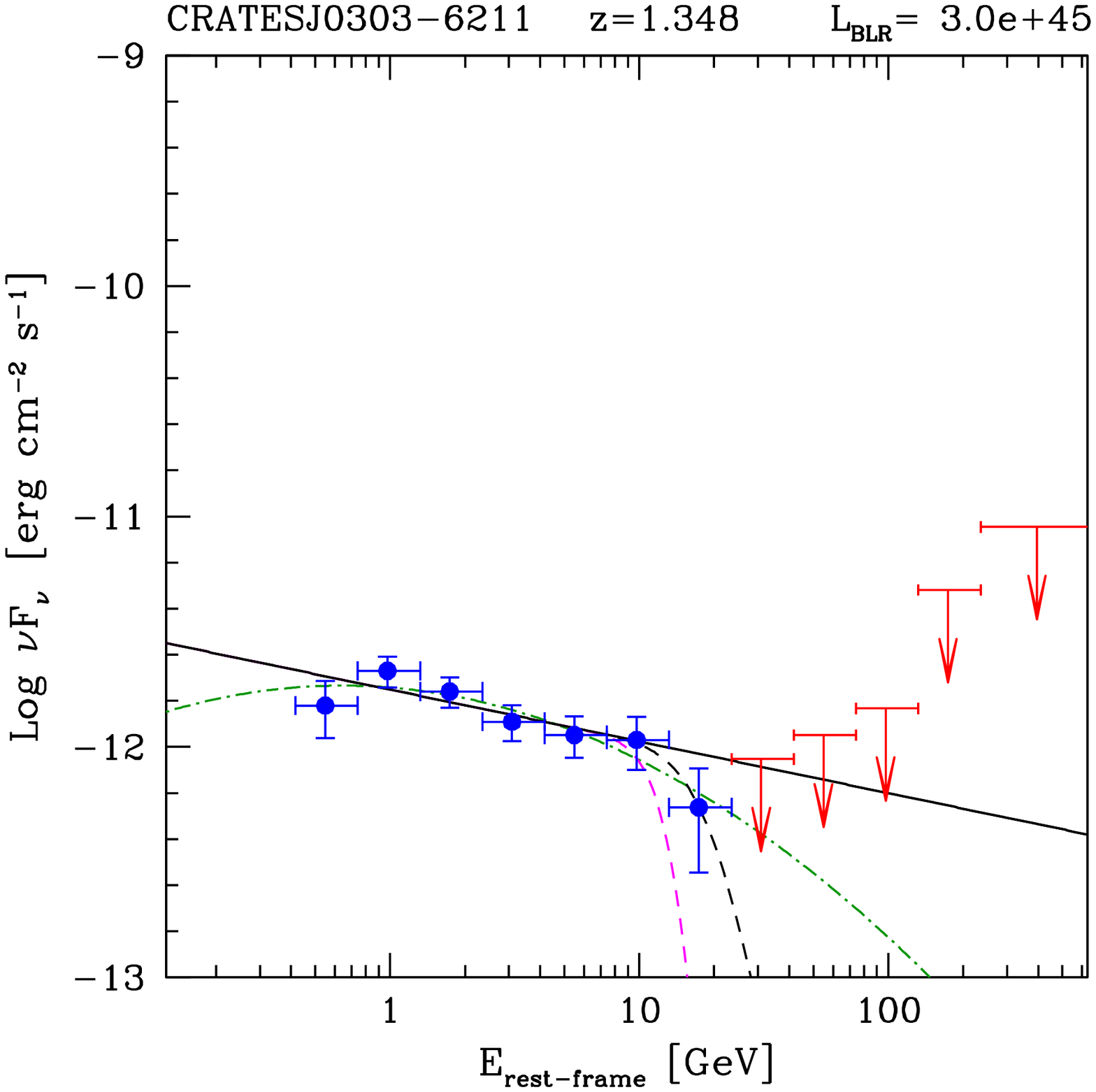}
\caption{Spectra of the additional FSRQ with large BLR. Upper row: data are constraining, showing a lower BLR absorption than expected. 
Lower row: data are not conclusive.}
\label{largeblr}
\end{figure*}


\begin{figure*}
\begin{tabular}{llrr}
 \psfig{file=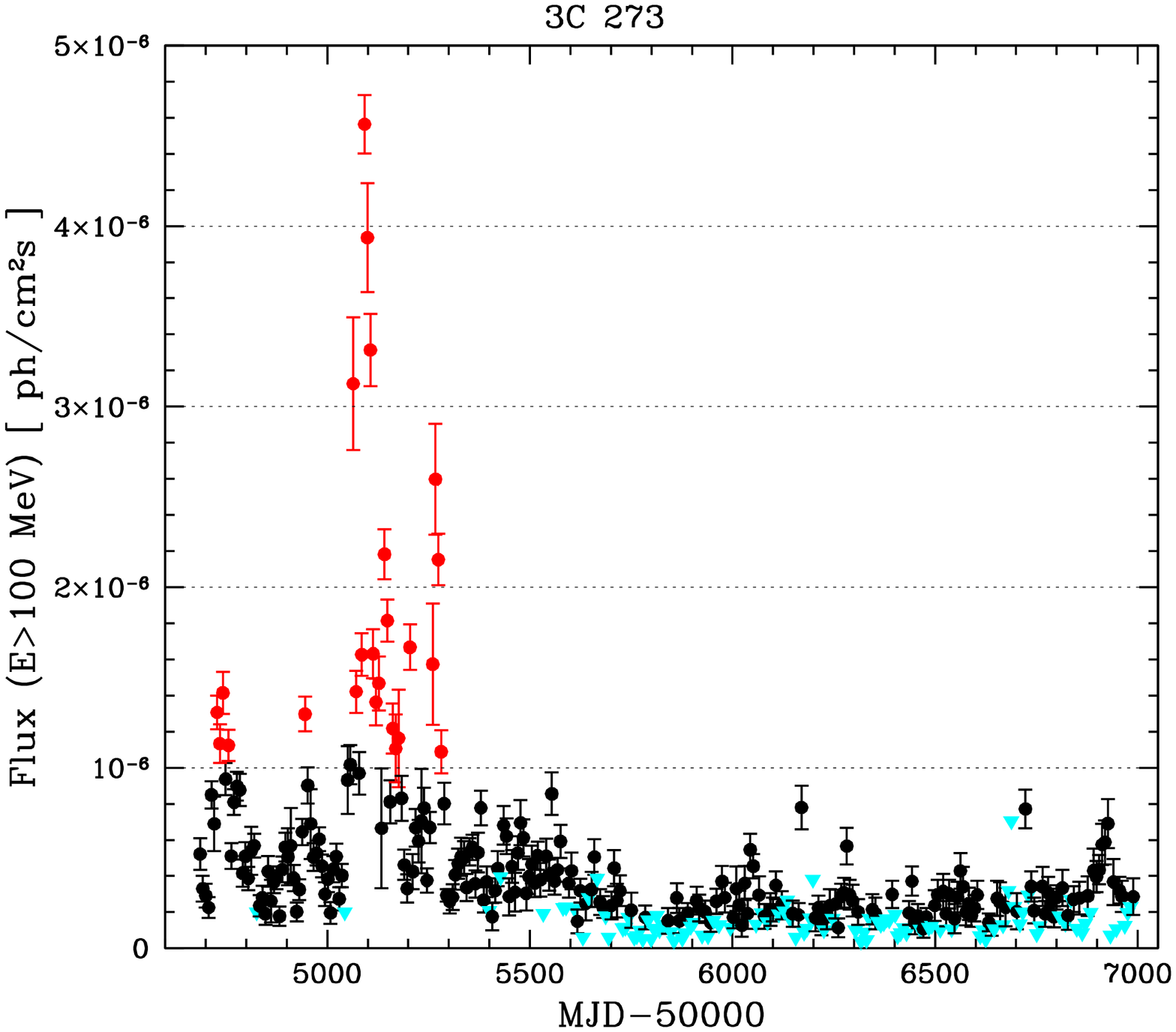,width=4.3cm,height=4.03cm}   & \psfig{file=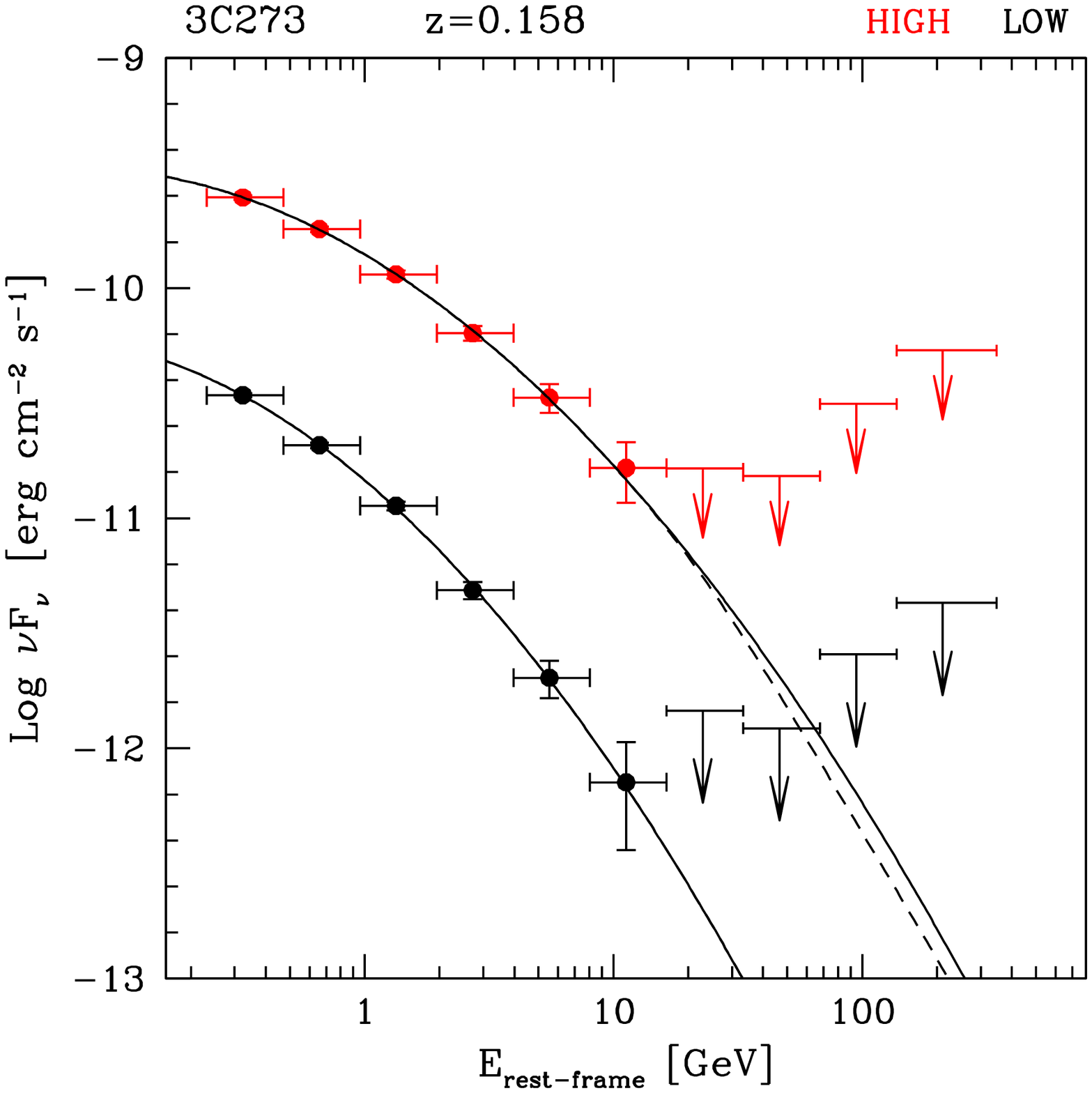,width=4.3cm,height=4.03cm} 
&\psfig{file=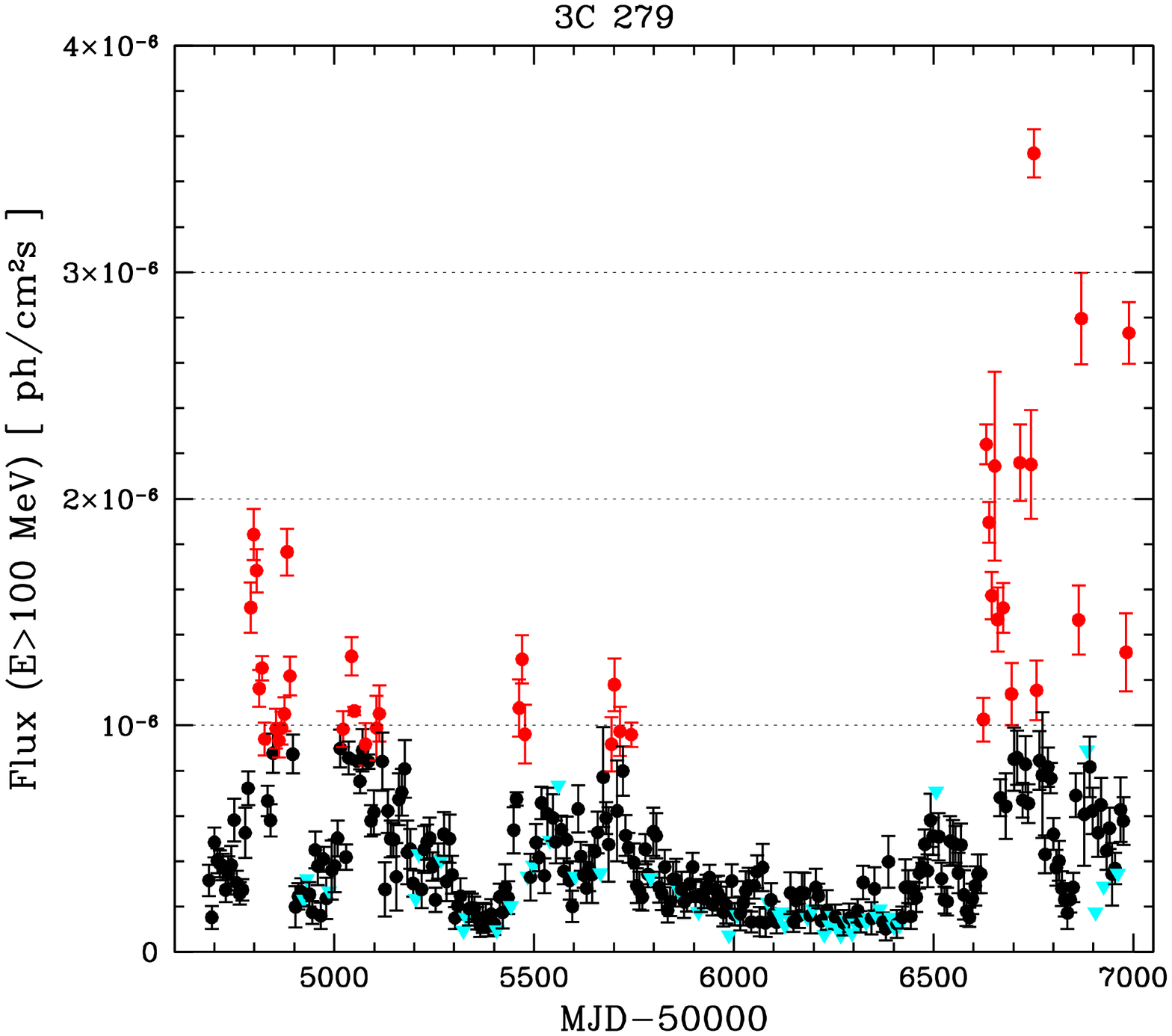,width=4.3cm,height=4.03cm}   & \psfig{file=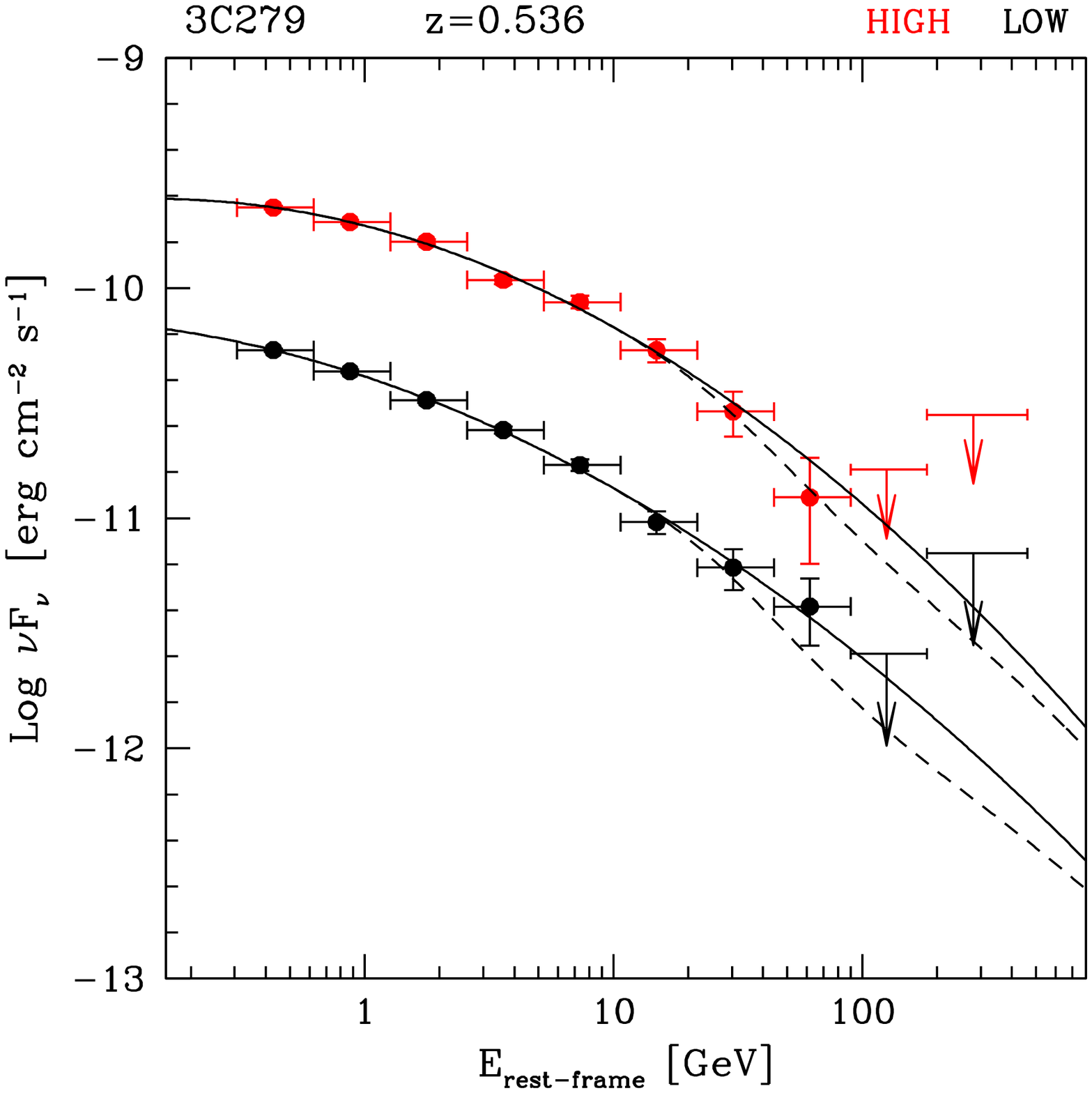,width=4.3cm,height=4.03cm} \\
 \psfig{file=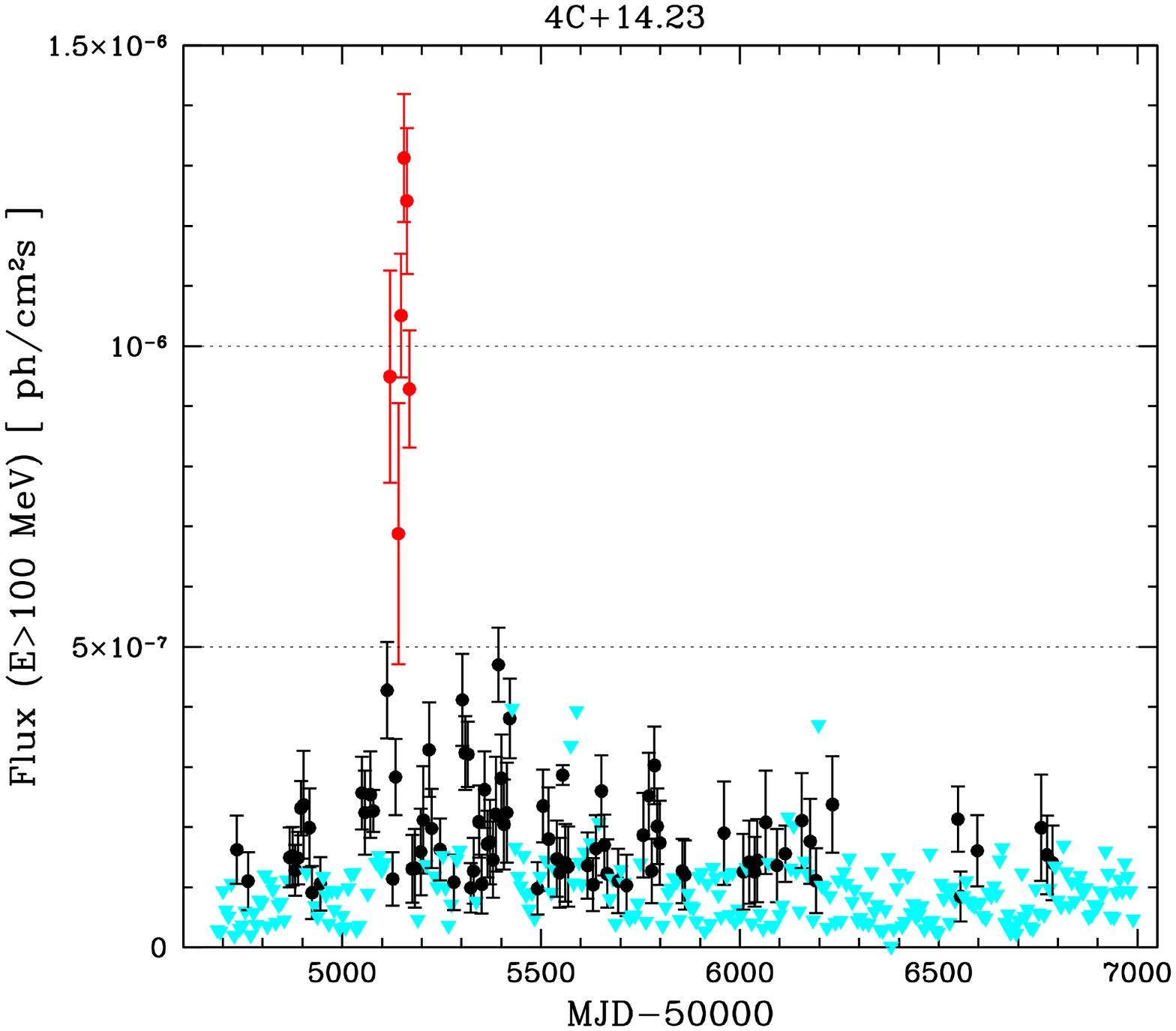,width=4.3cm,height=4.03cm}   & \psfig{file=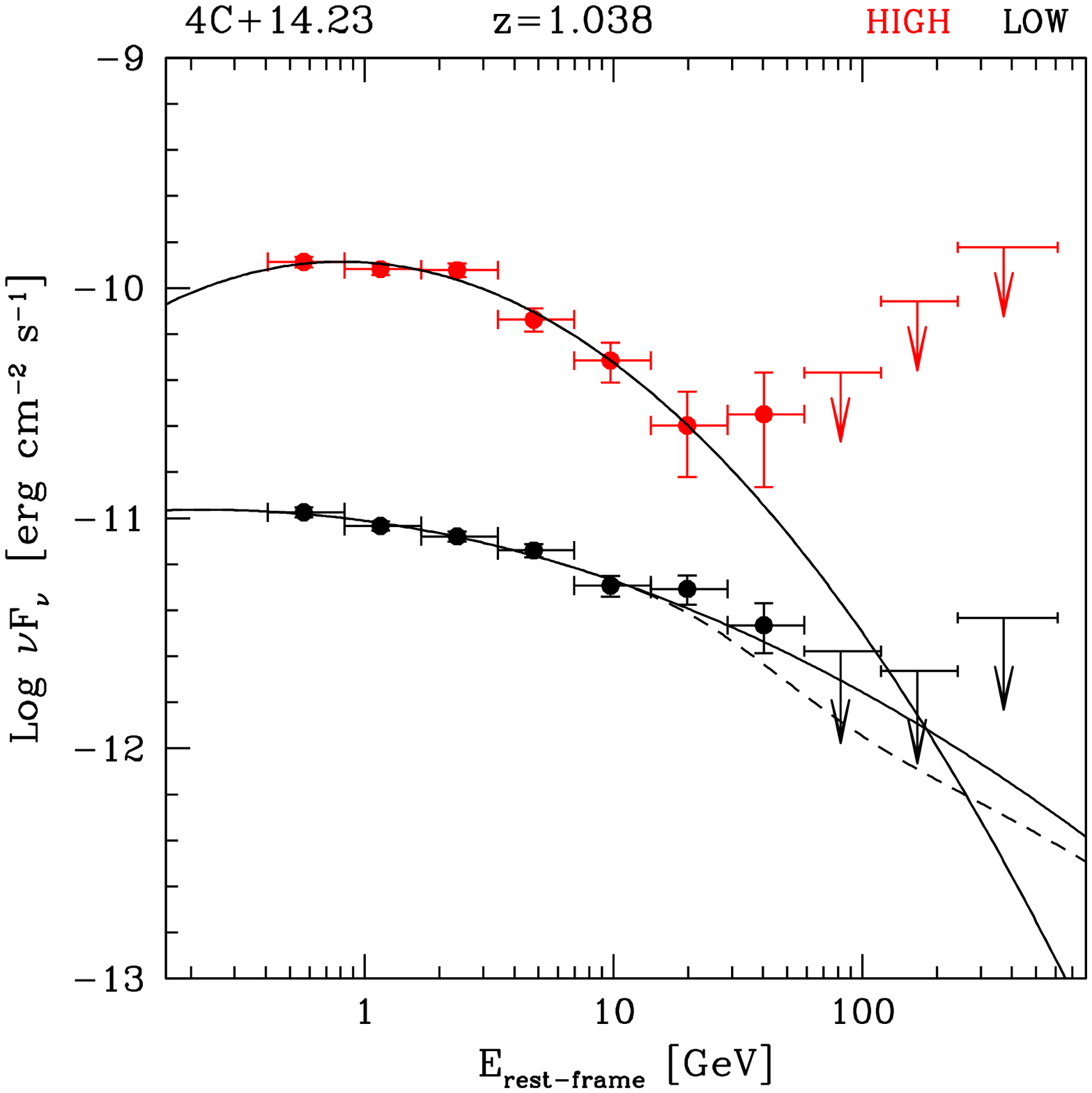,width=4.3cm,height=4.03cm} 
&\psfig{file=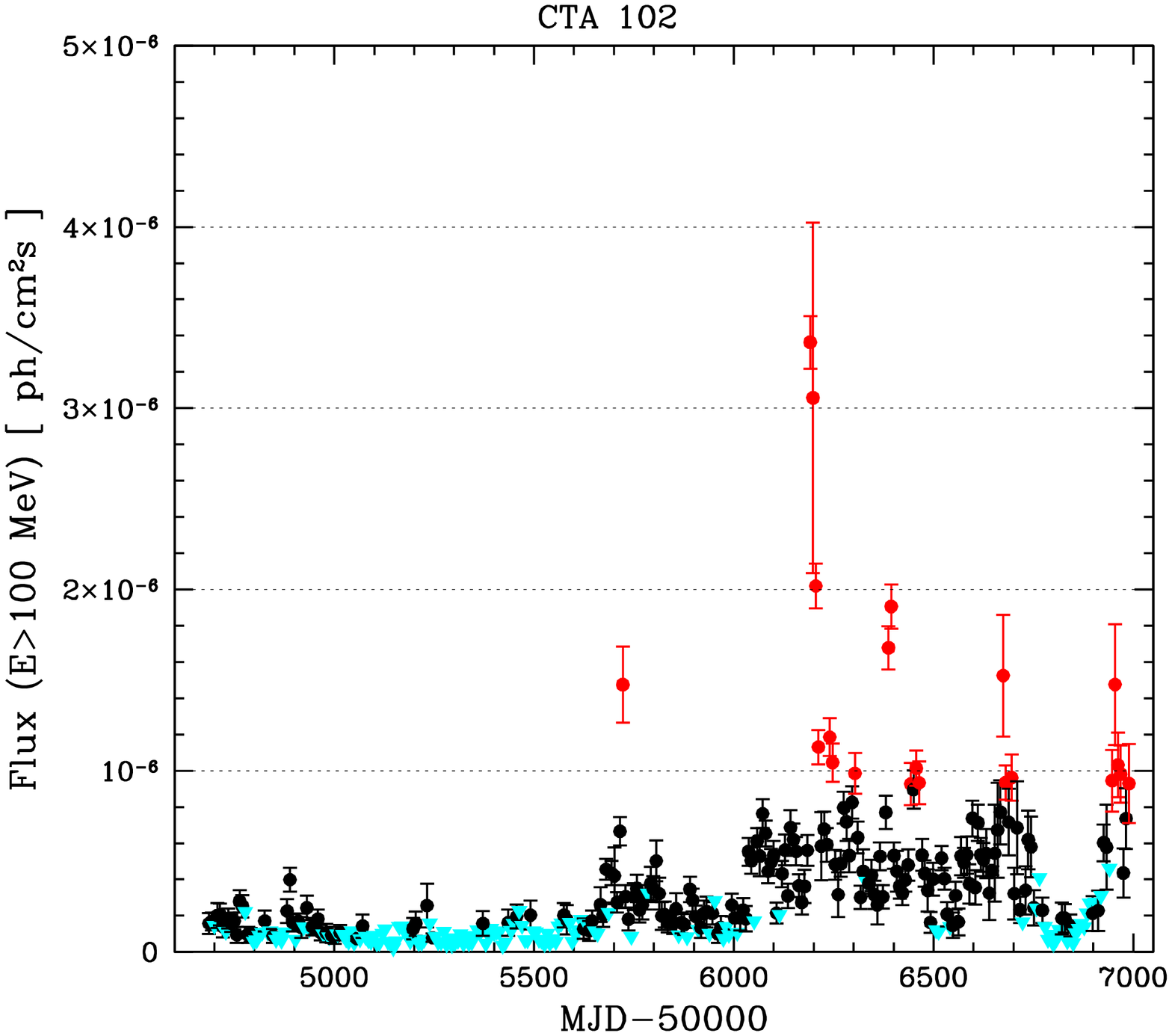,width=4.3cm,height=4.03cm }  & \psfig{file=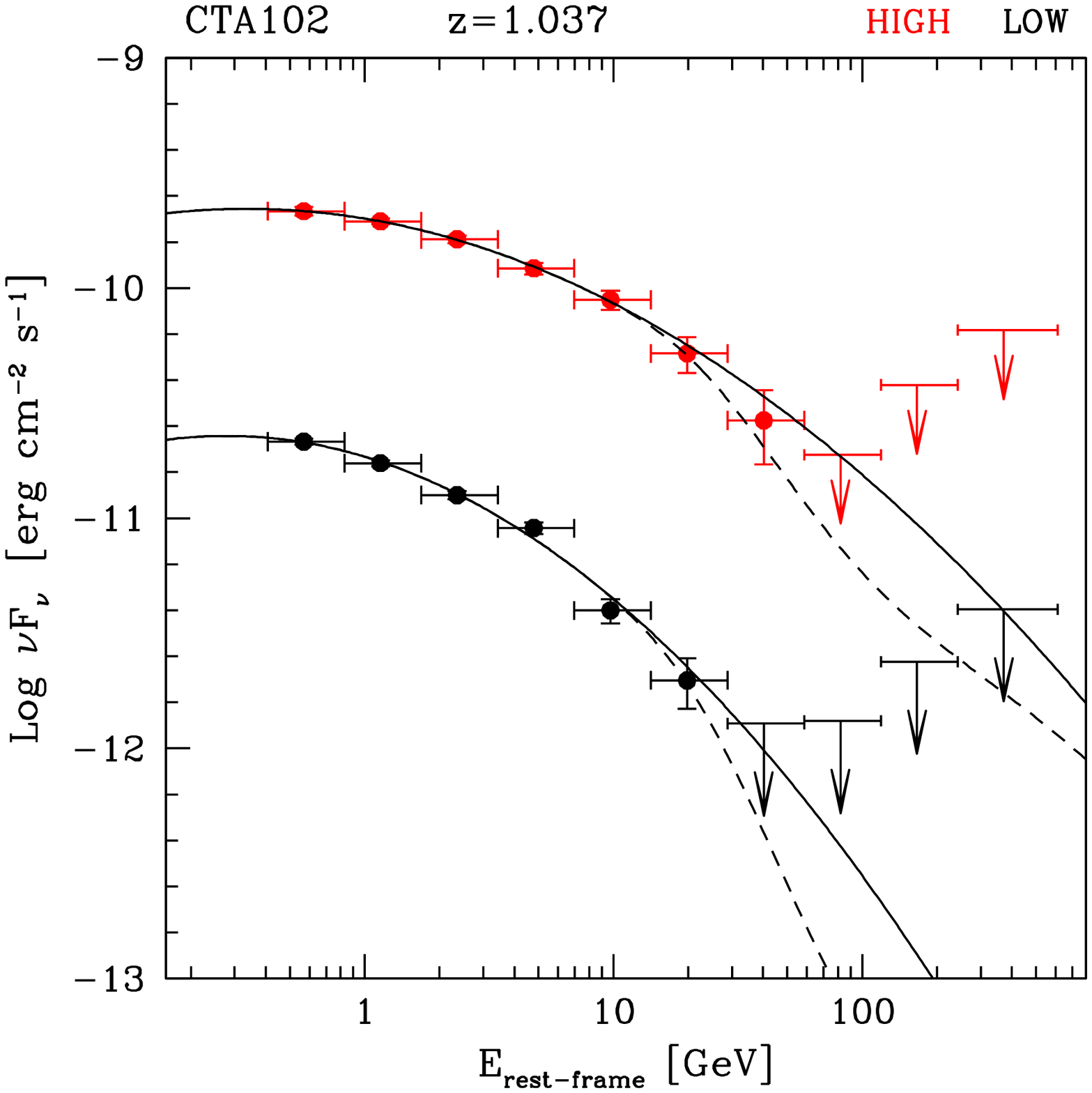,width=4.3cm,height=4.03cm } \\
 \psfig{file=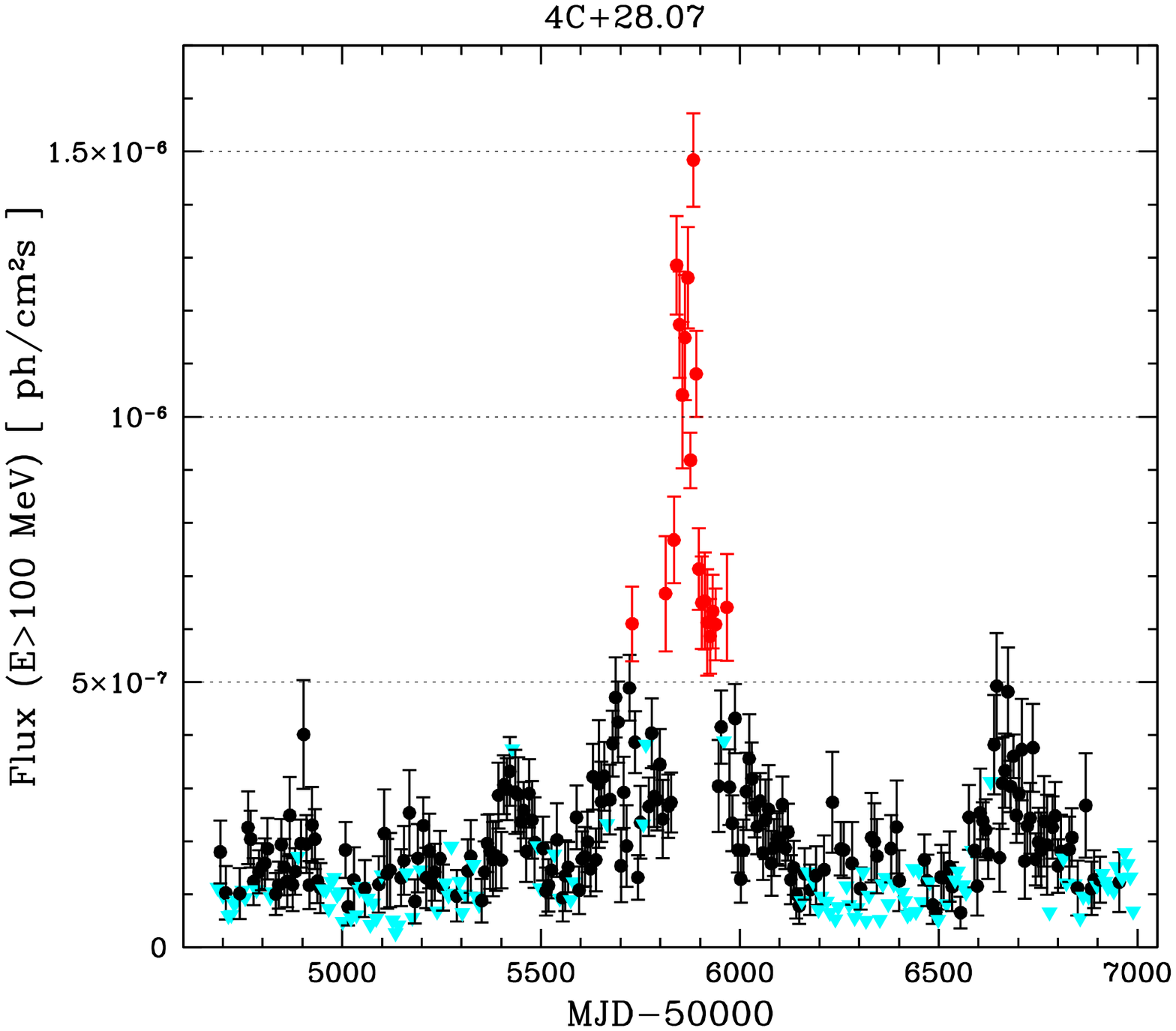,width=4.3cm,height=4.03cm}   & \psfig{file=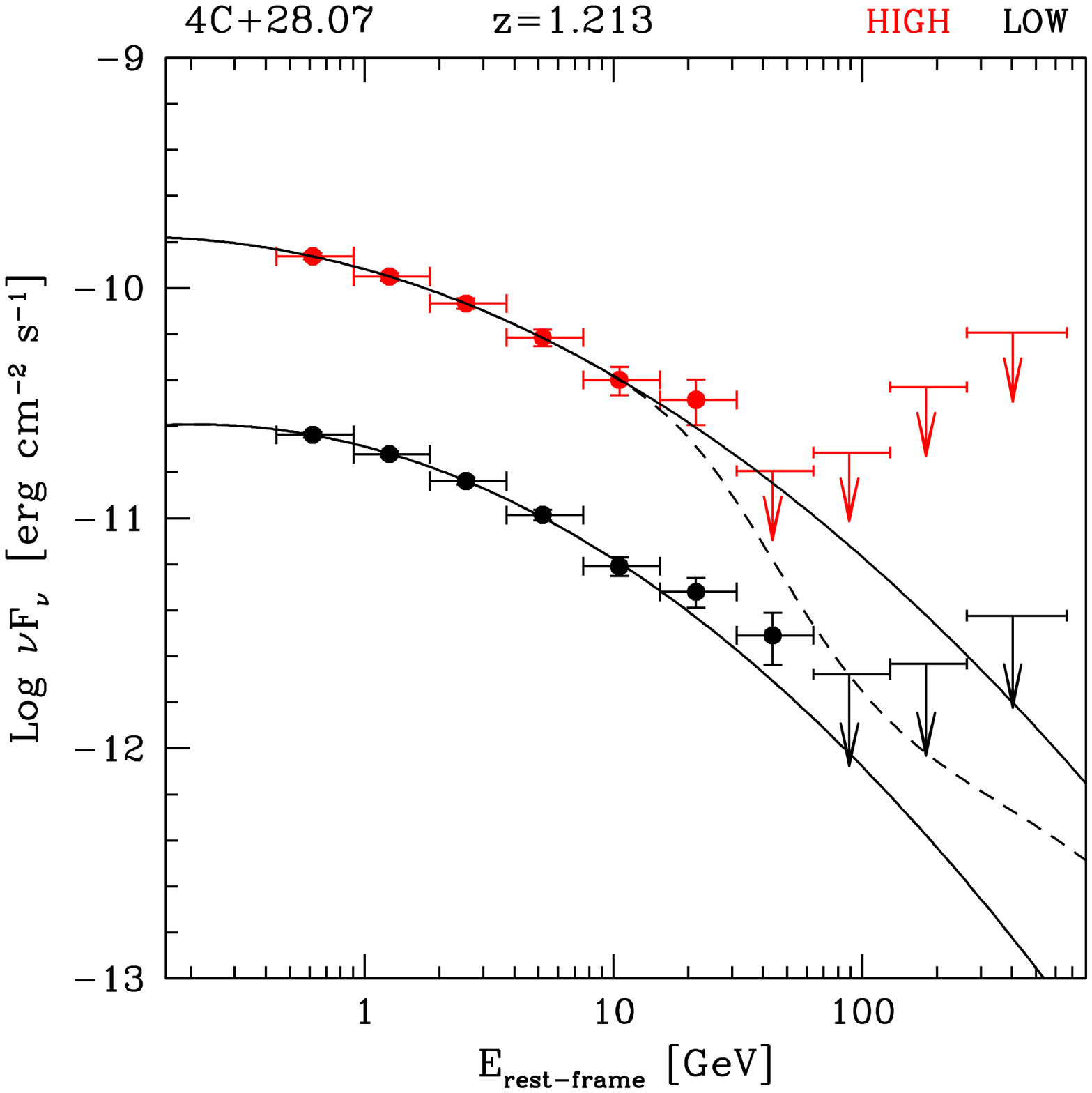,width=4.3cm,height=4.03cm} 
&\psfig{file=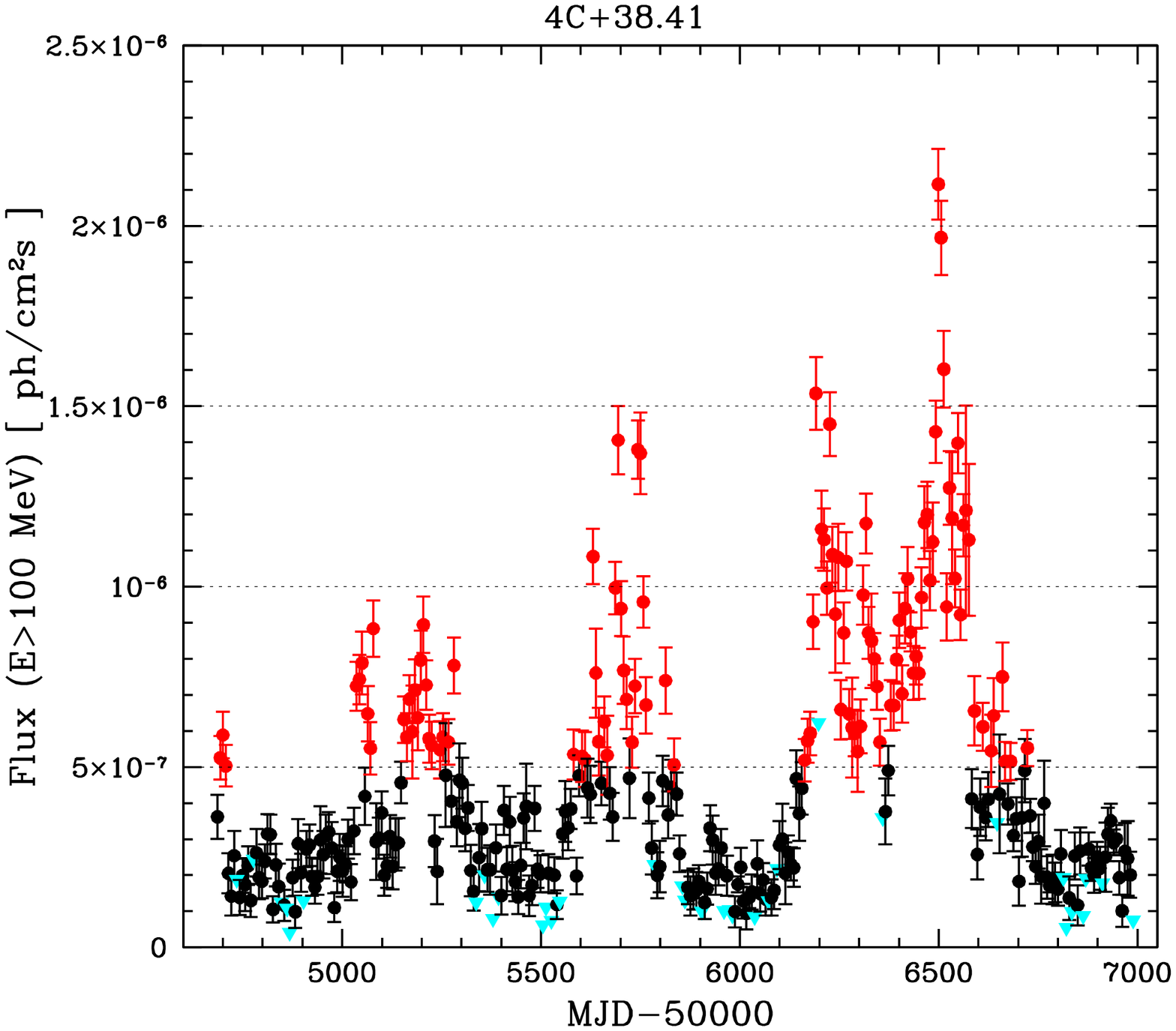,width=4.3cm,height=4.03cm}   & \psfig{file=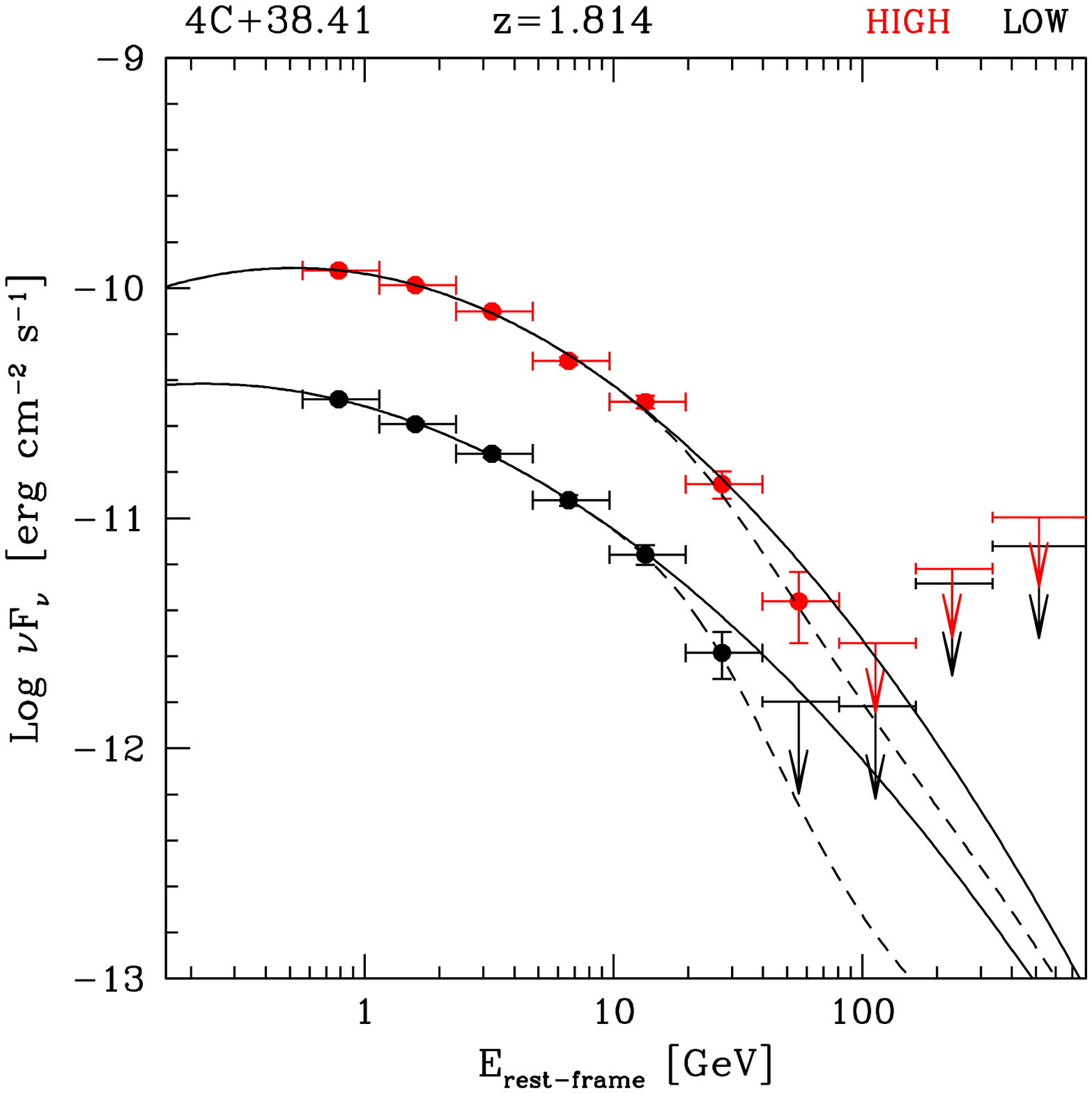,width=4.3cm,height=4.03cm} \\
 \psfig{file=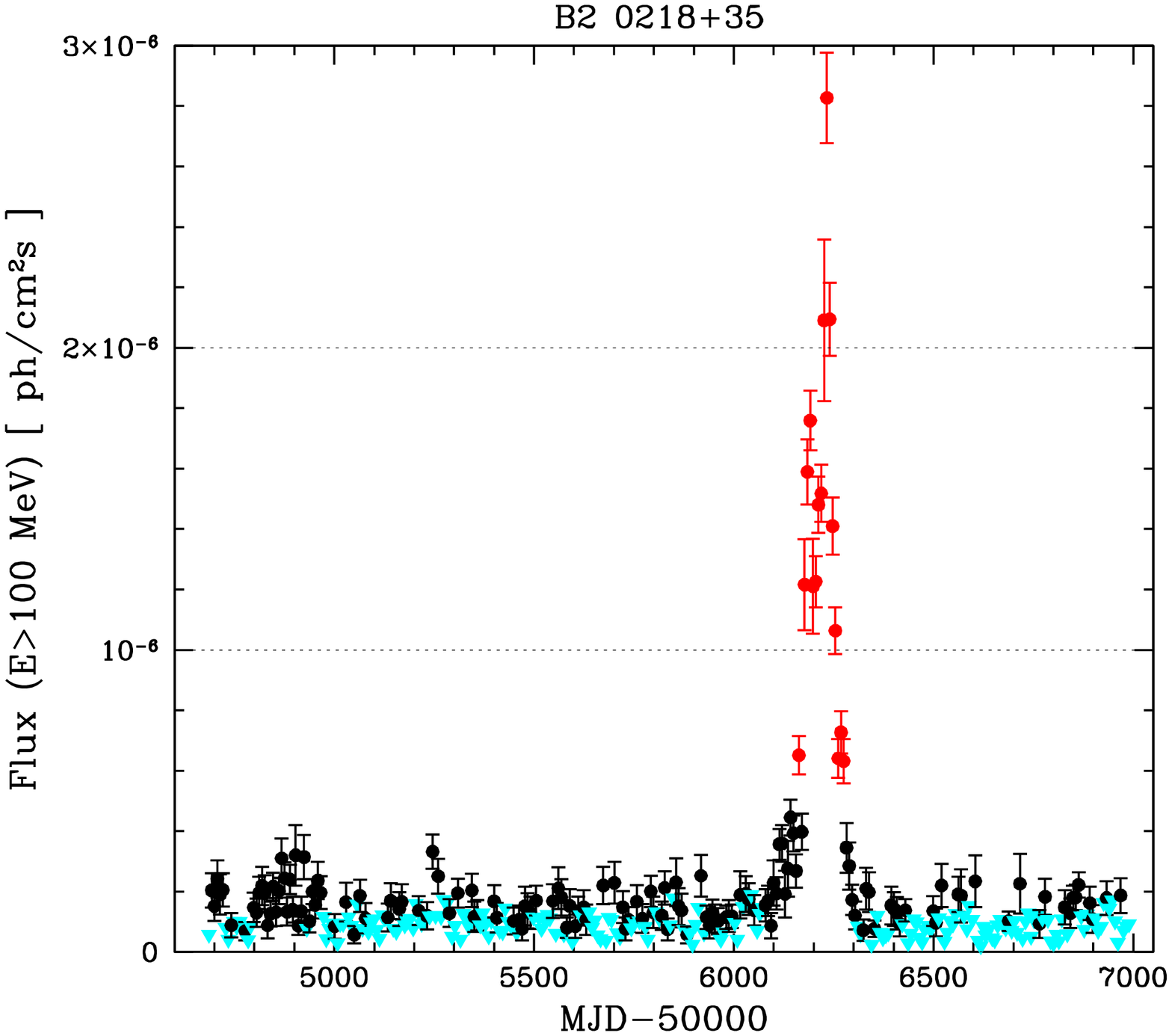,width=4.3cm,height=4.03cm}   & \psfig{file=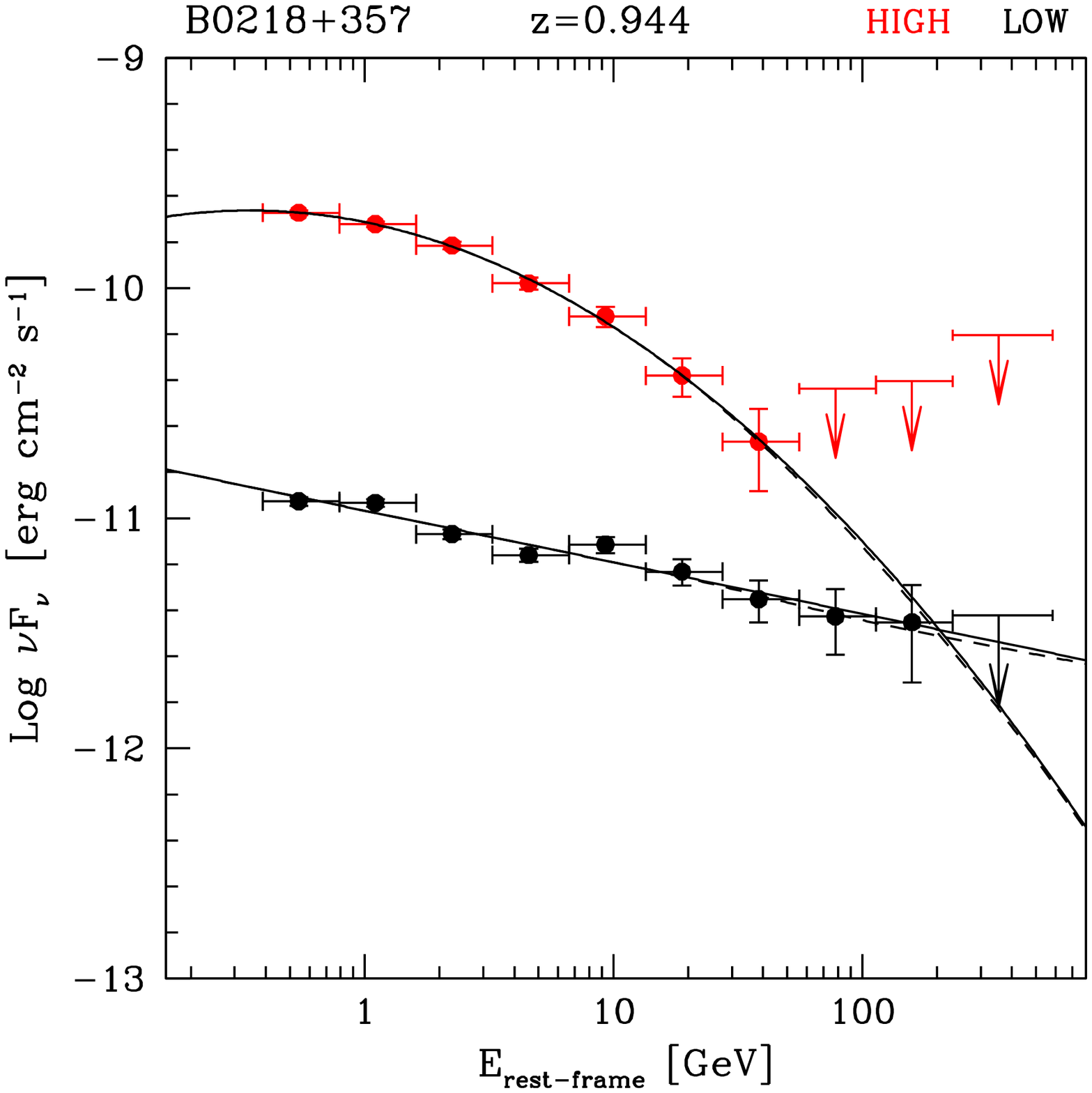,width=4.3cm,height=4.03cm} 
& \psfig{file=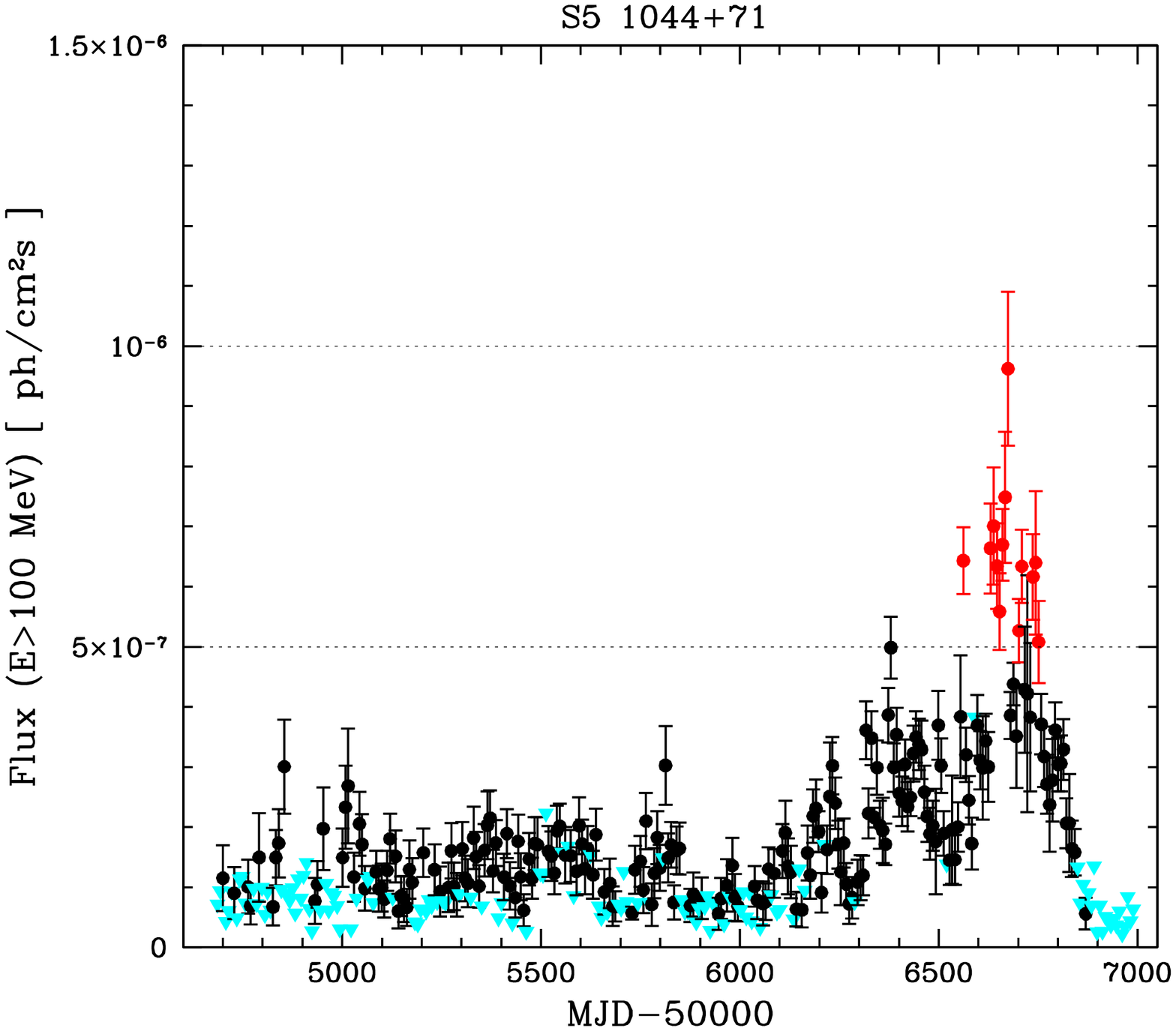,width=4.3cm,height=4.03cm }  & \psfig{file=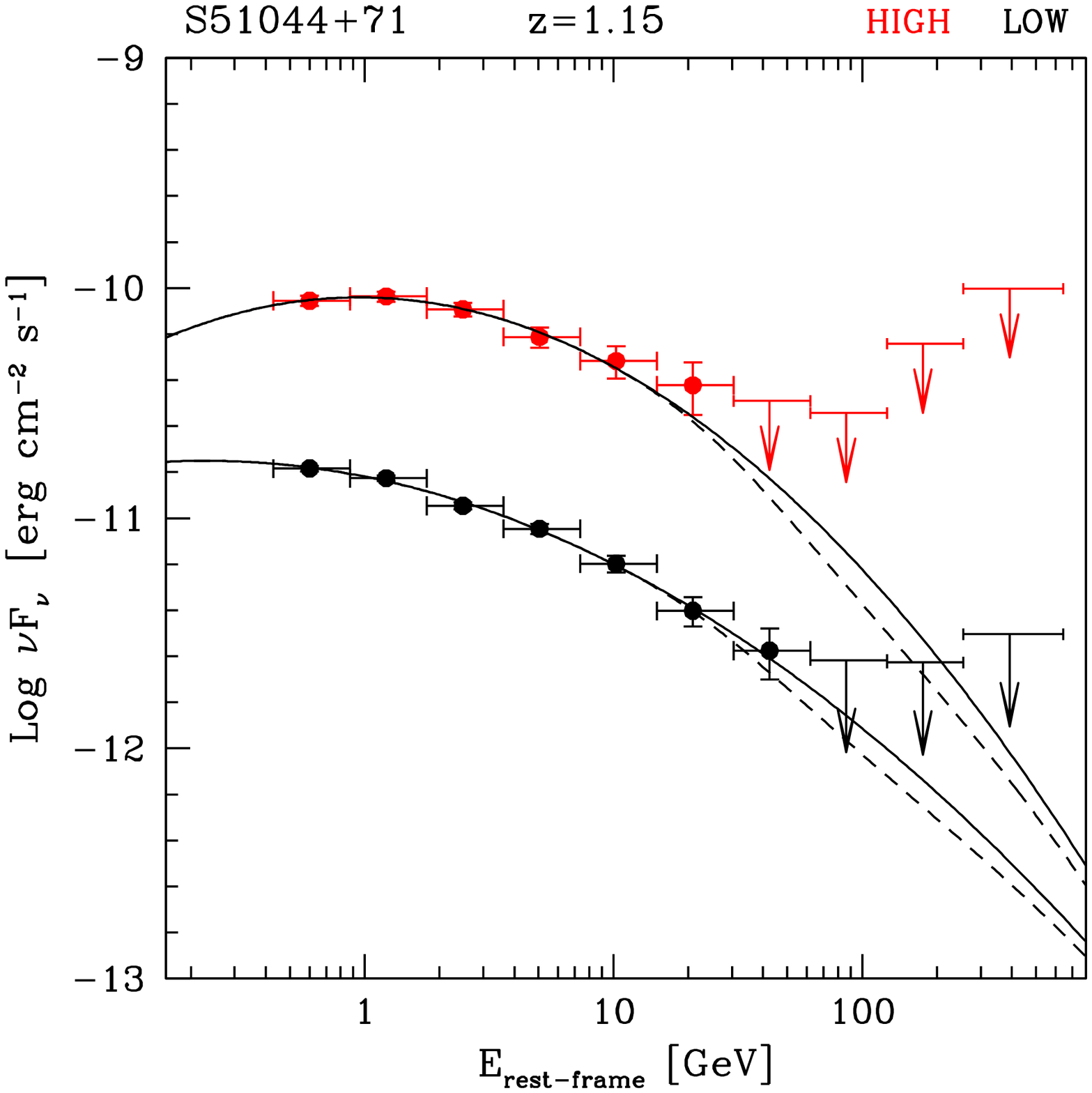,width=4.3cm,height=4.03cm } \\
 \psfig{file=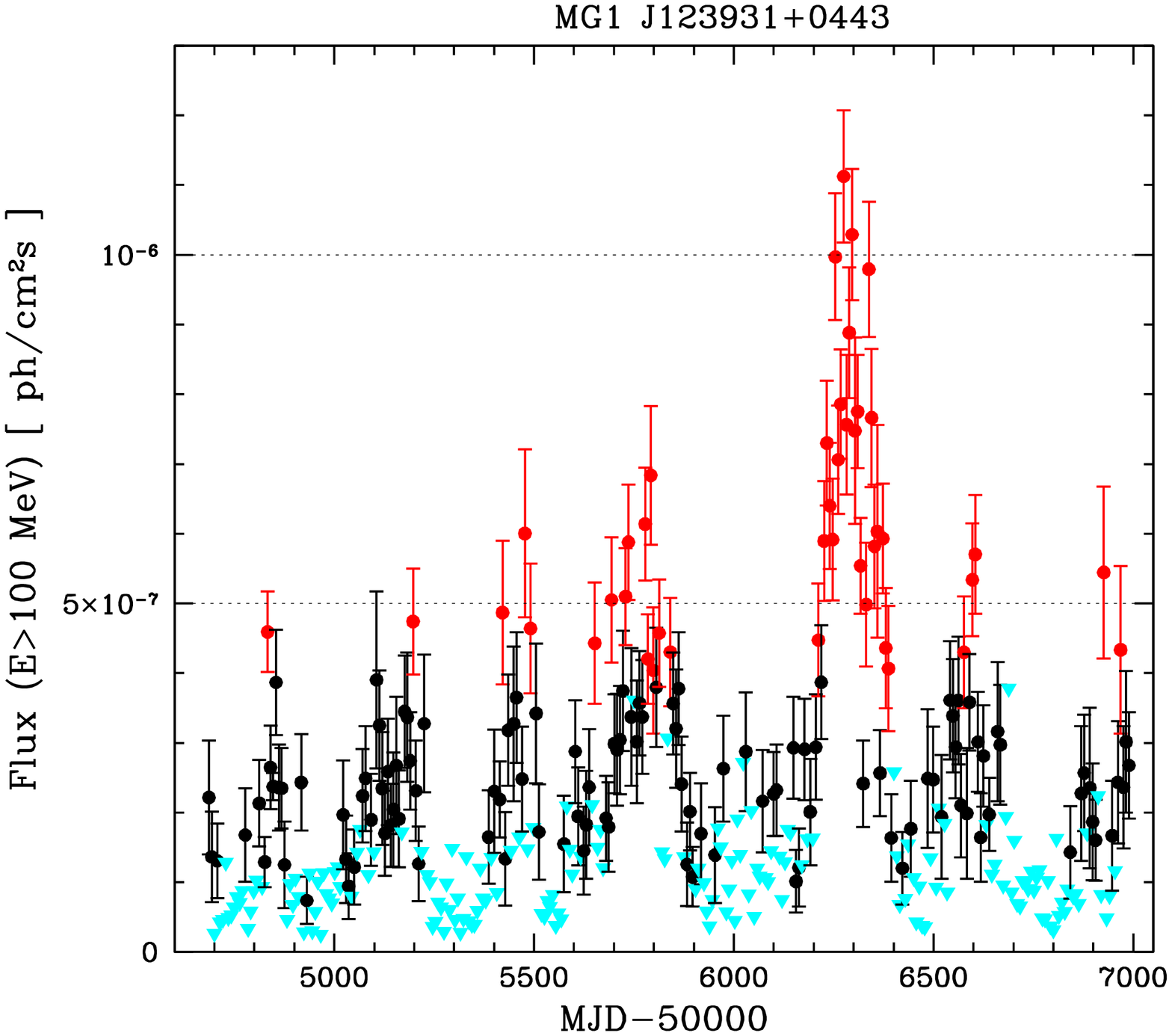,width=4.3cm,height=4.03cm }  & \psfig{file=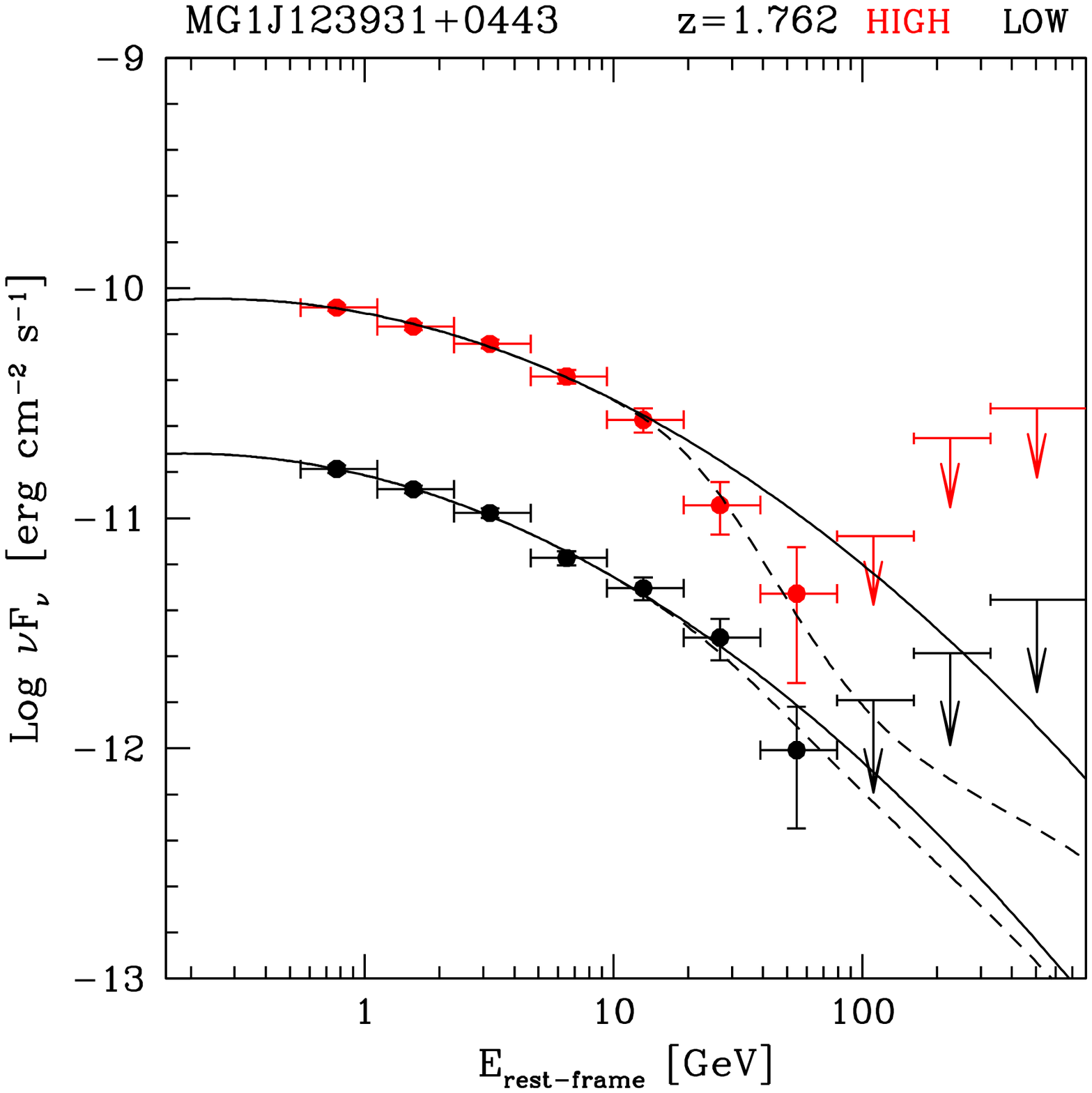,width=4.3cm,height=4.03cm } 
&\psfig{file=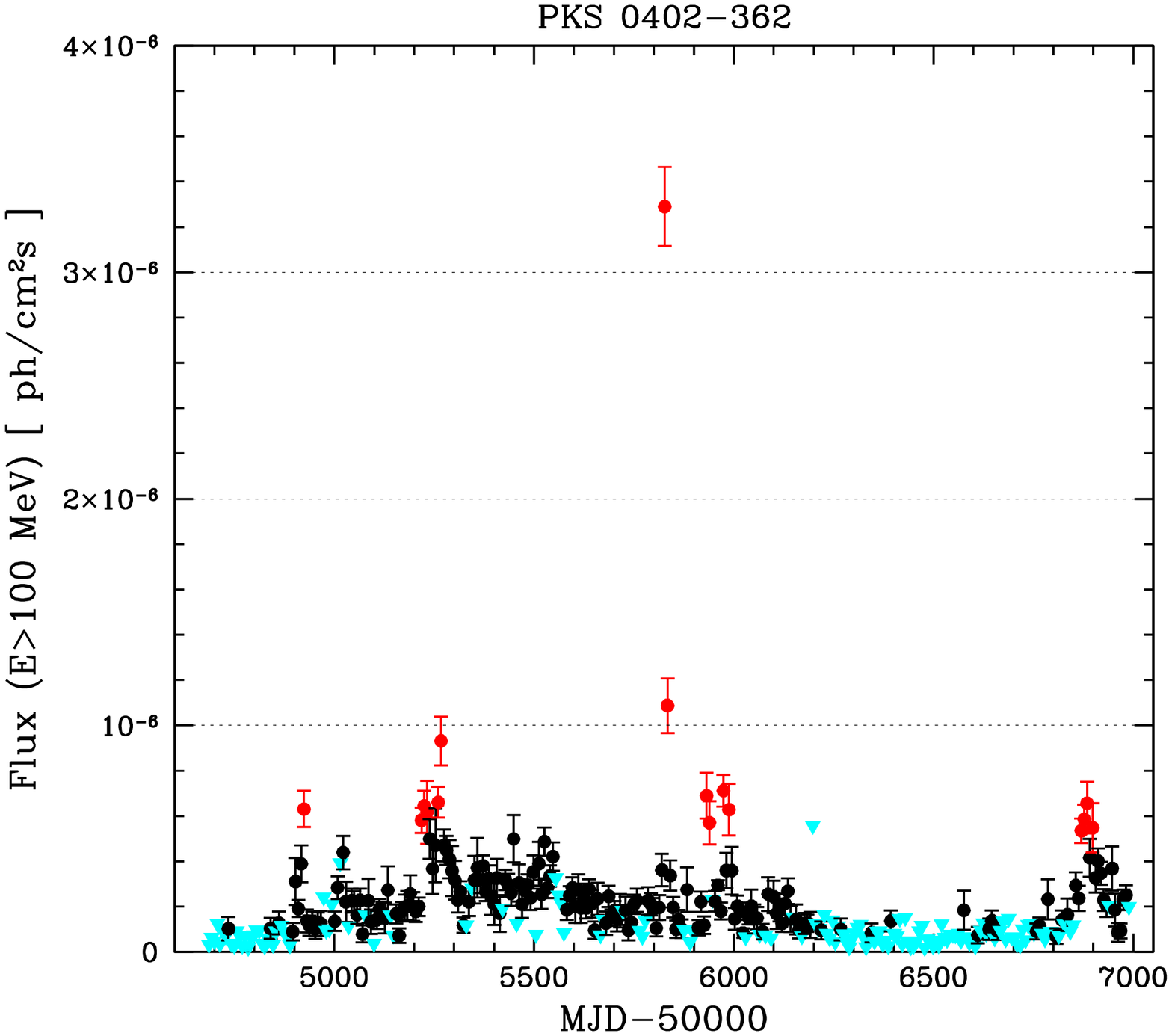,width=4.3cm,height=4.03cm }  & \psfig{file=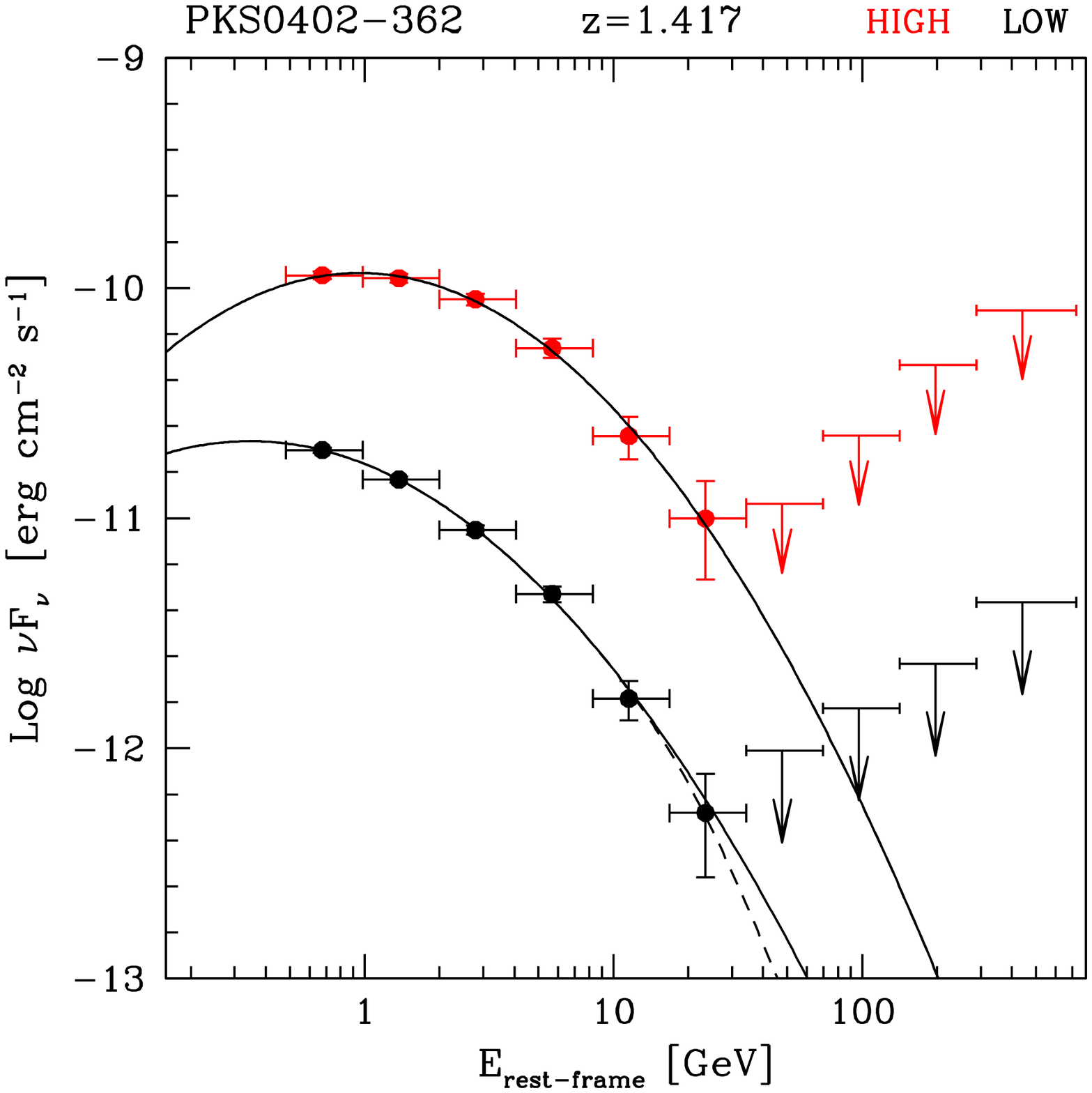,width=4.3cm,height=4.03cm } 
\end{tabular}
\caption{Lightcurves and gamma-ray SEDs of the FSRQ which have been divided in an average High (red) and Low-state (black) spectra. 
Lightcurves: time binning is 7-day bins. High-state bins are marked in red, Low-state bins in black or cyan. Cyan triangles 
correspond to upper limits, for the bins when the source is not detected in the 7-days timespan.
Spectra:  full lines show the intrinsic spectrum with parameters determined below 13 GeV,
dashed lines show the result of BLR absorption applied to the intrinsic spectrum and fitted to the data, 
with free optical depth.  }
\label{highlow1}
\end{figure*}

\begin{figure*}
\begin{tabular}{llrr}
 \psfig{file=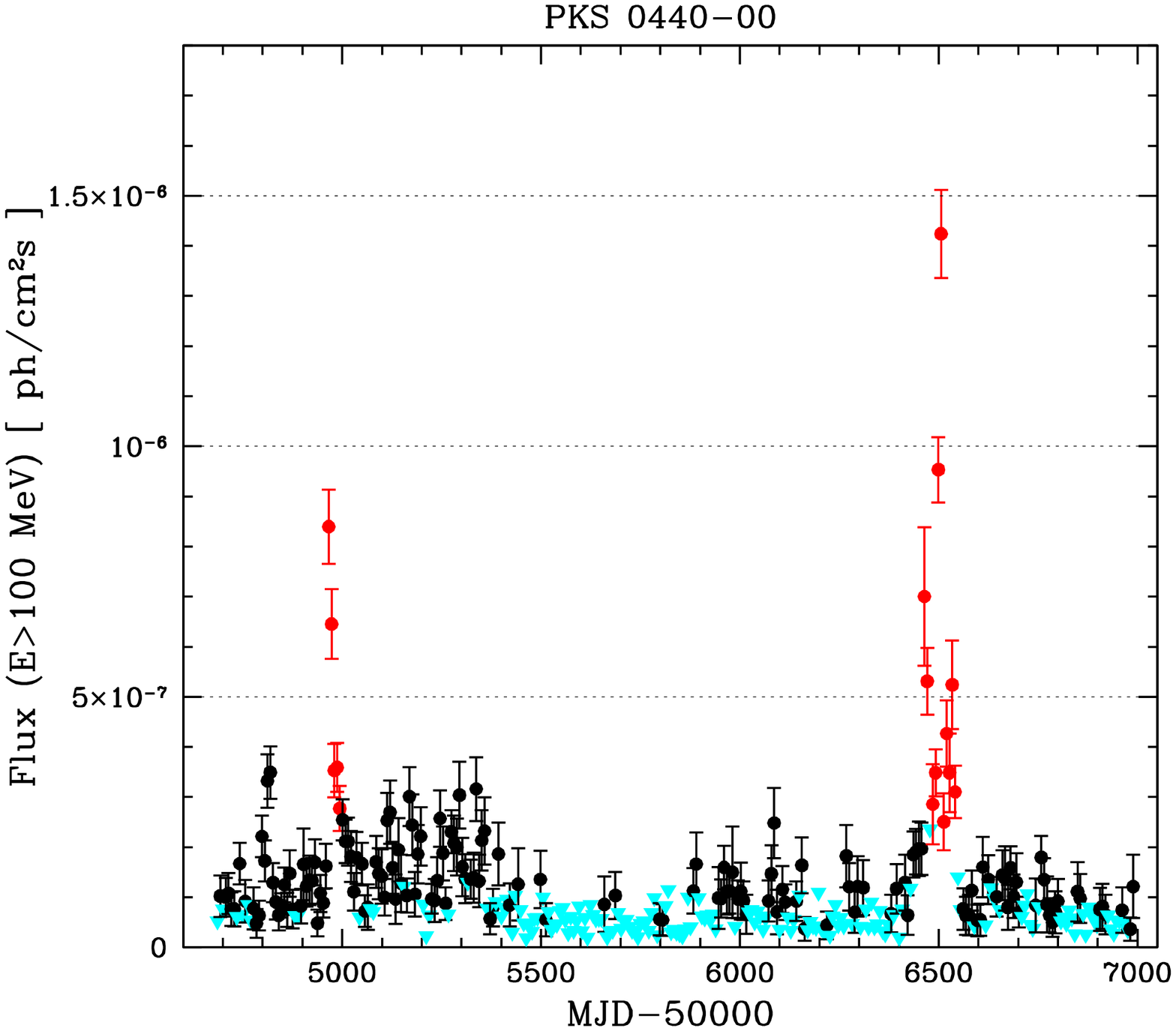,width=4.3cm,height=4.03cm }  & \psfig{file=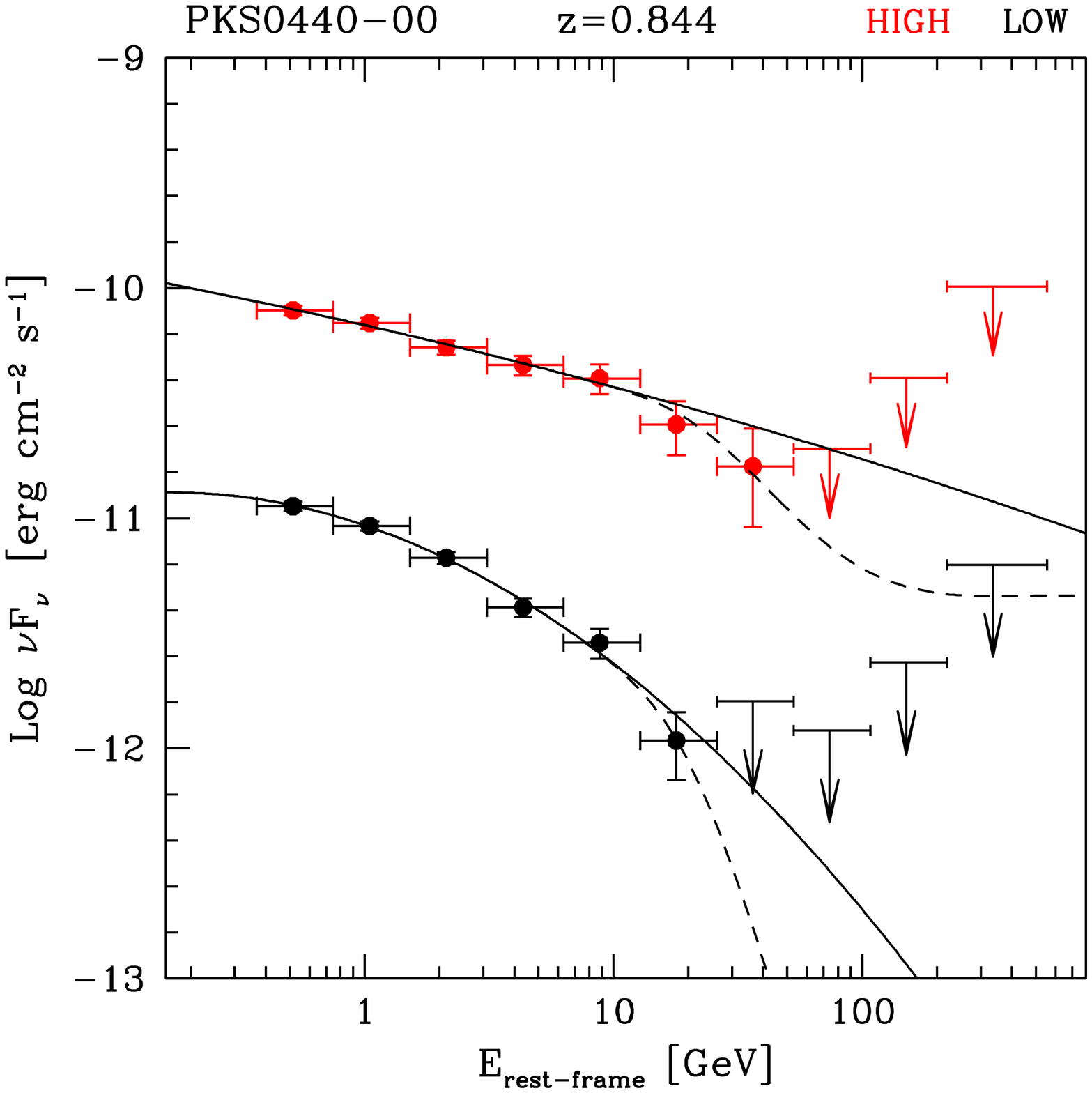,width=4.3cm,height=4.03cm } 
&\psfig{file=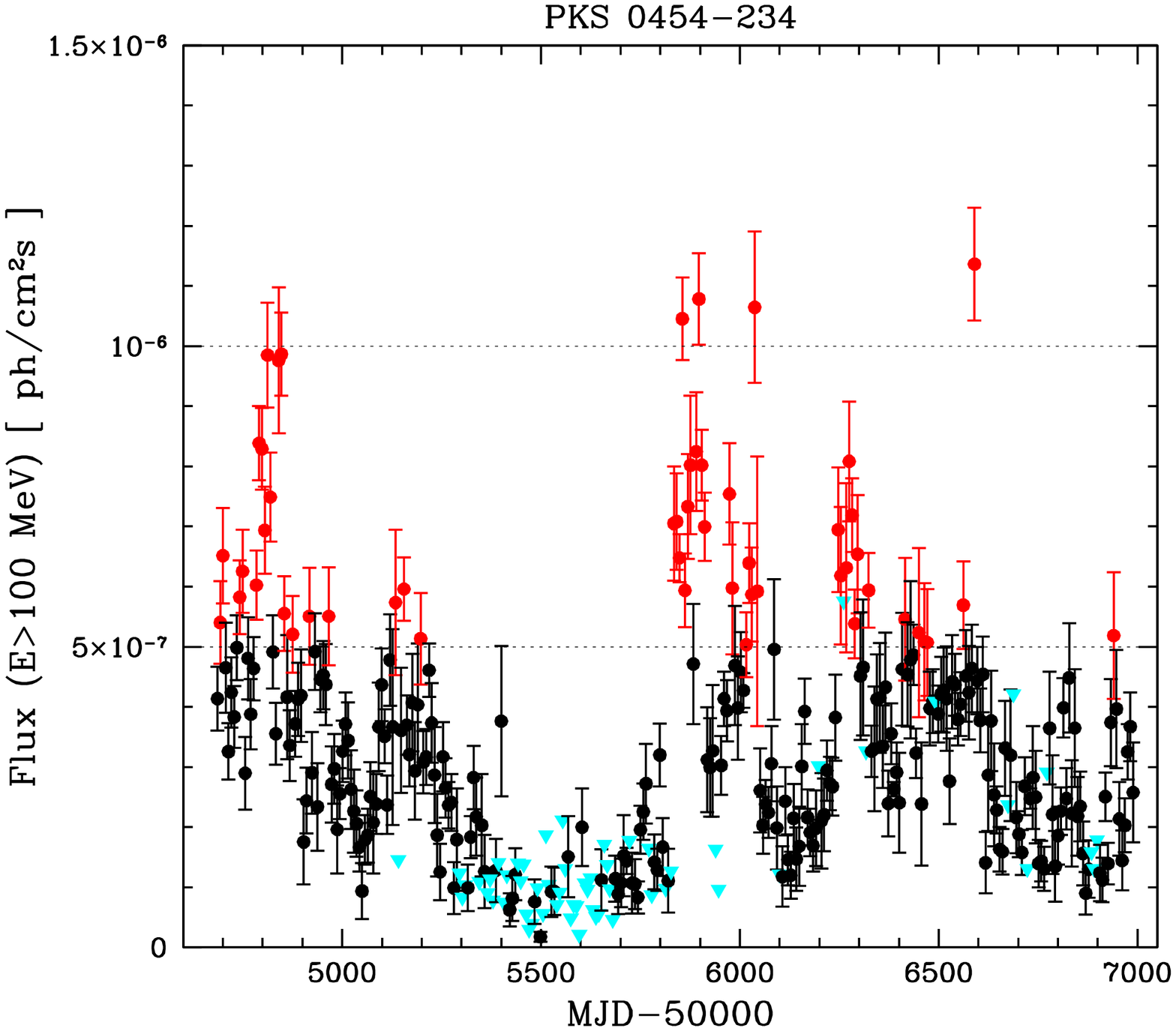,width=4.3cm,height=4.03cm }  & \psfig{file=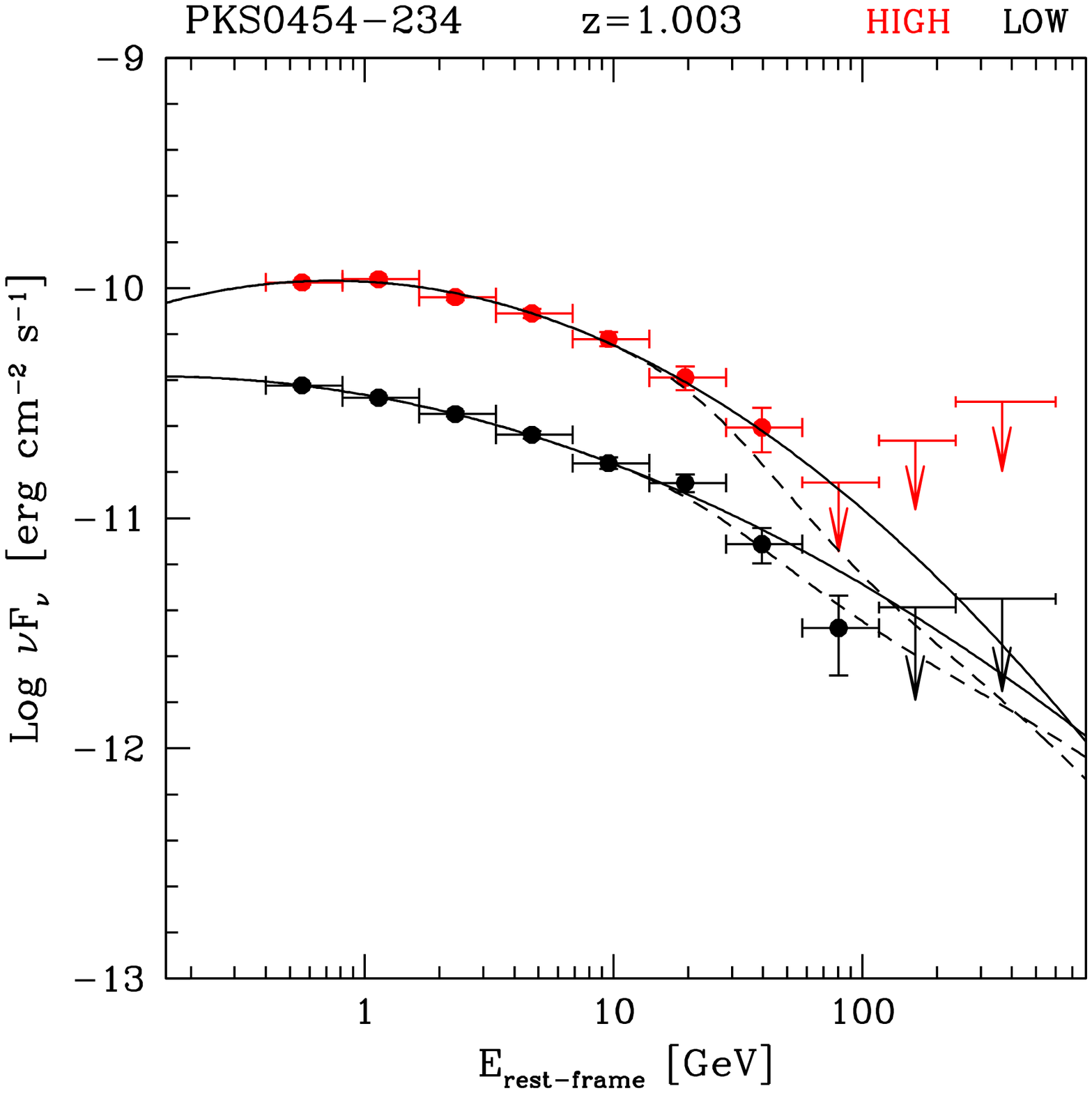,width=4.3cm,height=4.03cm } \\
 \psfig{file=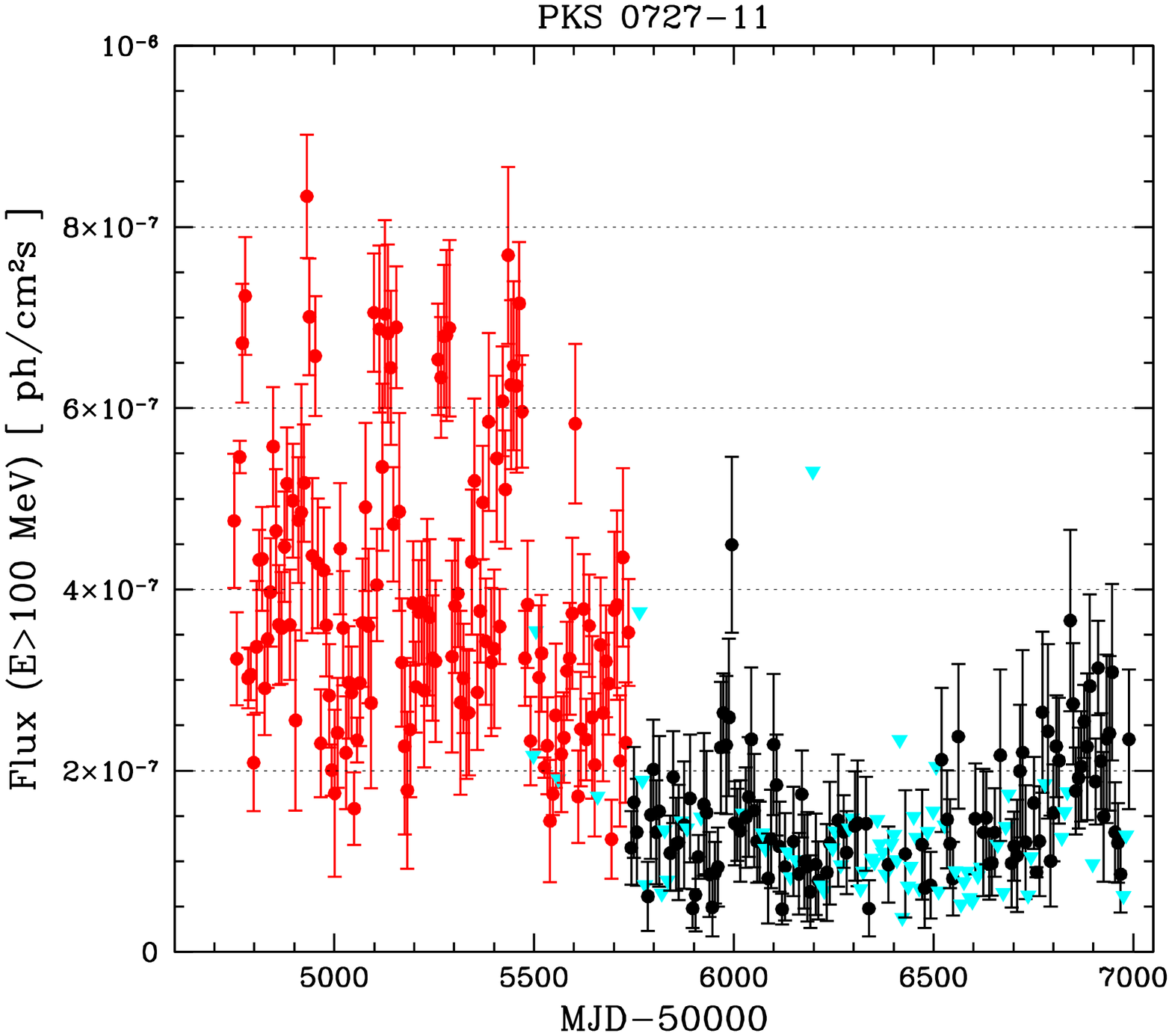,width=4.2cm,height=4.03cm }  & \psfig{file=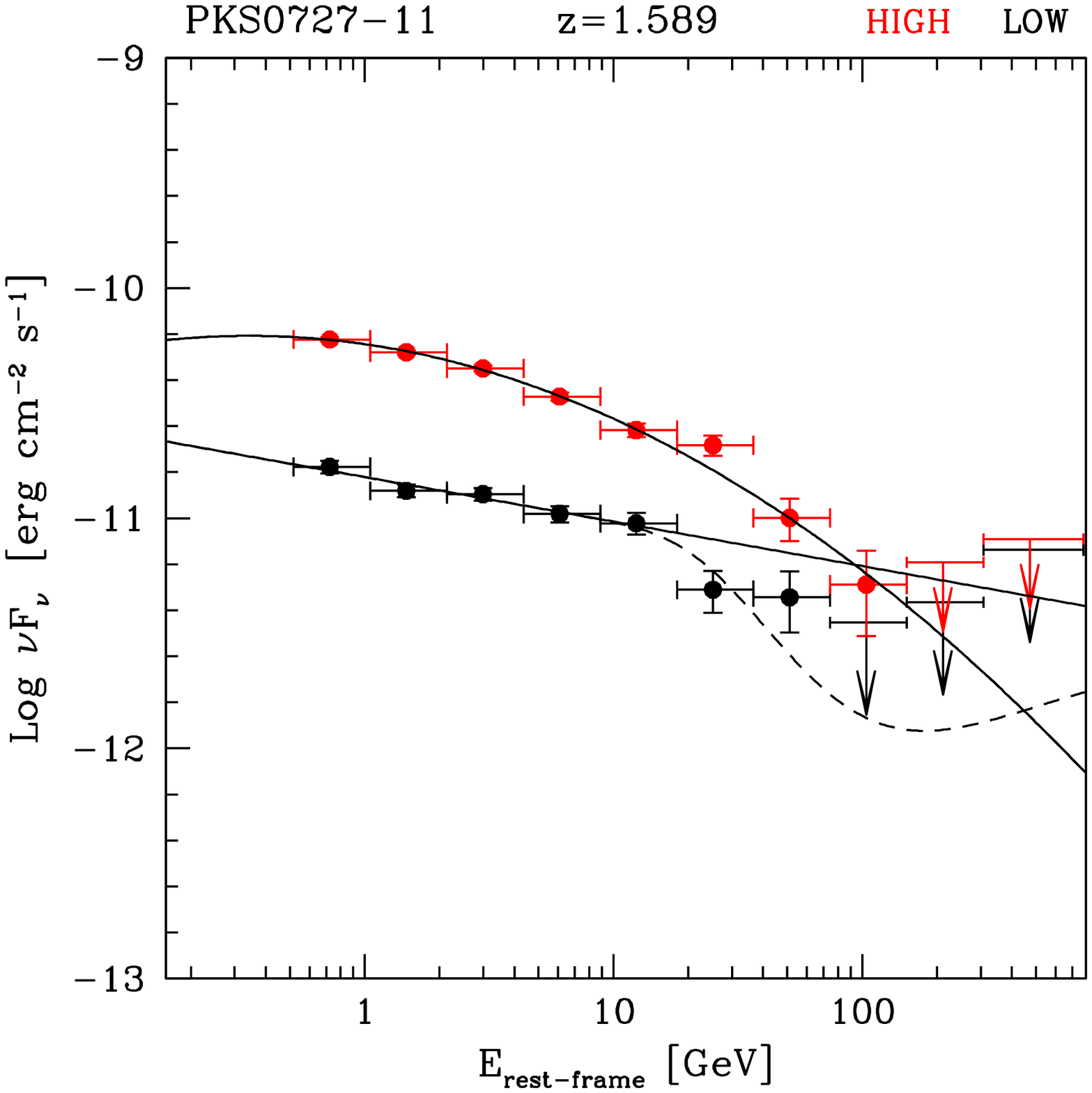,width=4.5cm,height=3.8cm } 
&\psfig{file=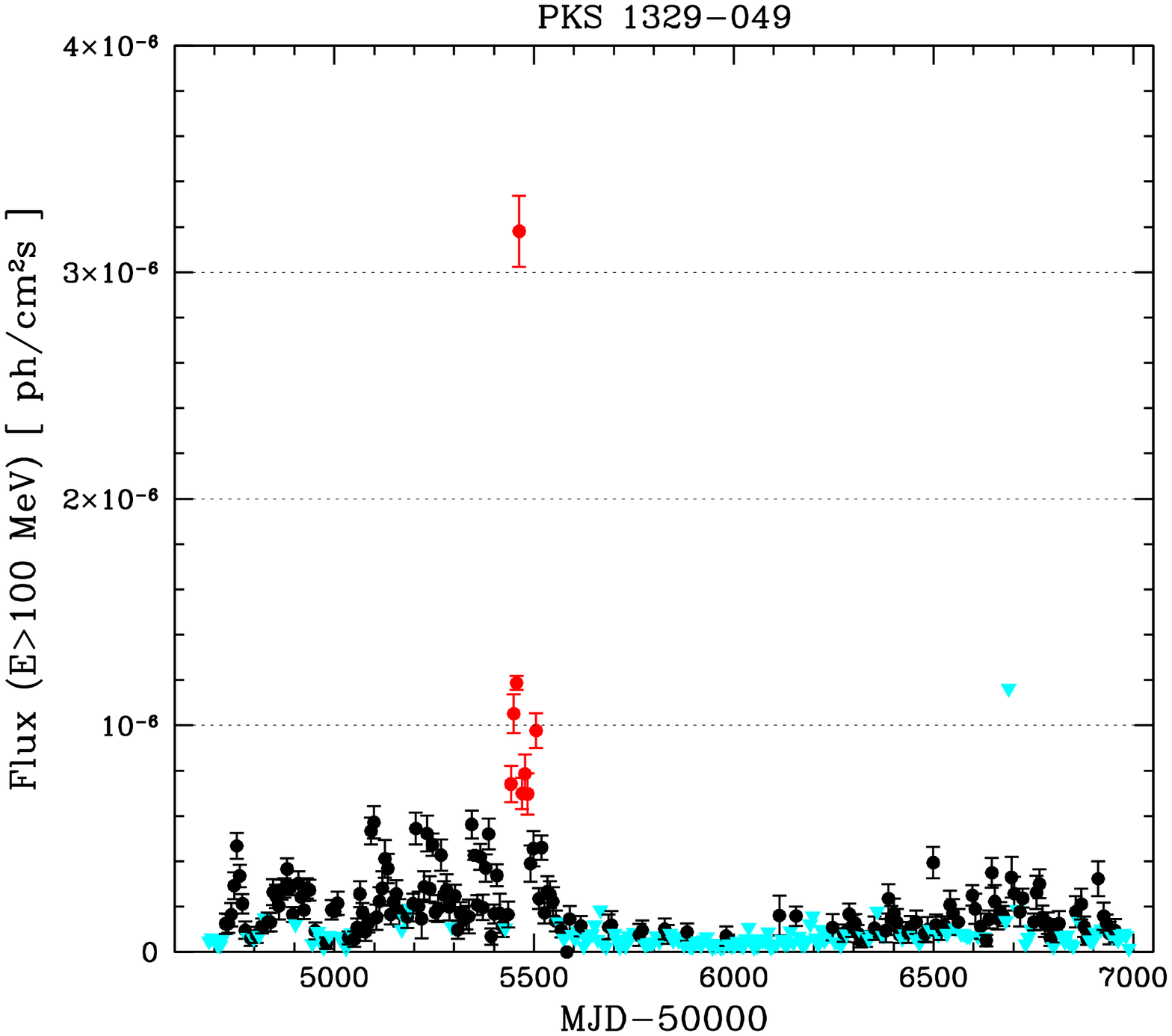,width=4.2cm,height=4.03cm }  & \psfig{file=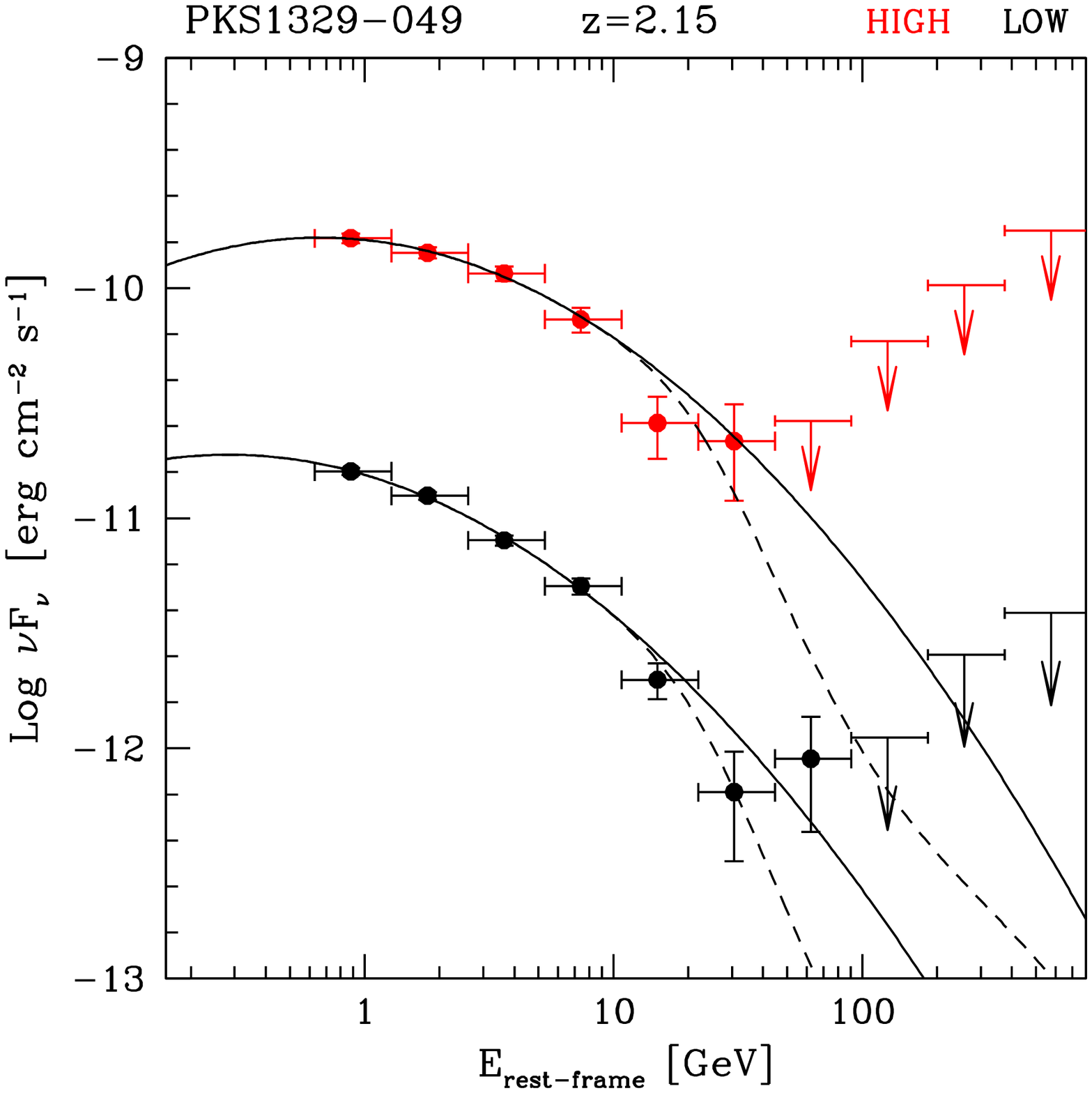,width=4.5cm,height=3.8cm } \\
 \psfig{file=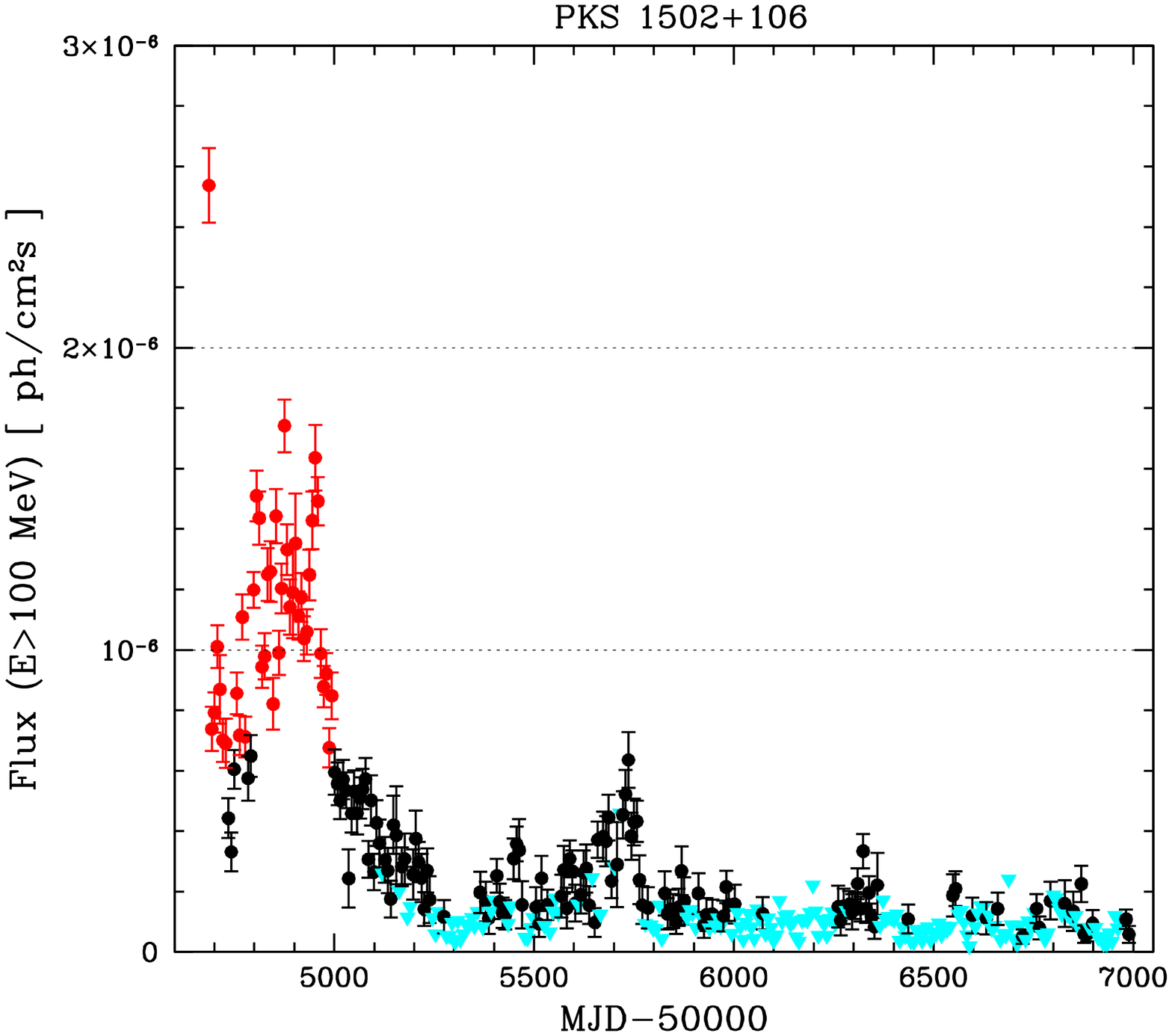,width=4.2cm,height=4.03cm }  & \psfig{file=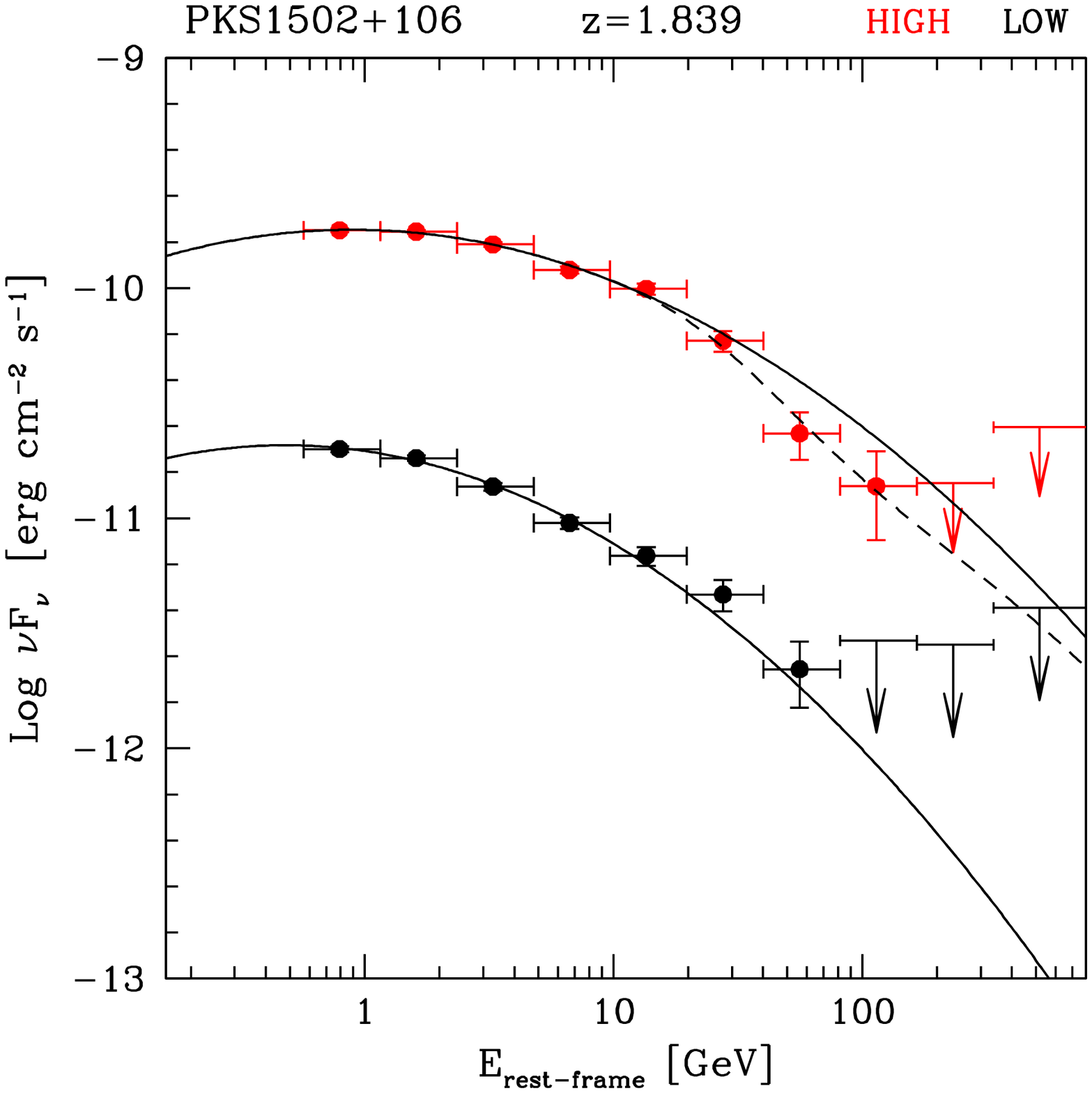,width=4.5cm,height=3.8cm } 
&\psfig{file=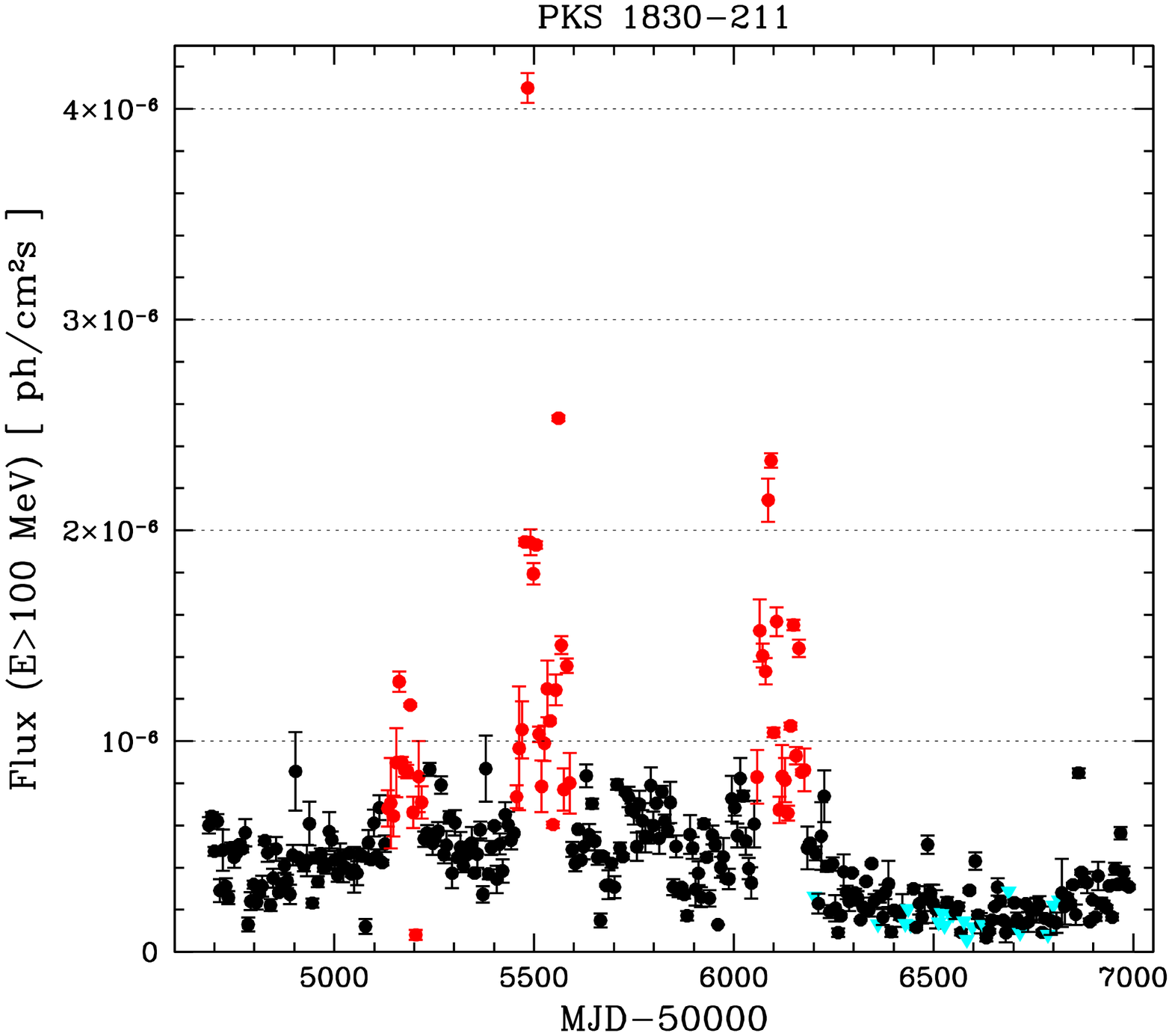,width=4.2cm,height=4.03cm }  & \psfig{file=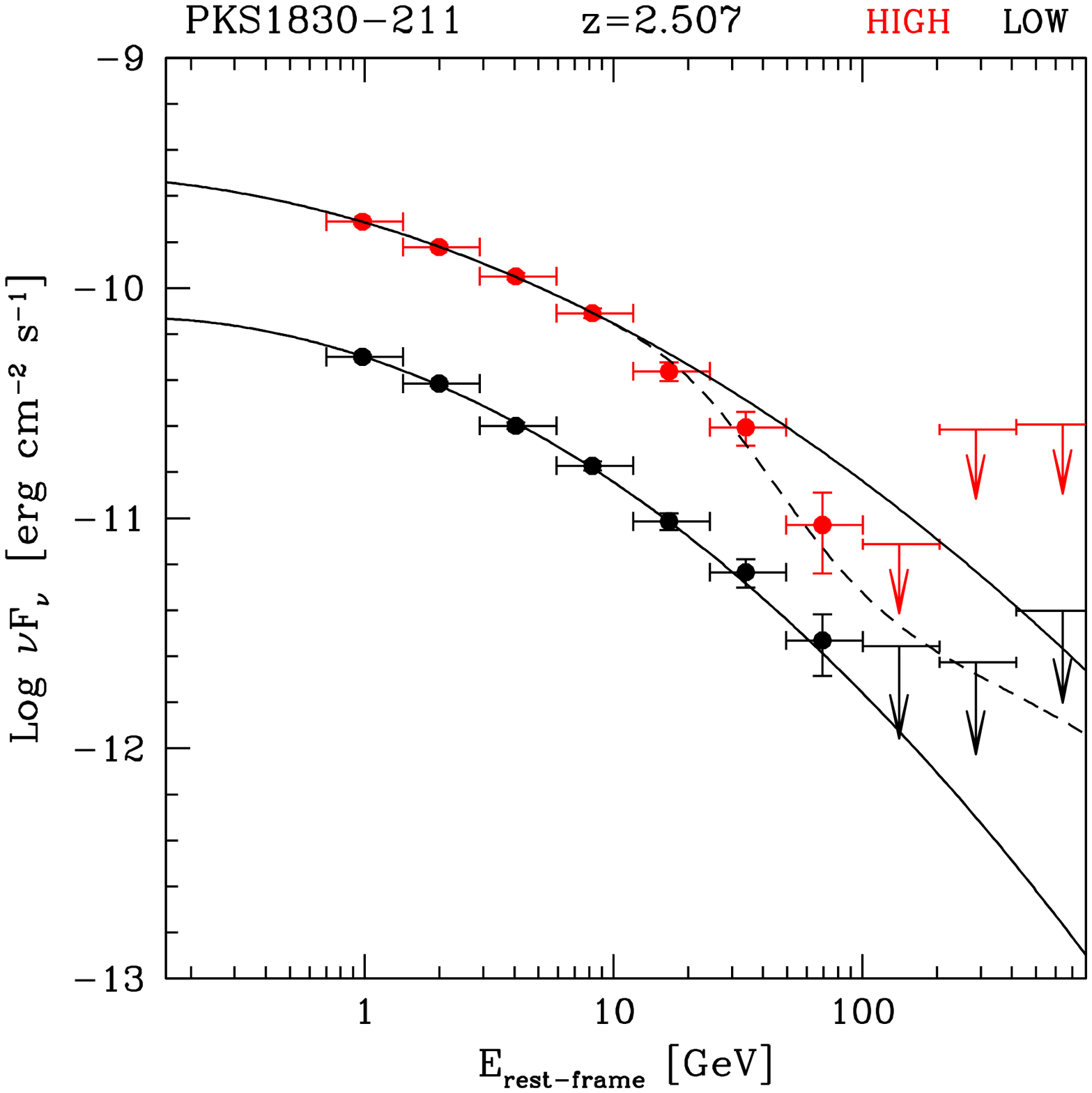,width=4.5cm,height=3.8cm } \\
 \psfig{file=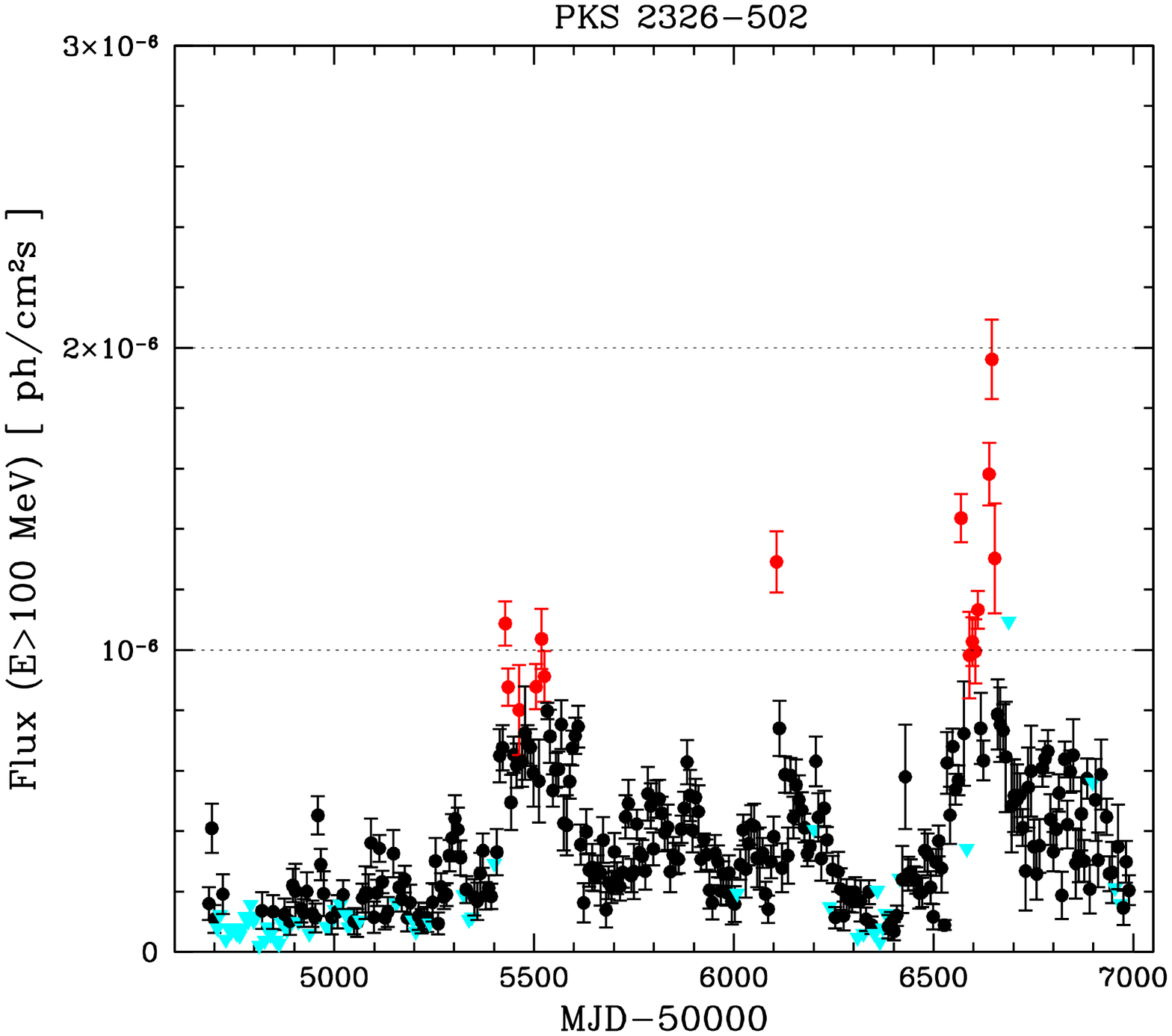,width=4.2cm,height=4.03cm }  & \psfig{file=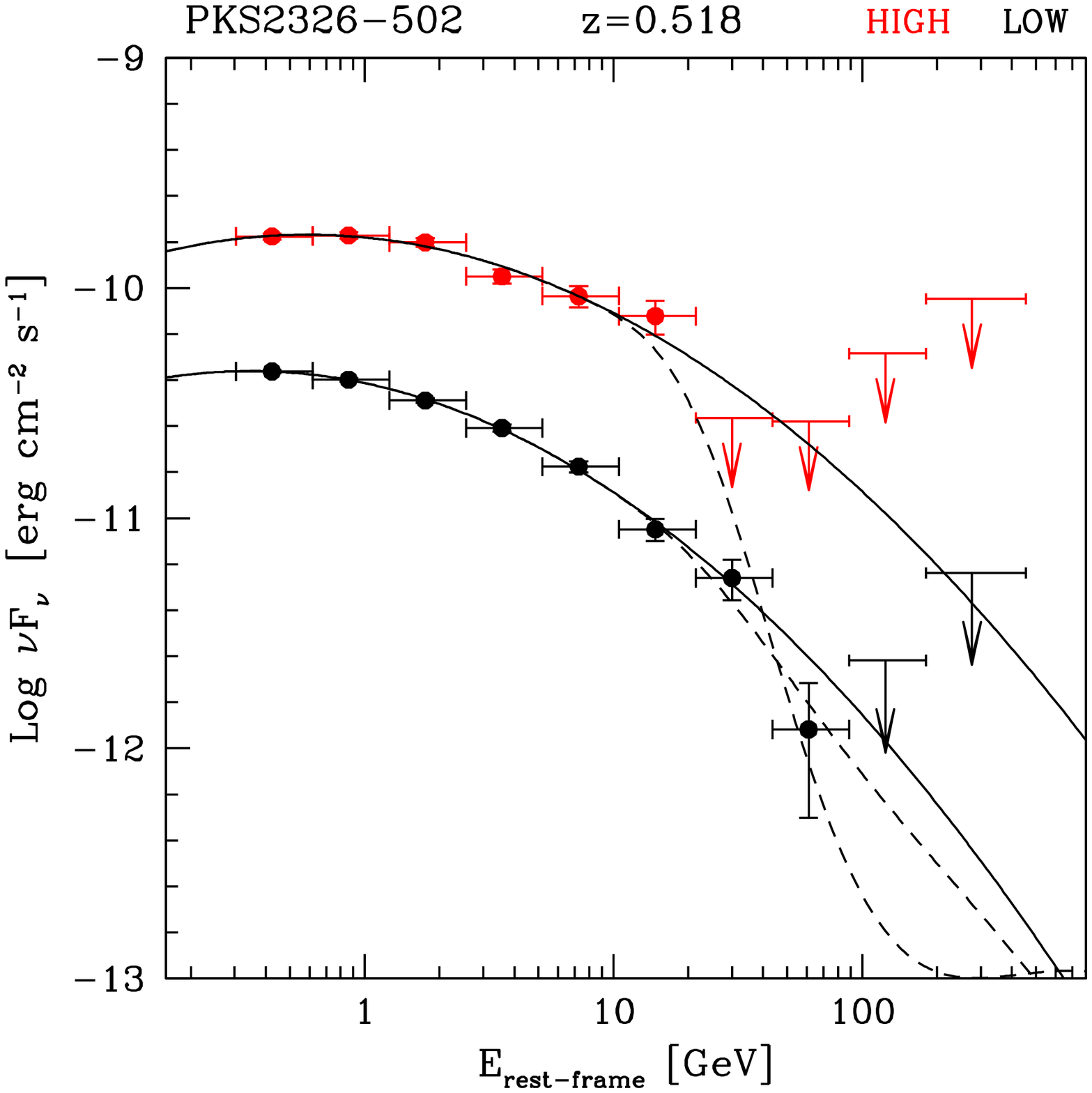,width=4.5cm,height=3.8cm } 
&\psfig{file=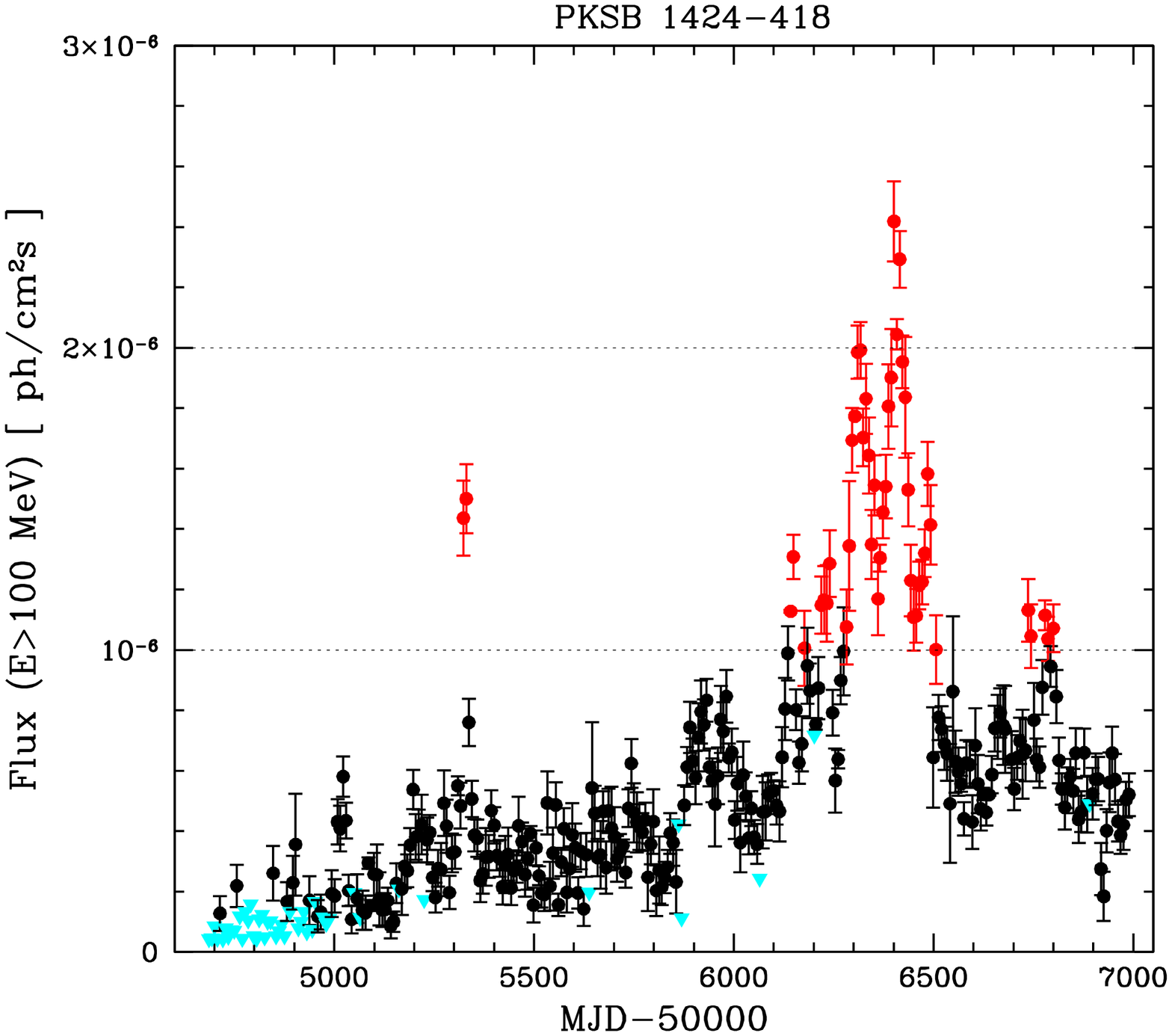,width=4.2cm,height=4.03cm }  & \psfig{file=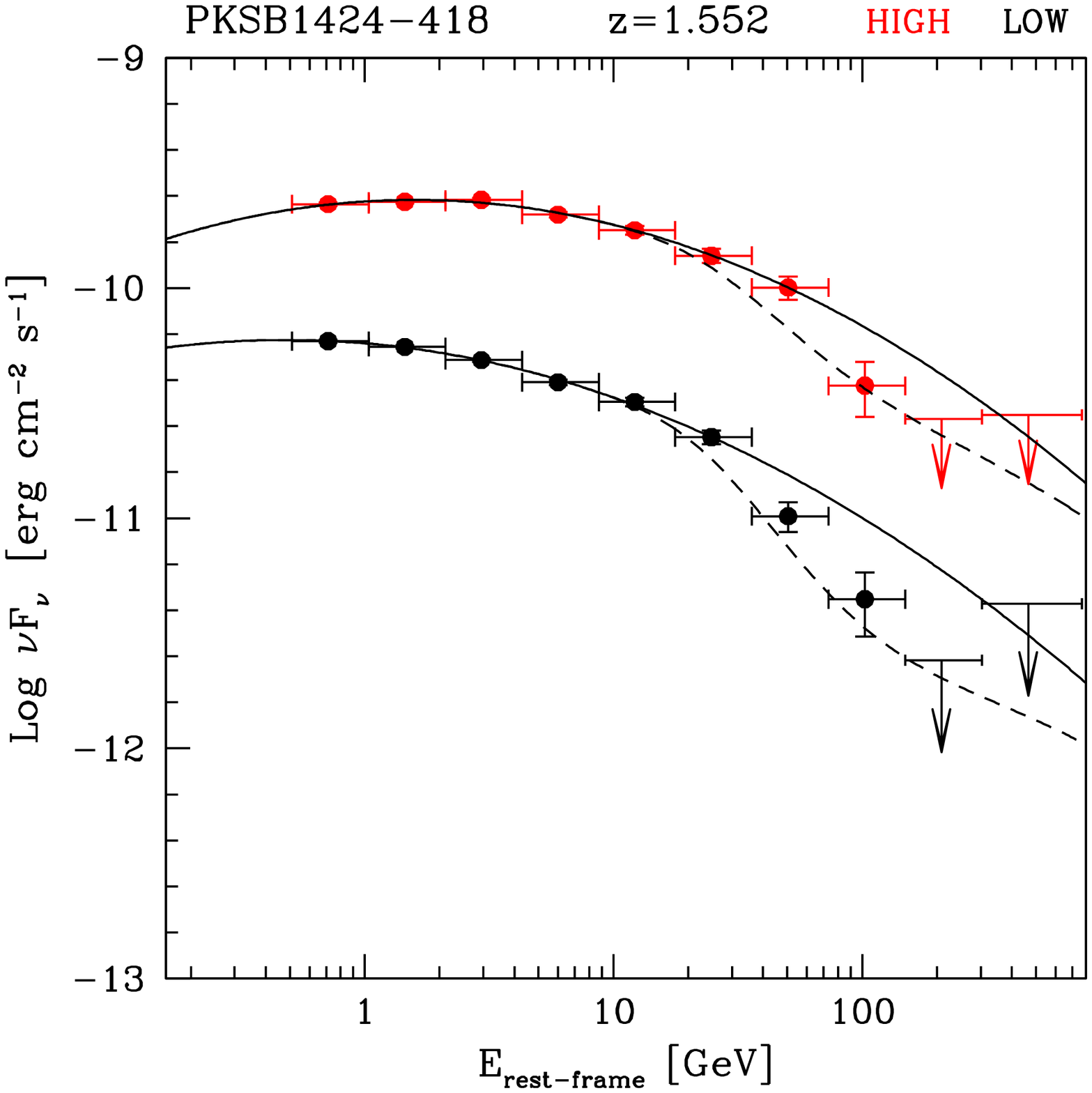,width=4.5cm,height=3.8cm } 
\end{tabular}
\contcaption{}  
\end{figure*}

\end{document}

%% file: tableAa.tex
\begin{table*}
\centering
\caption{
Sample of FSRQ blazars studied in this paper, sorted according to their overall significance in the 3LAC catalog (Col. 4).   
The upper table lists the 100 highest-significance objects in the 3LAC,  while the bottom table lists the six 
additional objects included for their large BLR.
Col. [1]: object name (from the association in the 3LAC).
Col. [2]: {\it Fermi}-LAT name in the 3LAC.
Col. [3]: redshift.
Col. [4]: significance of the detection in the 3LAC \citep[from][]{3LAC}. 
Col. [5]: luminosity of the Broad Line Region ($\equiv 0.1\,L_{\rm d}$), in  erg s$^{-1}$.
Col. [6]: associated  radius of the BLR, in cm, derived from the size-luminosity relation (see Sect. 2.1).
Col. [7]: maximum optical depth, at the peak of the \gg cross-section, for a path length $\ell=R_{\rm BLR}/2$ (see Sect. 2.1).  
Col. [8]: references for the BLR or emission lines luminosities.  
   X14: \citet{xiong14} from data in \citet{cao99,chai12,liu06};  GG10: \citet{gg2010}; GG11: \citet{gg2011highz}; 
   GG15: \citet{gg2015}; I15: \citet{isler15};  P14: \citet{pacciani14};  S12: \citet{Shaw12}; Sb14: \citet{sbarrato14};  
   T13: \citet{tavecchio1424}; To12: \citet{torrealba12}. 
}
\begin{tabular}{llcrrcrc}
\hline
\hline
Name  & 3LAC Name  &  $z$  &  $\sigma$  &  $L_{\rm BLR}$ & $R_{\rm BLR}$ & $\tau^{max}_{\rm BLR}$ & Refs   \\
~[1]     &[2]   &[3]   &[4]   &[5]   &[6]   &[7]   &[8]    \\ 
\hline 
	3C 454.3  & 3FGL J2254.0+1608 &   0.859 & 480.74  & 33.3e+44   & 5.8e+17 &   55.8 &  GG11  \\ 
       4C +21.35  & 3FGL J1224.9+2122 &   0.435 & 219.67  & 16.2e+44   & 4.0e+17 &   38.9 &  GG11  \\ 
     PKS 1510-08  & 3FGL J1512.8-0906 &    0.36 & 207.27  &  7.4e+44   & 2.7e+17 &   26.3 &  GG11  \\ 
      B2 1520+31  & 3FGL J1522.1+3144 &   1.487 & 192.68  &  4.5e+44   & 2.1e+17 &   20.5 &  GG15  \\ 
    PKS 1502+106  & 3FGL J1504.4+1029 &   1.839 & 189.47  & 19.8e+44   & 4.4e+17 &   43.0 &  GG11  \\ 
    PKS 0454-234  & 3FGL J0457.0-2324 &   1.003 & 161.08  & 18.8e+44   & 4.3e+17 &   41.9 &  GG10  \\ 
	  3C 279  & 3FGL J1256.1-0547 &   0.536 & 160.60  &  2.4e+44   & 1.5e+17 &   15.0 &  GG11  \\ 
	  3C 273  & 3FGL J1229.1+0202 &   0.158 & 149.01  & 33.8e+44   & 5.8e+17 &   56.2 &  GG11  \\ 
    PKS 2326-502  & 3FGL J2329.3-4955 &   0.518 & 130.01  &           &     --  &    --  &        \\  
   PKS B1424-418  & 3FGL J1427.9-4206 &   1.552 & 129.82  & 10.0e+44   & 3.2e+17 &   30.6 &  T13   \\ 
    PKS 1830-211  & 3FGL J1833.6-2103 &   2.507 & 129.72  &           &     --  &    --  &        \\  
       4C +55.17  & 3FGL J0957.6+5523 &   0.899 & 112.86  &  3.8e+44   & 1.9e+17 &   18.7 &  GG11  \\ 
     PKS 0727-11  & 3FGL J0730.2-1141 &   1.589 & 111.86  &           &     --  &    --  &  	  \\  
       4C +28.07  & 3FGL J0237.9+2848 &   1.213 & 105.19  & 18.0e+44$^a$   & 4.2e+17 &   41.0 &  I15   \\ 
    PKS 0402-362  & 3FGL J0403.9-3604 &   1.417 & 103.56  & 14.0e+44$^a$   & 3.7e+17 &   36.2 &  I15   \\ 
	 Ton 599  & 3FGL J1159.5+2914 &   0.725 &  93.40  &  5.1e+44  & 2.3e+17 &   21.9 &  GG11  \\ 
    PKS 1244-255  & 3FGL J1246.7-2547 &   0.635 &  92.64  & 1.2e+44$^a$  & 1.1e+17 &   10.5 &  To12   \\ 
       4C +38.41  & 3FGL J1635.2+3809 &   1.814 &  86.97  & 60.0e+44   & 7.7e+17 &   74.9 &  GG15  \\ 
  PMN J2345-1555  & 3FGL J2345.2-1554 &   0.621 &  82.32  &  2.1e+44   & 1.4e+17 &   14.0 &  GG15  \\ 
      S4 0917+44  & 3FGL J0920.9+4442 &    2.19 &  82.11  & 70.8e+44   & 8.4e+17 &   81.3 &  GG11  \\ 
     PKS 2052-47  & 3FGL J2056.2-4714 &   1.491 &  78.95  &  8.5e+44$^a$  & 2.9e+17 &   28.1 &  I15   \\  
    PKS 1124-186  & 3FGL J1127.0-1857 &   1.048 &  76.26  & 	      &     --  &    --  &         \\ 
     PKS 2142-75  & 3FGL J2147.3-7536 &   1.138 &  75.82  &  8.0e+44$^a$   & 2.8e+17 &   27.3 &  I15    \\  
     B3 1343+451  & 3FGL J1345.6+4453 &   2.534 &  75.77  & 10.5e+44   & 3.2e+17 &   31.3 &  GG15   \\  
     PKS 0805-07  & 3FGL J0808.2-0751 &   1.837 &  73.84  & 24.0e+44   & 4.9e+17 &   47.4 &  P14      \\  
    PKS 0244-470  & 3FGL J0245.9-4651 &   1.385 &  73.49  & 27.0e+44   & 5.2e+17 &   50.2 &  GG15   \\   
	 CTA 102  & 3FGL J2232.5+1143 &   1.037 &  72.20  & 41.4e+44   & 6.4e+17 &   62.2 &  GG11   \\ 
       4C +01.02  & 3FGL J0108.7+0134 &   2.099 &  71.65  & 75.0e+44   & 8.7e+17 &   83.7 &  GG11   \\ 
       4C +14.23  & 3FGL J0725.2+1425 &   1.038 &  69.90  &  9.0e+44   & 3.0e+17 &   29.0 &  GG15   \\ 
    PKS 0250-225  & 3FGL J0252.8-2218 &   1.427 &  69.69  &  3.2e+44  & 1.8e+17 &   17.2 &  GG15   \\ 
      B2 2308+34  & 3FGL J2311.0+3425 &   1.817 &  66.87  & 	      &     --  &    --  &  	   \\ 
      B2 0716+33  & 3FGL J0719.3+3307 &   0.779 &  66.55  &   4.5e+44  & 2.1e+17 &   20.5 &  GG10   \\ 	     	
MG1 J123931+0443  & 3FGL J1239.5+0443 &   1.762 &  65.00  &  9.1e+44  & 3.0e+17 &   29.2 &  GG11   \\ 
  PMN J1802-3940  & 3FGL J1802.6-3940 &   1.319 &  63.57  & 15.0e+44   & 3.9e+17 &   37.4 &  GG15   \\ 
      S4 1849+67  & 3FGL J1849.2+6705 &   0.657 &  63.28  &  3.4e+44$^a$  & 1.8e+17 &   17.8 &  To12    \\ 
     PKS 2023-07  & 3FGL J2025.6-0736 &   1.388 &  63.11  &  22.5e+44  & 4.7e+17 &   45.8 &  GG10   \\ 	     	 
    PKS 1454-354  & 3FGL J1457.4-3539 &   1.424 &  62.20  &  45.0e+44  & 6.7e+17 &   64.8 &  GG10   \\ 	     	
     B3 1708+433  & 3FGL J1709.6+4318 &   1.027 &  62.20  &  1.5e+44   & 1.2e+17 &   11.8 &  GG15   \\  
  GB6 J0742+5444  & 3FGL J0742.6+5444 &    0.72 &  61.62  & 	      &     --  &    --  &  	   \\ 
	  OX 169  & 3FGL J2143.5+1744 &   0.213 &  60.57  &  1.8e+44  & 1.3e+17 &   13.0 &  GG11   \\  
     PKS 2227-08  & 3FGL J2229.7-0833 &    1.56 &  58.57  & 46.1e+44  & 6.8e+17 &   65.7 &  GG11   \\
    PKS 2255-282  & 3FGL J2258.0-2759 &   0.926 &  56.77  & 18.5e+44$^a$  & 4.3e+17 &   41.6 &  To12   \\ 
      B2 2155+31  & 3FGL J2157.5+3126 &   1.488 &  56.67  &  5.4e+44   & 2.3e+17 &   22.5 &  GG15  \\   
    PKS 1622-253  & 3FGL J1625.7-2527 &   0.786 &  56.26  &  0.4e+44$^a$  & 6.6e+16 &    6.4 & X14  \\
     PKS 0420-01  & 3FGL J0423.2-0119 &   0.916 &  56.14  &  7.1e+44$^a$   & 2.7e+17 &   25.8 & X14  \\
	  OC 457  & 3FGL J0137.0+4752 &   0.859 &  55.94  &  4.6e+44$^a$   & 2.1e+17 &   20.7 & X14  \\
    PKS 2201+171  & 3FGL J2203.4+1725 &   1.076 &  54.14  &  12.0e+44  & 3.5e+17 &   33.5 &  GG10  \\  
	  OP 313  & 3FGL J1310.6+3222 &   0.997 &  51.97  &  8.4e+44  & 2.9e+17 &   28.0 &  GG11  \\  
    TXS 0059+581  & 3FGL J0102.8+5825 &   0.644 &  51.50  &  4.5e+44   & 2.1e+17 &   20.5 &  GG15  \\  
      S4 1030+61  & 3FGL J1033.8+6051 &   1.401 &  51.44  &  4.5e+44  & 2.1e+17 &   20.5 &  GG11  \\  
	  OG 050  & 3FGL J0532.7+0732 &   1.254 &  50.90  &  9.0e+44   & 3.0e+17 &   29.0 &  GG15  \\  
     PKS 0440-00  & 3FGL J0442.6-0017 &   0.844 &  50.84  &  6.0e+44   & 2.4e+17 &   23.7 &  GG15  \\  
\hline		 
\hline  	 
\end{tabular}	 
\label{lista}
\end{table*}

%% file: tableAb.tex
\begin{table*}
\centering
\contcaption{}
\begin{tabular}{llcrrcrc}
\hline
\hline
Name  & 3LAC Name  &  $z$  &  $\sigma$  &  $L_{\rm BLR}$ & $R_{\rm BLR}$ & $\tau^{max}_{\rm BLR}$ & Refs   \\
~[1]     &[2]   &[3]   &[4]   &[5]   &[6]   &[7]   &[8]    \\ 
\hline 
    PKS 1329-049  & 3FGL J1332.0-0508 &    2.15 &  49.71  & 67.0e+44   & 8.2e+17 &   79.1 &  GG11  \\   
    TXS 1700+685  & 3FGL J1700.1+6829 &   0.301 &  49.45  & 	      &     --  &    --  &  	  \\ 
      S4 1144+40  & 3FGL J1146.8+3958 &   1.089 &   48.70  &  11.7e+44 & 3.4e+17 &   33.1 &  GG11  \\   
       B0218+357  & 3FGL J0221.1+3556 &   0.944 &  48.44  &  0.6e+44   & 7.7e+16 &    7.5 &  GG10  \\   
    PKS 0215+015  & 3FGL J0217.8+0143 &   1.715 &  46.29  &  30.0e+44  & 5.5e+17 &   52.9 &  GG10  \\   
     PKS 0336-01  & 3FGL J0339.5-0146 &    0.85 &   45.00  &  10.0e+44$^a$  & 3.2e+17 &   30.6 &  X14 \\
	4C 31.03  & 3FGL J0112.8+3207 &   0.603 &  44.54  & 	      &     --  &    --  &  	  \\ 
	  OM 484  & 3FGL J1153.4+4932 &   0.334 &  42.49  &   2.7e+44  & 1.6e+17 &   15.9 &        \\   
    TXS 1920-211  & 3FGL J1923.5-2104 &   0.874 &  41.03  &  15.0e+44  & 3.9e+17 &   37.4 &  GG10  \\ 	
MG2 J101241+2439  & 3FGL J1012.6+2439 &   1.805 &  40.84  &  13.5e+44  & 3.7e+17 &   35.5 &  GG15  \\   
      S5 1044+71  & 3FGL J1048.4+7144 &    1.15 &  40.66  & 	      &     --  &    --  &        \\  
  PMN J0850-1213  & 3FGL J0850.2-1214 &   0.566 &  40.36  & 	      &     --  &    --  &  	  \\ 
     PKS 0601-70  & 3FGL J0601.2-7036 &   2.409 &  39.57  &  6.0e+44   & 2.4e+17 &   23.7 &  GG15  \\   
     B3 0650+453  & 3FGL J0654.4+4514 &   0.933 &  39.37  &  1.8e+44   & 1.3e+17 &   13.0 &  GG15  \\   
    PKS 2320-035  & 3FGL J2323.5-0315 &   1.393 &  39.35  & 	      &     --  &    --  &  	  \\ 
     PKS 0736+01  & 3FGL J0739.4+0137 & 0.18941 &  38.99  &  1.6e+44$^a$  & 1.2e+17 &   12.0 &  X14  \\
    PKS 1954-388  & 3FGL J1958.0-3847 &    0.63 &  37.66  &  1.6e+44$^a$  & 1.3e+17 &   12.1 &  X14  \\
    TXS 1206+549  & 3FGL J1208.7+5442 &   1.345 &  37.23  &  3.2e+44  & 1.8e+17 &   17.3 &  GG11  \\   
      B2 0200+30  & 3FGL J0203.6+3043 &   0.955 &   36.90  &  0.3e+44   & 5.5e+16 &    5.3 &  GG15  \\ 
     PKS 1730-13  & 3FGL J1733.0-1305 &   0.902 &  36.81  &  6.8e+44$^a$  & 2.6e+17 &   25.1 &  X14  \\
    TXS 0106+612  & 3FGL J0109.8+6132 &   0.783 &  36.55  & 	      &     --  &    --  &  	  \\ 
    PKS 0528+134  & 3FGL J0530.8+1330 &    2.07 &   35.90  &  75.0e+44  & 8.7e+17 &   83.7 &  GG11  \\  
     PKS 0446+11  & 3FGL J0449.0+1121 &   2.153 &  35.64  &  7.2e+44   & 2.7e+17 &   25.9 &  GG15  \\   
     B2 1732+38A  & 3FGL J1734.3+3858 &   0.976 &  34.65  &  2.0e+44   & 1.4e+17 &   13.7 &  GG15  \\   
     PKS 0130-17  & 3FGL J0132.6-1655 &    1.02 &  34.58  & 	      &     --  &    --  &  	  \\ 
    MRC 1659-621  & 3FGL J1703.6-6211 &   1.747 &  34.56  & 26.2e+44$^a$  & 5.1e+17 &   49.5 &  S12  \\ 
      S5 0836+71  & 3FGL J0841.4+7053 &   2.218 &  34.53  & 225.0e+44  & 1.5e+18 &  145.0 &  GG11  \\   
     PKS 0906+01  & 3FGL J0909.1+0121 &   1.024 &  34.06  &  18.8e+44  & 4.3e+17 &   41.9 &  GG11  \\   
    PKS 2123-463  & 3FGL J2126.5-4605 &    1.67 &  34.03  & 	      &     --  &    --  &  	  \\ 
MG2 J071354+1934  & 3FGL J0713.9+1933 &    0.54 &  33.61  &  0.6e+44   & 7.7e+16 &    7.5 &  GG15  \\   
  PMN J1344-1723  & 3FGL J1344.2-1724 &    2.49 &  33.15  &  10.8e+44  & 3.3e+17 &   31.8 &  GG15  \\   
    TXS 0529+483  & 3FGL J0533.2+4822 &   1.162 &   32.90  & 	      &     --  &    --  &        \\  
    PKS 1551+130  & 3FGL J1553.5+1256 &   1.308 &  32.53  &  15.9e+44  & 4.0e+17 &   38.5 &  GG11  \\ 	
      S3 0458-02  & 3FGL J0501.2-0157 &   2.291 &  32.49  &  37.0e+44  & 6.1e+17 &   58.8 &  GG11  \\   
   PKS B1908-201  & 3FGL J1911.2-2006 &   1.119 &  32.12  &  30.0e+44  & 5.5e+17 &   52.9 &  GG10  \\   
MG2 J201534+3710  & 3FGL J2015.6+3709 &   0.859 &   32.10  & 	      &     --  &    --  &  	  \\ 
	CTS 0490  & 3FGL J2325.3-3557 &    0.36 &  32.06  & 	      &     --  &    --  &  	  \\ 
     PKS 0451-28  & 3FGL J0453.2-2808 &   2.564 &  32.05  &  120.0e+44 & 1.1e+18 &  105.9 &  GG11  \\ 	
       4C +47.44  & 3FGL J1637.7+4715 &   0.735 &  31.98  &   1.8e+44  & 1.3e+17 &   13.0 &  GG15  \\ 	
    PKS 0235-618  & 3FGL J0236.7-6136 &   0.465 &  31.78  & 	      &     --  &    --  &  	  \\ 
       4C +04.42  & 3FGL J1222.4+0414 &   0.966 &  30.02  &   7.2e+44  & 2.7e+17 &   25.9 &  GG11  \\   
     PKS 0454-46  & 3FGL J0455.7-4617 &   0.858 &  30.01  & 	      &     --  &    --  &  	  \\ 
      S4 1726+45  & 3FGL J1727.1+4531 &   0.717 &  29.92  & 	      &     --  &    --  &  	  \\ 
    PKS 0142-278  & 3FGL J0145.1-2732 &   1.148 &  29.47  &   7.2e+44  & 2.7e+17 &   25.9 &  GG10  \\ 	
    PKS 2144+092  & 3FGL J2147.2+0929 &   1.113 &  29.42  &   15.0e+44 & 3.9e+17 &   37.4 &  GG10  \\ 	
      S4 1851+48  & 3FGL J1852.4+4856 &    1.25 &  29.34  &   	      &     --  &    --  &  	  \\ 
     PKS 1118-05  & 3FGL J1121.4-0554 &   1.297 &  29.33  & 	      &     --  &    --  &  	  \\ 
       4C +51.37  & 3FGL J1740.3+5211 &   1.381 &  29.27  & 	      &     --  &    --  &  	  \\ 
\hline  											  
PKS 0208-512$^b$   &   3FGL J0210.7-5101   & 	1.003   &   62.95$^b$ &   37e+44    &   6.1e+17   &	58.8 &  Sb14 \\
TXS 0322+222       &   3FGL J0325.5+2223   & 	2.066   &   22.18     &   60e+44    &   7.7e+17   &	74.9 &  GG15 \\
PKS B1127-145      &   3FGL J1129.9-1446   & 	1.184   &   22.00     &  112.5e+44   &	1.1e+18   &  102.5 &  GG10 \\
CRATES J0303-6211  &   3FGL J0303.7-6211   & 	1.348   &   16.76     &   30e+44    &   5.5e+17   &	52.9 &  GG15 \\
PMN J1617-5848     &   3FGL J1617.4-5846   & 	1.422   &   15.81     &  105e+44   &  1.0e+18     &	99.0 &  GG15 \\
CRATES J1613+3412  &   3FGL J1613.8+3410   & 	1.397   &   13.23     &   54e+44    &   7.3e+17   &	71.0 &  GG15 \\
\hline		 
\hline  	 
\multicolumn{8}{l}{$a$: our calculation from the line luminosities reported in the respective reference, following \citet{celotti97}.} \\
\multicolumn{8}{l}{$b$: this object was left out of the main FSRQ selection despite its LAT significance, because it was classified as BCU-I}\\
\multicolumn{8}{l}{\,\,\,\, in the 3LAC.} \\ 
\end{tabular}	 
\end{table*}

%% file: tableBa.tex
\begin{landscape}
\begin{table} 
\centering
\caption{Results of the BLR absorption fits to the gamma-ray SED data.	 																					  
Col. [1]: object Name.																				  
Col. [2]: redshift.																					  
Col. [3]: number of data points above 20 GeV rest-frame. It indicates how much farther the spectrum 
extends into the ``forbidden region" of strong BLR absorption.                    							  
Col. [4]: average restframe energy, in GeV, of the last ``good" datapoint in the spectrum at high energy (see text).  
Col. [5]: best fit model of the data below 13 GeV rest-frame (namely, with the lowest $\chi^2_{r}$). 
It is the model used to extrapolate the spectrum to estimate the optical depth.  
Col. [6]: path length $\ell$ inside the BLR, in units of $10^{16}$ cm. Obtained by extrapolating the model in Col. 5 to the full band.
Col. [7]: corresponding maximum optical depth, at the peak of the \gg~ cross-section with a Planckian spectrum.
Col. [8]: maximum optical depth assuming the dissipation region is located at $R_{\rm diss}$= max($2\times10^{16}$ cm, $R_{\rm BLR}/2$).
Col. [9]: corresponding path length inside the BLR, in units of $10^{16}$ cm, to be compared with Col. 6;
Col. [10]: logarithm of the ratio between the observed data flux of the highest-energy data point and the expected model with $\tau^{max}_{\rm BLR}$. 
It gives the separation in the SED (in orders of magnitudes) between the observed and expected fluxes if the assumptions of the EC(BLR) model were right.
Col. [11]: probability value (from F-test) between the model with free (measured) or fixed ($\ell_{\rm BLR}$) paths inside the BLR (see text). A 3-$\sigma$ exclusion corresponds to a p-value $\leq 2.7$e-03.           
Col. [12], [13]: path and corresponding $\tau^{max}$ for the fit with a logparabolic model plus BLR absorption allowing all parameters to be free.
Col. [14]: model of the fullband LAT spectrum yielding the lowest $\chi^2_{r}$, among: logparabola (logp), logparabola with free BLR absorption (logptau), 
power-law with under-exponential cutoff at high energy with  $\beta_{c}=1/3$~ (exp03). See Sect. 2.1.
Col. [15]: flag indicating if the upper limit is below the last ``good" datapoint in the SED, i.e.  if it is meaningful to constrain the spectral steepening (strong) or not.
If it is, the fits are checked that they comply to the UL constraints.  
}
\label{fitblr}
\begin{tabular}{llcrccrrrrccrll}
\hline
\hline
Name              &  $z$  &  n$>$20 & MaxBin & best$<$13    &  Path  & $\tau^{max}$  & $\tau^{max}_{BLR}$ & Path$_{\rm BLR}$ & ratio   &  p-value(BLR) &  Path$_{\rm F}$ & $\tau^{max}_{\rm F}$ & bestfitFull &   UL  \\
~[1]              &  [2]    &  [3] &   [4]  &     [5]       &  [6]   &     [7]       &      [8]           &  [9]             &   [10]  &     [11]  &     [12]  &     [13]   &   [14]  &  [15]           \\
\hline 
	3C 454.3  &   0.859 &	 3 &   77.3 &	    logp & 0.63$\pm$0.09 &   1.23 &   55.8 &	   28.9 &  21.0 & 5.8e-10 &  0.60$\pm$0.13 &	1.15 &       exp03 & strong   \\ 
       4C +21.35  &   0.435 &	 2 &   59.7 &	    logp & 0.16$\pm$0.11 &   0.31 &   38.9 &	   20.1 &  12.8 & 1.9e-07 &  0.16$\pm$0.17 &	0.31 &       exp03 & strong   \\ 
     PKS 1510-08  &    0.36 &	 3 &  100.6 &	    logp & 0.00$\pm$0.16 &   0.00 &   26.3 &	   13.6 &  11.2 & 3.3e-07 &  0.00$\pm$0.21 &	0.00 &        logp &	 no   \\ 
      B2 1520+31  &   1.487 &	 2 &   58.2 &	    logp & 0.17$\pm$0.13 &   0.33 &   20.5 &	   10.6 &   6.3 & 9.7e-07 &  0.21$\pm$0.23 &	0.40 &       exp03 &	 no   \\ 
    PKS 1502+106  &   1.839 &	 3 &   66.4 &	    logp & 0.37$\pm$0.11 &   0.72 &   43.0 &	   22.2 &  14.9 & 2.6e-07 &  0.42$\pm$0.23 &	0.81 &       exp03 & strong   \\ 
    PKS 0454-234  &   1.003 &	 3 &   83.3 &	    logp & 0.20$\pm$0.13 &   0.40 &   41.9 &	   21.7 &  16.0 & 1.0e-07 &  0.25$\pm$0.22 &	0.48 &       exp03 &	 no   \\ 
	  3C 279  &   0.536 &	 3 &   63.9 &	    logp & 0.36$\pm$0.16 &   0.69 &   15.0 &	    7.7 &   4.8 & 1.7e-06 &  0.32$\pm$0.25 &	0.62 &     logptau &	 no   \\ 
	  3C 273  &   0.158 &	 0 &   15.2 &	    logp & 0.00$\pm$0.47 &   0.00 &   56.2 &	   29.1 &   1.1 & 7.8e-05 &  0.00$\pm$0.67 &	0.00 &        logp & strong   \\ 
    PKS 2326-502  &   0.518 &	 1 &   35.5 &	    logp & 0.88$\pm$0.17 &   1.69 &    --  &	    --  &   --  &     --  &  0.96$\pm$0.25 &	1.85 &     logptau & strong   \\ 
   PKS B1424-418  &   1.552 &	 3 &  106.1 &	    logp & 0.37$\pm$0.08 &   0.71 &   30.6 &	   15.8 &  12.6 & 1.4e-08 &  0.35$\pm$0.18 &	0.68 &       exp03 & strong   \\ 
    PKS 1830-211  &   2.507 &	 3 &   82.0 &	    logp & 0.48$\pm$0.09 &   0.92 &    --  &	    --  &   --  &     --  &  0.24$\pm$0.15 &	0.47 &       exp03 & strong   \\ 
       4C +55.17  &   0.899 &	 4 &  140.5 &	    logp & 0.04$\pm$0.07 &   0.08 &   18.7 &	    9.7 &   8.0 & 9.0e-09 &  0.09$\pm$0.16 &	0.17 &       exp03 &	 no   \\ 
     PKS 0727-11  &   1.589 &	 2 &   60.6 &	    logp & 0.29$\pm$0.10 &   0.56 &    --  &	    --  &   --  &     --  &  0.34$\pm$0.23 &	0.66 &       exp03 & strong   \\ 
       4C +28.07  &   1.213 &	 2 &   51.8 &	    logp & 0.00$\pm$0.15 &   0.00 &   41.0 &	   21.2 &  12.1 & 6.5e-07 &  0.00$\pm$0.21 &	0.00 &        logp & strong   \\ 
    PKS 0402-362  &   1.417 &	 0 &   17.9 &	    logp & 0.11$\pm$0.44 &   0.21 &   36.2 &	   18.7 &   1.3 & 1.0e-03 &  0.24$\pm$0.73 &	0.46 &        logp & strong   \\ 
	 Ton 599  &   0.725 &	 3 &   71.7 &	    logp & 0.00$\pm$0.35 &   0.00 &   21.9 &	   11.3 &   8.1 & 9.6e-05 &  0.00$\pm$0.49 &	0.00 &        logp &	 no   \\ 
    PKS 1244-255  &   0.635 &	 2 &   38.2 &	    logp & 0.14$\pm$0.21 &   0.27 &   10.5 &	    5.5 &   2.0 & 3.0e-06 &  0.21$\pm$0.33 &	0.40 &        logp &	 no   \\ 
       4C +38.41  &   1.814 &	 3 &   65.8 &	    logp & 0.44$\pm$0.17 &   0.84 &   74.9 &	   38.7 &  25.5 & 7.3e-08 &  0.46$\pm$0.27 &	0.88 &       exp03 &	 no   \\ 
  PMN J2345-1555  &   0.621 &	 3 &   67.4 &	    logp & 0.31$\pm$0.11 &   0.60 &   14.0 &	    7.2 &   4.7 & 1.8e-07 &  0.28$\pm$0.20 &	0.54 &       exp03 &	 no   \\ 
      S4 0917+44  &    2.19 &	 2 &   42.0 &	    logp & 0.46$\pm$0.13 &   0.89 &   81.3 &	   42.1 &  18.3 & 1.9e-06 &  0.25$\pm$0.25 &	0.47 &       exp03 & strong   \\ 
     PKS 2052-47  &   1.491 &	 2 &   58.3 &	    logp & 0.00$\pm$0.31 &   0.00 &   28.1 &	   14.6 &   9.5 & 7.4e-05 &  0.00$\pm$0.47 &	0.00 &        logp &	 no   \\ 
    PKS 1124-186  &   1.048 &	 3 &   85.2 &	    logp & 0.00$\pm$0.07 &   0.00 &    --  &	    --  &   --  &     --  &  0.00$\pm$0.16 &	0.00 &        logp &	 no   \\ 
     PKS 2142-75  &   1.138 &	 1 &   28.1 &	    logp & 0.54$\pm$0.08 &   1.05 &   27.3 &	   14.1 &   3.1 & 5.1e-08 &  0.65$\pm$0.13 &	1.25 &     logptau & strong   \\ 
     B3 1343+451  &   2.534 &	 4 &  147.0 &	    logp & 0.17$\pm$0.04 &   0.34 &   31.3 &	   16.2 &  13.3 & 2.2e-10 &  0.19$\pm$0.11 &	0.37 &       exp03 &	 no   \\ 
     PKS 0805-07  &   1.837 &	 3 &   66.4 &	    logp & 0.24$\pm$0.10 &   0.45 &   47.4 &	   24.5 &  16.4 & 4.7e-07 &  0.37$\pm$0.25 &	0.72 &       exp03 & strong   \\ 
    PKS 0244-470  &   1.385 &	 1 &   31.4 &	    logp & 0.29$\pm$0.16 &   0.57 &   50.2 &	   26.0 &   7.1 & 1.7e-05 &  0.21$\pm$0.29 &	0.41 &       exp03 & strong   \\  
	 CTA 102  &   1.037 &	 2 &   47.6 &	    logp & 0.26$\pm$0.29 &   0.50 &   62.2 &	   32.2 &  16.2 & 1.3e-05 &  0.18$\pm$0.45 &	0.34 &        logp &	 no   \\ 
       4C +01.02  &   2.099 &	 3 &   72.5 &	    logp & 0.32$\pm$0.18 &   0.61 &   83.7 &	   43.3 &  30.5 & 6.0e-07 &  0.17$\pm$0.35 &	0.34 &        logp &	 no   \\ 
       4C +14.23  &   1.038 &	 2 &   47.7 &	    logp & 0.00$\pm$0.12 &   0.00 &   29.0 &	   15.0 &   7.9 & 2.0e-06 &  0.13$\pm$0.21 &	0.25 &       exp03 & strong   \\ 
    PKS 0250-225  &   1.427 &	 2 &   56.8 &	    logp & 0.02$\pm$0.15 &   0.05 &   17.2 &	    8.9 &   5.3 & 1.3e-06 &  0.02$\pm$0.27 &	0.04 &        logp &	 no   \\ 
      B2 2308+34  &   1.817 &	 3 &   65.9 &	    logp & 0.10$\pm$0.16 &   0.19 &    --  &	    --  &   --  &     --  &  0.28$\pm$0.30 &	0.54 &        logp &	 no   \\ 
      B2 0716+33  &   0.779 &	 2 &   41.6 &	    logp & 0.56$\pm$0.09 &   1.07 &   20.5 &	   10.6 &   4.4 & 1.9e-08 &  0.59$\pm$0.16 &	1.15 &       exp03 &	 no   \\ 
MG1 J123931+0443  &   1.762 &	 3 &   64.6 &	    logp & 1.22$\pm$0.14 &   2.37 &   29.2 &	   15.1 &   9.6 & 9.5e-06 &  1.25$\pm$0.24 &	2.42 &     logptau &	 no   \\ 
  PMN J1802-3940  &   1.319 &	 2 &   54.2 &	    logp & 0.24$\pm$0.30 &   0.46 &   37.4 &	   19.4 &  11.0 & 3.7e-05 &  0.21$\pm$0.59 &	0.40 &        logp &	 no   \\ 
      S4 1849+67  &   0.657 &	 2 &   38.8 &	    logp & 0.38$\pm$0.09 &   0.74 &   17.8 &	    9.2 &   3.4 & 3.9e-06 &  0.43$\pm$0.14 &	0.83 &        logp &	 no   \\ 
     PKS 2023-07  &   1.388 &	 2 &   55.9 &	    logp & 0.00$\pm$0.24 &   0.00 &   45.8 &	   23.7 &  14.4 & 2.2e-05 &  0.00$\pm$0.40 &	0.00 &        logp &	 no   \\ 
\hline		 																					  
\hline  	 																					  
\end{tabular}	 																					  
\end{table}
\end{landscape}

%% file: tableBb.tex
\begin{landscape}
\begin{table} 
\centering
\contcaption{}
\begin{tabular}{llcrccrrrrccrll}
\hline
\hline
Name              &  $z$  &  n$>$20 & MaxBin & best$<$13    &  Path  & $\tau^{max}$  & $\tau^{max}_{BLR}$ & Path$_{\rm BLR}$ & ratio   &  p-value(BLR) &  Path$_{\rm F}$ & $\tau^{max}_{\rm F}$ & bestfitFull &   UL  \\
~[1]              &  [2]    &  [3] &   [4]  &     [5]       &  [6]   &     [7]       &      [8]           &  [9]             &   [10]  &     [11]  &     [12]  &     [13]   &   [14]  &  [15]           \\
\hline 
    PKS 1454-354  &   1.424 &	 2 &   56.7 &	    logp & 0.41$\pm$0.27 &   0.80 &   64.8 &	   33.5 &  20.0 & 1.4e-05 &  0.47$\pm$0.49 &	0.91 &        logp &	 no   \\ 
     B3 1708+433  &   1.027 &	 2 &   47.4 &	    logp & 0.00$\pm$0.54 &   0.00 &   11.8 &	    6.1 &   3.6 & 3.0e-03 &  0.00$\pm$0.79 &	0.00 &        logp &	 no   \\ 
  GB6 J0742+5444  &    0.72 &	 2 &   40.2 &	    logp & 0.00$\pm$0.50 &   0.00 &    --  &	    --  &   --  &     --  &  0.00$\pm$0.79 &	0.00 &        logp &	 no   \\ 
	  OX 169  &   0.213 &	 1 &   28.4 &	    logp & 0.00$\pm$0.81 &   0.00 &   13.0 &	    6.7 &   1.8 & 1.8e-03 &  0.00$\pm$1.16 &	0.00 &        logp &	 no   \\ 
     PKS 2227-08  &    1.56 &	 0 &   18.9 &	    logp & 0.00$\pm$0.26 &   0.00 &   65.7 &	   34.0 &   2.6 & 7.7e-05 &  0.05$\pm$0.45 &	0.09 &        logp & strong   \\ 
    PKS 2255-282  &   0.926 &	 1 &   25.3 &	    logp & 0.00$\pm$0.99 &   0.00 &   41.6 &	   21.5 &   3.8 & 2.4e-03 &  0.00$\pm$1.63 &	0.00 &        logp &	 no   \\ 
      B2 2155+31  &   1.488 &	 2 &   58.2 &	    logp & 0.79$\pm$0.20 &   1.52 &   22.5 &	   11.6 &   6.9 & 9.6e-05 &  0.90$\pm$0.35 &	1.74 &       exp03 &	 no   \\ 
    PKS 1622-253  &   0.786 &	 2 &   41.8 &	    logp & 0.37$\pm$0.20 &   0.72 &    6.4 &	    3.3 &   1.3 & 7.6e-05 &  0.49$\pm$0.33 &	0.95 &        logp &	 no   \\ 
     PKS 0420-01  &   0.916 &	 2 &   44.8 &	    logp & 0.00$\pm$0.43 &   0.00 &   25.8 &	   13.3 &   6.5 & 1.8e-04 &  0.00$\pm$0.70 &	0.00 &        logp &	 no   \\ 
	  OC 457  &   0.859 &	 2 &   43.5 &	    logp & 0.13$\pm$0.13 &   0.24 &   20.7 &	   10.7 &   4.9 & 4.9e-08 &  0.23$\pm$0.22 &	0.45 &       exp03 &	 no   \\ 
    PKS 2201+171  &   1.076 &	 2 &   48.6 &	    logp & 0.00$\pm$0.81 &   0.00 &   33.5 &	   17.3 &   9.4 & 1.1e-03 &  0.00$\pm$1.07 &	0.00 &        logp &	 no   \\ 
	  OP 313  &   0.997 &	 3 &   83.1 &	     pow & 0.69$\pm$0.23 &   1.34 &   28.0 &	   14.5 &  10.7 & 8.9e-05 &  0.30$\pm$0.32 &	0.58 &       exp03 &	 no   \\ 
    TXS 0059+581  &   0.644 &	 2 &   38.5 &	    logp & 0.00$\pm$0.88 &   0.00 &   20.5 &	   10.6 &   4.2 & 3.1e-03 &  0.00$\pm$1.39 &	0.00 &        logp &	 no   \\ 
      S4 1030+61  &   1.401 &	 2 &   56.2 &	    logp & 0.44$\pm$0.11 &   0.84 &   20.5 &	   10.6 &   6.1 & 8.1e-08 &  0.32$\pm$0.20 &	0.62 &       exp03 &	 no   \\ 
	  OG 050  &   1.254 &	 2 &   52.7 &	    logp & 0.00$\pm$0.18 &   0.00 &   29.0 &	   15.0 &   8.6 & 1.4e-05 &  0.08$\pm$0.30 &	0.16 &        logp & strong   \\ 
     PKS 0440-00  &   0.844 &	 2 &   43.1 &	    logp & 1.08$\pm$0.30 &   2.08 &   23.7 &	   12.2 &   5.4 & 9.6e-04 &  0.84$\pm$0.45 &	1.63 &       exp03 &	 no   \\ 
    PKS 1329-049  &    2.15 &	 3 &   73.7 &	    logp & 0.00$\pm$0.42 &   0.00 &   79.1 &	   40.9 &  29.7 & 3.4e-04 &  0.00$\pm$0.70 &	0.00 &        logp &	 no   \\ 
    TXS 1700+685  &   0.301 &	 1 &   30.4 &	    logp & 0.38$\pm$0.75 &   0.74 &    --  &	    --  &   --  &     --  &  0.04$\pm$1.07 &	0.08 &        logp &	 no   \\ 
      S4 1144+40  &   1.089 &	 3 &   86.9 &	    logp & 0.00$\pm$0.22 &   0.00 &   33.1 &	   17.1 &  13.6 & 8.8e-06 &  0.00$\pm$0.32 &	0.00 &        logp &	 no   \\ 
       B0218+357  &   0.944 &	 3 &   80.9 &	    logp & 0.03$\pm$0.20 &   0.07 &    7.5 &	    3.9 &   2.7 & 2.0e-04 &  0.00$\pm$0.37 &	0.00 &        logp &	 no   \\ 
    PKS 0215+015  &   1.715 &	 3 &   63.5 &	     pow & 0.92$\pm$0.23 &   1.78 &   52.9 &	   27.4 &  17.2 & 8.9e-06 &  0.59$\pm$0.41 &	1.13 &       exp03 &	 no   \\ 
     PKS 0336-01  &    0.85 &	 2 &   43.3 &	    logp & 0.20$\pm$0.42 &   0.38 &   30.6 &	   15.8 &   7.2 & 5.6e-04 &  0.35$\pm$0.70 &	0.68 &        logp &	 no   \\ 
	4C 31.03  &   0.603 &	 2 &   37.5 &	    logp & 1.95$\pm$1.04 &   3.77 &    --  &	    --  &   --  &     --  &  1.56$\pm$1.52 &	3.01 &        logp &	 no   \\ 
	  OM 484  &   0.334 &	 0 &   17.5 &	     pow & 7.46$\pm$1.76 &  14.42 &   15.9 &	    8.2 &   0.0 & 6.1e-01 &  7.35$\pm$2.69 &   14.20 &       exp03 &	 no   \\ 
    TXS 1920-211  &   0.874 &	 1 &   24.6 &	    logp & 0.72$\pm$0.32 &   1.39 &   37.4 &	   19.4 &   3.1 & 7.8e-04 &  1.09$\pm$0.63 &	2.10 &     logptau & strong   \\ 
MG2 J101241+2439  &   1.805 &	 1 &   20.7 &	     pow & 1.82$\pm$0.44 &   3.52 &   35.5 &	   18.4 &   1.7 & 3.7e-03 &  2.11$\pm$0.94 &	4.07 &     logptau & strong   \\ 
      S5 1044+71  &    1.15 &	 2 &   50.3 &	    logp & 0.16$\pm$0.12 &   0.30 &    --  &	    --  &   --  &     --  &  0.28$\pm$0.19 &	0.55 &       exp03 &	 no   \\ 
  PMN J0850-1213  &   0.566 &	 2 &   36.6 &	    logp & 0.00$\pm$0.81 &   0.00 &    --  &	    --  &   --  &     --  &  0.00$\pm$1.17 &	0.00 &        logp &	 no   \\ 
     PKS 0601-70  &   2.409 &	 2 &   44.8 &	    logp & 0.00$\pm$1.06 &   0.00 &   23.7 &	   12.2 &   6.3 & 4.2e-03 &  0.00$\pm$1.72 &	0.00 &        logp &	 no   \\ 
     B3 0650+453  &   0.933 &	 2 &   45.2 &	    logp & 0.00$\pm$0.91 &   0.00 &   13.0 &	    6.7 &   3.6 & 8.5e-03 &  0.00$\pm$0.82 &	0.00 &        logp &	 no   \\ 
    PKS 2320-035  &   1.393 &	 1 &   31.5 &	     pow & 0.94$\pm$0.21 &   1.82 &    --  &	    --  &   --  &     --  &  0.82$\pm$0.41 &	1.59 &     logptau & strong   \\ 
     PKS 0736+01  & 0.18941 &	 1 &   27.8 &	    logp & 0.00$\pm$1.14 &   0.00 &   12.0 &	    6.2 &   1.4 & 5.1e-02 &  0.00$\pm$1.74 &	0.00 &        logp &	 no   \\ 
    PKS 1954-388  &    0.63 &	 1 &   21.4 &	    logp & 1.21$\pm$0.28 &   2.34 &   12.1 &	    6.3 &   0.6 & 1.6e-03 &  1.25$\pm$0.51 &	2.42 &     logptau & strong   \\ 
    TXS 1206+549  &   1.345 &	 0 &   17.3 &	     pow & 2.10$\pm$0.55 &   4.07 &   17.3 &	    9.0 &   0.3 & 1.5e-02 &  1.43$\pm$1.01 &	2.77 &     logptau & strong   \\ 
      B2 0200+30  &   0.955 &	 2 &   45.7 &	    logp & 0.00$\pm$0.27 &   0.00 &    5.3 &	    2.7 &   1.5 & 9.3e-05 &  0.07$\pm$0.46 &	0.14 &        logp &	 no   \\ 
     PKS 1730-13  &   0.902 &	 2 &   44.5 &	    logp & 0.22$\pm$0.08 &   0.43 &   25.1 &	   13.0 &   6.0 & 8.9e-09 &  0.28$\pm$0.14 &	0.54 &       exp03 &	 no   \\ 
    TXS 0106+612  &   0.783 &	 0 &   13.2 &	    logp & 3.40$\pm$9.70 &   6.58 &    --  &	    --  &   --  &     --  & 3.52$\pm$16.50 &	6.80 &        logp &	 no   \\ 
    PKS 0528+134  &    2.07 &	 0 &   12.8 &	    logp & 2.22$\pm$1.08 &   4.30 &   83.7 &	   43.3 &   0.7 & 3.5e-03 &  2.21$\pm$2.29 &	4.27 &       exp03 & strong   \\ 
     PKS 0446+11  &   2.153 &	 2 &   41.5 &	    logp & 0.00$\pm$0.30 &   0.00 &   25.9 &	   13.4 &   5.9 & 1.2e-05 &  0.00$\pm$0.54 &	0.00 &        logp &	 no   \\ 
     B2 1732+38A  &   0.976 &	 1 &   26.0 &	    logp & 0.00$\pm$0.31 &   0.00 &   13.7 &	    7.1 &   1.4 & 3.4e-06 &  0.00$\pm$0.42 &	0.00 &       exp03 &	 no   \\ 
     PKS 0130-17  &    1.02 &	 1 &   26.6 &	    logp & 0.00$\pm$2.35 &   0.00 &    --  &	    --  &   --  &     --  &  0.00$\pm$3.53 &	0.00 &        logp &	 no   \\ 
    MRC 1659-621  &    1.75 &	 1 &   20.3 &	    logp & 3.33$\pm$1.35 &   6.44 &   49.5 &	   25.6 &   2.1 & 9.6e-03 &  3.47$\pm$2.57 &	6.72 &       exp03 &	 no   \\ 
      S5 0836+71  &   2.218 &	 1 &   23.8 &	    logp & 0.00$\pm$2.70 &   0.00 &  145.0 &	   75.0 &  11.6 & 2.7e-02 &  0.00$\pm$4.34 &	0.00 &        logp &	 no   \\ 
     PKS 0906+01  &   1.024 &	 1 &   26.6 &	    logp & 0.00$\pm$0.78 &   0.00 &   41.9 &	   21.7 &   4.4 & 8.8e-05 &  0.00$\pm$1.22 &	0.00 &        logp &	 no   \\ 
\hline		 																					  
\hline  	 																					  
\end{tabular}	 																					  
\end{table}
\end{landscape}

%% file: tableBc.tex
\begin{landscape}
\begin{table} 
\centering
\contcaption{}
\begin{tabular}{llcrccrrrrccrll}
\hline
\hline
Name              &  $z$  &  n$>$20 & MaxBin & best$<$13    &  Path  & $\tau^{max}$  & $\tau^{max}_{BLR}$ & Path$_{\rm BLR}$ & ratio   &  p-value(BLR) &  Path$_{\rm F}$ & $\tau^{max}_{\rm F}$ & bestfitFull &   UL  \\
~[1]              &  [2]    &  [3] &   [4]  &     [5]       &  [6]   &     [7]       &      [8]           &  [9]             &   [10]  &     [11]  &     [12]  &     [13]   &   [14]  &  [15]           \\
\hline 
    PKS 2123-463  &    1.67 &	 0 &   19.7 &	     pow & 3.69$\pm$2.08 &   7.13 &    --  &	    --  &   --  &     --  &  2.47$\pm$3.67 &	4.77 &       exp03 &	 no  \\ 
MG2 J071354+1934  &    0.54 &	 0 &   11.4 &	    logp & 3.50$\pm$1.48 &   6.77 &    7.5 &	    3.9 &   0.2 & 8.1e-01 &  3.38$\pm$2.80 &	6.53 &       exp03 & strong  \\ 
  PMN J1344-1723  &    2.49 &	 3 &   81.6 &	     pow & 0.68$\pm$0.36 &   1.31 &   31.8 &	   16.4 &  11.7 & 1.3e-03 &  0.00$\pm$0.71 &	0.00 &       exp03 &	 no  \\ 
    TXS 0529+483  &   1.162 &	 1 &   28.4 &	    logp & 0.00$\pm$0.31 &   0.00 &    --  &	    --  &   --  &     --  &  0.00$\pm$0.59 &	0.00 &        logp & strong  \\ 
    PKS 1551+130  &   1.308 &	 1 &   30.4 &	     pow & 0.32$\pm$0.86 &   0.61 &   38.5 &	   19.9 &   5.1 & 3.3e-03 &  0.00$\pm$1.62 &	0.00 &       exp03 &	 no  \\ 
      S3 0458-02  &   2.291 &	 2 &   43.3 &	    logp & 0.00$\pm$0.11 &   0.00 &   58.8 &	   30.4 &  14.0 & 4.2e-07 &  0.00$\pm$0.23 &	0.00 &        logp & strong  \\ 
   PKS B1908-201  &   1.119 &	 1 &   27.9 &	    logp & 0.09$\pm$0.12 &   0.17 &   52.9 &	   27.4 &   6.0 & 1.6e-06 &  0.04$\pm$0.25 &	0.07 &        logp & strong  \\ 
MG2 J201534+3710  &   0.859 &	 2 &   43.5 &	    logp & 0.00$\pm$0.16 &   0.00 &    --  &	    --  &   --  &     --  &  0.00$\pm$0.22 &	0.00 &        logp & strong  \\ 
	CTS 0490  &    0.36 &	 0 &   10.1 &	    logp & 1.10$\pm$1.44 &   2.13 &    --  &	    --  &   --  &     --  &  1.14$\pm$2.66 &	2.21 &        logp & strong  \\ 
     PKS 0451-28  &   2.564 &	 0 &   14.8 &	    logp & 2.99$\pm$0.74 &   5.77 &  105.9 &	   54.8 &   1.4 & 4.2e-03 &  2.80$\pm$1.56 &	5.41 &     logptau & strong  \\ 
       4C +47.44  &   0.735 &	 1 &   22.8 &	    logp & 0.00$\pm$1.45 &   0.00 &   13.0 &	    6.7 &   1.1 & 3.2e-03 &  0.00$\pm$2.30 &	0.00 &       exp03 &	 no  \\ 
    PKS 0235-618  &   0.465 &	 0 &   10.8 &	     pow & 2.63$\pm$2.12 &   5.09 &    --  &	    --  &   --  &     --  &  0.14$\pm$3.94 &	0.27 &        logp & strong  \\ 
       4C +04.42  &   0.966 &	 0 &   14.5 &	    logp & 0.00$\pm$6.57 &   0.00 &   25.9 &	   13.4 &   0.6 & 1.1e-01 & 0.00$\pm$10.33 &	0.00 &        logp &	 no  \\ 
     PKS 0454-46  &   0.858 &	 1 &   24.4 &	    logp & 0.00$\pm$1.94 &   0.00 &    --  &	    --  &   --  &     --  &  0.00$\pm$3.02 &	0.00 &        logp &	 no  \\ 
      S4 1726+45  &   0.717 &	 0 &   12.7 &	    logp & 0.00$\pm$1.22 &   0.00 &    --  &	    --  &   --  &     --  &  0.00$\pm$2.01 &	0.00 &        logp & strong  \\ 
    PKS 0142-278  &   1.148 &	 0 &   15.9 &	    logp & 0.00$\pm$9.78 &   0.00 &   25.9 &	   13.4 &   0.7 & 1.5e-01 & 0.00$\pm$15.76 &	0.00 &        logp &	 no  \\ 
    PKS 2144+092  &   1.113 &	 0 &   15.6 &	    logp & 2.12$\pm$3.50 &   4.11 &   37.4 &	   19.4 &   0.6 & 3.6e-02 &  2.24$\pm$6.07 &	4.32 &        logp &	 no  \\ 
      S4 1851+48  &    1.25 &	 1 &   29.6 &	    logp & 0.00$\pm$0.62 &   0.00 &    --  &	    --  &   --  &     --  &  0.00$\pm$0.97 &	0.00 &        logp & strong  \\ 
     PKS 1118-05  &   1.297 &	 0 &   17.0 &	    logp & 0.00$\pm$0.71 &   0.00 &    --  &	    --  &   --  &     --  &  0.00$\pm$1.25 &	0.00 &        logp & strong  \\ 
       4C +51.37  &   1.381 &	 2 &   55.7 &	    logp & 0.00$\pm$0.51 &   0.00 &    --  &	    --  &   --  &     --  &  0.00$\pm$0.84 &	0.00 &        logp &	 no  \\ 
\hline
     PKS 0208-512 &  1.003 &	1 &   26.3 &	   logp & 0.76$\pm$0.30 &   1.47 &   58.8 &	  30.4 &   5.8 & 1.8e-03 & 0.86$\pm$0.54 &    1.67 &	   exp03 & strong   \\
     TXS 0322+222 &  2.066 &	0 &   12.8 &	    pow & 2.97$\pm$1.00 &   5.75 &   74.9 &	  38.7 &   0.4 & 1.2e-01 & 3.10$\pm$2.17 &    5.98 &	   exp03 & strong   \\
    PKS B1127-145 &  1.184 &	0 &   16.2 &	   logp & 0.00$\pm$3.44 &   0.00 &  102.5 &	  53.0 &   2.3 & 1.5e-02 & 0.00$\pm$6.00 &    0.00 &	 logptau &     no   \\
CRATES J0303-6211 &  1.348 &	0 &   17.4 &	    pow & 4.18$\pm$2.84 &   8.09 &   52.9 &	  27.4 &   1.3 & 3.4e-02 & 1.09$\pm$5.18 &    2.10 &	    logp &     no   \\
   PMN J1617-5848 &  1.422 &	0 &   17.9 &	    pow & 0.00$\pm$3.47 &   0.00 &   99.0 &	  51.2 &   3.2 & 5.0e-02 & 0.20$\pm$8.00 &    0.39 &	   exp03 &     no   \\
CRATES J1613+3412 &  1.397 &	1 &   31.5 &	   logp & 0.00$\pm$2.87 &   0.00 &   71.0 &	  36.7 &  11.0 & 4.2e-02 & 0.00$\pm$4.43 &    0.00 &	    logp &     no   \\
\hline		 																					  
\hline  	 																					  
\end{tabular}	 																					  
\end{table}
\end{landscape}

%% file: tableD.tex
\begin{table*} 
\centering
\caption{Results of the BLR absorption fits to separate ``High" and ``Low" states, for a sub-sample of FSRQs.
The division of the data is done with a cut in flux or epoch (see Fig. \ref{tevs} and Fig. \ref{highlow1}).
Col. [1]: object name.																			  
Col. [2]: state of the source, as defined in Col. 3.																		  
Col. [3]: flux $N_{cut}$ (in cm$^{-2}$ s$^{-1}$) or time interval separating the High from Low-state spectra. 
Namely,  the High (Low) datasets are obtained summing all intervals in the 7-day lightcurve with Flux $\geq N_{cut}$ ($<N_{cut}$),
or the intervals specified by MJD.
Col. [4]: number of data points above 20 GeV rest-frame.                  							  
Col. [5]: average rest-frame energy, in GeV, of the last ``good" datapoint in the spectrum at high energy (namely with TS$>4$ and npred$>3$ and flux error $<50\%$, see Sect. 2); 				  
Col. [6]: best fit model of the data below 13 GeV rest-frame (i.e. with the lowest $\chi^2_{r}$). It is the model used to extrapolate the spectrum to estimate the optical depth.  
Col. [7]: path length inside the BLR, in units of $10^{16}$ cm, obtained by extrapolating the model in Col. 6 to the full LAT band.
Col. [8]: corresponding maximum optical depth, at the peak of the \gg cross-section (see text).
Col. [9]: model of the fullband LAT spectrum yielding the lowest overall $\chi^2_{r}$, among:  logparabola (logp), logparabola with BLR absoprtion (logptau) 
or power-law with under-exponential cutoff with $\beta_{c}=1/3$ (exp03).
Col. [10]:  flag indicating if the upper limit in the SED is below the last ``good" datapoint at high energy, 
i.e.  if it is  meaningful to constrain the spectral steepening (strong) or not.  
}
\begin{tabular}{lcccrcccll}
\hline
\hline
Name            &  State  &  Cut  &  n$>$20 & MaxBin  & best$<$13   &  Path  & $\tau^{max}$  &  bestfitFull &   UL  \\
~[1]            &  [2]    &  [3]  &   [4]   &   [5]   &  [6]        &  [7]   &      [8]      &  [9]         &   [10]  \\
\hline 
         3C 273 & High  & 1.0e-6 &     0 &   11.2 &    logp &  0.16$\pm$0.37 &   0.31 &        logp & strong   \\ 
       	        & Low	&	 &     0 &   11.2 &    logp &  0.0$\pm$11.7  &   0.00 &       exp03 &	  no   \\
         3C 279 & High  & 0.9e-6 &     2 &   61.7 &    logp &  0.20$\pm$0.21 &   0.38 &        logp &	  no   \\
                & Low	&	 &     2 &   61.7 &    logp &  0.27$\pm$0.11 &   0.52 &       exp03 & strong   \\
      4C +14.23 & High  & 0.6e-6 &     1 &   40.2 &    logp &  0.00$\pm$0.99 &   0.00 &        logp &	  no   \\
                & Low	&	 &     1 &   40.2 &    logp &  0.24$\pm$0.16 &   0.46 &       exp03 & strong   \\
      4C +21.35 & High  & 1.0e-6 &     2 &   57.6 &    logp &  0.07$\pm$0.10 &   0.14 &     logptau & strong   \\
                & Low	&	 &     2 &   57.6 &    logp &  0.24$\pm$0.15 &   0.47 &       exp03 & strong   \\
      4C +28.07 & High  & 0.5e-6 &     1 &   21.5 &    logp &  0.72$\pm$0.26 &   1.40 &       exp03 & strong   \\
                & Low	&	 &     2 &   43.7 &    logp &  0.00$\pm$0.15 &   0.00 &       exp03 & strong   \\
      4C +38.41 & High  & 0.5e-6 &     2 &   55.6 &    logp &  0.34$\pm$0.14 &   0.65 &     logptau & strong   \\
                & Low	&	 &     1 &   27.3 &    logp &  0.84$\pm$0.11 &   1.62 &     logptau & strong   \\
      B0218+357 & High  & 0.5e-6 &     1 &   38.4 &    logp &  0.03$\pm$0.16 &   0.05 &        logp &	  no   \\
                & Low	&	 &     3 &  158.8 &	pow &  0.03$\pm$0.17 &   0.07 &       exp03 &	  no   \\
        CTA 102 & High  & 0.9e-6 &     1 &   40.2 &    logp &  0.53$\pm$0.07 &   1.02 &     logptau & strong   \\
                & Low	&	 &     0 &   19.8 &    logp &  0.86$\pm$0.43 &   1.67 &       exp03 & strong   \\
MG1 J123931+0443& High  & 0.4e-6 &     1 &   26.8 &    logp &  0.76$\pm$0.16 &   1.47 &     logptau & strong   \\
                & Low	&	 &     2 &   54.5 &    logp &  0.16$\pm$0.17 &   0.31 &        logp &	  no   \\
   PKS 0402-362 & High  & 0.5e-6 &     1 &   23.5 &    logp &  0.00$\pm$0.58 &   0.00 &       exp03 &	  no   \\
                & Low	&	 &     1 &   23.5 &    logp &  0.51$\pm$0.53 &   0.98 &       exp03 &	  no   \\
    PKS 0440-00 & High  & mjd$^a$ &    1 &   36.4 &	pow &  0.59$\pm$0.23 &   1.14 &       exp03 &	  no   \\
                & Low	&	 &     0 &   17.9 &    logp &  1.71$\pm$1.09 &   3.30 &        logp &	  no   \\
   PKS 0454-234 & High  & 0.5e-6 &     1 &   39.6 &    logp &  0.36$\pm$0.18 &   0.69 &     logptau & strong   \\
                & Low	&	 &     2 &   80.4 &    logp &  0.20$\pm$0.08 &   0.38 &     logptau &	  no   \\
    PKS 0727-11 & High  & mjd$^b$ &    3 &  104.0 &    logp &  0.00$\pm$0.10 &   0.00 &       exp03 &	  no   \\
                & Low	&	 &     2 &   51.1 &	pow &  0.81$\pm$0.16 &   1.56 &       exp03 & strong   \\
   PKS 1329-049 & High  & 0.6e-6 &     1 &   30.6 &    logp &  0.93$\pm$1.13 &   1.79 &       exp03 &	  no   \\
                & Low	&	 &     2 &   62.2 &    logp &  0.96$\pm$0.67 &   1.86 &       exp03 &	  no   \\
   PKS 1502+106 & High  & 0.65e-6&     3 &  114.0 &    logp &  0.28$\pm$0.09 &   0.54 &     logptau &	  no   \\
                & Low	&	 &     2 &   56.1 &    logp &  0.00$\pm$0.21 &   0.00 &        logp &	  no   \\
    PKS 1510-08 & High  & 1.7e-6 &     2 &   54.6 &    logp &  0.00$\pm$0.19 &   0.00 &        logp &	  no   \\
                & Low	&	 &     3 &  111.1 &    logp &  0.00$\pm$0.31 &   0.00 &        logp &	  no   \\
   PKS 1830-211 & High  & mjd$^c$ &    2 &   69.3 &    logp &  0.60$\pm$0.09 &   1.16 &       exp03 & strong   \\
                & Low	&	 &     2 &   69.3 &    logp &  0.00$\pm$0.08 &   0.00 &        logp & strong   \\
   PKS 2326-502 & High  & 0.8e-6 &     0 &   14.7 &    logp &  2.17$\pm$0.73 &   4.20 &       exp03 & strong   \\
                & Low	&	 &     1 &   30.0 &    logp &  0.32$\pm$0.14 &   0.62 &     logptau & strong   \\
  PKS B1424-418 & High  & 1.0e-6 &     3 &  102.5 &    logp &  0.33$\pm$0.12 &   0.63 &       exp03 & strong   \\
                & Low	&	 &     3 &  102.5 &    logp &  0.58$\pm$0.13 &   1.12 &     logptau & strong   \\
     S5 1044+71 & High  & 0.5e-6 &     1 &   20.9 &    logp &  0.19$\pm$0.24 &   0.36 &       exp03 & strong   \\
                & Low	&	 &     2 &   42.5 &    logp &  0.14$\pm$0.12 &   0.27 &     logptau & strong   \\
\hline		 																		    
\hline  	 					  
\multicolumn{10}{l}{$a$: MJD 54960--55000 and 56460--56545}\\
\multicolumn{10}{l}{$b$: MJD $\geq$55740}\\ 
\multicolumn{10}{l}{$c$: MJD 55130--55221, 55452--55592 and 56054--56180} \\ 
\end{tabular}	 																		  
\label{tabHL}
\end{table*}

%% file: tableCa.tex
\begin{landscape}
\begin{table}
\centering
\caption{Spectral parameters of the fits of the gamma-ray SED data in Table 2.  
Col. [1]: object name.
Col. [2]: redshift.
Col. [3]: integral flux over the 0.2--10 GeV band, in cm$^{-2}$ s$^{-1}$, from the logparabolic full-band model.
Col. [4], [5], [6]: photon index, curvature and $\chi^2_r$ of the best-fit model below 13 GeV rest-frame (Unabsorbed band).
Col. [7], [8], [9]: photon index, curvature and $\chi^2_r$ of the log-parabolic fit with BLR absorption (logptau), over the full LAT band and with all parameters free.
Col. [10], [11]: $\chi^2_r$ of the following fits over the full band:  pure log-parabola with no BLR absorption (logp), 
power-law with high-energy cutoff with $\beta_{c}=1/3$~ (exp03).
}
\label{fitparam}
\begin{tabular}{llccccccccc}
\hline
\hline
        Name    &  $z$   &  Flux 0.2-10 GeV  &  GammaU  &   betaU     & $\chi^2_r$/d.o.f.  &   GammaF &  betaF  & $\chi^2_r$(logptau)/d.o.f. & $\chi^2_r$(logp)/d.o.f. & $\chi^2_r$(exp03)/d.o.f.   \\  
~[1]            &[2]     &    [3]            &   [4]    &   [5]       &  [6]       &[7]        &[8]       &[9]     &[10]       &[11]        \\
\hline 
	3C 454.3   &   0.859 &  8.49e-07 & 2.41$\pm$0.01 & 0.20$\pm$0.02 &    2.00/3 & 2.41$\pm$0.01 &  0.20$\pm$0.01 &       1.73/7 &    8.40/8 &     0.84/8	\\ 
       4C +21.35   &   0.435 &  1.94e-07 & 2.34$\pm$0.01 & 0.09$\pm$0.02 &    1.22/4 & 2.34$\pm$0.01 &  0.09$\pm$0.02 &       1.47/7 &    1.45/8 &     1.20/8	\\
     PKS 1510-08   & 	0.36 &  3.47e-07 & 2.44$\pm$0.01 & 0.16$\pm$0.01 &    0.55/4 & 2.44$\pm$0.01 &  0.16$\pm$0.02 &       1.90/8 &    1.40/9 &     3.18/9	\\
      B2 1520+31   &   1.487 &  1.33e-07 & 2.42$\pm$0.02 & 0.18$\pm$0.04 &    1.81/3 & 2.42$\pm$0.01 &  0.17$\pm$0.03 &       1.48/6 &    1.83/7 &     0.92/7	\\
    PKS 1502+106   &   1.839 &  1.22e-07 & 2.26$\pm$0.02 & 0.20$\pm$0.05 &    2.89/3 & 2.25$\pm$0.02 &  0.20$\pm$0.05 &       2.47/6 &    3.47/7 &     2.07/7	\\
    PKS 0454-234   &   1.003 &  1.36e-07 & 2.22$\pm$0.01 & 0.18$\pm$0.03 &    1.35/4 & 2.22$\pm$0.01 &  0.17$\pm$0.03 &       2.07/7 &    2.47/8 &     1.82/8	\\
	  3C 279   &   0.536 &  2.01e-07 & 2.37$\pm$0.01 & 0.12$\pm$0.02 &    1.26/4 & 2.38$\pm$0.01 &  0.13$\pm$0.03 &       1.21/7 &    1.62/8 &     1.22/8	\\
	  3C 273   &   0.158 &  1.19e-07 & 2.86$\pm$0.01 & 0.24$\pm$0.03 &    0.35/4 & 2.86$\pm$0.01 &  0.24$\pm$0.03 &       0.42/5 &    0.32/6 &     0.71/6	\\
    PKS 2326-502   &   0.518 &  1.33e-07 & 2.29$\pm$0.01 & 0.24$\pm$0.03 &    1.09/4 & 2.29$\pm$0.01 &  0.23$\pm$0.03 &       1.45/6 &    6.11/7 &     2.73/7	\\
   PKS B1424-418   &   1.552 &  2.28e-07 & 2.13$\pm$0.01 & 0.16$\pm$0.02 &    1.41/3 & 2.13$\pm$0.02 &  0.17$\pm$0.04 &       3.56/7 &    4.84/8 &     1.44/8	\\
    PKS 1830-211   &   2.507 &  1.63e-07 & 2.46$\pm$0.01 & 0.09$\pm$0.02 &    0.10/2 & 2.48$\pm$0.01 &  0.14$\pm$0.03 &       0.80/6 &    1.02/7 &     0.45/7	\\
       4C +55.17   &   0.899 &  4.57e-08 & 1.91$\pm$0.02 & 0.17$\pm$0.07 &    2.08/3 & 1.91$\pm$0.02 &  0.16$\pm$0.04 &       1.26/8 &    1.01/9 &     0.90/9	\\
     PKS 0727-11   &   1.589 &  9.91e-08 & 2.27$\pm$0.02 & 0.12$\pm$0.04 &    1.06/3 & 2.26$\pm$0.02 &  0.11$\pm$0.05 &       1.89/6 &    2.27/7 &     1.43/7	\\
       4C +28.07   &   1.213 &  7.39e-08 & 2.40$\pm$0.01 & 0.27$\pm$0.02 &    0.30/3 & 2.40$\pm$0.02 &  0.26$\pm$0.03 &       0.76/6 &    0.53/7 &     1.25/7	\\
    PKS 0402-362   &   1.417 &  5.97e-08 & 2.70$\pm$0.04 & 0.45$\pm$0.07 &    1.50/3 & 2.69$\pm$0.03 &  0.43$\pm$0.07 &       1.45/4 &    1.19/5 &     1.40/5	\\
	 Ton 599   &   0.725 &  4.61e-08 & 2.26$\pm$0.03 & 0.21$\pm$0.07 &    2.29/4 & 2.26$\pm$0.03 &  0.19$\pm$0.07 &       1.86/7 &    1.50/8 &     1.86/8	\\
    PKS 1244-255   &   0.635 &  6.78e-08 & 2.30$\pm$0.01 & 0.27$\pm$0.03 &    0.48/4 & 2.29$\pm$0.01 &  0.26$\pm$0.03 &       0.45/6 &    0.42/7 &     0.49/7	\\
       4C +38.41   &   1.814 &  1.59e-07 & 2.48$\pm$0.02 & 0.27$\pm$0.04 &    1.82/3 & 2.47$\pm$0.02 &  0.27$\pm$0.04 &       1.47/6 &    2.41/7 &     1.21/7	\\
  PMN J2345-1555   &   0.621 &  6.98e-08 & 2.04$\pm$0.01 & 0.17$\pm$0.03 &    1.10/4 & 2.04$\pm$0.01 &  0.18$\pm$0.03 &       0.75/7 &    0.87/8 &     0.71/8	\\
      S4 0917+44   & 	2.19 &  3.13e-08 & 2.29$\pm$0.02 & 0.17$\pm$0.05 &    0.31/2 & 2.31$\pm$0.03 &  0.22$\pm$0.06 &       0.84/5 &    0.85/6 &     0.59/6	\\
     PKS 2052-47   &   1.491 &  5.32e-08 & 2.48$\pm$0.01 & 0.32$\pm$0.02 &    0.10/3 & 2.48$\pm$0.02 &  0.32$\pm$0.05 &       1.38/6 &    0.96/7 &     1.69/7	\\
    PKS 1124-186   &   1.048 &  6.36e-08 & 2.18$\pm$0.01 & 0.14$\pm$0.02 &    0.15/3 & 2.18$\pm$0.01 &  0.14$\pm$0.03 &       0.69/7 &    0.55/8 &     0.63/8	\\
     PKS 2142-75   &   1.138 &  6.21e-08 & 2.43$\pm$0.01 & 0.16$\pm$0.02 &    0.08/3 & 2.43$\pm$0.01 &  0.14$\pm$0.02 &       0.15/5 &    0.92/6 &     0.42/6	\\
     B3 1343+451   &   2.534 &  6.42e-08 & 2.20$\pm$0.03 & 0.16$\pm$0.05 &    0.87/2 & 2.20$\pm$0.01 &  0.15$\pm$0.03 &       0.64/7 &    0.59/8 &     0.34/8	\\
     PKS 0805-07   &   1.837 &   4.8e-08 & 2.13$\pm$0.01 & 0.15$\pm$0.04 &    0.41/3 & 2.11$\pm$0.03 &  0.12$\pm$0.06 &       1.59/6 &    1.90/7 &     1.10/7	\\
    PKS 0244-470   &   1.385 &  3.39e-08 & 2.43$\pm$0.03 & 0.14$\pm$0.07 &    1.01/3 & 2.44$\pm$0.02 &  0.15$\pm$0.05 &       0.70/5 &    0.65/6 &     0.48/6	\\
	 CTA 102   &   1.037 &  8.41e-08 & 2.41$\pm$0.03 & 0.29$\pm$0.06 &    1.64/3 & 2.42$\pm$0.02 &  0.31$\pm$0.04 &       1.31/6 &    1.03/7 &     1.39/7	\\
       4C +01.02   &   2.099 &  6.19e-08 & 2.40$\pm$0.02 & 0.11$\pm$0.04 &    0.61/3 & 2.42$\pm$0.03 &  0.14$\pm$0.07 &       1.46/6 &    1.28/7 &     1.30/7	\\
       4C +14.23   &   1.038 &   3.3e-08 & 2.18$\pm$0.01 & 0.23$\pm$0.02 &    0.07/3 & 2.17$\pm$0.02 &  0.19$\pm$0.05 &       0.72/6 &    0.66/7 &     0.64/7	\\
    PKS 0250-225   &   1.427 &  4.31e-08 & 2.35$\pm$0.03 & 0.18$\pm$0.06 &    0.90/3 & 2.35$\pm$0.02 &  0.19$\pm$0.04 &       0.59/6 &    0.49/7 &     0.60/7	\\
      B2 2308+34   &   1.817 &  4.08e-08 & 2.36$\pm$0.02 & 0.21$\pm$0.04 &    0.49/3 & 2.34$\pm$0.02 &  0.17$\pm$0.06 &       1.11/6 &    0.91/7 &     1.06/7	\\
      B2 0716+33   &   0.779 &  2.81e-08 & 2.10$\pm$0.01 & 0.15$\pm$0.02 &    0.11/3 & 2.10$\pm$0.01 &  0.15$\pm$0.02 &       0.20/6 &    0.46/7 &     0.18/7	\\
MG1 J123931+0443   &   1.762 &  6.34e-08 & 2.35$\pm$0.01 & 0.17$\pm$0.02 &    0.17/3 & 2.35$\pm$0.01 &  0.16$\pm$0.03 &       1.16/6 &    2.06/7 &     1.45/7	\\
  PMN J1802-3940   &   1.319 &  3.75e-08 & 2.11$\pm$0.05 & 0.30$\pm$0.11 &    3.18/4 & 2.11$\pm$0.04 &  0.31$\pm$0.11 &       2.40/6 &    2.01/7 &     2.61/7	\\
      S4 1849+67   &   0.657 &  3.18e-08 & 2.25$\pm$0.01 & 0.20$\pm$0.01 &    0.05/4 & 2.24$\pm$0.01 &  0.20$\pm$0.01 &       0.45/6 &    0.21/7 &     0.31/7	\\
     PKS 2023-07   &   1.388 &  4.86e-08 & 2.36$\pm$0.02 & 0.38$\pm$0.03 &    0.42/4 & 2.36$\pm$0.02 &  0.38$\pm$0.05 &       1.29/6 &    1.05/7 &     1.87/7	\\
\hline		 
\hline  	 
\end{tabular}	 
\end{table}
\end{landscape}

%% file: tableCb.tex
\begin{landscape}
\begin{table}
\centering
\contcaption{}
\begin{tabular}{llccccccccc}
\hline
\hline
        Name    &  $z$   &  Flux 0.2-10 GeV  &  GammaU  &   betaU     & $\chi^2_r$/d.o.f.  &   GammaF &  betaF  & $\chi^2_r$(logptau)/d.o.f. & $\chi^2_r$(logp)/d.o.f. & $\chi^2_r$(exp03)/d.o.f.   \\  
~[1]            &[2]     &    [3]            &   [4]    &   [5]       &  [6]       &[7]        &[8]       &[9]     &[10]       &[11]        \\
\hline 
    PKS 1454-354   & 	1.424 &  3.64e-08 & 2.26$\pm$0.04 & 0.23$\pm$0.10 &    2.15/3 & 2.26$\pm$0.03 &  0.22$\pm$0.07 &       1.56/6 &    1.25/7 &	1.50/7  \\
     B3 1708+433   & 	1.027 &  2.51e-08 & 2.27$\pm$0.03 & 0.36$\pm$0.07 &    0.73/3 & 2.27$\pm$0.03 &  0.34$\pm$0.08 &       1.83/6 &    1.33/7 &	1.97/7  \\
  GB6 J0742+5444   & 	 0.72 &  2.12e-08 & 2.22$\pm$0.04 & 0.29$\pm$0.08 &    1.49/4 & 2.22$\pm$0.03 &  0.28$\pm$0.07 &       1.57/6 &    1.25/7 &	1.52/7  \\
	  OX 169   & 	0.213 &  5.27e-08 & 2.53$\pm$0.02 & 0.18$\pm$0.04 &    0.50/4 & 2.53$\pm$0.02 &  0.18$\pm$0.04 &       0.92/6 &    0.69/7 &	0.95/7  \\
     PKS 2227-08   & 	 1.56 &  4.91e-08 & 2.80$\pm$0.03 & 0.24$\pm$0.06 &    0.55/3 & 2.79$\pm$0.03 &  0.22$\pm$0.05 &       0.52/4 &    0.42/5 &	0.45/5  \\
    PKS 2255-282   & 	0.926 &  3.63e-08 & 2.49$\pm$0.04 & 0.33$\pm$0.09 &    1.40/3 & 2.49$\pm$0.04 &  0.35$\pm$0.08 &       1.35/5 &    1.10/6 &	1.38/6  \\
      B2 2155+31   & 	1.488 &  3.22e-08 & 2.24$\pm$0.01 & 0.26$\pm$0.03 &    0.17/3 & 2.23$\pm$0.02 &  0.25$\pm$0.05 &       1.31/6 &    1.12/7 &	1.10/7  \\
    PKS 1622-253   & 	0.786 &  7.28e-08 & 2.42$\pm$0.02 & 0.16$\pm$0.04 &    0.41/3 & 2.41$\pm$0.01 &  0.15$\pm$0.03 &       0.64/6 &    0.45/7 &	0.45/7  \\
     PKS 0420-01   & 	0.916 &  3.62e-08 & 2.29$\pm$0.05 & 0.23$\pm$0.12 &    2.54/3 & 2.29$\pm$0.04 &  0.22$\pm$0.08 &       1.61/6 &    1.27/7 &	1.63/7  \\
	  OC 457   & 	0.859 &  3.83e-08 & 2.26$\pm$0.01 & 0.17$\pm$0.03 &    0.31/3 & 2.25$\pm$0.01 &  0.15$\pm$0.03 &       0.23/6 &    0.30/7 &	0.22/7  \\
    PKS 2201+171   & 	1.076 &  2.55e-08 & 2.25$\pm$0.03 & 0.41$\pm$0.07 &    0.62/3 & 2.24$\pm$0.05 &  0.39$\pm$0.11 &       2.32/6 &    1.12/7 &	1.82/7  \\
	  OP 313   & 	0.997 &   3.1e-08 & 2.26$\pm$0.01 &	      0.0 &    0.19/4 & 2.29$\pm$0.02 &  0.06$\pm$0.05 &       0.81/7 &    0.72/8 &	0.67/8  \\
    TXS 0059+581   & 	0.644 &  3.05e-08 & 2.14$\pm$0.06 & 0.37$\pm$0.12 &    4.18/5 & 2.14$\pm$0.06 &  0.37$\pm$0.14 &       3.67/6 &    3.07/7 &	4.30/7  \\
      S4 1030+61   & 	1.401 &  2.52e-08 & 2.15$\pm$0.02 & 0.13$\pm$0.04 &    0.43/3 & 2.16$\pm$0.01 &  0.15$\pm$0.03 &       0.30/6 &    0.40/7 &	0.27/7  \\
	  OG 050   & 	1.254 &  4.43e-08 & 2.32$\pm$0.04 & 0.19$\pm$0.09 &    1.79/3 & 2.30$\pm$0.03 &  0.14$\pm$0.07 &       1.26/6 &    1.09/7 &	1.14/7  \\
     PKS 0440-00   & 	0.844 &  3.72e-08 & 2.39$\pm$0.01 & 0.12$\pm$0.01 &    0.04/3 & 2.40$\pm$0.02 &  0.14$\pm$0.03 &       0.92/6 &    0.41/7 &	0.37/7  \\
    PKS 1329-049   & 	 2.15 &  4.94e-08 & 2.65$\pm$0.04 & 0.33$\pm$0.10 &    1.74/3 & 2.64$\pm$0.04 &  0.33$\pm$0.09 &       2.14/6 &    1.72/7 &	2.02/7  \\
    TXS 1700+685   & 	0.301 &  2.62e-08 & 2.32$\pm$0.03 & 0.20$\pm$0.06 &    1.07/4 & 2.33$\pm$0.03 &  0.22$\pm$0.06 &       0.89/6 &    0.75/7 &	0.89/7  \\
      S4 1144+40   & 	1.089 &  3.83e-08 & 2.39$\pm$0.01 & 0.19$\pm$0.02 &    0.23/4 & 2.38$\pm$0.02 &  0.18$\pm$0.04 &       0.95/7 &    0.56/8 &	0.97/8  \\
       B0218+357   & 	0.944 &   5.7e-08 & 2.26$\pm$0.03 & 0.11$\pm$0.07 &    3.15/4 & 2.26$\pm$0.03 &  0.12$\pm$0.06 &       2.45/7 &    2.12/8 &	2.47/8  \\
    PKS 0215+015   & 	1.715 &  1.87e-08 & 2.06$\pm$0.03 &	      0.0 &    0.51/3 & 2.08$\pm$0.04 &  0.08$\pm$0.09 &       1.04/6 &    1.39/7 &	1.04/7  \\
     PKS 0336-01   & 	 0.85 &   3.4e-08 & 2.41$\pm$0.04 & 0.22$\pm$0.09 &    1.28/3 & 2.40$\pm$0.03 &  0.20$\pm$0.06 &       1.36/6 &    0.90/7 &	1.06/7  \\
	4C 31.03   & 	0.603 &  1.74e-08 & 2.24$\pm$0.02 & 0.25$\pm$0.05 &    0.42/4 & 2.24$\pm$0.03 &  0.27$\pm$0.08 &       1.91/6 &    1.15/7 &	1.34/7  \\
	  OM 484   & 	0.334 &  1.41e-08 & 2.31$\pm$0.02 &	      0.0 &    0.27/5 & 2.31$\pm$0.02 &  0.01$\pm$0.05 &       1.76/5 &    1.73/6 &	1.23/6  \\
    TXS 1920-211   & 	0.874 &  1.57e-08 & 2.00$\pm$0.07 & 0.43$\pm$0.19 &    2.47/3 & 1.98$\pm$0.06 &  0.35$\pm$0.12 &       1.71/5 &    2.38/6 &	1.92/6  \\
MG2 J101241+2439   & 	1.805 &  1.55e-08 & 2.32$\pm$0.05 &	      0.0 &    2.13/5 & 2.27$\pm$0.07 &  0.00$\pm$0.16 &       2.23/4 &    5.68/5 &	4.61/5  \\
      S5 1044+71   & 	 1.15 &  5.09e-08 & 2.30$\pm$0.01 & 0.18$\pm$0.03 &    0.38/3 & 2.29$\pm$0.01 &  0.16$\pm$0.02 &       1.10/6 &    1.47/7 &	0.69/7  \\
  PMN J0850-1213   & 	0.566 &  1.54e-08 & 2.14$\pm$0.05 & 0.33$\pm$0.11 &    1.39/4 & 2.14$\pm$0.05 &  0.31$\pm$0.10 &       1.27/6 &    0.96/7 &	1.22/7  \\
     PKS 0601-70   & 	2.409 &  2.94e-08 & 2.66$\pm$0.12 & 0.34$\pm$0.22 &    2.66/2 & 2.66$\pm$0.10 &  0.33$\pm$0.20 &       3.05/5 &    1.32/6 &	1.50/6  \\
     B3 0650+453   & 	0.933 &  1.52e-08 & 2.32$\pm$0.05 & 0.26$\pm$0.12 &    1.01/3 & 2.28$\pm$0.05 &  0.17$\pm$0.11 &       1.32/6 &    0.95/7 &	1.09/7  \\
    PKS 2320-035   & 	1.393 &  3.82e-08 & 2.47$\pm$0.04 &	      0.0 &    1.66/4 & 2.48$\pm$0.04 &  0.00$\pm$0.08 &       1.52/5 &    3.96/6 &	3.21/6  \\
     PKS 0736+01   &  0.18941 &  3.53e-08 & 2.54$\pm$0.03 & 0.24$\pm$0.08 &    1.05/4 & 2.54$\pm$0.03 &  0.25$\pm$0.07 &       1.57/6 &    1.29/7 &	1.59/7  \\
    PKS 1954-388   & 	 0.63 &  2.64e-08 & 2.37$\pm$0.03 & 0.18$\pm$0.07 &    1.04/4 & 2.37$\pm$0.03 &  0.17$\pm$0.06 &       0.81/5 &    1.65/6 &	0.89/6  \\
    TXS 1206+549   & 	1.345 &  1.36e-08 & 2.48$\pm$0.06 &	      0.0 &    1.67/4 & 2.51$\pm$0.06 &  0.12$\pm$0.13 &       1.32/4 &    1.63/5 &	1.40/5  \\
      B2 0200+30   & 	0.955 &   1.4e-08 & 2.09$\pm$0.03 & 0.28$\pm$0.07 &    0.34/3 & 2.08$\pm$0.03 &  0.25$\pm$0.07 &       0.46/6 &    0.38/7 &	0.53/7  \\
     PKS 1730-13   & 	0.902 &  4.39e-08 & 2.27$\pm$0.01 & 0.17$\pm$0.03 &    0.12/3 & 2.27$\pm$0.01 &  0.16$\pm$0.02 &       0.14/6 &    0.35/7 &	0.12/7  \\
    TXS 0106+612   & 	0.783 &  2.85e-08 & 2.62$\pm$0.07 & 0.49$\pm$0.15 &    1.45/3 & 2.62$\pm$0.07 &  0.49$\pm$0.15 &       1.89/4 &    0.89/5 &	0.91/5  \\
    PKS 0528+134   & 	 2.07 &  2.51e-08 & 2.59$\pm$0.08 & 0.57$\pm$0.19 &    1.75/3 & 2.59$\pm$0.09 &  0.57$\pm$0.21 &       1.93/3 &    1.94/4 &	1.55/4  \\
     PKS 0446+11   & 	2.153 &   2.7e-08 & 2.58$\pm$0.05 & 0.33$\pm$0.10 &    0.46/2 & 2.57$\pm$0.03 &  0.33$\pm$0.08 &       0.38/5 &    0.28/6 &	0.44/6  \\
     B2 1732+38A   & 	0.976 &  1.57e-08 & 2.37$\pm$0.01 & 0.22$\pm$0.03 &    0.09/3 & 2.36$\pm$0.01 &  0.20$\pm$0.03 &       0.31/5 &    0.15/6 &	0.13/6  \\
     PKS 0130-17   & 	 1.02 &  1.54e-08 & 2.55$\pm$0.04 & 0.29$\pm$0.10 &    0.59/3 & 2.55$\pm$0.06 &  0.29$\pm$0.14 &       1.72/5 &    0.94/6 &	1.08/6  \\
    MRC 1659-621   & 	 1.75 &  1.84e-08 & 2.31$\pm$0.05 & 0.16$\pm$0.13 &    1.11/3 & 2.31$\pm$0.06 &  0.15$\pm$0.16 &       2.36/4 &    1.61/5 &	1.32/5  \\
      S5 0836+71   & 	2.218 &  2.59e-08 & 3.12$\pm$0.09 & 0.56$\pm$0.15 &    1.15/2 & 3.12$\pm$0.09 &  0.56$\pm$0.14 &       2.40/4 &    1.84/5 &	2.34/5  \\
     PKS 0906+01   & 	1.024 &  2.65e-08 & 2.48$\pm$0.04 & 0.24$\pm$0.10 &    1.06/3 & 2.47$\pm$0.04 &  0.23$\pm$0.08 &       0.71/5 &    0.57/6 &	0.73/6  \\
\hline		 
\hline  	 
\end{tabular}	 
\end{table}
\end{landscape}

%% file: tableCc.tex
\begin{landscape}
\begin{table}
\centering
\contcaption{}
\begin{tabular}{llccccccccc}
\hline
\hline
        Name    &  $z$   &  Flux 0.2-10 GeV  &  GammaU  &   betaU     & $\chi^2_r$/d.o.f.  &   GammaF &  betaF  & $\chi^2_r$(logptau)/d.o.f. & $\chi^2_r$(logp)/d.o.f. & $\chi^2_r$(exp03)/d.o.f.   \\  
~[1]            &[2]     &    [3]            &   [4]    &   [5]       &  [6]       &[7]        &[8]       &[9]     &[10]       &[11]        \\
\hline 
    PKS 2123-463   &   1.67 &  1.37e-08 & 2.45$\pm$0.06 &	    0.0 &    1.16/4 & 2.48$\pm$0.09 &  0.11$\pm$0.19 &       2.12/4 &	 1.13/5 &     1.11/5  \\
MG2 J071354+1934   &   0.54 &  1.75e-08 & 2.38$\pm$0.05 & 0.27$\pm$0.12 &    1.44/4 & 2.38$\pm$0.06 &  0.28$\pm$0.13 &       1.57/4 &	 1.75/5 &     1.51/5  \\
  PMN J1344-1723   &   2.49 &	1.5e-08 & 2.08$\pm$0.03 &	    0.0 &    0.20/2 & 2.21$\pm$0.08 &  0.11$\pm$0.23 &       2.49/6 &	 2.10/7 &     2.08/7  \\
    TXS 0529+483   &  1.162 &  1.81e-08 & 2.30$\pm$0.05 & 0.30$\pm$0.13 &    1.22/3 & 2.30$\pm$0.05 &  0.29$\pm$0.11 &       1.31/5 &	 1.08/6 &     1.35/6  \\
    PKS 1551+130   &  1.308 &  1.72e-08 & 2.43$\pm$0.06 &	    0.0 &    2.38/5 & 2.43$\pm$0.08 &  0.10$\pm$0.18 &       2.19/5 &	 1.82/6 &     1.81/6  \\
      S3 0458-02   &  2.291 &  2.95e-08 & 2.36$\pm$0.03 & 0.26$\pm$0.06 &    0.29/2 & 2.35$\pm$0.02 &  0.24$\pm$0.06 &       0.49/5 &	 0.41/6 &     0.51/6  \\
   PKS B1908-201   &  1.119 &  3.07e-08 & 2.32$\pm$0.03 & 0.27$\pm$0.07 &    0.54/3 & 2.33$\pm$0.02 &  0.28$\pm$0.05 &       0.36/5 &	 0.31/6 &     0.33/6  \\
MG2 J201534+3710   &  0.859 &  8.76e-08 & 2.46$\pm$0.02 & 0.24$\pm$0.05 &    0.41/3 & 2.46$\pm$0.02 &  0.22$\pm$0.04 &       0.43/6 &	 0.30/7 &     0.55/7  \\
	CTS 0490   &   0.36 &  1.16e-08 & 2.13$\pm$0.04 & 0.53$\pm$0.08 &    0.59/4 & 2.13$\pm$0.04 &  0.53$\pm$0.09 &       0.59/4 &	 0.50/5 &     0.57/5  \\
     PKS 0451-28   &  2.564 &  1.84e-08 & 2.70$\pm$0.13 & 0.30$\pm$0.25 &    2.45/2 & 2.72$\pm$0.09 &  0.33$\pm$0.19 &       1.67/3 &	 2.88/4 &     1.86/4  \\
       4C +47.44   &  0.735 &  1.35e-08 & 2.36$\pm$0.04 & 0.26$\pm$0.08 &    0.74/4 & 2.36$\pm$0.04 &  0.26$\pm$0.10 &       1.07/5 &	 0.78/6 &     0.70/6  \\
    PKS 0235-618   &  0.465 &  1.03e-08 & 2.32$\pm$0.06 &	    0.0 &    1.72/5 & 2.33$\pm$0.07 &  0.16$\pm$0.17 &       1.79/4 &	 1.43/5 &     1.54/5  \\
       4C +04.42   &  0.966 &  2.02e-08 & 2.93$\pm$0.03 & 0.37$\pm$0.06 &    0.16/3 & 2.93$\pm$0.05 &  0.37$\pm$0.10 &       1.81/4 &	 1.26/5 &     1.42/5  \\
     PKS 0454-46   &  0.858 &  1.46e-08 & 2.60$\pm$0.07 & 0.16$\pm$0.14 &    1.03/3 & 2.60$\pm$0.06 &  0.15$\pm$0.13 &       1.52/5 &	 0.96/6 &     1.00/6  \\
      S4 1726+45   &  0.717 &  1.48e-08 & 2.51$\pm$0.04 & 0.33$\pm$0.10 &    0.79/4 & 2.50$\pm$0.05 &  0.32$\pm$0.11 &       0.94/4 &	 0.72/5 &     0.86/5  \\
    PKS 0142-278   &  1.148 &  1.18e-08 & 2.72$\pm$0.11 & 0.45$\pm$0.23 &    1.96/3 & 2.72$\pm$0.12 &  0.45$\pm$0.25 &       2.94/4 &	 2.15/5 &     2.21/5  \\
    PKS 2144+092   &  1.113 &  2.18e-08 & 2.56$\pm$0.05 & 0.32$\pm$0.12 &    1.03/3 & 2.56$\pm$0.05 &  0.32$\pm$0.12 &       1.92/4 &	 0.75/5 &     0.86/5  \\
      S4 1851+48   &   1.25 &  9.28e-09 & 2.07$\pm$0.06 & 0.39$\pm$0.14 &    1.32/4 & 2.06$\pm$0.08 &  0.36$\pm$0.18 &       2.04/5 &	 1.38/6 &     1.73/6  \\
     PKS 1118-05   &  1.297 &  1.59e-08 & 2.37$\pm$0.08 & 0.29$\pm$0.19 &    2.21/3 & 2.36$\pm$0.06 &  0.26$\pm$0.15 &       1.73/4 &	 1.37/5 &     1.52/5  \\
       4C +51.37   &  1.381 &  7.93e-09 & 2.15$\pm$0.04 & 0.41$\pm$0.09 &    0.29/3 & 2.15$\pm$0.05 &  0.40$\pm$0.12 &       1.73/6 &	 1.32/7 &     1.59/7  \\ 
\hline
     PKS 0208-512 &   1.003 &  3.45e-08 & 2.35$\pm$0.05 & 0.18$\pm$0.12 &    3.29/3 & 2.34$\pm$0.04 & 0.16$\pm$0.08 &	    2.60/4 &	2.83/5 &     2.08/5 \\
     TXS 0322+222 &   2.066 &  1.84e-08 & 2.71$\pm$0.06 &	    0.0 &    1.22/4 & 2.70$\pm$0.09 & 0.01$\pm$0.19 &	    2.02/2 &	2.12/3 &     1.95/3 \\
    PKS B1127-145 &   1.184 &  9.13e-09 & 2.70$\pm$0.06 & 0.19$\pm$0.13 &    0.34/3 & 2.70$\pm$0.06 & 0.19$\pm$0.14 &	    0.36/3 &	0.45/4 &     0.48/4 \\
CRATES J0303-6211 &   1.348 &  5.18e-09 & 2.22$\pm$0.08 &	    0.0 &    0.86/4 & 2.25$\pm$0.10 & 0.22$\pm$0.25 &	    0.89/3 &	0.69/4 &     0.75/4 \\
   PMN J1617-5848 &   1.422 &  1.01e-08 & 2.46$\pm$0.13 &	    0.0 &    1.82/4 & 2.46$\pm$0.17 & 0.00$\pm$0.44 &	    2.45/3 &	1.83/4 &     1.83/4 \\
CRATES J1613+3412 &   1.397 &  6.97e-09 & 2.43$\pm$0.07 & 0.40$\pm$0.17 &    0.89/4 & 2.43$\pm$0.11 & 0.40$\pm$0.28 &	    1.99/4 &	1.86/5 &     2.09/5 \\
\hline		 
\hline  	 
\end{tabular}	 
\end{table}
\end{landscape}